\documentclass[10pt]{article}
\usepackage{amsmath,appendix,color,amssymb,graphicx,bm,stmaryrd,mathabx,amssymb,mathdots,esint,amsfonts}
\DeclareGraphicsExtensions{.pdf}
\UseRawInputEncoding
\DeclareMathAlphabet{\mathpzc}{OT1}{pzc}{m}{it}

\makeatletter
\@addtoreset{equation}{section}
\makeatother
\newtheorem{theorem}{Theorem}[section]
\newtheorem{conjecture}[theorem]{Conjecture}
\newtheorem{proposition}[theorem]{Proposition}
\newtheorem{lemma}[theorem]{Lemma}
\newtheorem{corollary}[theorem]{Corollary}

\newtheorem{example}{Example}[section]
\newtheorem{definition}[example]{Definition}
\newtheorem{remark}[example]{Remark}
\newtheorem{hypothesis}[example]{Hypothesis}
\def\br{\begin{remark}\rm\small}
\def\er{\end{remark}}
\def\bt{\begin{theorem}}
\def\et{\end{theorem}}
\def\bcj{\begin{conjecture}}
\def\ecj{\end{conjecture}}
\def\bd{\begin{definition}}
\def\ed{\end{definition}}
\def\bp{\begin{proposition}}
\def\ep{\end{proposition}}
\def\bl{\begin{lemma}}
\def\el{\end{lemma}}
\def\bc{\begin{corollary}}
\def\ec{\end{corollary}}
\def\bh{\begin{hypothesis}}
\def\eh{\end{hypothesis}}
\def\beaq{\begin{eqnarray}}
\def\eeaq{\end{eqnarray}}

\newcommand{\dd}{\mathrm{d}}
\newcommand{\beq}{\begin{equation}}
\newcommand{\eeq}{\end{equation}}
\newcommand{\bea}{\begin{eqnarray}}
\newcommand{\eea}{\end{eqnarray}}

\newcommand{\ra}{\rightarrow}

\newcommand{\p}{{\parallel}}

\newcommand{\Tr}{{\,\rm Tr}\:}

\newcommand{\Res}{\mathop{\,\rm Res\,}}

\renewcommand{\Re}{{\mathrm{Re}}\:}

\newcommand{\Vref}{V^{\mathrm{ref}}}
\definecolor{rouge}{rgb}{0.84,0.18,0.07}
\definecolor{bleu}{rgb}{0.22,0.41,0.74}
\definecolor{vertf}{rgb}{0.08,0.46,0.07}
\usepackage[pdftex]{hyperref}
\hypersetup{colorlinks,urlcolor=vertf,citecolor=rouge,linkcolor=bleu,filecolor=black}

\begin{document}

\pagestyle{empty}
\addtolength{\baselineskip}{0.20\baselineskip}
\begin{center}
\vspace{26pt}
{\large \bf {Asymptotic expansion of $\beta$ matrix models
in the multi-cut regime}}
\end{center}

\vspace{26pt}
 
\begin{center}
{\sl Ga\"etan Borot}\hspace*{0.05cm}\footnote{The work has been conducted at Section de Math\'ematiques, Universit\'e de Gen\`eve, at MIT, Department of Mathematics, at MPIM Bonn, and at (current address) Humboldt-Universit\"at zu Berlin, Institut f\"ur Mathematik und Institut f\"ur Physik, Unter den Linden 6, 10099 Berlin, Germany. Email: \href{mailto:gborot@mpim-bonn.mpg.de}{gaetan.borot@hu-berlin.de}},
{\sl Alice Guionnet}\hspace*{0.05cm}\footnote{The work has been conducted at MIT, Department of Mathematics and (current address) UMPA, CNRS  UMR 5669, ENS Lyon, 46 all\'ee d'Italie, 69007 Lyon, France. Email: \href{mailto:alice.guionnet@ens-lyon.fr}{alice.guionnet@ens-lyon.fr}}
%
%
%

\vspace{0.2cm}

\end{center}

\vspace{20pt}
\begin{center}
{\bf Abstract}
\end{center}

%

\vspace{0.5cm}

We establish the asymptotic expansion in $\beta$  matrix models with a confining, off-critical potential, in the regime where the support of the equilibrium measure is a finite union of segments. We first address the case where the filling fractions of these segments are fixed, and show the existence of a $\frac{1}{N}$ expansion. We then study the asymptotics of the sum over the  filling fractions, to obtain the full asymptotic expansion for the initial problem in the multi-cut regime. In particular, we identify the fluctuations of  the linear statistics and show that they are approximated in law by the sum of a Gaussian random variable and an independent  Gaussian discrete random variable with oscillating center. Fluctuations of filling fractions are also described by an oscillating discrete Gaussian random variable. We apply our results to study the all-order small dispersion asymptotics of solutions of the Toda chain associated with the one   Hermitian matrix model ($\beta = 2$) as well as orthogonal ($\beta = 1$) and skew-orthogonal ($\beta = 4$) polynomials outside the bulk.
\vspace{0.5cm}

\noindent \textbf{MSC Classification:} 60B20, 15B52, 60F05

\newpage


\vspace{26pt}
\pagestyle{plain}
\setcounter{page}{1}
\addtocounter{footnote}{-2}

\tableofcontents

\newpage

\section{Introduction}
\label{sec:intro}

This paper is concerned with the asymptotic expansion  for the partition function and  the multilinear statistics of $\beta$ matrix models. These laws represent a generalisation of the joint distribution of the $N$ eigenvalues of the Gaussian Unitary Ensemble \cite{Mehtabook}. The convergence of the empirical measure of the eigenvalues is well-known (see, \textit{e.g.} \cite{Boutet}), and we are interested in the all-order finite size corrections to the moments of this empirical measure. Much attention  has been paid to this problem 
 in the regime when the eigenvalues condense on a single segment, usually referred to as a  one-cut regime. In this case, a central limit theorem for linear statistics was proved by Johansson \cite{Johansson98}, while a full $\frac{1}{N}$ expansion was derived first for $\beta = 2$ \cite{APS01,ErcMcL,BleherItsZ}, then for any $\beta > 0$ in \cite{BG11}. On the other hand, the multi-cut regime was  until recently poorly understood at the rigorous level, except for $\beta = 2$ which is related to integrable systems, and can be treated with the powerful asymptotic analysis techniques for Riemann--Hilbert problems, see \textit{e.g.} \cite{DKMcLVZ2}. Nevertheless, a heuristic derivation of the asymptotic expansion for the multi-cut regime has  been  proposed to leading order by Bonnet, David and Eynard \cite{BDE}, and extended to all orders in \cite{Ecv}, in terms of Theta functions and their derivatives. It features oscillatory behaviour, whose origin lies in the tunneling of eigenvalues between the different connected components of the support. This heuristic, originally written for $\beta = 2$, can be trivially extended to $\beta > 0$, see \textit{e.g.} \cite{GBThese}. 

More recently, M.~Shcherbina has established this asymptotic expansion up to terms of order $1$ \cite{Schuniv,Smulticut}. This allows us to observe,  for instance,  that linear statistics do not always satisfy a central limit theorem (this fact was already noticed for $\beta = 2$ in \cite{Pasturs}). In this work, we go beyond the $O(1)$ and put the heuristics of \cite{Ecv} to all orders on a firm mathematical ground. Our strategy is to first study the asymptotics in the model with fixed filling fractions, and then reconstruct the asymptotics in the original model via a finite-dimensional analysis. As a consequence we obtain a replacement for the central limit theorem for linear statistics and for filling fractions. Besides, we treat uniformly soft and hard edges, while \cite{Smulticut} assumed soft edges.

For $\beta = 2$, we can establish the full asymptotic expansion outside of the bulk for the orthogonal polynomials with real-analytic potentials, and the all-order asymptotic expansion of certain solutions of the Toda lattice in the continuum limit. The same method allows  us  to  rigorously establish  the asymptotics of skew-orthogonal polynomials ($\beta = 1$ and $4$) away from the bulk, derived heuristically in \cite{Eskew}. To our knowledge,  Riemann--Hilbert analysis of  skew-orthogonal polynomials is possible in principle, but is cumbersome and has not been done before, so our method provides the first proof of these asymptotics. After this work was released, this method was extended to treat more general Coulomb-like interactions in \cite{BGK15}. We also note that a proof of the asymptotics up to $o(1)$ with $\beta = 2$ was obtained by the Riemann--Hilbert approach in the two-cuts situation in \cite{CGMcL15}, and in $k$-cut situation with $k \geq 2$ in \cite{CFWW}.

Since the first release of this work, several authors have considered asymptotic questions in the multi-cut regime of $\beta$-ensembles. A recent approach to central limit theorems inspired by Stein's method  was proposed in \cite{LLW19}, but it  is restricted to the one-cut regime. The transport method  introduced in \cite{BGF15}
 allowed to establish rigidity of eigenvalues \cite{Li16} and universality \cite{B18} in the multi-cut regime. In \cite{BLS18}, the validity of central limit theorems for linear fluctuations has also  been extended to include test functions   with weaker regularity assumptions  and to  critical cases (and  then test functions in the range of the so-called ''master operator''). Beyond being a source of inspiration for these works, and the first rigorous article where Dyson--Schwinger equations  were used to derive central limit theorem in the multi-cut regime,
 the present article contains results that  still did not appear anywhere else, such as  the  asymptotics of  to (skew) orthogonal polynomials and integrable systems (see Section \ref{S2}),  a discussion about the relation  with Chekhov--Eynard--Orantin topological recursion (see Section \ref{CEO}), and  the detailed use of precise estimates of beta ensembles with fixed filling fractions to estimate the free energy in multi-cut models and the reconstruction of the Theta function (see Section \ref{musqo}). Besides, Shcherbina derives in \cite{Smulticut} via operator methods and for soft edges an expression of the order $N$ in the free energy in terms of the entropy of the equilibrium measure and a universal constant. Our work proves a similar formula both with soft and hard edges and with a different method based on complex analysis.

 Our results on the asymptotics of the partition function have been used e.g. to study the asymptotics of the determinant of T\"oplitz matrices in \cite{Mar20,Mar21}. 
The ideas that we introduce to handle the multi-cut regime are extended in a work in progress \cite{BGG} to study the fluctuations of discrete $\beta$-ensembles appearing in random tiling models  in non-simply connected domains (with holes and/or frozen regions).

For Coulomb gases in dimension $d > 1$, carrying out the asymptotic analysis when the support of the equilibrium measure has several connected components remains in general an open problem. Some specific $d = 2$, $\beta = 2$ situations have been treated in \cite{ACC,ACCL} relying on the determinantal structure of these models. In general, probabilistic methods in the spirit of this article that do not rely on integrability, and therefore could address arbitrary $\beta > 0$ (where integrability is absent), are still insufficiently developed.

\subsection{Definitions}
\label{S11}

\subsubsection{Model and empirical measure}

We consider the probability measure $\mu_{N,\beta}^{V;\mathsf{B}}$ on $\mathsf{B}^N$
given by:
\beq
\label{eqmes}\mathrm{d}\mu_{N,\beta}^{V;\mathsf{B}}(\lambda) = \frac{1}{Z_{N,\beta}^{V;\mathsf{B}}}\prod_{i = 1}^N \mathrm{d}\lambda_i\,\mathbf{1}_{\mathsf{B}}(\lambda_i)\,e^{-\frac{\beta N}{2} \,V(\lambda_i)}\,\prod_{1 \leq i < j \leq N} |\lambda_i - \lambda_j|^{\beta}.
\eeq
$\mathsf{B}$ is a finite disjoint union of closed intervals of $\mathbb{R}$ possibly with infinite endpoints, $\beta$ is a positive number, and $Z_{N,\beta}^{V;\mathsf{B}}$ is the partition function so that \eqref{eqmes} has total mass $1$. This model is usually called the $\beta$-ensemble \cite{Mehtabook,DE02,Loggas}. We introduce the unnormalised empirical measure $M_N$ of the eigenvalues:
$$
M_N=\sum_{i=1}^N \delta_{\lambda_i},
$$
and we consider several types of statistics for $M_N$. We sometimes denote $\mathbb{L} = \mathrm{diag}(\lambda_1,\ldots,\lambda_N)$.

\subsubsection{Correlators}

We introduce the Stieltjes transform of the $n$-th order moments of the empirical measure, called \emph{disconnected correlators}:
$$
\widetilde{W}_n(x_1,\ldots,x_n) = \mu_{N,\beta}^{V;\mathsf{B}}\Big[\Big(\int_{\mathbb{R}}\frac{\dd M_N(\xi_1)}{x_1-\xi_1}\times \cdots\times\int_{\mathbb{R}}\frac{\dd M_N(\xi_n)}{x_n - \xi_n}\Big)\Big]. 
$$
They are holomorphic functions of $x_i \in \mathbb{C}\setminus \mathsf{B}$. It is more convenient to consider the \emph{correlators} to study large $N$ asymptotics:
\begin{equation} 
\label{defcor}
\begin{split}
W_n(x_1,\ldots, x_n) & = \partial_{t_1}\cdots\partial_{t_n}\Big(\ln Z_{N,\beta}^{V-\frac{2}{\beta N}\sum_{i = 1}^n \frac{t_i}{x_i - \bullet};\mathsf{B}}\Big)\Big|_{t_i = 0} \\
& = \mu_{N,\beta}^{V;\mathsf{B}}\Big[\prod_{i = 1}^n \Tr\,\frac{1}{x_j - \mathbb{L}}\Big]_{c}. 
\end{split}
\end{equation}
By construction, the coefficients of their expansions as a Laurent series in the variables $x_i $ (sufficiently large) give the $n$-th order cumulants of $M_N$. If $I$ is a set, we introduce the notation $x_I = (x_i)_{i \in I}$ for a set of variables indexed by $I$; their order will not matter as we insert them only  in symmetric functions of their variables (like $W_n$, $\widetilde{W}_n$, etc.). The two types of correlators are related by:
$$
\widetilde{W}_n(x_1,\ldots,x_n) = \sum_{s = 1}^n \sum_{\substack{J_1\dot{\cup}\cdots\dot{\cup}J_s = \llbracket 1,n \rrbracket}} \prod_{i = 1}^s W_{|J_i|}(x_{J_i}).
$$
where $\dot{\cup}$ stands for the disjoint union. If $\varphi_n$ is an analytic (symmetric) function in $n$ variables in a neighbourhood of $\mathsf{B}^n$, the $n$-linear statistics can be deduced as contour integrals of the disconnected correlators:
\beq
\label{repw}\mu_{N,\beta}^{V;\mathsf{B}}\Big[ \sum_{1 \leq i_1,\ldots,i_n \leq N} \varphi_n(\lambda_{i_1},\ldots,\lambda_{i_n})\Big] = \oint_{\mathsf{B}} \frac{\dd \xi_1}{2{\rm i}\pi} \cdots \oint_{\mathsf{B}} \frac{\dd \xi_n}{2{\rm i}\pi}\,\varphi_n(\xi_1,\ldots,\xi_n)\,\widetilde{W}_n(\xi_1,\ldots,\xi_n).
\eeq
We remark that the knowledge of the correlators for an analytic family of potentials $(V_{t})_{t}$ determines the partition function up to an integration constant, since:
$$
\partial_t \ln Z_{N,\beta}^{V_t;\mathsf{B}} = -\frac{\beta N}{2}\,\mu_{N,\beta}^{V_t;\mathsf{B}}\Big[\sum_{i = 1}^N \partial_t V_t(\lambda_i)\Big] = -\frac{\beta N}{2}\,\oint_{\mathsf{B}} \frac{\dd\xi}{2{\rm i}\pi}\,\partial_t V_t(\xi)\,W_1^t(\xi).
$$
where $W_1^{t}$ is the first correlator in the model with potential $V_t$, and the notation $\oint_{\mathsf{B}} \dd \xi \cdots$ means integration along a contour in $\mathbb{C} \setminus \mathsf{B}$ surrounding $\mathsf{B}$ with positive orientation. If the integrand has poles in $\mathbb{C} \setminus \mathsf{B}$ (e.g. it depends on extra variables $x_i \in \mathbb{C} \setminus \mathsf{B}$ that are not integrated upon and has poles at $\xi = x_i$), the contour should be chosen (unless stated otherwise) so that the poles remain outside. The notation should not be confused with $\int_{\mathsf{B}} \dd \xi \cdots$, which is the Lebesgue integral on $\mathsf{B} \subseteq \mathbb{R}$.

\subsubsection{Kernels}
\label{Kdefsec}
Let $\mathbf{c}$ be a $n$-uple of non zero complex numbers. We introduce the $n$-point kernels:
\begin{equation}
\label{kerde}
\begin{split}
\mathsf{K}_{n,\mathbf{c}}(x_1,\ldots,x_n) & =  \mu_{N,\beta}^{V;\mathsf{B}}\left[\prod_{j = 1}^n \mathrm{det}^{c_j}(x_j - \mathbb{L})\right]  \\
& =  \frac{Z_{N,\beta}^{V - \frac{2}{\beta N}\sum_{j = 1}^n c_j\ln(x_j - \bullet);\mathsf{B}}}{Z_{N,\beta}^{V;\mathsf{B}}}.
\end{split}
\end{equation}
When $c_j$ are integers, the kernels are holomorphic functions of $x_j \in \mathbb{C}\setminus \mathsf{B}$. When $c_j$ are not integers, the kernels are multivalued holomorphic functions of $x_j$ in $\mathbb{C}\setminus \mathsf{B}$, with monodromies around the connected components of $\mathsf{B}$ and around $\infty$. The right-hand side of \eqref{kerde}, where we used $\ln$, has the same multivalued nature. Alternatively, both sides of \eqref{kerde} can be defined as   single-valued functions of $x_1,\ldots,x_n$ by choosing a determination of the logarithm in a domain $\mathsf{D}$ of the form $\mathbb{C} \setminus \ell$, where $\ell$ is a smooth path in $\mathbb{C}$  from $0$ to $\infty$, and using $z^{c} = e^{c \ln z}$ for the left-hand side.

In particular, for $\beta = 2$, $\mathsf{K}_{1,(1)}(x)$ is the monic $N$-th orthogonal polynomial associated to the weight $\mathbf{1}_{\mathsf{B}}(x)\,e^{-N\,V(x)}\dd x$ on the real line, and $\mathsf{K}_{2,(1,-1)}(x,y)$ is the $N$-th Christoffel--Darboux kernel associated to those orthogonal polynomials, see Section~\ref{S2}.

\subsection{Equilibrium measure and multi-cut regime}

By standard results of potential theory and large deviations, see \cite{Johansson98, BAG} or the textbooks \cite[Theorem 6]{Deift} or \cite[Theorem 2.6.1 and Corollary
2.6.3]{AGZ} (note there that $\mathsf{B}=\mathbb R$, but the generalisation to integration over general sets $\mathsf{B}$ is straightforward), we have:
\begin{theorem}
\label{th:1} Assume that $V\,:\, \mathsf{B} \rightarrow \mathbb{R}$ is a continuous function,  and if $V$ depends on $N$, assume also that $V $ converges towards $V^{\{0\}}$ when $N$ goes to infinity  in the space of continuous functions over $\mathsf{B}$ for the sup norm. Moreover, for $\tau \in \{\pm 1\}$ with $\tau\infty \in \mathsf{B}$, assume that:
$$
\liminf_{x \rightarrow \tau\infty} \frac{V^{\{0\}}(x)}{2\ln|x|} > 1.
$$
 We consider the  normalised empirical measure $L_N=N^{-1}\,M_N$ in the space $\mathcal{P}(\mathsf{B})$ of probability measures on $\mathsf{B}$ equipped with its weak topology. Then, the law of $L_N$ under
$\mu_{N,\beta}^{V;\mathsf{B}}$  satisfies a large deviation principle with scale $N^2$ and good rate function $J$ given by:
\beq
\label{Enf} J[\mu]= E[\mu]-\inf_{\nu\in\mathcal{P}(\mathsf{B})} E[\nu],\qquad E[\mu] =\frac{\beta}{2} \iint_{\mathsf{B}^2} \dd\mu(\xi)\dd\mu(\eta)\Big(\frac{V^{\{0\}}(\xi) + V^{\{0\}}(\eta)}{2}  -\ln|\xi-\eta|\Big).
\eeq
As a consequence, $L_N$ converges almost surely and in expectation
to the unique probability measure $\mu_{\mathrm{eq}}^{V}$ on $\mathsf{B}$ which minimises $E$.
$\mu_{\mathrm{eq}}^{V}$ has compact support, denoted $\mathsf{S}$. It is characterised by the existence of a constant $C^V$ such that:
\beq
\label{ina} \forall x \in \mathsf{B},\qquad 2\int_{\mathsf{B}}\dd \mu_{\mathrm{eq}}^{V}(\xi)\ln|x - \xi| - V^{\{0\}}(x) \leq C^V,
\eeq
with equality realised $\mu_{\mathrm{eq}}^V$ almost surely.
\end{theorem}

The goal of this article is to establish an all-order expansion of the partition function, the correlators and the kernels, in all such situations.

\subsection{Assumptions}

We will refer throughout the text to the following set of assumptions. An integer number $g\ge 0$ is fixed. 

\begin{hypothesis}
\label{hmainc1}
\begin{itemize} \item[] \phantom{sss}
\item[$\bullet$] (Regularity) $V\,:\,\mathsf{B} \rightarrow \mathbb{R}$ is continuous, and if $V$ depends on $N$, it has a limit $V^{\{0\}}$ in the space of continuous functions on $\mathsf{B}$ for the sup norm.
\item[$\bullet$] (Confinement) For $\tau \in \{\pm 1\}$ so that $\tau\infty \in \mathsf{B}$, $\liminf_{x \rightarrow \tau\infty} \frac{V(x)}{2\ln|x|} > 1$. If $V$ depends on $N$, we require its limit $V^{\{0\}}$ to satisfy this condition. 
\item[$\bullet$] ($(g + 1)$-cut regime) The support of $\mu_{\mathrm{eq}}^{V}$ is of the form $\mathsf{S} = \bigcup_{h = 0}^{g} \mathsf{S}_h$ where $\mathsf{S}_h = [\alpha_{h}^{-},\alpha_{h}^{+}]$ are pairwise disjoint and $\alpha_{h}^{-} < \alpha_{h}^+$ for any $h \in \ldbrack 0,g \rdbrack$.
\item[$\bullet$] (Control of large deviations) The effective potential $U^{V;\mathsf{B}}_{{\rm eq}}(x) = V(x)- 2\int_{\mathsf{B}} \ln|x-\xi|\dd\mu_{\mathrm{eq}}^{V}(\xi)$ for $x \in \mathsf{B}$ achieves its minimum value for $x \in \mathsf{S}$ only.
\item[$\bullet$] (Off-criticality) $\mu_{\mathrm{eq}}^{V}$ has a density of the form:
\beq
\label{eqns}\frac{\dd\mu_{\mathrm{eq}}^{V}}{\dd x} = \frac{S(x)}{\pi}\,\prod_{h = 0}^{g} (\alpha_h^{+} - x)^{\rho_h^{+}/2}(x - \alpha_h^{-})^{\rho_h^{-}/2},
\eeq
where $\rho_{h}^{\bullet}$ is $+1$ (resp. $-1$) if the corresponding edge is soft (resp. hard), and $S(x) > 0$ for $x \in \mathsf{S}$. Hard edges must be boundary points of $\mathsf{B}$.
\end{itemize}
\end{hypothesis}
Hard edges are boundary points of $\mathsf{B}$.
Note that  if $V^{\{0\}}$ is real-analytic in a neighbourhood of $\mathsf{B}$, the  $(g + 1)$-cut regime hypothesis is always satisfied (the support consists of a finite disjoint union of segments) and $S$ is analytic in a neighbourhood of $\mathsf{S}$. We will hereafter say that $V$ is regular and confining in $\mathsf{B}$ if it satisfies the two first assumptions above.

We will also require regularity of the potential:
\begin{hypothesis}
\label{hmainc3}
\begin{itemize}
\item[\phantom{sss}]
\item[$\bullet$] (Analyticity) $V$ extends to a holomorphic function in some open neighbourhood $\mathsf{U}$ of $\mathsf{S}$.
\item[$\bullet$] ($\frac{1}{N}$ expansion of the potential) There exists a sequence $(V^{\{k\}})_{k \geq 0}$ of holomorphic functions in $\mathsf{U}$ and constants $(v^{\{k\}})_{k \geq 1}$ such that, for any $K \geq 0$,
\beq\label{expV}
\sup_{\xi \in \mathsf{U}} \Big|V(\xi) - \sum_{k = 0}^{K} N^{-k}\,V^{\{k\}}(\xi)\Big| \leq v^{\{K + 1\}}\,N^{-(K + 1)}.
\eeq
\end{itemize}
\end{hypothesis}
In Section~\ref{refine}, we shall weaken Hypothesis~\ref{hmainc3} by allowing complex perturbations of order $\frac{1}{N}$ and harmonic functions instead of analytic functions:
\begin{hypothesis}
\label{hmainc4}
$V\,:\,\mathsf{B} \rightarrow \mathbb{C}$ can be decomposed as $V = \mathcal{V}_1 + \overline{\mathcal{V}_2}$ where:
\begin{itemize}
\item[$\bullet$] For $j = 1,2$, $\mathcal{V}_j$ extends to a holomorphic function in some neighbourhood $\mathsf{U}$ of $\mathsf{B}$. There exists a sequence of holomorphic functions $(\mathcal{V}_{j}^{\{k\}})_{k \geq 0}$ and constants $(v_{j}^{\{k\}})_{k \geq 1}$ so that, for any $K \geq 0$:
$$
\sup_{\xi \in \mathsf{U}} \Big|\mathcal{V}_j(\xi) - \sum_{k = 0}^{K} N^{-k}\,\mathcal{V}_j^{\{k\}}(\xi)\Big| \leq v_j^{\{K + 1\}}\,N^{-(K + 1)}.
$$
\item[$\bullet$] $V^{\{0\}} = \mathcal{V}_1^{\{0\}} + \overline{\mathcal{V}_2^{\{0\}}}$ is real-valued on $\mathsf{B}$.
\end{itemize}
\end{hypothesis}
The topology for which we study the large $N$ expansion of correlators is described in \S~\ref{S3}, and amounts to controlling the (moments of order $p$)$\times C^p$ uniformly in $p$ for a constant $C > 0$. We now describe our strategy and announce our results.

\subsection{Main result with fixed filling fractions: partition function and correlators}
\label{fixfil}
Before coming to the multi-cut regime, we analyse a different model where the number of $\lambda$s in a small enlargement of $\mathsf{S}_h$ is fixed. Let $\mathsf{A} = \bigcup_{h = 0}^{g} \mathsf{A}_h$ where $\mathsf{A}_h = [a_{h}^{-},a_h^{+}]$ are pairwise disjoint segments such that $a_{h}^{-} \leq \alpha_{h}^{-} < \alpha_{h}^+ \leq a_{h}^+$, where the inequalities are equalities if the corresponding edge is hard, and are strict if the corresponding edge is soft. We introduce the set:
\beq
\label{Egcaldef} \mathcal{E} = \Big\{\bm{\epsilon} \in (0,1)^{g}\quad\Big| \quad \sum_{h = 1}^{g} \epsilon_h < 1\Big\}.
\eeq
If $\bm{N}=(N_1,\ldots,N_g)$ is an integer vector such that $\bm{\epsilon}=\frac{\bm{N}}{N}
\in \mathcal{E}$, we denote $N_0 = N - \sum_{h = 1}^{g} N_h$, and  consider the probability measure on $\prod_{h = 0}^{g} \mathsf{A}_h^{N_h}$:
\begin{equation}
\label{eqmes2}
\begin{split}
\dd\mu_{N,\beta;\bm{\epsilon}}^{V;\mathsf{A}}(\bm{\lambda}) & = \frac{1}{Z_{N,\beta;\bm{\epsilon}}^{V;\mathsf{A}}}\prod_{h = 0}^g \Big[\prod_{i = 1}^{N_h} \mathrm{d}\lambda_{h,i}\,\mathbf{1}_{\mathsf{A}_{h}}(\lambda_{h,i})\,e^{-\frac{\beta N}{2}\,V(\lambda_{h,i})}\,\prod_{1 \leq i < j \leq N} |\lambda_{h,i} - \lambda_{h,j}|^{\beta}\Big] \\
& \quad \times \prod_{0 \leq h < h' \leq g} \prod_{\substack{1 \leq i \leq N_h \\ 1 \leq i' \leq N_{h'}}} |\lambda_{h,i} - \lambda_{h',i'}|^{\beta}.
\end{split}
\end{equation}
The empirical measure $M_{N}$ and the correlators $W_{n;\bm{N}/N}(x_1,\ldots,x_n)$ for this model are defined as in \S~\ref{S11} with $\mu_{N,\beta}^{V;\mathsf{A}}$ replaced by $\mu_{N,\beta;\bm{N}/N}^{V;\mathsf{A}}$. We call $\epsilon_h=\frac{N_h}{N}$ the \emph{filling fraction} of $\mathsf{A}_h$. It follows from the definitions that:
\beq
\label{nrm}\oint_{\mathsf{A}_h} \frac{\dd\xi}{2{\rm i}\pi}\,W_{n;\bm{N}/N}(\xi,x_2,\ldots,x_n) = \delta_{n,1}\,N_{h}=\delta_{n,1}\,N \epsilon_{h}.
\eeq
for $x_2,\ldots,x_n \in \mathbb{C} \setminus \mathsf{A}$. Indeed, from the definition of the correlators \eqref{defcor} $W_{n;\bm{N}/N}(x_{1},x_2,\ldots,x_n) $ for $n \geq 2$ can be expressed as a sum of  products of moments of products of the $n$-tuple of random variables $\big(\sum_{i = 1}^{N} \frac{1}{x_j - \lambda_{i}}-\mu_{N,\beta;\bm{\epsilon}}^{V;\mathsf{A}}[\sum_{i = 1}^{N} \frac{1}{x_j - \lambda_{i}}]
\big)_{j = 1}^{n}$ which are linear in each of these variables. Therefore we can integrate over the variable $x_{1}$ in each of these terms by Fubini's theorem. The key  observation is that 
$\oint_{\mathsf{A}_h} \sum_{i = 1}^{N} \frac{\dd \xi}{2{\rm i}\pi}\,\frac{1}{\xi - \lambda_i}$ is the number $N_{h}$ of $\lambda_i$s belonging to $\mathsf{A}_h$. Since $N_h$ is deterministic in the fixed filling fraction model, it is equal to its expectation and therefore each of these terms vanish which implies \eqref{nrm} for $n\ge 2$. When $n=1$, the cumulant is simply equal to the expectation of $ \sum_{i = 1}^{N} \frac{1}{\xi - \lambda_i}$ and the previous remark proves \eqref{nrm}.

We will refer to \eqref{eqmes} as the initial model, and to \eqref{eqmes2} as the model with fixed filling fractions. Standard results from potential theory or a straightforward generalisation of  \cite[Theorem 2.6.1 and Corollary
2.6.3]{AGZ}  imply:
\begin{theorem}
\label{th:10} Assume $V$ regular and confining on $\mathsf{A}$. We consider the normalised empirical measures
 $L_{N,h}=\frac{1}{N_h}\sum_{i=1}^{N_h} \delta_{\lambda_{h,i}}\in \mathcal{P}(\mathsf{A}_h)$ for $h\in \ldbrack 0,g \rdbrack$.
Take a sequence $\mathbf{N} = (N_1,\ldots,N_g)$ of $g$-uple of integers, indexed by $N$, such that $\sum_{h = 1}^g N_h \leq N$, and such that $\bm{N}/N$ converges to a given $\bm{\epsilon}\in \mathcal{E}$ when $N \rightarrow \infty$. Then, the law of $(L_{N,h})_{0\le h\le g}$ under $\mu_{N,\beta;\bm{N}/N}^{V;\mathsf{A}}$  satisfies a large deviation principle with scale $N^2$ and good rate function $$J_{\bm{\epsilon}}[\mu_0,\ldots,\mu_g]=E\Big[\sum_{h=0}^g\epsilon_h \mu_h\Big]-\inf_{\nu_h\in\mathcal{P}
(\mathsf{A}_h)} E\Big[\sum_{h=0}^g\epsilon_h \nu_h\Big]\,,$$
where $\epsilon_0=1-\sum_{h=1}^g \epsilon_h$, $N_0=N-\sum_{h=1}^g N_h$ and $E$ is defined in Equation~\eqref{Enf}. 
 As a consequence, the  empirical measure $L_{N;\bm{\epsilon}}=\sum_{h=0}^g\frac{N_h}{N} L_{N,h}$ 
converges almost surely and in expectation towards the unique probability measure $\mu_{\mathrm{eq};\bm{\epsilon}}^{V;\mathsf{A}}$ on $\mathsf{A}$ which minimises $E$ among probability measures with fixed mass $\epsilon_h$ on $\mathsf{A}_h$ for any $h \in \ldbrack 0,g \rdbrack$. It is characterised by the existence of constants $C_{\bm{\epsilon},h}^{V,\mathsf{A}}$ such that:
\beq
\label{ina0} \forall h \in \llbracket 0,g \rrbracket,\quad \forall x \in \mathsf{A}_h,\qquad 2\int_{\mathsf{B}}\dd \mu_{\mathrm{eq};\bm{\epsilon}}^{V;\mathsf{A}}(\xi)\ln|x - \xi| - V^{\{0\}}(x) \leq C_{\bm{\epsilon},h}^{V;\mathsf{A}},
\eeq
with equality realised $\mu_{\mathrm{eq};\bm{\epsilon}}^{V;\mathsf{A}}$ almost surely. $\mu_{\mathrm{eq};\bm{\epsilon}}^{V;\mathsf{A}}$ can be decomposed as a sum of positive measures $\mu_{\mathrm{eq};\bm{\epsilon},h}^{V}$ having compact support in $\mathsf{A}_h$, denoted $\mathsf{S}_{\bm{\epsilon},h}$. Moreover, if $V^{\{0\}}$ is real-analytic in a neighbourhood of $\mathsf{A}$, the support $\mathsf{S}_{\bm{\epsilon},h}$ consists of a finite union of segments.
\end{theorem}
Later in the text, we shall consider $\mu_{\mathrm{eq};\bm{N}/N}^{V;\mathsf{A}}$ with $\bm{N}=(N_1,\ldots,N_g)$ a vector of positive integers so that $\sum_{h = 1}^{g} N_h < N$: this will denote the unique solution of \eqref{ina0} with $\bm{\epsilon}=\bm{N}/N$. 
$\mu_{\mathrm{eq}}^{V;\mathsf{A}}$ appearing in Theorem~\ref{th:1} coincides with $\mu_{\mathrm{eq};\bm{\epsilon}_{\star}}^{V}$ for the optimal value $\bm{\epsilon}_{\star} = (\mu_{\mathrm{eq}}^{V;\mathsf{A}}(\mathsf{A}_{h}))_{1 \leq h \leq g}$, and in this case $\mathsf{S}_{\bm{\epsilon}_{\star},h}$ is actually the segment $[\alpha^{-}_h,\alpha_{h}^+]$. The key point --- justified in Appendix \ref{appA} ---  is that, for $\bm{\epsilon}$ close enough to $\bm{\epsilon}_{\star}$, the support $\mathsf{S}_{\bm{\epsilon},h}$ remains connected, and the model with fixed filling fractions enjoys a $\frac{1}{N}$ expansion.

\begin{theorem}
\label{th:3}If $V$ satisfies Hypotheses~\ref{hmainc1} and \ref{hmainc4} on $\mathsf{A}$, there exists $t > 0$ such that, uniformly for integers $\bm{N}=(N_1,\ldots,N_g)$ such that $\bm{N} /N\in \mathcal{E}$  and $|\bm{N}/N- \bm{\epsilon}_{\star}|_1 < t$, we have an expansion for the correlators, for any $K \geq 0$
\beq
\label{expco}W_{n;\bm{N}/N}(x_1,\ldots,x_n) = \sum_{k = n - 2}^{K} N^{-k}\,W_{n;\bm{N}/N}^{\{k\}}(x_1,\ldots,x_n) + O(N^{-(K + 1)}).
\eeq
Up to a fixed $O(N^{-(K + 1)})$ and for a fixed $n$, Equation~\eqref{expco} holds uniformly for $x_1,\ldots,x_n$ in compact regions of $\mathbb{C}\setminus \mathsf{A}$ . The $W_{n;\bm{\epsilon}}^{\{k\}}$ can be extended into  smooth functions of $\bm{\epsilon}\in\mathcal{E}$ close enough to $\bm{\epsilon}_{\star}$.
\end{theorem}
We prove this theorem, independently of the nature soft/hard of the edges, in Section~\ref{S3} for real-analytic potential (\textit{i.e.} Hypothesis~\ref{hmainc3} instead of \ref{hmainc4}). For $\beta = 2$ and potential $V$ independent of $N$, the coefficients of expansion $W_{n;\bm{N}/N}^{\{k\}} = 0$ are zero for $k = (n + 1) \,\,{\rm mod} 2$, as is well-known for hermitian random matrix models (see \eqref{Wndom} and  the remarks on $\beta$-dependence in Section~\ref{CEO}).  The result is extended to harmonic potentials (\textit{i.e.}, Hypothesis~\ref{hmainc4}) in Section~\ref{S62}. In Proposition~\ref{expoqq}, we provide an explicit control of the errors in terms of the distance of $x_1,\ldots,x_k$ to $\mathsf{A}$, and its proof makes clear that the expansion of the correlators is  not  expected to be uniform for $x_1,\ldots,x_n$ chosen in a compact of $\mathbb{C}\setminus \mathsf{A}$ independently of $n$ and $K$ (namely it is uniform only for $K$ fixed). Note that we will sometimes omit to specify the dependence in $\mathsf{A},V$, etc. in the notations (e.g. for the equilibrium measure, for the correlators and their coefficient of expansions), but we will at least include it when this dependence is of particular importance.

We then compute in Section~\ref{PartitionSec} the expansion of the partition function thanks to the expansion of $W_{1;\bm{N}/N}$ and $W_{2;\bm{N}/N}$, by an interpolation that reduces the strength of pairwise interactions between eigenvalues in different segments while preserving the equilibrium measure. At the end of the interpolation, we are left with a product of $(g + 1)$ partition functions in a one-cut regime, for which the asymptotic expansion was established in \cite{BG11}.

\begin{theorem}
\label{th:2}If $V$ satisfies Hypotheses \ref{hmainc1} and \ref{hmainc4} on $\mathsf{A}$, there exists $t > 0$ such that, uniformly for $g$-dimensional vectors of positive integers $\bm{N}$  such that $\bm{N} /N\in \mathcal{E}$ and $|\bm{N}/N- \bm{\epsilon}_{\star}|_1 < t$, we have for any $K \geq 0$:
\beq 
\label{sqiqq}\frac{N!\,Z_{N,\beta;\bm{N}/N}^{V;\mathsf{A}}}{\prod_{h = 0}^{g} N_h!}= N^{\frac{\beta}{2}N + \varkappa}\exp\Big(\sum_{k = -2}^{K} N^{-k}\,F^{\{k\};V}_{\beta;\bm{N}/N} + O(N^{-(K + 1)})\Big),
\eeq
with
$$
\varkappa = \frac{1}{2} + (\# {\rm soft} + 3\#†{\rm hard})\frac{-3 + \beta/2 + 2/\beta}{24}.
$$
Besides, $F^{\{k\};V}_{\beta;\bm{\epsilon}}$ extends to a smooth function of $\bm{\epsilon}$ close enough to $\bm{\epsilon}_{\star}$, and at the value $\bm{\epsilon} = \bm{\epsilon}_{\star}$, the first derivatives of $F^{\{-2\};V}_{\beta;\bm{\epsilon}}$ vanish and its Hessian is negative definite.
\end{theorem}
We can identify explicitly:
\begin{equation}
\label{F1ent} \begin{split}
F^{\{-2\};V}_{\beta;\bm{\epsilon}} & =  \frac{\beta}{2}\bigg(\iint_{\mathsf{A}^2} \ln|x - y|\,\dd\mu_{{\rm eq};\bm{\epsilon}}^{V}(x)\dd\mu_{{\rm eq};\bm{\epsilon}}^{V}(y) -  \int_{\mathsf{A}} V^{\{0\}}(x)\dd\mu_{{\rm eq};\bm{\epsilon}}^{V}(x)\bigg) = -\frac{\beta}{2}\,\inf_{\nu_h \in \mathcal{P}(\mathsf{A}_h)} E\Big[\sum_{h = 0}^{g} \epsilon_h\nu_h\Big], \\
F^{\{-1\};V}_{\beta;\bm{\epsilon}} & =  - \frac{\beta}{2} \int_{\mathsf{A}} V^{\{1\}}(x)\dd\mu_{{\rm eq};\bm{\epsilon}}^{V}(x) + \Big(1 - \frac{\beta}{2}\Big)\Big({\rm Ent}[\mu_{{\rm eq};\bm{\epsilon}}^{V}] - \ln\big(\tfrac{\beta}{2}\big)\Big) +\frac{\beta}{2}\ln\big(\tfrac{2\pi}{e}\big) - \ln\Gamma\big(\tfrac{\beta}{2}\big),
\end{split}
\end{equation}
where
$$
{\rm Ent}[\mu] = -\int_{\mathbb{R}} \ln\Big(\frac{\dd\mu}{\dd x}\Big) \dd \mu(x)
$$
is the entropy. The formula for $F^{\{-2\};V}_{\beta;\bm{\epsilon}}$ is obvious from potential theory, while the formula for $F^{\{-1\};V}_{\beta;\bm{\epsilon}}$ is established in Proposition~\ref{LemEntV} (the first term comes from the fact that we let the potential  depend on $N$). The appearance of the entropy in the term of order $N$ in the free energy is well known in the one-cut case, and here we prove that it appears in the same way for the multi-cut case with fixed filling fractions and we determine the 
additional constant. The term  $\frac{\beta}{2} N \ln N$ is universal, while the term $\varkappa \ln N$  only depends on the nature of the endpoints of the support. These logarithmic corrections can already be observed in the asymptotic expansion of Selberg integrals for  large $N$ computing the partition function of the classical Jacobi, Laguerre or Gaussian $\beta$-ensembles, corresponding to a one-cut regime \cite{BG11}. The fact that the coefficient of $\ln N$ shadows in some way the geometry of the support was observed in other contexts, see e.g. \cite{Cardy}, and is not specific to two-dimensional Coulomb gases living on a line. Their identification in the multi-cut regime and fixed filling fractions results from an interpolation with a product of one such  model for each cut, which changes only  the coefficients of powers of $N$. Up to a given $O(N^{-K})$, all expansions are uniform with respect to the parameters of the potential and of $\bm{\epsilon}$ chosen in a compact set so that  the assumptions hold.  Theorems~\ref{th:3}--\ref{th:2} are the generalisations to the fixed filling fractions model of our earlier results about existence of the $\frac{1}{N}$ expansion in the one-cut regime \cite{BG11} (see also \cite{Johansson98,APS01,ErcMcL,BleherItsZ,GMS,Shch2} for earlier results concerning the one-cut regime in $\beta = 2$ or general $\beta$-ensembles).

\subsection{Relation with Chekhov--Eynard--Orantin topological recursion}\label{CEO}

Once these asymptotic expansions are shown to exist, by consistency their coefficients $W_{n;\bm{\epsilon}}^{\{k\}}$ are computed by the $\beta$ topological recursion of Chekhov and Eynard \cite{CE06}. As a matter of fact, the asymptotic expansion
$$
W_{n;\bm{\epsilon}}(x_1,\ldots,x_n) = \sum_{k \geq -1} N^{-k}\,W_{n;\bm{\epsilon}}^{\{k\}}(x_1,\ldots,x_n)
$$
has a finer structure so that for  $n \geq 1$ and $k \geq -1$ we can write:
\begin{equation}
\label{Wndom}W_{n;\bm{\epsilon}}^{\{k\}}(x_1,\ldots,x_n) = \sum_{G = 0}^{\lfloor \frac{k - n}{2} \rfloor + 1} \Big(\frac{\beta}{2}\Big)^{1 - n - G}\Big(1 - \frac{2}{\beta}\Big)^{k + 2 - 2G - n}\,\mathcal{W}_{n;\bm{\epsilon}}^{[G,k + 2 - 2G - n]}(x_1,\ldots,x_n),
\end{equation}
where $\mathcal{W}_{n;\bm{\epsilon}}^{[G,l]}$ are the quantities computed by the topological recursion of \cite{CE06}. The initial data consists of the non-decaying terms in the correlators, namely:
\begin{equation*}
\begin{split}
W_{1;\bm{\epsilon}}^{\{-1\}}(x) & = \mathcal{W}_{1;\bm{\epsilon}}^{[0,0]}(x) ,\\
W_{1;\bm{\epsilon}}^{\{0\}}(x) & = \Big(1 - \frac{2}{\beta}\Big)\mathcal{W}_{1;\bm{\epsilon}}^{[0,1]}(x), \\
W_{2;\bm{\epsilon}}^{\{0\}}(x_1,x_2) & = \frac{2}{\beta}\,\mathcal{W}_{2;\bm{\epsilon}}^{[0,0]}(x_1,x_2).
\end{split}
\end{equation*}
All these quantities have an analytic continuation in the variables $x_i$ on the same Riemann surface $\mathcal{C}_{\bm{\epsilon}}$ called \emph{spectral curve}. The curve $\mathcal{C}_{\bm{\epsilon}}$ can in fact be defined as the maximal Riemann surface on which $W_{1;\bm{\epsilon}}^{\{-1\}}(x)$, initially defined for $x \in \mathbb{C} \setminus \mathsf{A}$, admits an analytic continuation (cf. Section~\ref{S17} for a continued discussion on geometry of spectral curves). The information carried by the decomposition \eqref{Wndom} is that, if $V$ is chosen independent of $\beta$ and $N$, all the $\mathcal{W}_{n;\bm{\epsilon}}^{[G,K]}$ are also independent of $\beta$ and $N$ (except perhaps through the implicit dependence in $N$ of $\bm{\epsilon}$), and thus the coefficients of  the expansions of the correlators display a remarkable structure of Laurent polynomial in $\frac{\beta}{2}$. This property comes from the structure of the Dyson--Schwinger equations.

From the same initial data, Chekhov and Eynard also define numbers $W_{0;\bm{\epsilon}}^{[G,K]} = \mathcal{F}_{\bm{\epsilon}}^{[G,K]}$, which give the coefficients of the asymptotic expansion of the free energy $\ln Z_{N,\beta;\bm{N}/N}^{V;\mathsf{A}}$ up to an integration constant independent of the potential, and which are independent of $\beta$ provided $V$ is chosen independent of $\beta$. More precisely, we mean that for any two potentials $V$ and $\tilde{V}$ satisfying the assumptions of Theorem~\ref{th:2} and leading to a $(g + 1)$-cut regime, we must have for $k \geq -2$, by consistency with \cite{CE06}:
$$
F^{\{k\};V}_{\beta;\bm{\epsilon}} - F^{\{k\};\tilde{V}}_{\beta;\bm{\epsilon}} = \sum_{G = 0}^{\lfloor \frac{k}{2} \rfloor + 1} \Big(\frac{\beta}{2}\Big)^{1 - G}\Big(1 - \frac{2}{\beta}\Big)^{k + 2 - 2G}\,\big(\mathcal{F}^{[G,k + 2 - 2G];V}_{\bm{\epsilon}} - \mathcal{F}^{[G,k + 2 - 2G];\tilde{V}}_{\bm{\epsilon}}\big).
$$
In particular, the topological recursion defines $\mathcal{F}^{[0,0];V}_{\bm{\epsilon}} = E[\mu_{{\rm eq};\bm{\epsilon}}^V]$ and $\mathcal{F}^{[0,1];V}_{\bm{\epsilon}} = -{\rm Ent}[\mu_{{\rm eq};\bm{\epsilon}}^{V}]$. By comparison with \eqref{F1ent}, we arrive to an absolute comparison (here assume the potential to be independent of $N$, \textit{i.e.} $V = V^{\{0\}}$):
\begin{equation}
\label{Fcomu}
\begin{split}
F^{\{-2\};V}_{\beta;\bm{\epsilon}} & = \frac{\beta}{2}\,\mathcal{F}^{[0,0];V}_{\bm{\epsilon}}, \\
F^{\{-1\};V}_{\beta;\bm{\epsilon}} & = \frac{\beta}{2}\Big(1 - \frac{2}{\beta}\Big)\Big(\mathcal{F}^{[0,1];V}_{\bm{\epsilon}} + \ln\big(\tfrac{\beta}{2}\big)\Big)  + \frac{\beta}{2}\ln\big(\tfrac{2\pi}{e}\big) - \ln \Gamma\big(\tfrac{\beta}{2}\big).
\end{split}
\end{equation}
The constant in the second line was not computed in \cite{CE06}. To our knowledge, the absolute --- including a $\beta$-dependent, possibly $g$-dependent but otherwise $V$-independent constant --- comparison between the coefficients $F_{\beta;\bm{\epsilon}}^{\{k\};V}$ of the asymptotic expansion of the $\beta$-ensembles and the invariants $\mathcal{F}^{[G,m]}$ for $(G,m) \neq (0,0),(0,1)$ produced by the topological recursion has not been performed in full generality. It is only known for $\beta = 2$ for all $G$ in the one-cut regime, see \cite[Proposition 2.5]{Mar17}.

When $\beta = 2$, only $\mathcal{W}_{n;\bm{\epsilon}}^{[G]} = \mathcal{W}_{n;\bm{\epsilon}}^{[G,0]}$ and $\mathcal{F}_{\bm{\epsilon}}^{[G]} = \mathcal{F}^{[G,0]}_{\bm{\epsilon}}$ appear. These are the quantities defined by the Chekhov--Eynard--Orantin topological recursion \cite{EO07}, and we retrieve the usual asymptotic expansions
\begin{equation*}
\begin{split}
\mathcal{W}_{n;\bm{\epsilon}}(x_1,\ldots,x_n) & = \sum_{G \geq 0} N^{2 - 2G - n}\,\mathcal{W}_{n;\bm{\epsilon}}^{[G]}(x_1,\ldots,x_n), \\
\ln\bigg(\frac{Z_{N,\beta;\bm{\epsilon}}^{V;\mathsf{A}}}{Z_{N,\beta;\bm{\epsilon}}^{\tilde{V};\mathsf{A}}}\bigg)  & = \sum_{G \geq 0} N^{2 - 2G}\big(\mathcal{F}^{[G];V}_{\bm{\epsilon}} - \mathcal{F}^{[G];\tilde{V}}_{\bm{\epsilon}}\big),
\end{split}
\end{equation*}
involving only powers of $\frac{1}{N}$ with parity $(-1)^n$ in the $n$-point correlators, and powers of $\frac{1}{N^2}$ in the free energy.

\subsection{Main results in the multi-cut regime: partition function}
\label{msqqu}
Let us come back to the initial model \eqref{eqmes}. We can always take $\mathsf{A} = \bigcup_{h = 0}^{g} \mathsf{A}_h \subseteq \mathsf{B}$ to be a small enlargement of the support $\mathsf{S}$ respecting the setup of \S~\ref{fixfil}. It is  indeed well-known that the partition function $Z_{N,\beta}^{V;\mathsf{B}}$ can be replaced by $Z_{N,\beta}^{V;\mathsf{A}}$ up to exponentially small corrections when $N$ is large (see \cite{PSbook,BG11} for results in this direction, and we give a proof for completeness in \S~\ref{S31as} below). The latter can be decomposed as a sum over all possible ways of distributing the $\lambda$s between the segments $\mathsf{A}_h$, namely:
\beq
\label{sum1}Z_{N,\beta}^{V;\mathsf{A}} = \sum_{\substack{ N_0,\ldots,N_{g} \geq 0 \\ \sum_{h=0}^g N_h= N}} \frac{N!}{\prod_{h = 0}^{g} N_h!}\,Z_{N,\beta;\bm{N}/N}^{V;\mathsf{A}},
\eeq
where we have denoted $N_0 = N - \sum_{h = 1}^{g} N_h$ the number of $\lambda$s put in the segment $\mathsf{A}_0$. So, we can use our results for the model with fixed filling fractions to analyse the asymptotic behaviour of each term in the sum, and then find the asymptotic expansion of the sum taking into account the interference of all contributions. This is carried out in Section~\ref{S8111}.

Before stating the results, we need two ingredients. First, we let $\mathfrak{Z}_{N,\beta;\bm{\epsilon}}^{V;\mathsf{A}}$ be the (truncated at an arbitrary order $K$) asymptotic series depending on a $g$-dimensional vector with positive entries, at least when its coefficients are defined:
\beq
\label{ildeZ} \mathfrak{Z}_{N,\beta;\bm{\epsilon}}^{V;\mathsf{A}}= N^{\frac{\beta}{2}N + \varkappa}\exp\Big(\sum_{k = -2}^{K} N^{-k}\,F^{\{k\};V}_{\beta;\bm{\epsilon}} + O(N^{-(K + 1)})\Big)\,.
\eeq
If we substitute $\bm{\epsilon} = \bm{N}/N$ as in Theorem~\ref{th:2}, it gives the asymptotic expansion of the partition function of the fixed filling fractions model with unordered eigenvalues, and we recall that $F^{\{k\};V}_{\beta;\bm{\epsilon}}$ exists as a smooth function of $\bm{\epsilon}$ in some non-empty open set. We shall denote $(F^{\{k\};V}_{\beta;\bm{\epsilon}})^{(j)}$ the tensor of $j$-th derivatives with respect to $\bm{\epsilon}$.

Second, we introduce the Siegel Theta function with characteristics $\bm{\mu},\bm{\nu} \in \mathbb{C}^{g}$. If $\bm{\tau}$ is a symmetric $g \times g$ matrix of complex numbers such that $\mathrm{Im}\,\bm{\tau} > 0$,  the Siegel Theta function
 is the entire function of $\bm{v} \in \mathbb{C}^{g}$ defined by the exponentially fast converging series:
\begin{equation}
\label{thetadeff}
\vartheta\!\left[\begin{array}{@{\hspace{-0.02cm}}l@{\hspace{-0.02cm}}} \bm{\mu} \\ \bm{\nu} \end{array}\right]\!\!(\bm{v}|\bm{\tau}) = \sum_{\bm{m} \in \mathbb{Z}^{g}} \exp\Big({\rm i}\pi (\bm{m} + \bm{\mu})\cdot\bm{\tau}\cdot(\bm{m} + \bm{\mu}) + 2{\rm i}\pi(\bm{v} + \bm{\nu})\cdot(\bm{m} + \bm{\mu})\Big).
\end{equation}
Among its essential properties, we mention:
\begin{itemize}
\item[$\bullet$] for any characteristics $\bm{\mu},\bm{\nu}$, it satisfies the diffusion-like equation $4{\rm i}\pi\partial_{\tau_{h,h'}}\vartheta = \partial_{v_h}\partial_{v_{h'}}\vartheta$.
\item[$\bullet$] it is a quasi-periodic function with lattice $\mathbb{Z}^{g} \oplus \bm{\tau}(\mathbb{Z}^{g})$: for any $\bm{m}_0,\bm{n}_0 \in \mathbb{Z}^{g}$,
$$
\vartheta\!\left[\begin{array}{@{\hspace{-0.02cm}}l@{\hspace{-0.02cm}}} \bm{\mu} \\ \bm{\nu} \end{array}\right]\!\!(\bm{v} + \bm{m}_0 + \bm{\tau}\cdot\bm{n}_0|\bm{\tau}) = \exp\big(2{\rm i}\pi\bm{m}_0\cdot\bm{\mu} - 2{\rm i}\pi \bm{n_0}\cdot(\bm{v} + \bm{\nu}) - {\rm i}\pi\bm{n}_0\cdot\bm{\tau}\cdot\bm{n}_0\big)\,\vartheta\!\left[\begin{array}{@{\hspace{-0.02cm}}l@{\hspace{-0.02cm}}} \bm{\mu} \\ \bm{\nu} \end{array}\right]\!\!(\bm{v}|\bm{\tau}).
$$
\item[$\bullet$] it has a nice transformation law under $\bm{\tau} \rightarrow (\bm{A\tau} + \bm{B})(\bm{C\tau} + \bm{D})^{-1}$ where $\bm{A},\bm{B},\bm{C},\bm{D}$ are the $g \times g$ blocks of a $2g \times 2g$ symplectic matrix \cite{TataMumford}.
\item[$\bullet$] when $\bm{\tau}$ is the matrix of periods of a genus $g$ Riemann surface, it satisfies the Fay identity \cite{Fay}.
\end{itemize}
We define the gradient operator $\nabla_{\bm{v}}$ acting on the variable $\bm{v}$ of this function. For instance, the diffusion equation takes the form $4{\rm i}\pi\partial_{\bm{\tau}}\vartheta = \nabla^{\otimes 2}_{\bm{v}}\vartheta$.

\begin{theorem}
\label{th:22}Assume Hypotheses~\ref{hmainc1} and \ref{hmainc4}. Let $\bm{\epsilon}_{\star} = (\mu_{\mathrm{eq}}^V[\mathsf{S}_h])_{1 \leq h \leq g}$ --- we shall replace all indices $\bm{\epsilon}$ by $\star$ in our notations to indicate a specialisation at $\bm{\epsilon} = \bm{\epsilon}_\star$. Then, the partition function has an asymptotic expansion of the form, with $\mathsf{C}=\mathsf{B}$ or $\mathsf{A}$, for any $K \geq -2$:
\beq
\label{300} Z_{N,\beta}^{V;\mathsf{C}} = \mathfrak{Z}_{N,\beta;\star}^{V;\mathsf{A}}\left\{\Big(\sum_{k = 0}^{K} N^{-k}\,T_{\beta;\star}^{\{k\}}\big[\tfrac{\nabla_{\bm v}}{2{\rm i}\pi}\big]\Big)\vartheta\!\left[\begin{array}{@{\hspace{-0.03cm}}c@{\hspace{-0.03cm}}} -N\bm{\epsilon}_{\star}\, \\ \bm{0} \end{array}\right]\!\!(\bm{v}_{\beta;\star}|\bm{\tau}_{\beta;\star}) + O(N^{-(K + 1)})\right\}.
\eeq
In this expression, $\mathfrak{Z}_{N,\beta;\star}^{V;\mathsf{A}}$ is the asymptotic series defined in Equation~\eqref{ildeZ} and evaluated at $\bm{\epsilon} = \bm{\epsilon}_{\star}$. If $\bm{X}$ is a vector with $g$ components, we set $T^{\{0\}}_{\beta;\bm{\epsilon}}[\bm{X}] = 1$, and for $k \geq 1$:
\beq
\label{TMde}T_{\beta;\bm{\epsilon}}^{\{k\}}[\bm{X}] = \sum_{r = 1}^{k} \frac{1}{r!} \sum_{\substack{k_1,\ldots,k_r \geq -2 \\ j_1,\ldots,j_r > 0 \\ k_i + j_i > 0 \\ \sum_{i = 1}^{r} k_i + j_i = k}} \Big(\bigotimes_{i = 1}^{r} \frac{(F_{\beta;\bm{\epsilon}}^{\{k_i\};V})^{(j_i)}}{j_i!}\Big)\cdot\bm{X}^{\otimes(\sum_{i = 1}^r j_i)},
\eeq
where $\cdot$ denotes the standard scalar product on the tensor space. We have also introduced:
$$
\bm{v}_{\beta;\star} = \frac{(F^{\{-1\};V}_{\beta;\star})'}{2{\rm i}\pi},\qquad \bm{\tau}_{\beta;\star} = \frac{(F^{\{-2\};V}_{\beta;\star})''}{2{{\rm i}\pi}}.
$$
\end{theorem}
Being more explicit but less compact, we may rewrite:
\begin{equation}
\label{Tmsa}
\begin{split}
 \quad T_{\beta;\star}^{\{k\}}\big[\tfrac{\nabla_{\bm{v}}}{2{\rm i}\pi}\big]\vartheta\!\left[\begin{array}{@{\hspace{-0.03cm}}c@{\hspace{-0.03cm}}} -N\bm{\epsilon}_{\star}\, \\ \bm{0} \end{array}\right]\!\!(\bm{v}_{\beta;\star}|\bm{\tau}_{\beta;\star})  & = \sum_{r = 1}^{k} \frac{1}{r!} \sum_{\substack{k_1,\ldots,k_r \geq -2 \\ j_1,\ldots,j_r > 0 \\ k_i + j_i > 0 \\ \sum_{i = 1}^{r} k_i + j_i = k}}\!\!\!\! \Big(\bigotimes_{i = 1}^{r} \frac{(F_{\beta;\star}^{\{k_i\};V})^{(j_i)}}{j_i!}\Big) \\
 & \quad  \cdot \Big(\sum_{\bm{m} \in \mathbb{Z}^{g}}(\bm{m} - N\bm{\epsilon}_{\star})^{\otimes(\sum_{i = 1}^r j_i)}\,e^{{\rm i}\pi \cdot\bm{\tau_{\beta;\star}}\cdot(\bm{m} - N\bm{\epsilon}_{\star})^{\otimes 2} + 2{\rm i}\pi \bm{v}_{\beta;\star}\cdot(\bm{m} - N\bm{\epsilon}_{\star})}\Big).
\end{split}
\end{equation}

For $\beta = 2$, this result has been derived heuristically to leading order in \cite{BDE}, and to all orders in \cite{Ecv}. These heuristic arguments can be extended straightforwardly to all values of $\beta$, see \textit{e.g.} \cite{GBThese}. Our work justifies their heuristic argument. To prove this result, we exploit the Dyson--Schwinger equations for the $\beta$-ensemble with fixed filling fractions taking advantage of a rough control on the large $N$ behaviour of the correlators. The result of Theorem~\ref{th:22} has been derived up to $o(1)$ by Shcherbina \cite{Smulticut} for real-analytic potentials, with different techniques, based on the representation of $\prod_{h < h'} \prod_{i,j} |\lambda_{h,i} - \lambda_{h',j}|^{\beta}$, which is the exponential of a quadratic statistic, as expectation value of a linear statistics coupled to a Brownian motion. The rough \textit{a priori} controls on the correlators do not allow at present the description of the $o(1)$ by such methods. The results in \cite{Smulticut} were also written in a different form: $F^{\{0\};V}_{\beta;\bm{\epsilon}}$ appearing in $\mathfrak{Z}$ was identified with a combination of Fredholm determinants (see also the physics paper \cite{WiegZ}), while this representation does not come naturally in our approach. Also, the steps undertaken in Section~\ref{musqo} where we replace the sum over nonnegative integers such that $N_0 + \cdots + N_g = N$ in Equation~\eqref{sum1}, by a sum over $\bm{N} \in \mathbb{Z}^{g}$, thus reconstructing the Siegel Theta function, was not performed in \cite{Smulticut}.

The $2{\rm i}\pi$ appears because we used the standard definition of the Siegel Theta function, and should not hide the fact that all terms in Equation~\eqref{Tmsa} are real-valued. Here, the matrix:
\beq
\label{30}\bm{\tau}_{\beta;\star} = \frac{\mathrm{Hessian}(F^{\{-2\};V}_{\beta;\bm{\epsilon}})\big|_{\bm{\epsilon} = \bm{\epsilon}_{\star}}}{2{\rm i}\pi}
\eeq
involved in the Theta function has purely imaginary entries, and $\mathrm{Im}\,\bm{\tau}_{\beta;\star}$ is definite positive according to Theorem~\ref{th:2}, hence the Theta function in the right-hand side makes sense. Notice also that for it is $\mathbb{Z}^g$-periodic with respect to $\bm{\mu}$, hence we can replace $-N\bm{\epsilon}_{\star}$ by $-N\bm{\epsilon}_{\star} + \lfloor N\bm{\epsilon}_{\star} \rfloor$, and this is responsible for modulations in the asymptotic expansion, and thus breakdown of the $\frac{1}{N}$ expansion. Still, the model has ``subsequential" asymptotic expansions in $\frac{1}{N}$. For instance, for an even potential with two cuts ($g = 1$) model, we have $\epsilon_{\star} = \frac{1}{2}$ so $(-N\epsilon_{\star}\,\,{\rm mod}\,\,\mathbb{Z})$ appearing as characteristic in the Theta function only depends on the parity of $N$, and for each fixed parity we get an asymptotic expansion in $\frac{1}{N}$. In fact, having an even potential implies that the fixed-filling fraction model is invariant under $\epsilon \rightarrow 1 - \epsilon$, so only the terms with even numbers $j_i$ of derivatives with respect to filling fractions contribute in $T^{\{k\}}_{\beta;\bm{\epsilon}}$. If furthermore $\beta = 2$, only the $(F^{\{k\};V}_{\beta = 2;\star})^{(j)}$ with $k$ even survive, and we deduce that the same is true for $T^{\{k\}}_{\beta =2;\star}$, so that the logarithm of the partition function has an asymptotic expansion in $\frac{1}{N^2}$ for $N$ odd, and different asymptotic expansion in $\frac{1}{N^2}$ for $N$ even (of course up to the universal logarithmic corrections $\frac{\beta}{2}N\ln N + \varkappa \ln N$).

Let us give the two first orders of Equation~\eqref{Tmsa}:
$$
T_{\beta;\star}^{\{1\}}[\bm{X}] = \frac{1}{6}\,(F_{\beta;\star}^{\{-2\};V})'''\cdot\bm{X}^{\otimes 3} + \frac{1}{2}\,(F_{\beta;\star}^{\{-1\};V})''\cdot\bm{X}^{\otimes 2} + (F_{\beta;\star}^{\{0\};V})'\cdot\bm{X},
$$
and:
\begin{equation*}
\begin{split}
T_{\beta;\star}^{\{2\}}[\bm{X}] & = \frac{1}{72}\,\big[(F_{\beta;\star}^{\{-2\};V})'''\big]^{\otimes 2}\cdot\bm{X}^{\otimes 6} + \frac{1}{12}\,\big[(F_{\beta;\star}^{\{-2\};V})'''\otimes (F_{\beta;\star}^{\{-1\};V})''\big]\cdot \bm{X}^{\otimes 5} \\
& \quad + \Big(\frac{1}{6}\,\big[(F_{\beta;\star}^{\{-2\};V})'''\otimes (F_{\beta;\star}^{\{0\};V})'\big] + \frac{1}{8}\,\big[(F_{\beta;\star}^{\{-1\};V})''\big]^{\otimes 2} +  \frac{1}{24}\,(F_{\beta;\star}^{\{-2\};V})^{(4)}\Big)\cdot\bm{X}^{\otimes 4} \\
& \quad  + \Big(\frac{1}{2}\,\big[(F_{\beta;\star}^{\{-1\};V})''\otimes(F_{\beta;\star}^{\{0\};V})'\big] + \frac{1}{6}\,(F_{\beta;\star}^{\{-1\};V})'''\Big)\cdot\bm{X}^{\otimes 3} \\
& \quad + \Big(\frac{1}{2}\,\big[(F_{\beta;\star}^{\{0\};V})'\big]^{\otimes 2} + \frac{1}{2}\,(F_{\beta;\star}^{\{0\};V})''\Big)\cdot\bm{X}^{\otimes 2} + (F_{\beta;\star}^{\{1\};V})'\cdot \bm{X}.
\end{split}
\end{equation*}

For $\beta = 2$, unlike the one-cut regime where the asymptotic expansion was in $\frac{1}{N^2}$ up to constants independent of the potential, the multi-cut regime features an asymptotic expansion with non-trivial terms in powers of $\frac{1}{N}$. For instance we have a contribution at order $\frac{1}{N}$ of 
$$
T_{\beta = 2;\star}^{\{1\}}[\bm{X}] = \frac{1}{6}\,(F_{\beta = 2;\star}^{\{-2\};V})'''\cdot\bm{X}^{\otimes 3} + (F_{\beta = 2;\star}^{\{0\};V})'\cdot\bm{X}.
$$
 In a two-cuts regime ($g = 1$), a sufficient condition for all terms of order $N^{-(2k + 1)}$ to vanish (again, up to integration constants already present in $\mathfrak{Z}$) is that $\epsilon_{\star} = \frac{1}{2}$ and $Z_{N,\beta = 2;\epsilon}^{V;\mathsf{A}} = Z_{N,\beta = 2;1 - \epsilon}^{V;\mathsf{A}}$, for the same reasons that we mentioned for the case of an even potential with two cuts. In such a case, we have an expansion in powers of $\frac{1}{N^2}$ for the partition function, whose coefficients depend on the parity of $N$. In general, we also observe that $\bm{v}_{\beta = 2;\star} = \bm{0}$, \textit{i.e.} Thetanullwerten appear in the expansion.

Using the fact that the $n$-th correlator is the $n$-derivative of the free energy of the partition function for a perturbed potential of order $N$, and our asymptotic results are uniform for small perturbations of this kind, it is pure algebra to derive from \eqref{th:22} an asymptotic expansion for the correlators $W_n$ for the initial model in the multi-cut regime. For $\beta = 2$, the resulting expression can be found for instance in \cite[Section 6.2]{BEInt} up to $O(\frac{1}{N})$  and a systematic diagrammatic for all orders is given in \cite[Appendix A]{BEknots}. This can be straightforwardly extended to the $\beta \neq 2$ case, simply by including half-integer genera $g$ (in our conventions, $k$ not having fixed parity)

\subsection{Comments relative to the geometry of the spectral curve}
\label{S17}
We stress now facts from the theory of the topological recursion \cite{CE06,EO07} which are relevant in the present case --- for further details on the geometry compact Riemann surfaces, see for instance \cite{Eyn18}. When $V$ is a polynomial and $\bm{\epsilon}$ is close enough to $\bm{\epsilon}_{\star}$, the density of the equilibrium measure can be analytically continued to a hyperelliptic curve of genus $g$, denoted $\mathcal{C}_{\bm{\epsilon}}$ (the spectral curve). Its equation is:
\beq
\label{hype}y^2 = \prod_{h = 0}^{g} (x - \alpha^{-}_{\bm{\epsilon},h})(x - \alpha^{+}_{\bm{\epsilon},h}).
\eeq
and $\mathcal{C}_{\bm{\epsilon}}$ is the compactification of the locus of such $(x,y)$ obtained by adding the two points at $\infty$ where $y \sim x^{g +1}$ (first sheet) and $y \sim -x^{g+1}$ (second sheet). Let $\mathcal{A}_h$ be the cycle in $\mathcal{C}_{\bm{\epsilon}}$ surrounding $\mathsf{A}_{\bm{\epsilon},h} = [\alpha^{-}_{\bm{\epsilon},h},\alpha^{+}_{\bm{\epsilon},h}]$. The family $\bm{\mathcal{A}} = (\mathcal{A}_h)_{1 \leq h \leq g}$ can be completed by a family of cycles $\bm{\mathcal{B}}$ so that $(\bm{\mathcal{A}},\bm{\mathcal{B}})$ is a symplectic basis of homology of $\mathcal{C}_{\bm{\epsilon}}$. More precisely, the cycle $\mathcal{B}_h$ travels from $\alpha_{\bm{\epsilon},h}^{-}$ to $\alpha_{\bm{\epsilon},h - 1}^+$ in the first sheet, and $\alpha_{\bm{\epsilon},h - 1}^+$ to $\alpha_{\bm{\epsilon},h}^-$ in the second sheet. The correlators $W_{n;\bm{\epsilon}}^{[G,K]}$ are meromorphic functions on $\mathcal{C}^n_{\bm{\epsilon}}$, computed recursively by a residue formula on $\mathcal{C}_{\bm{\epsilon}}$.

In particular, the analytic continuation of
\begin{equation}
\label{Bebl} \bigg(\frac{\beta}{2} W_{2;\bm{\epsilon}}^{\{0\}}(x_1,x_2) + \frac{1}{(x_1 - x_2)^2}\bigg)\dd x_1\dd x_2 = \bigg(\mathcal{W}_{2;\bm{\epsilon}}^{[0,0]}(x_1,x_2) + \frac{1}{(x_1 - x_2)^2}\bigg)\dd x_1\dd x_2
\end{equation}
is the unique meromorphic bidifferential, denoted $\Omega$, on $\mathcal{C}_{\bm{\epsilon}}$, which has vanishing $\bm{\mathcal{A}}$-periods, and has for only singularity a double pole at coinciding point with leading coefficient $1$ and without residue. This $\Omega$ plays an important role for the geometry of the spectral curve and is called \emph{fundamental bidifferential of the second kind}. It sometimes appear under the name of ``Bergman kernel''  although it does not coincide with (but it is related to) the kernel introduced by Bergman in \cite{BS53}. It can be explicitly computed by the formula:
\begin{equation} 
\label{W2Beb}
\Omega(z_1,z_2) = \dd_{z_1} \dd_{z_2} \ln \theta\Big(\int_{z_1}^{z_2} \bm{\varpi}\dd x + \mathbf{c}\,\Big|\,\bm{\tau}^{\mathcal{C}_{\bm{\epsilon}}}\Big),
\end{equation} 
where: 
\begin{itemize}
\item $\theta = \vartheta\big[\begin{smallmatrix} \mathbf{0} \\ \mathbf{0} \end{smallmatrix}\big]$ is the Riemann Theta function.
\item $\bm{\varpi}(z)\dd x(z)$ is the basis of holomorphic one-forms dual to the $\bm{\mathcal{A}}$-cycles, \textit{i.e.} characterised by:
\begin{equation}
\label{varpian}
\forall h,h' \in \ldbrack 1,g \rdbrack,\qquad \oint_{\mathcal{A}_{h}} \varpi_{h'}\dd x = \delta_{h,h'}.
\end{equation}
\item $\bm{\tau}^{\mathcal{C}_{\bm{\epsilon}}}$ is the Riemann matrix of periods of the spectral curve $\mathcal{C}_{\bm{\epsilon}}$:
$$
\forall h,h' \in \ldbrack 1,g \rdbrack,\qquad \oint_{\mathcal{B}_{h}} \varpi_{h'}\dd x = \tau^{\mathcal{C}_{\bm{\epsilon}}}_{h,h'}.
$$
\item $\mathbf{c} = \frac{1}{2}(\mathbf{r} + \bm{\tau}(\mathbf{s}))$ with $\mathbf{r},\mathbf{s} \in \mathbb{Z}^{g}$ such that $\mathbf{r} \cdot \mathbf{s}$ is odd, is a non singular characteristics for the Theta function, \textit{i.e.} such that $\theta\big(\int_{z_1}^{z_2} \bm{\varpi}\dd x + \mathbf{c}\,\big|\,\bm{\tau}\big)$ is not identically $0$ when $z_1,z_2 \in \mathcal{C}_{\bm{\epsilon}}$. Such a $\mathbf{c}$ exist and the result then does not depend on which such $\mathbf{c}$ is chosen.
\end{itemize}

It is a property of the topological recursion that the derivatives of $F_{\beta;\bm{\epsilon}}^{\{k\};V}$ can be computed as $\bm{\mathcal{B}}$-cycle integrals of the correlators: 
\begin{equation}
\label{Fdedin}
(F_{\beta;\bm{\epsilon}}^{\{k\};V})^{(j)} = \Big(\frac{\beta}{2}\Big)^{j} \oint_{\bm{\mathcal{B}}} \dd \xi_1 \cdots \oint_{\bm{\mathcal{B}}} \dd \xi_j\,W_{j;\bm{\epsilon}}^{\{k + j\}}(\xi_1,\ldots,\xi_j).
\end{equation}
This relation extends as well to derivatives of correlators:
$$
\big(W_{n;\bm{\epsilon}}^{\{k\}}(x_1,\ldots,x_n)\big)^{(j)} = \Big(\frac{\beta}{2}\Big)^{j} \oint_{\bm{\mathcal{B}}} \dd \xi_1 \cdots \oint_{\bm{\mathcal{B}}} \dd \xi_{j}\,W_{n + j;\bm{\epsilon}}^{\{k + j\}}(x_1,\ldots,x_k,\xi_1,\ldots,\xi_{j}).
$$
where it is understood that we differentiate keeping $x$ fixed. In particular:
\begin{equation}
\label{varpinu} \big(W_{1;\bm{\epsilon}}^{\{-1\}}(x)\big)' \dd x = 2{\rm i}\pi\,\bm{\varpi}(x) \dd x = \oint_{\bm{\mathcal{B}}} \Omega(x,\bullet) = \frac{\beta}{2} \oint_{\bm{\mathcal{B}}} \dd \xi\,W_{2;\bm{\epsilon}}^{\{0\}}(x,\xi),
\end{equation} 
Besides, the matrix to use in the Theta function appearing in Theorem~\ref{th:22} is
$$
\tau_{\beta;\star} = \frac{\beta}{2}\, \bm{\tau}^{\mathcal{C}_{\bm{\epsilon}}}.
$$
This simple dependence in $\beta$ of $W_{2;\bm{\epsilon}}^{\{0\}}$ can be traced back to the fact that, as a consequence of the Dyson--Schwinger equations, we have
$$
\forall h \in \llbracket 0,g \rrbracket\quad \forall x \in \mathsf{A}_h,\qquad \frac{\beta}{2}\big(W_{2;\bm{\epsilon}}^{\{0\}}(x_1 + {\rm i}0,x_2) + W_{2;\bm{\epsilon}}^{\{0\}}(x_1 - {\rm i}0,x_2)\big) = - \frac{1}{(x_1 - x_2)^2},
$$
and this equation (together with the properties of the analytic continuation of $W_{2;\bm{\epsilon}}^{\{0\}}$ on $\mathcal{C}_{\bm{\epsilon}}$ and the constraint of vanishing $\bm{\mathcal{A}}$-periods) fully characterises $W_{2;\bm{\epsilon}}^{\{0\}}$.

This relation has a long history, and follows from the identification of $F^{\{-2\};V}_{\beta;\bm{\epsilon}} = \frac{\beta}{2} \mathcal{F}^{[0,0];V}_{\bm{\epsilon}}$ (\textit{cf.} Equation~\eqref{Fcomu}) with the prepotential of the Hurwitz space associated to the family of curves \eqref{hype} --- considered as a Frobenius manifold --- computed by Dubrovin \cite{DubF0}, as well as with the tau function of the Whitham hierarchy as shown by Krichever \cite{KricheverF0}. A derivation in the context of matrix model is for instance given in \cite{CheM}. Although \textit{a priori} differentiability of $F^{\{-2\};V}_{\beta;\bm{\epsilon}}$ is not justified in \cite{CheM}, it is guaranteed by our results of Section~\ref{Sumo}.

Equation~\ref{Fdedin} at $\bm{\epsilon} = \bm{\epsilon}_{\star}$ can be used to compute $T_{\beta;\star}^{\{k\}}[\bm{X}]$ appearing in Equation~\eqref{TMde}. The derivation with respect to $\bm{\epsilon}$ is not a natural operation in the initial model when $N$ is finite, since $N\epsilon_h$ are forced to be integers in Equation~\eqref{eqmes2}. Yet, we show that the coefficients of expansion themselves are smooth functions of $\bm{\epsilon}$, and thus $\partial_{\bm{\epsilon}}$ makes sense.

\subsection{Central limit theorems for fluctuations and their breakdown}

In Section~\ref{Lare}, we describe the fluctuation of the number of particles $N_h$ in each segment $\mathsf{A}_h$: when $N \rightarrow \infty$, its law is approximated by the law of a Gaussian conditioned to live in a shifted integer lattice. The shift of the lattice oscillates with $N$, by an amount $\lfloor N\epsilon_{\star,h} \rfloor$. Note that, since $N \epsilon_{\star,h}$ is for general $N$ not an integer, strictly speaking one cannot say that it converges in law to a discrete Gaussian random variable. This is however true along subsequences of $N$ in case $\epsilon_{\star,h} = \mu_{{\rm eq}}^{V}(\mathsf{A}_h)$ is a rational number.

\begin{theorem}
\label{th:33t}
Assume Hypotheses~\ref{hmainc1} and \ref{hmainc4}, and let $\bm{N} = (N_1,\ldots,N_g)$ be the vector of filling fractions as above. If $\bm{P}$ is a $g$-tuple of integers depending on $N$ and such that $\bm{P} - N\bm{\epsilon}_{\star} = o(N^{\frac{1}{3}})$ when $N \rightarrow \infty$, we have:
\begin{equation}
\label{eq130eq}
\mu_{N,\beta}^{V;\mathsf{A}}\big(\bm{N} = \bm{P}\big) \sim \frac{e^{\frac{1}{2}\,(F^{\{-2\}}_{\beta;\star})^{''}\cdot(\bm{P} - N\bm{\epsilon}_{\star})^{\otimes 2} + (F^{\{-1\}}_{\beta;\star})'\cdot(\bm{P} - N\bm{\epsilon}_{\star})}}{\vartheta\big[\begin{smallmatrix} -N\bm{\epsilon}_{\star}\\ \bm{0} \end{smallmatrix}\big](\bm{v}_{\beta;\star}|\bm{\tau}_{\beta;\star})}.
\end{equation}
\end{theorem}

In Section~\ref{S833}, we describe the fluctuations of linear statistics in the multi-cut regime.

\begin{theorem}
Assume Hypotheses~\ref{hmainc1} and \ref{hmainc4}. Let $\varphi$ be an analytic test function in a neighbourhood of $\mathsf{A}$, and $s \in \mathbb{R}$. We have when $N \rightarrow \infty$:
\begin{equation}
\label{unedzn}\begin{split}
& \quad 
 \mu_{N,\beta}^{V;\mathsf{A}}\big(e^{{\rm i}s\big(\sum_{i = 1}^N \varphi(\lambda_i) - N\int_{\mathsf{S}} \varphi(\xi)\dd\mu_{{\rm eq}}^{V}(\xi)\big)}\big) \\ 
 & \mathop{\sim}_{N \rightarrow \infty} \exp\Big({\rm i}s\,M_{\beta;\star}[\varphi] - \frac{s^2}{2}\,Q_{\beta;\star}[\varphi,\varphi]\Big)\,\frac{\vartheta\!\left[\begin{smallmatrix} -N\bm{\epsilon}_{\star} \\ \bm{0} \end{smallmatrix}\right]\!\!\big(\bm{v}_{\beta;\star} + {\rm i}s\,\bm{u}_{\beta;\star}[\varphi]\big|\bm{\tau}_{\beta;\star}\big)}{\vartheta\!\left[\begin{smallmatrix} -N\bm{\epsilon}_{\star} \\ \bm{0} \end{smallmatrix}\right]\!\!\big(\bm{v}_{\beta;\star}\big|\bm{\tau}_{\beta;\star}\big)},
\end{split}
\end{equation}
where:
\begin{equation*}
\begin{split}
\bm{u}_{\beta;\star}[\varphi] & =  \Big(\oint_{\mathsf{S}_h} \frac{\dd \xi}{2{\rm i}\pi}\,\varphi(\xi)\,\varpi_h(\xi)\Big)_{1 \leq h \leq g}, \\
 M_{\beta;\star}[\varphi] & = \oint_{\mathsf{A}} \frac{\dd\xi}{2{\rm i}\pi}\,\varphi(\xi)\,W_{1;\star}^{\{0\}}(\xi), \\
Q_{\beta;\star}[\varphi,\varphi] & = \oiint_{\mathsf{A}} \frac{\dd\xi_1\,\dd\xi_2}{(2{\rm i}\pi)^2}\,\varphi(\xi_1)\varphi(\xi_2)\,W_{2;\star}^{\{0\}}(\xi_1,\xi_2).
\end{split}
\end{equation*}
We recall that the $\varpi_h(x)\dd x$ are the holomorphic one-forms from Equations~\eqref{varpian}-\eqref{varpinu}, while $W_{1;\bm{\epsilon}}^{\{0\}}$ and $W_{2;\bm{\epsilon}}^{\{0\}}$ appear in  the asymptotic expansion of the correlators in the model with fixed filling fractions (Theorem~\ref{th:3}), and here they must be specialised at $\bm{\epsilon} = \bm{\epsilon}_{\star}$.
\end{theorem}
\begin{remark}\label{remgaussian}
In particular, $\bm{u}_{\beta;\star}$ is a linear map associating to a  test function $\varphi$ a $g$-dimensional vector. When $\varphi$ is such that $\bm{u}_{\beta;\star}[\varphi] = 0$, the Theta functions cancel out and we deduce that the random variable
$$
\Phi_N[\varphi] := \sum_{i = 1}^N \varphi(\lambda_i) - N\int_{\mathsf{S}} \varphi(\xi)\dd\mu_{{\rm eq}}^{V}(\xi)
$$
converges in law to a Gaussian random variable with mean $M_{\beta;\star}[\varphi]$ and covariance $Q_{\beta;\star}[\varphi,\varphi]$. We remark that we have the alternative formula from \eqref{linearterm}:
$$
 \bm{u}_{\beta;\star}[\varphi] = \Big(\frac{1}{2{\rm i}\pi} \partial_{{\epsilon}_h}\int_{\mathsf{S}} \varphi(\xi)\,\dd\mu_{{\rm eq};\bm{\epsilon}}^V(\xi)\Big)_{1 \leq h \leq g}\Big|_{\bm{\epsilon} = \bm{\epsilon}_{\star}} $$
showing that $\bm{u}_{\beta;\star}[\varphi] $ vanishes when $ \bm{\epsilon}_{\star}$ is a critical point of $\int_{\mathsf{S}} \varphi(\xi)\,\dd\mu_{{\rm eq};\bm{\epsilon}}^V(\xi)$. Even though our results are obtained for analytic potentials and test functions, this condition clearly makes sense with less regularity. In fact, it is possible to generalize our  results and techniques  to consider sufficiently smooth potential and test functions instead of analytic ones. We refer the interested reader to \cite[Sections 4 and 6]{GColumbia} to such a generalization in the one-cut case.

\end{remark}
When $\bm{u}_{\beta;\star}[\varphi] \neq 0$, the central limit theorem does not hold anymore. Instead, from the shape of the right-hand side, $\Phi_N[\varphi]$ is approximated when $N \rightarrow \infty$ by the sum of two independent random variables: the first one is a Gaussian random variable with mean $M_{\beta;\star}[\varphi]$ and covariance $Q_{\beta;\star}[\varphi,\varphi]$, and the second one is the scalar product with $2{\rm i}\pi \bm{u}_{\beta;\star}[\varphi]$ (which is a vector in $\mathbb{R}^g$ when $\varphi$ is real-valued) of a random Gaussian vector conditioned to live on the lattice $-\lfloor N\bm{\epsilon}_{\star} \rfloor + \mathbb{Z}^{g}$. This also display $N$-dependent oscillations. These oscillations can be interpreted in physical terms from tunnelling of particles between different segments. One sees indeed than moving a single $\lambda_i$ from $\mathsf{A}_h$ to $\mathsf{A}_{h'}$ changes $\Phi_N[\varphi]$ by a quantity of order $1$, which is already the typical order of fluctuation of linear statistics when filling fractions are fixed.

The next term in the asymptotic expansion of the left-hand side of \eqref{unedzn} is of relative order $O(\frac{1}{N})$, which therefore gives the speed of convergence of the associated linear statistics of the empirical measure.

\subsection{Asymptotic expansion of kernels and correlators}

Once the result on large $N$ expansion of the partition function is obtained, we can easily infer the asymptotic expansion of the correlators and the kernels by perturbing the potential by terms of order $\frac{1}{N}$, maybe complex-valued, as allowed by Hypothesis~\ref{hmainc4}.

\subsubsection{Leading behaviour of the correlators}

Although we could write down the expansion for the correlators as a corollary of Theorem~\ref{th:22}, we bound ourselves to point out their leading behaviour. Whereas $W_n$ behaves as $O(N^{2 - n})$ in the one-cut regime or in the model with fixed filling fractions, $W_n$ for $n \geq 3$ does not decay when $N$ is large in a $(g + 1)$-cut regime with $g \geq 1$.  More precisely:
\begin{theorem}
Assume Hypothesis~\ref{hmainc1}, \ref{hmainc4} and that the number of cuts $(g + 1)$ is greater or equal to  $2$. When $N \rightarrow \infty$, we have, uniformly when $x_1,\ldots,x_n$ belongs to any compact of  $(\mathbb{C}\setminus\mathsf{A})^n$:
$$
W_2(x_1,x_2) = W_{2;\star}^{\{0\}}(x_1,x_2) + \Big(\bm{\varpi}(x_1)\otimes\bm{\varpi}(x_2)\Big)\cdot\nabla_{\bm{v}}^{\otimes 2}\ln\vartheta\!\left[\begin{array}{@{\hspace{-0.03cm}}c@{\hspace{-0.03cm}}} -N\bm{\epsilon}_{\star}\, \\ \bm{0} \end{array}\right]\!\!\big(\bm{v}_{\beta;\star}\big|\bm{\tau}_{\beta;\star}\big) + o(1)\,,
$$
and for any $n \geq 3$:
$$
W_n(x_1,\ldots,x_n) = \Big(\bigotimes_{i = 1}^n \bm{\varpi}(x_i)\Big)\cdot\nabla_{\bm{v}}^{\otimes n}\ln\vartheta\!\left[\begin{array}{@{\hspace{-0.03cm}}c@{\hspace{-0.03cm}}} -N\bm{\epsilon_{\star}}\, \\ \bm{0} \end{array}\right]\!\!\big(\bm{v}_{\beta;\star}\big|\bm{\tau}_{\beta;\star}\big) + o(1)\,.
$$
\end{theorem}
Integrating this result over $\bm{\mathcal{A}}$-cycles provide the leading order behaviour of $n$-th order moments of the filling fractions $\bm{N}$ and the result agrees with Theorem~\ref{th:33t}.

\subsubsection{Kernels}

We explain in \S~\ref{K63} that the following result concerning the kernel --- defined in Equation~\eqref{kerde} --- is a consequence of Theorem~\ref{th:3}:
\begin{corollary}
\label{th:4}Assume Hypothesis~\ref{hmainc1} and \ref{hmainc4}. There exists $t > 0$ such that, for any sequence of vectors of positive integers $\mathbf{N} = (N_1,\ldots,N_{g})$ such that \mbox{$|\bm{N}/N - \bm{\epsilon}_{\star}|_1 < t$}, the $n$-point kernels in the model with fixed filling fractions have an asymptotic expansion when $N \rightarrow \infty$ of the form, for any $K \geq 0$:
\beq
\label{ineq} \mathsf{K}_{n,\bm{c};\bm{\epsilon}}(x_1,\ldots,x_n)  = \exp\bigg[\sum_{j = 1}^{n} Nc_j\big(\ln(x_j) + 2{\rm i}\pi \chi_j\big) + \sum_{k = -1}^{K} N^{-k}\Big(\sum_{r = 1}^{k + 2} \frac{1}{r!}\mathcal{L}_{\bm{x},\bm{c}}^{\otimes r}[W_{r;\bm{\epsilon}}^{\{k\}}]\Big) + O(N^{-(K + 1)})\bigg],
\eeq
where $\mathcal{L}_{\bm{x},\bm{c}}$ is the linear form 
:
\beq
\label{linf}
\mathcal{L}_{\bm{x},\bm{c}}[f] = \sum_{j = 1}^{n} c_j\int_{\infty}^{x_j} \check{f}(x)\dd x,\qquad {\rm where}\,\,\check{f}(x) = f(x) + \frac{1}{x} \Res_{x = \infty} f(\xi)\dd \xi,
\eeq 
 The error terms in this expansion are uniform for $x_1,\ldots,x_n$ in any compact of $\mathbb{C}\setminus \mathsf{A}$.
\end{corollary}
The $(r,k) = 1$ term in \eqref{ineq} depends on choices for the path of integration from $\infty$ to $x_j$ (the other terms do not, and are also unaffected by the difference between $f$ and $\check{f}$ in \eqref{linf}), and $\chi_j \in \mathbb{Z}$. These two features are a manifestation of the fact that the definition of the kernel depends on a choice of determination for the complex logarithm; resolving them by the choice of suitable determinations and domain of definition leads to specific integer values for $\chi_j$. These subtleties are explained in details in \S~\ref{K63} and can be ignored if all $c_j \in \mathbb{Z}$ (in that case the definition of the kernel does not depend on choices).

Hereafter, if $\gamma$ is a smooth path in $\mathbb{C}\setminus\mathsf{S}_{\bm{\epsilon}}$, we set $\mathcal{L}_{\gamma} = \int_{\gamma}$, and $\mathcal{L}_{\gamma}^{\otimes r}$ is given by:
$$
\mathcal{L}_{\gamma}^{\otimes r}[W_{r;\bm{\epsilon}}^{\{k\}}] = \int_{\gamma}\dd x_1\cdots\int_{\gamma}\dd x_r\,W_{r;\bm{\epsilon}}^{\{k\}}(x_1,\ldots,x_r).
$$
\textit{A priori}, the integrals in the right-hand side of Equation~\eqref{ineq} depend on the relative homology class in $\mathbb{C}\setminus \mathsf{A}$ of paths between $\infty$ to  $x_i$. A basis of homology cycles in $\mathbb{C}\setminus \mathsf{A}$ is given by $\overline{\bm{\mathcal{A}}} = (\mathcal{A}_h)_{0 \leq h \leq g}$, and we deduce from Equation~\eqref{nrm} that:
\beq
\label{sqq}\forall h \in \ldbrack 0,g \rdbrack,\qquad \oint_{\mathcal{A}_{h}} \frac{\dd\xi}{2{\rm i}\pi}\,W_{n;\bm{N}/N}^{\{k\}}(\xi,x_2,\ldots,x_n) = \delta_{n,1}\delta_{k,-1}\,\frac{N_h}{N}.
\eeq
Therefore, the only multivaluedness of the right-hand side comes from the first term $N \mathcal{L}_{\bm{x},\bf{c}}[W_{1;\bm{\epsilon}}^{\{-1\}}]$, and given Equation~\eqref{sqq} and observing that $N_h = N\epsilon_h$ are integers, we see that it exactly reproduces the monodromies of the kernels depending on $c_j$.

We now come to the multi-cut regime of the initial model. If $\bm{X}$ is a vector with $g$ components, and $\mathcal{L}$ is a linear form on the space of holomorphic functions on $\mathbb{C}\setminus\mathsf{S}_{\bm{\epsilon}}$, let us define:
$$
\tilde{T}^{\{k\}}_{\beta;\bm{\epsilon}}[\mathcal{L};\bm{X}] = \sum_{r = 1}^{k} \frac{1}{r!} \sum_{\substack{j_1,\ldots,j_r \geq 1 \\ k_1,\ldots,k_r \geq -2 \\ n_1,\ldots,n_r \geq 0 \\  k_i + j_i + n_i > 0 \\ \sum_{i = 1}^{r} k_i + j_i + n_i = k}} \Big(\bigotimes_{i = 1}^{r} \frac{\mathcal{L}^{\otimes n_i}[(W_{n_i;\bm{\epsilon}}^{\{k_i\}})^{(j_i)}]}{n_i!\,j_i!}\Big)\cdot\bm{X}^{\otimes(\sum_{i = 1}^r j_i)},
$$
where we took as convention $W_{n = 0;\bm{\epsilon}}^{\{k\}} = F_{\beta;\bm{\epsilon}}^{\{k\}}$ and the derivatives are computed for fixed $x$s.
Then, as a consequence of Theorem~\ref{th:22}:
\begin{corollary}
\label{th:33}Assume Hypothesis \ref{hmainc1} and \ref{hmainc4}. With the notations of Corollary~\ref{th:4}, the $n$-point kernels have an asymptotic expansion, for any $K \geq 0$:
$$
\mathsf{K}_{n,\bf{c}}(\bm{x}) = \mathsf{K}_{n,\bf{c};\star}(\bm{x}) \frac{\Big(\sum_{k = 0}^K N^{-k}\,\tilde{T}_{\beta;\star}^{\{k\}}\big[\mathcal{L}_{\bm{x},\bm{c}},\frac{\nabla_{\bm{v}}}{2{\rm i}\pi}\big]\Big)\vartheta\!\left[\begin{smallmatrix} -N\bm{\epsilon}_{\star} \\ \bm{0} \end{smallmatrix}\right]\!\!\big(\bm{v}_{\beta;\star} + \mathcal{L}_{\bm{x},\bm{c}}[\bm{\varpi}]\big|\bm{\tau}_{\beta;\star}\big)}{\Big(\sum_{k = 0}^K N^{-k}\,T_{\beta;\star}^{\{k\}}\big[\frac{\nabla_{\bm{v}}}{2{\rm i}\pi}\bigr]\Big)\vartheta\!\left[\begin{smallmatrix} -N\bm{\epsilon}_{\star} \\ \bm{0} \end{smallmatrix}\right]\!\!\big(\bm{v}_{\beta;\star}\big|\bm{\tau}_{\beta;\star}\big)}\big(1 + O(N^{-(K + 1)})\big).
$$
The first factor comes from evaluation of the right-hand side of Equation~\eqref{ineq} at $\bm{\epsilon} = \bm{\epsilon}_{\star}$, $\mathcal{L}_{\bm{x},\bm{c}} = \sum_{j = 1}^{n} c_j\int_{\infty}^{x_j}$ and $\bm{\varpi}\dd x$ is the basis of holomorphic one-forms.
\end{corollary}
A diagrammatic representation for the terms of such expansion was proposed in \cite[Appendix A]{BEknots}.

\subsection{Strategy of the proof}
The key idea of this article is to establish an asymptotic expansion for the partition functions of our models for fixed filling fractions:
\beq 
\label{sqiqq2}\frac{N!\,Z_{N,\beta;\bm{N}/N}^{V;\mathsf{A}}}{\prod_{h = 0}^{g} N_h!}= N^{\frac{\beta}{2}N + \varkappa}\exp\Big(\sum_{k = -2}^{K} N^{-k}\,F^{\{k\};V}_{\beta;\bm{N}/N} + O(N^{-(K + 1)})\Big),
\eeq
for any $K \geq 0$. Indeed, such an expansion allows to estimate the free energy of the original model $\ln Z_{N,\beta}^{V;\mathsf{A}}$ up to errors of order $O(N^{-K-1 + \epsilon})$, see \eqref{sum1} and Theorem \ref{th:22}.  It also allows  to analyse the asymptotic distribution of the filling fractions $\bm{\epsilon} = \bm{N}/N$, see Theorem \ref{th:33t}, since this distribution is given as the following ratio of partition functions:
\begin{equation}
\label{mufill}\mu_{N,\beta}^{V;\mathsf{A}}\big(\bm{N} \big)= \frac{N!}{\prod_{h = 0}^g N_h!}\,\frac{Z_{N,\beta;\bm{N}/N}^{V;\mathsf{A}}}{Z_{N,\beta}^{V;\mathsf{A}}}\,.
\end{equation}
In particular, if \eqref{sqiqq2} is known up to $o(1)$, the leading behaviour of \eqref{mufill} when $N \rightarrow \infty$ can be computed. This analysis is detailed in Section  \ref{Lare}. 

To handle fluctuations of linear statistics, we use the well-known approach of considering  the free energy for perturbations of order $\frac{1}{N}$ of the potential. In fact, if we denote by $\Phi_N[\varphi] := \sum_{i = 1}^N \varphi(\lambda_i) - N\int_{\mathsf{S}} \varphi(\xi)\dd\mu_{{\rm eq}}^{V}(\xi)
$, as in  Remark \ref{remgaussian}, we see that for any real number $s$
$$
\mu_{N,\beta}^{V;\mathsf{A}}\big[e^{s\Phi_{N}[\varphi]}\big] = e^{-sN\int_{\mathsf{S}} \varphi(\xi)\dd\mu_{{\rm eq}}^{V}(\xi) } \frac{Z_{N,\beta}^{V-\frac{2s}{N\beta}\varphi;\mathsf{A}}}{Z_{N,\beta}^{V;\mathsf{A}}}\,.$$
Again, the expansion of the free energies up to $o(1)$ allows to derive the asymptotics of the Laplace transform of $\Phi_{N}[\varphi]$ and hence the central limit theorem, see Section \ref{S833}. Note in passing that another way to study these fluctuations is to first condition the law $\mu_{N,\beta}^{V;\mathsf{A}}$ by fixing its filling fractions to be equal to some $\bm{N}$. Indeed,  we can also recover the fluctuations of the linear statistics from those under  the conditioned law (that can be deduced from the ratio of the partition functions of Theorem~\ref{th:2} and lead to classical central limit theorems with Gaussian limits), together with the fluctuations of the filling fractions. Then, one easily sees that the term  $\bm{u}_{\beta;\star}$ comes from the fluctuations of the filling fractions and more precisely from the difference of centerings
$N(\int_{\mathsf{S}} \varphi(\xi)\dd\mu_{{\rm eq}}^{V}(\xi)-\int_{\mathsf{S}} \varphi(\xi)\dd\mu_{{\rm eq};\bm{N}/N}^{V}(\xi))$ for varying $\bm{N}/N$.

Therefore, the central result  of this article  is Theorem \ref{th:22}. To prove this theorem, we shall as in \cite{BG11} interpolate between the partition functions we are interested in and explicitly computable reference partition functions. For the latter we take a product of partition functions of one-cut models with Gaussian, Laguerre or Jacobi weight (depending on the nature of edges, soft or hard, of the equilibrium measure one wishes to match) that are evaluated as Selberg integrals. Such reference partition functions were already used in \cite{BG11}. One important new element of the present analysis  is the interpolation from a model with several cuts to independent one-cut models. This is realised by considering the $s$-dependent model
\begin{equation*}
\begin{split}
& \quad Z_{N,\beta;\bm{\epsilon}}^{V;\mathsf{A}}(s) \\
& = \! \int_{\prod_{h = 0}^{g} \mathsf{A}_h^{N_h}} \Bigg[\prod_{h = 0}^{g} \prod_{i = 1}^{N_h} \dd\lambda_{h,i}\,e^{-N\frac{\beta}{2}V_h(\lambda_{h,i})}\Bigg] \Bigg[\prod_{0 \leq h < h' \leq g} \prod_{\substack{1 \leq i \leq N_h \\ 1 \leq i' \leq N_{h'}}}\!\! |\lambda_{h,i} - \lambda_{h',i'}|^{s\beta}\Bigg] \Bigg[\prod_{h = 0}^{g} \prod_{1 \leq i < j \leq N_{h}} \!\! |\lambda_{h,i} - \lambda_{h,j}|^{\beta}\Bigg]
\end{split}
\end{equation*}
for $s \in [0,1]$. We choose to take the $s$-dependent potential $V_h(x) = T^s_h(x)$ on the $h$-segment
$$
T^s_h(x) = V(x) - 2(1 - s)\sum_{h' \neq h} \int_{\mathsf{A}_{h'}} \dd\mu_{{\rm eq};\bm{\epsilon}}^{V}(\xi)\,\ln|x - \xi|,\qquad {\rm for}\,\,x \in \mathsf{A}_{h}\,,
$$
where $V$ is the potential of the original model. This choice is such that the equilibrium measure associated with the model $Z_{N,\beta;\bm{\epsilon}}^{T_{s};\mathsf{A}}(s) $ is the equilibrium measure of the original model, see Section \ref{decoupleproof}.
Moreover $Z_{N,\beta;\bm{\epsilon}}^{V;\mathsf{A}}=Z_{N,\beta;\bm{\epsilon}}^{T_{1};\mathsf{A}}(1) $ whereas $Z_{N,\beta;\bm{\epsilon}}^{T_{0};\mathsf{A}}(0)$ is a product of models whose equilibrium measure has only one-cut (they are the restriction of the equilibrium measure of the original model to each of the connected pieces of its support) which we can compute by  \cite{BG11} (see Section \ref{firstsection}). Interpolating along this family yields 
\begin{equation} 
\label{lnaun}
\begin{split}
& \quad \ln\bigg(\frac{Z_{N,\beta;\bm{\epsilon}}^{V;\mathsf{A}}}{Z_{N,\beta;\bm{\epsilon}}^{T_0;\mathsf{A}}(s = 0)}\bigg) =\int_{0}^{1}\partial_{s}\ln Z_{N,\beta;\bm{\epsilon}}^{T_{s};\mathsf{A}}(s) \dd s  \\ 
&= \beta \int_{0}^{1} \dd s\mu_{N,\beta;\bm{\epsilon}}^{T_{s};\mathsf{A}}(s) \bigg[\sum_{0 \leq h < h' \leq g}\sum_{{\substack{1 \leq i \leq N_h \\ 1 \leq i' \leq N_{h'}}}}\ln|\lambda_{h,i} - \lambda_{h',i'}| 
-N\sum_{0 \leq h' \neq h \leq g}\sum_{i = 1}^{N_h} \int_{\mathsf S_{h'}} \ln |\lambda_{h,i}-x| \dd\mu_{{\rm eq};\bm{\epsilon}}^{V}(x)\bigg] \\
& =   -N\beta \sum_{0 \leq h \neq h' \leq g} \oint_{\mathsf{A}_{h}}\oint_{\mathsf{A}_{h'}} \frac{\dd x\,\dd x'}{(2{\rm i}\pi)^2}\,\ln[(x - x'){\rm sgn}(h - h')]\,W_{1;\bm{\epsilon}}^{\{-1\}}(x)\bigg(\int_{0}^{1} \dd s\,W_{1;\bm{\epsilon}}^{s}(x')\bigg) \\
&\quad + \sum_{0 \leq h' \neq h \leq g} \frac{\beta}{2} \oint_{\mathsf{A}_{h}}\oint_{\mathsf{A}_{h'}} \frac{\dd x\,\dd x'}{(2{\rm i}\pi)^2} \ln[(x - x'){\rm sgn}(h - h')]\bigg(\int_{0}^{1} \dd s \big[W_{2;\bm{\epsilon}}^{s}(x,x') + W_{1;\bm{\epsilon}}^{s}(x)W_{1;\bm{\epsilon}}^{s}(x')\big]\bigg),
\end{split}
\end{equation} 
It is important to note that in the first equality, the singularity of the logarithm is away from the range of integration as it involves variables in distinct segments, so we could express \eqref{lnaun} in terms of analytic linear and quadratic statistics, which in turn can be expressed in terms of the correlators $W_{n;\bm{\epsilon}}^{s}$  of the model associated with $Z_{N,\beta;\bm{\epsilon}}^{T_{s};\mathsf{A}}(s)$. Lemma \ref{htem} gives the large $N$ expansion of these correlators.
  
These expansions are  based on the so-called Dyson--Schwinger equations \eqref{dystheo}, see also \eqref{SDmoda} for the correlators of the interpolating models. These equations are exact equations satisfied by the correlators for any fixed $N$ and obtained simply by  integration by parts. They are a priori not closed but the idea is to show that they are asymptotically closed so that if we can show that the correlators have a large $N$ expansion of topological type, their coefficients will satisfy a closed system of equations. The latter is based on the fact that coefficients beyond the leading order satisfy a inhomogeneous linear equation, with inhomogeneous term involving coefficients of lower order only. Hence, solving the linear equation allows to define uniquely and recursively all the coefficients in the expansion of the correlators. The linear equation is described  by a linear operator, called the master-operator, that we denote $\mathcal K$, see \eqref{Kdeff} and which is the same for all orders. An inversion of this operator (continuously on some function space) precisely allows to solve the linear equation.

The central point of our approach is therefore to invert the operator $\mathcal K$. In fact, the operator is not invertible, but rather has a kernel of dimension at least $g$, where $(g + 1)$ is the number of cuts (i.e. connected components of the support of the equilibrium measure). However, its extension $\hat{\mathcal K}$ where we also record the periods around the cuts, is invertible in an off-critical situation --- see Section \ref{opK222}. Fixing the filling fractions exactly amount to use the extended operator $\widehat{\mathcal{K}}$ instead of $\mathcal{K}$, and this is why we first consider the model with fixed filling fractions. The invertibility of the extended operator indeed allows not only to solve formally the Dyson--Schwinger equations but also to  show the existence of this asymptotic expansion to all orders in $\frac{1}{N}$. To this end, it is necessary to use a priori rough estimates on the correlators, which we obtain by  classical methods of concentration of measure and large deviations, see Section \ref{concea}. These estimates can be improved iteratively with the Dyson--Schwinger equations, see e.g Section \ref{S523}, to obtain optimal estimates and eventually reach the all order asymptotic expansion. This bootstrap strategy was first introduced in \cite{BG11} for the one-cut model. We detail these computations in the case where $s=1$ in Section \ref{S3}. We also need to carry this out for the interpolating $s$-dependent model in order to have asymptotic expansions to insert in \eqref{lnaun}. In that case the extended operator does not have an explicit inverse, but we can nevertheless show by Fredholm arguments that it is invertible. Then we indicate in Section~\ref{PartitionSec} the modifications to take into account the previous bootstrap argument for $s\in [0,1]$.
 
We stress again  that  we cannot use the inversion and bootstrap strategy in the Dyson--Schwinger equations for the correlators of the original model in the multi-cut regime, because the relevant master operator is not invertible. This is the reason why we need the detour through the partition function with fixed filling fractions  (via \eqref{sqiqq2}), from which  any desired expansion of the correlators of the original model can be obtained by looking at $\frac{1}{N}$-perturbations of the potential.

\section{Application to (skew) orthogonal polynomials and integrable systems}
\label{S2}

The one-hermitian matrix model (\textit{i.e.} $\beta = 2$) is related to the Toda chain and orthogonal polynomials (see \textit{e.g.} \cite{Deift}). Similarly, the one-symmetric (resp. quaternionic self-dual) matrix model corresponds to $\beta = 1$ (resp. $\beta = 4$), and is related to the Pfaff lattice and skew-orthogonal polynomials \cite{Eskew,AMVTP,HAMVTP}. Therefore, our results establish the all-order asymptotics of certain solutions (those related to matrix integrals) of the Toda chain and the Pfaff lattice in the continuum limit, and the all-order asymptotics of (skew) orthogonal polynomials away from the bulk. We illustrate it for orthogonal polynomials with respect to an analytic weight defined on the whole real line. It could be applied equally well to orthogonal polynomials with respect to an analytic weight on a finite union of segments of the real axis. We review with less details in \S~\ref{S444a} the definition of skew-orthogonal polynomials and the way to obtain them from Corollary~\ref{th:33}.

The leading order asymptotic of orthogonal polynomials is well-known since the work of Deift et al. \cite{DKMcLVZ1,DKMcLVZ2,DKMcLVZ3}, using the asymptotic analysis of Riemann--Hilbert problems which was pioneered in \cite{DZ}. In principle, it is possible to push the Riemann--Hilbert analysis beyond leading order, but this approach being very cumbersome, it has not been performed yet to our knowledge. Notwithstanding, the all-order expansion has a nice structure, and was heuristically derived by Eynard \cite{EHouches} based on the general works \cite{BDE,Ecv}. In this article, we provide a proof of those heuristics.

Unlike the Riemann--Hilbert technique which becomes cumbersome to study the asymptotics of skew-orthogonal polynomials (\textit{i.e.} $\beta = 1$ and $4$) and thus has not been performed up to now, our method could be applied without difficulty to those values of $\beta$, and would allow to justify the heuristics of Eynard \cite{Eskew} formulated for the leading order, and describe all subleading orders. In other words, it provides a purely probabilistic approach to address asymptotic problems in integrable systems. It also suggests that the appearance of Theta functions is not intrinsically related to integrability. In particular, we see in Theorem~\ref{thepo} that for $\beta = 2$, the Theta function appearing in the leading order is associated to the matrix of periods of the hyperelliptic curve $\mathcal{C}_{\bm{\epsilon}_{\star}}$ defined by the equilibrium measure. Actually the Theta function is just the basic block to construct analytic functions on this curve, and this is the reason why it pops up in the Riemann--Hilbert analysis. However, for $\beta \neq 2$, the Theta function comes is associated to $\frac{\beta}{2}$ times the matrix of periods of $\mathcal{C}_{\bm{\epsilon}_{\star}}$, which might be or not the matrix of period of a curve, and anyway is not that of $\mathcal{C}_{\bm{\epsilon}_{\star}}$. So, the monodromy problem solved by this Theta function is not directly related to the equilibrium measure, which makes for instance for $\beta = 1$ or $4$ its construction via Riemann--Hilbert techniques \textit{a priori} more involved.
 
Contrarily to Riemann--Hilbert techniques however, we are not yet in position within our method to consider the asymptotic in the bulk, at the edges, or the double-scaling limit for varying weights close to a critical point, or the case of complex-values weights which has been studied in \cite{BertMo}. It would be very interesting to find a way out of these technical restrictions within our method.

\subsection{Setting}

We first review the standard relations between orthogonal polynomials on the real line, random matrices and integrable systems, see \textit{e.g.} \cite[Section 5]{GraCla}. In this section, $\beta = 2$ and we omit to precise it in the notations. Let $V_{\bf{t}}(\lambda) = V(\lambda) + \sum_{k = 1}^{d} t_k\lambda^{k}$. Let $(P_{n,N}(x))_{n \geq 0}$ be the monic orthogonal polynomials associated to the weight $\dd w(x) = \dd x\,e^{-NV_{\bf{t}}(x)}$ on $\mathsf{B} = \mathbb{R}$. We choose $V$ and restrict in consequence $t_k$ so that the weight decreases quickly at $\pm \infty$. If we denote $h_{n,N}$ the $L^2(\dd w)$ norm of $P_{n,N}$, the polynomials $\hat{P}_{n,N} = P_{n,N}/\sqrt{h_{n,N}}$ are orthonormal. They satisfy a three-term recurrence relation:
$$
x\hat{P}_{n,N}(x) = \sqrt{h_{n,N}}\hat{P}_{n + 1,N}(x) + \beta_{n,N}\hat{P}_{n,N}(x) + \sqrt{h_{n - 1,N}}\hat{P}_{n - 1,N}(x).
$$
The recurrence coefficients are solutions of a Toda chain: if we set
$$
u_{n,N} = \ln h_{n,N},\qquad v_{n,N} = -\beta_{n,N},
$$
we have:
\beq
\label{Toda}\partial_{t_1} u_{n,N} = v_{n,N} - v_{n - 1,N},\qquad \partial_{t_1} v_{n,N} = e^{u_{n + 1,N}} - e^{u_{n,N}},
\eeq
and the coefficients $t_{k}$ generate higher Toda flows. The recurrence coefficients also satisfy the string equations:
\beq
\label{string}\sqrt{h_{n,N}} [V'(\mathbf{Q}_{N})]_{n,n - 1} = \frac{n}{N},\qquad [V'(\mathbf{Q}_{N})]_{n,n} = 0,
\eeq
where $\mathbf{Q}_N$ is the semi-infinite matrix:
$$
\mathbf{Q}_N = \left(\begin{array}{ccccc} \sqrt{h_{1,N}} & \beta_{1,N} &  & & \\ \beta_{1,N} & \sqrt{h_{2,N}} & \beta_{2,N} & & \\ & \beta_{2,N} & \sqrt{h_{3,N}} & \beta_{3,N} & \\ & \ddots & \ddots & \ddots & \\ & & & & \end{array}\right) .
$$
The equations \ref{string} determine in terms of $V$ the initial condition for the system~\eqref{Toda}. The partition function $\mathcal{T}(\mathbf{t}) = Z_{N}^{V_{\mathbf{t}};\mathbb{R}}$ is the Tau function associated to the solution $(u_{n,N}(\mathbf{t}),v_{n,N}(\mathbf{t}))_{n \geq 1}$ of Equation~\eqref{Toda}. The partition function itself can be computed as \cite{Mehtabook,PSbook}:
$$
Z_{N}^{V;\mathbb R }
= N!\,\prod_{j = 0}^{N - 1} h_{j,N}.
$$
We insist on the dependence on $N$ and $V$ by writing $h_{j,N} = h_{j}(NV)$.
Therefore, the norms can be retrieved as:
\beq
\label{squ}h_{n}(NV) = \frac{\prod_{j = 1}^{n} h_{j}(NV)}{\prod_{j = 1}^{n - 1} h_{j}(NV)} = \frac{1}{n + 1}\,\frac{Z_{n + 1}^{NV/(n + 1);\mathbb R}}{Z_{n}^{NV/n; \mathbb R}} = \frac{1}{n + 1}\,\frac{Z_{n + 1}^{\frac{V}{s(1 + 1/n)};\mathbb{R}}}{Z_{n}^{V/s;\mathbb{R}}},\qquad s = \frac{n}{N}.
\eeq
The regime where $n,N \rightarrow \infty$ but $s = \frac{n}{N}$ remains fixed and positive corresponds to the small dispersion regime in the Toda chain, where $\frac{1}{n}$ plays the role of the dispersion parameter.

\subsection{Small dispersion asymptotics of \texorpdfstring{$h_{n,N}$}{hnN}}

When $V_{\mathbf{t}_0}/s_0$ satisfies Hypotheses~\ref{hmainc1} and \ref{hmainc3} for a given set of times $(s_0,\mathbf{t}_0)$, $V_{\mathbf{t}}/s$  satisfies the same assumptions at least for $(s,\mathbf{t})$ in some neighbourhood $\mathcal{U}$ of $(s_0,\mathbf{t}_0)$, and Theorem~\ref{th:22} determines the asymptotic expansion of  $\mathcal{T}_{N}(\mathbf{t})=Z_N^{V_{\mathbf{t}};\mathbb R}$ up to $O(N^{-\infty})$. Besides, we can apply Theorem~\ref{th:22} to study the ratio in the right-hand side of Equation~\eqref{squ} when $n \rightarrow \infty$ up to $o(n^{-\infty})$. For instance, we record below the expansion up to order $O(n^{-2})$.
\begin{theorem}
In the regime $n,N \rightarrow \infty$, $s = \frac{n}{N} > 0$ fixed, and Hypotheses~\ref{hmainc1} and \ref{hmainc3} are satisfied with soft edges, we have the asymptotic expansion:
\begin{equation}
\label{spAnun}\begin{split}
u_{n,N}& = n\big(2\mathcal{F}^{[0]}_{\star} - \mathcal{L}_{\frac{V_{\mathbf{t}}}{s}}[\mathcal{W}_{1;\star}^{[0]}]\big) + 1 + \mathcal{F}^{[0]}_{\star} - \mathcal{L}_{\frac{V_{\mathbf{t}}}{s}}\big[\mathcal{W}_{1;\star}^{[0]}\big] + \frac{1}{2}\mathcal{L}^{\otimes 2}_{\frac{V_{\mathbf{t}}}{s}}\big[\mathcal{W}_{2;\star}^{[0]}\big]  + \ln\Big(\frac{\tilde{\Theta}_{n}}{\Theta_{n}}\Big) \\ 
& \quad + \frac{1}{n}\Bigg\{\varkappa - \frac{1}{2} + \mathcal{L}_{\frac{\bm{\mathcal{B}}}{2{\rm i}\pi}}\big[\mathcal{W}_{1;\star}^{[1]}\big] \cdot \nabla\ln\Big(\frac{\tilde{\Theta}_n}{\Theta_n}\Big) - \mathcal{L}_{\frac{V_{\mathbf{t}}}{s}}\big[\mathcal{W}_{1;\star}^{[1]}\big] \\
& \quad \qquad + \frac{1}{6}\,\mathcal{L}_{\frac{\bm{\mathcal{B}}}{2{\rm i}\pi}}^{\otimes 3}\big[\mathcal{W}_{3;\star}^{[0]}\big] \cdot \Big(\frac{\nabla^{\otimes 3}\tilde{\Theta}_{n}}{\tilde{\Theta}_{n}} - \frac{\nabla^{\otimes 3} \Theta_{n}}{\Theta_n}\Big)  
- \frac{1}{2}\,\mathcal{L}_{\frac{\bm{\mathcal{B}}}{2{\rm i}\pi}}^{\otimes 2} \otimes \mathcal{L}_{\frac{V_{\mathbf{t}}}{s}}\big[W_{3;\star}^{[0]}\big] \cdot \frac{\nabla^{\otimes 2} \tilde{\Theta}_{n}}{\tilde{\Theta}_{n}} \\
& \quad
\qquad + \frac{1}{2} \mathcal{L}_{\frac{\bm{\mathcal{B}}}{2{\rm i}\pi}} \otimes \mathcal{L}_{\frac{V_{\mathbf{t}}}{s}}^{\otimes 2}\big[\mathcal{W}_{3;\star}^{[0]}\big] \cdot \nabla \ln \tilde{\Theta}_n - \frac{1}{6}  \mathcal{L}_{\frac{V_{\mathbf{t}}}{s}}^{\otimes 3}\big[\mathcal{W}_{3;\star}^{[0]}\big]\bigg\}  + O(n^{-2}).
\end{split} 
\end{equation}
We used the shortcut notations:
$$
\tilde{\Theta}_{n}  = \vartheta\!\left[\begin{array}{@{\hspace{-0.03cm}}c@{\hspace{-0.03cm}}} -(n + 1)\,\bm{\epsilon}_{\star}\, \\ \bm{0} \end{array}\right]\!\!\big(\bm{v} -\mathcal{L}_{V_{\mathbf{t}}/s}[\bm{\varpi}]\big|\bm{\tau}_{\star}\big),\qquad 
\Theta_{n} = \vartheta\!\left[\begin{array}{@{\hspace{-0.03cm}}c@{\hspace{-0.03cm}}} -n\,\bm{\epsilon}_{\star}\, \\ \bm{0} \end{array}\right]\!\!\big(\bm{v}\big|\bm{\tau}_{\star}\big).
$$
and in Equation~\eqref{spAnun} it is understood that the argument $\bm{v}$ is specialised to $\bm{0}$, after application of the $\nabla = \nabla_{\bm{v}}$. Besides:
$$  
\mathcal{L}_{\frac{V_{\mathbf{t}}}{s}}[f] = \oint_{\mathsf{S}} \frac{\dd\xi}{2{\rm i}\pi}\,\frac{V_{\mathbf{t}}(\xi)}{s}\,f(\xi),\qquad \mathcal{L}_{\frac{\bm{\mathcal{B}}}{2{\rm i}\pi}}[f] = \oint_{\bm{\mathcal{B}}} \frac{\dd \xi }{2{\rm i}\pi}\,f(\xi).
$$  
\end{theorem}
When $V_{\mathbf{t}}/s$ leads to a multi-cut regime, this asymptotic expansion features oscillations. Numerical evidence for such oscillations first appeared in \cite{Jurkiewicz}, where plots of $h_{n - 1,N}/h_{n,N}$ displaying the phase transitions from a one-cut to a multi-cut regime can be found for a sextic potential.

We recall that all the quantities $\mathcal{W}_{m;\star}^{[G]}$ can be computed from the equilibrium measure associated to the potential $V_{\mathbf{t}}$, so making those asymptotic explicit just requires to solve the scalar Riemann--Hilbert problem for $\mu_{\mathrm{eq}}^{sV_{\mathbf{t}}}$. Notice that the number $(g + 1)$ of cuts \textit{a priori} depends on $(s_0,\mathbf{t}_0)$, and we do not address the issue of transitions between regimes with different number of cuts (because we cannot relax at present our off-criticality assumption), which are expected to be universal \cite{Dubuniv}.

\subsection{Asymptotic expansion of orthogonal polynomials away from the bulk}

The orthogonal polynomials can be computed thanks to Heine formula \cite{Szego}:
$$
P_n(x) = \mu_{n}^{V_{\mathbf{t}}/s;\mathbb{R}}\bigg[\prod_{i = 1}^n (x - \lambda_i)\bigg] = \mathsf{K}_{1,1}(x).
$$
Hence, as a consequence of Corollary~\ref{th:33} we obtain their asymptotic expansion away from the bulk. We first collect some notations that appeared throughout the introduction, specialised to the case $\beta = 2$ relevant here:
$$
\mathcal{W}_{0;\star}^{[G]} = \mathcal{F}_{\star}^{[G]} = F_{\beta = 2;\bm{\epsilon}_{\star}}^{\{2G - 2\}},\qquad 
\mathcal{W}_{n;\star}^{[G]} = W_{n;\bm{\epsilon}_{\star}}^{\{2G - 2 + n\}},\qquad  \bm{\tau}_{\star} = \frac{(\mathcal{F}^{[0]}_{\beta=2;\star})''}{2{\rm i}\pi},
$$
and
\begin{equation*}
\begin{split}
T_{\star}^{\{k\}}[\bm{X}] & = \sum_{r = 1}^{k} \frac{1}{r!} \sum_{\substack{j_1,\ldots,j_r \geq 1 \\ G_1,\ldots,G_r \geq 0 \\  2G_i - 2 + j_i > 0 \\ \sum_{i = 1}^{r} (2G_i - 2 + j_i) = k}} \Big(\bigotimes_{i = 1}^{r} \frac{(\mathcal{F}_{\star}^{[G_i]})^{(j_i)}}{j_i!}\Big)\cdot\bm{X}^{\otimes(\sum_{i = 1}^r j_i)}, \\
\tilde{T}_{\star}^{\{k\}}[\mathcal{L};\bm{X}] & = \sum_{r = 1}^{k} \frac{1}{r!} \sum_{\substack{j_1,\ldots,j_r \geq 1 \\ G_1,\ldots,G_r \geq 0 \\ n_1,\ldots,n_r \geq 0 \\ 2G_i - 2 + n_i + j_i > 0 \\ \sum_{i = 1}^{r} (2G_i - 2 + n_i + j_i) = k}} \Big(\bigotimes_{i = 1}^{r} \frac{\mathcal{L}^{\otimes n_i}[(\mathcal{W}_{n_i;\star}^{[G_i]})^{(j_i)}]}{n_i!\,j_i!}\Big)\cdot\bm{X}^{\otimes(\sum_{i = 1}^r j_i)},
\end{split}
\end{equation*}
where:
$$
(\mathcal{W}_{n;\star}^{[G]})^{(j)}(x_1,\ldots,x_n) = \oint_{\bm{\mathcal{B}}} \cdots \oint_{\bm{\mathcal{B}}}  \mathcal{W}_{n + j;\star}^{[G]}(x_1,\ldots,x_n,\xi_1,\ldots,\xi_{j}) \dd \xi_1 \cdots \dd \xi_{j}.
$$

\begin{theorem}
\label{thepo}In the regime $n,N \rightarrow \infty$, $s = \frac{n}{N} > 0$ fixed, and Hypotheses~\ref{hmainc1} and \ref{hmainc3} are satisfied, for $x \in \mathbb{C}\setminus \mathsf{S}$, we have the asymptotic expansion, for any $K \geq 0$:
\begin{equation*}
\begin{split}
P_n(x) & = \exp\Big(\sum_{\substack{m \geq 1,\,\ G \geq 0 \\ 2G - 2 + m \leq K}} n^{2 - 2G - m} \frac{\mathcal{L}_{x}^{\otimes m}[\mathcal{W}_{m;\star}^{[G]}]}{m!}\Big)\big(1 + O(n^{-(K + 1)})\big) \\
& \quad \times \frac{\Big(\sum_{k = 0}^K n^{-k}\,\tilde{T}^{\{k\}}\big[\mathcal{L}_{x}\,;\,\frac{\nabla_{\bm{v}}}{2{\rm i}\pi}\big]\Big) \vartheta\!\left[\begin{array}{@{\hspace{-0.03cm}}c@{\hspace{-0.03cm}}} -n\,\bm{\epsilon}_{\star}\, \\ \bm{0} \end{array}\right]\!\!\big(\mathcal{L}_x[\bm{\varpi}]\big|\bm{\tau}_{\star}\big)}{\Big(\sum_{k = 0}^K n^{-k}\,T^{\{k\}}\big[\frac{\nabla_{\bm{v}}}{2{\rm i}\pi}\big]\Big) \vartheta\!\left[\begin{array}{@{\hspace{-0.03cm}}c@{\hspace{-0.03cm}}} -n\,\bm{\epsilon}_{\star}\, \\ \bm{0} \end{array}\right]\!\!\big(\bm{0}\big|\bm{\tau}_{\star}\big)},
\end{split}
\end{equation*}
where $\mathcal{L}_{x} = \int_{\infty}^{x}$. For a given $K$, this expansion is uniform for $x$ in any compact of $\mathbb{C}\setminus\mathsf{S}$.
\end{theorem}
We remark that $\mathcal{L}_{x}[\bm{\varpi}] = \int_{\infty}^{x} \bm{\varpi}$ is the Abel map evaluated between the points $x$ and $\infty$. The variable $s = \frac{n}{N}$ rescales the potential, and therefore the equilibrium measure and all the coefficient of expansion depend on $s$.

As such, the results presented in this article do not allow the study of the asymptotic expansion of orthogonal polynomials in the bulk, \textit{i.e.} for $x \in \mathsf{S}$. Indeed, this requires to perturb the potential $V(\lambda)$ by a term $-\frac{1}{n}\,\ln(\lambda - x)$ having a singularity at $x \in \mathsf{S}$, a case going beyond our Hypothesis~\ref{hmainc4}. Similarly, we cannot address at present the regime of transitions between a $g$-cut regime and a $g'$-cut regime with $g \neq g'$, because off-criticality was a key assumption in our derivation. Although it is the most interesting in regard of universality, the question of deriving uniform asymptotics, even at the leading order, valid for the crossover around a critical point is still open from the point of view of our methods. 

\subsection{Asymptotic expansion of skew-orthogonal polynomials}
\label{S444a}
The expectation values of $\prod_{i = 1}^N (x - \lambda_i)$ in the $\beta$-ensembles for $\beta = 1$ and $4$ are skew-orthogonal polynomials. Let us review this point, and just mention that the application of Corollary~\ref{th:33} implies all-order asymptotic for skew-orthogonal polynomials away from the bulk. Here, the relevant skew-symmetric bilinear products are:
\begin{equation}
\label{sek2}
\begin{split}
\langle f,g \rangle_{n,\beta = 1} & =  \int_{\mathbb{R}^2} \dd x\dd y\,e^{-n(V(x) + V(y))}\,\mathrm{sgn}(y - x)\,f(x)g(y), \\
\langle f,g \rangle_{n,\beta = 4} & = \int_{\mathbb{R}} \dd x\,e^{-n\,V(x)}\big(f(x)g'(x) - f'(x)g(x)\big).
\end{split}
\end{equation}
A family of polynomials $(P_{N}(x))_{N \geq 0}$ is \emph{skew-orthogonal} if:
$$
\forall j,k \geq 0,\qquad \big\langle P_{j},P_{k} \rangle_{n,\beta} = \big(\delta_{j,k - 1} - \delta_{j -1,k}\big)h_{j;n,\beta}.
$$
For a given skew-symmetric product, the family of skew-orthogonal polynomials is not unique, since one can add to $P_{2N + 1}$ any multiple of $P_{2N}$, and this does not change the skew-norms $h_N$. If we add the requirement that the degree $2N$ term in $P_{2N + 1}$ vanishes, the skew-orthogonal polynomials are then unique. The generalisation of Heine formula was proved in \cite{Eskew}:  
\begin{theorem}
Let $P_{N;n,\beta}$ be a set of monic skew-orthogonal polynomials associated to \eqref{sek2}. We can take
\begin{equation*}
\begin{split}
P_{2N;n,\beta = 1}(x) & =  \mu_{2N,\beta = 1}^{nV/N;\mathbb{R}}\Big[\prod_{i = 1}^{2N} (x - \lambda_i)\Big], \\
P_{2N + 1;n,\beta = 1}(x) & =  \mu_{2N,\beta=1}^{nV/N;\mathbb{R}}\Bigg[\Big(x + \sum_{i = 1}^{2N} \lambda_i\Big)\prod_{i = 1}^{2N}(x - \lambda_i)\Bigg], \\
P_{2N;n,\beta = 4}(x) & = \mu_{N,\beta = 4}^{nV/2N;\mathbb{R}}\Big[\prod_{i = 1}^{N} (x - \lambda_i)^2\Big], \\
P_{2N;n,\beta = 4}(x) & = \mu_{N,\beta = 4}^{nV/2N;\mathbb{R}}\Bigg[\Big(x + \sum_{i = 1}^{N} 2\lambda_i\Big)\prod_{i = 1}^{N} (x - \lambda_i)^2\Bigg].
\end{split}
\end{equation*}
\end{theorem}
Corollary~\ref{th:33} then determines the asymptotics of the right-hand side. The partition function itself can be deduced from the skew-norms
\cite{Mehtabook}
\begin{equation*}
\begin{split}
Z_{2N,\beta = 1}^{nV/2N;\mathbb{R}} & = (2N)! \prod_{j = 0}^{N - 1} h_{j;n,\beta =1} \\
Z_{2N + 1,\beta = 1}^{nV/(2N + 1);\mathbb{R}} & = (2N + 1)! \prod_{j = 0}^{N - 1} h_{j;n,\beta = 1} \cdot \int_{\mathbb{R}} e^{-nV(x)} P_{N-1;n,\beta}(x)\dd x \\
 \qquad Z_{N,\beta = 4}^{nV/2N;\mathbb{R}} & = N! \prod_{j = 0}^{N - 1} h_j,
\end{split}
\end{equation*}
and conversely
$$
h_{N;n,\beta = 1} = \frac{1}{(2N + 2)(2N + 1)}\,\frac{Z_{2N + 2,\beta = 1}^{nV/(2N + 2);\mathbb{R}}}{Z_{2N,\beta}^{nV/2N;\mathbb{R}}},\qquad h_{N;n,\beta = 4} = \frac{1}{N + 1} \frac{Z_{N + 1,\beta = 4}^{nV/(2N + 2);\mathbb{R}}}{Z_{N,\beta = 4}^{nV/2N;\mathbb{R}}}
$$
It has been shown that this partition function for $\beta = 1$ is a tau-function of the Pfaff lattice \cite{HAMVTP,AMVTP}. Here we obtain its asymptotic expansion from Theorem~\ref{th:22}.

\section{Large deviations and concentration of measure}
\label{concea}
\subsection{Restriction to a vicinity of the support}
\label{S31as}
Our first step is to show that the interval of integration in Equation~\eqref{eqmes} can be restricted to a vicinity of the support of the equilibrium measure, up to exponentially small corrections when $N$ is large. The proofs are very similar to the one-cut case \cite{BG11}, and we remind briefly their idea in \S~\ref{Spr}. Let $V$ be a regular and confining potential, and $\mu_{\mathrm{eq}}^{V;\mathsf{B}}$  the equilibrium measure determined by Theorem~\ref{th:1}. We denote by $\mathsf{S}$ its (compact) support. We define the effective potential by:
\beq
\label{eq31}U^{V;\mathsf{B}}_{{\rm eq}}(x)  = V^{\{0\}}(x) - 2\int_{\mathsf{B}} \dd \mu_{\mathrm{eq}}^{V}(\xi)\,\ln|x - \xi|,\qquad \tilde{U}_{{\rm eq}}^{V;\mathsf{B}}(x) = U_{{\rm eq}}^{V;\mathsf{B}}(x) - \inf_{\xi \in \mathsf{B}} U_{{\rm eq}}^{V;\mathsf{B}}(\xi),
\eeq
when $x \in \mathsf{B}$, and $+\infty$ otherwise.
\begin{lemma}
\label{uuu3} If $V$ is regular, confining, and converges uniformly to $V^{\{0\}}$ on $\mathsf{B}$,  then we have large deviation estimates: for any $\mathsf{F} \subseteq \overline{\mathsf{B}\backslash \mathsf{S}}$ closed in $\mathsf{B}$ and $\mathsf{O} \subseteq \mathsf{B}\backslash \mathsf{S}$ open in $\mathsf{B}$,
\begin{equation*}
\begin{split}
\limsup_{N\ra\infty}\frac{1}{N}\ln \mu^{V;\mathsf{B}}_{N,\beta}\left[\exists i\quad\lambda_i \in \mathsf{F}\right] & \leq  -\frac{\beta}{2}\,\inf_{x \in \mathsf{F}} \tilde{U}_{{\rm eq}}^{V;\mathsf{B}}(x), \\
\liminf_{N\ra\infty}\frac{1}{N}\ln \mu^{V;\mathsf{B}}_{N,\beta}\left[\exists i\quad\lambda_i \in \mathsf{O}\right] & \geq  -\frac{\beta}{2}\,\inf_{x \in \mathsf{O}} \tilde{U}_{{\rm eq}}^{V;\mathsf{B}}(x).
\end{split}
\end{equation*}
\end{lemma}

\begin{definition}
We say that $V$ satisfies a control of large deviations on $\mathsf{B}$ if $\tilde{U}_{{\rm eq}}^{V;\mathsf{B}}$ is positive on $\mathsf{B}\setminus\mathsf{S}$.
\end{definition}

Note that $\tilde{U}_{{\rm eq}}^{V;\mathsf{B}}$ vanishes at the boundary of $ \mathsf{S}$.
According to Lemma~\ref{uuu3}, such a property implies that large deviations outside $\mathsf{S}$ are exponentially small when $N$ is large.

\begin{corollary}
\label{uuu2} Let $V$ be regular, confining, satisfying a control of large deviations on $\mathsf{B}$. Let $ \mathsf{A} \subseteq \mathsf{B}$ be a finite union of segments which contains $\{x\in \mathsf{B}: d(x,\mathsf{S}) \le\epsilon\}$ for some positive $\epsilon$. There exists $\eta(\mathsf{A}) > 0$ so that:
\beq
Z_{N,\beta}^{V;\mathsf{B}} = Z_{N,\beta}^{V;\mathsf{A}}\big(1 + O(e^{-N\eta(\mathsf{A})})\big),
\eeq
and for any $n \geq 1$, there exists a universal constant $\gamma_n > 0$ so that, for any $x_1,\ldots,x_n \in (\mathbb{C}\setminus \mathsf{B})^{n}$:
\beq
\label{eq:coinW} \big|W_n^{V;\mathsf{B}}(x_1,\ldots,x_n) - W_n^{V;\mathsf{A}}(x_1,\ldots,x_n)\big| \leq \frac{\gamma_{n}\,e^{-N\eta(\mathsf{A})}}{\prod_{i = 1}^{n} d(x_i,\mathsf{B})}.
\eeq
\end{corollary}
Note that if all edges are hard we have $\mathsf{B} = \mathsf{S}$ and Lemma~\ref{uuu3} and Corollary~\ref{uuu2} are useless.

It is useful to have a local version of this result, saying that we can vary endpoints of the segments which are not hard edges for the equilibrium measure, up to exponentially small corrections.
\begin{corollary}
\label{pro} Let $V$ be regular, confining, satisfying a control of large deviations on $\mathsf{B}$. Let $\mathsf{A} \subseteq \mathsf{B}$ be a finite union of segments 
which contains $\{x\in \mathsf{B}: d(x,\mathsf{S}) \le\epsilon\}$ for some positive $\epsilon$.
 If $a_0$ is the left edge of a connected component of $\mathsf{A}$ and $a < a_0$ and is not in $\mathsf{S}$, let us define $\mathsf{A}_{a} = \mathsf{A} \cup [a,a_0]$. For any $\varepsilon > 0$ small enough, there exists $\eta_{\varepsilon} > 0$ so that, for $N$ large enough and any $a \in (a_0 - \varepsilon,a_0) \subseteq \mathsf{B}$, we have:
\beq
\label{36}\big|\partial_{a} \ln Z_{N,\beta}^{V;\mathsf{A}_{a}}\big| \leq e^{-N\eta_{\varepsilon}},
\eeq
and, for $N$ large enough and any $n \geq 1$ and $x_1,\ldots,x_n \in (\mathbb{C}\setminus \mathsf{A}_{a})$:
\beq
\label{37}\left| \partial_{a} W_{n}^{V;\mathsf{A}_a}(x_1,\ldots,x_n)\right| \leq \frac{\gamma_{n}\,e^{-N\eta_{\varepsilon}}}{\prod_{i = 1}^{n} d(x_i,\mathsf{A}_{a})}.
\eeq
A similar result holds at the right endpoint of a connected component of $\mathsf{A}$.
\end{corollary}

From now on, even though we initially want to study the model on $\mathsf{B}^N$, we are going first to study the model on $\mathsf{A}^N$, where $\mathsf{A}$ is a small (but fixed) enlargement of $\mathsf{S}$ within $\mathsf{B}$, as allowed above. In particular, when $\mathsf{S}$ is a disjoint union of finite segments $(\mathsf{S}_{h})_{h = 0}^g$, we can take $\mathsf{A}$ to be a disjoint union of finite segments $(\mathsf{A}_{h})_{h = 0}^g$ such that $\mathsf{A}_h$ is a neighborhood of $\mathsf{S}_h$ in $\mathsf{B}$. More precisely, we can take as endpoints of $\mathsf{A}$ points close enough to the soft edges of the equilibrium measure but outside of its support, while the hard edges must remain endpoints common to $\mathsf{S},\mathsf{A}$ and $\mathsf{B}$. We next state similar results for the fixed filling fractions model  of Section \ref{fixfil}. Recall that part of the data defining this model is a sequence (indexed by $N$) of $g$-uple of positive integers $\mathbf{N} = (N_1,\ldots,N_g)$ such that $N_0 = N - \sum_{h = 1}^{g} N_h \geq 0$ and such that $\bm{\epsilon} = \mathbf{N}/N$ converges to a point in 
$$
\mathcal{E}_{g} = \Big\{\bm{\epsilon} \in (0,1)^{g}\,\,\Big|\,\,\,\sum_{h = 1}^{g} \epsilon_h < 1\Big\}.
$$
In this context, the effective potential is defined, for $x \in \mathsf{A}_{h}$ by the formula
$$
U^{V;\mathsf{A}}_{{\rm eq};\bm{\epsilon}}(x)  = V^{\{0\}}(x) - 2\int_{\mathsf{A}} \dd \mu_{\mathrm{eq};\bm{\epsilon}}^{V}(\xi)\,\ln|x - \xi|,\qquad \tilde{U}^{V;\mathsf{A}}_{{\rm eq};\bm{\epsilon}}(x) = U^{V;\mathsf{A}}_{{\rm eq};\bm{\epsilon}}(x)-\inf_{\xi\in \mathsf{A}_h} U^{V;\mathsf{A}}_{{\rm eq};\bm{\epsilon}}(\xi),
$$
and for $x \notin \mathsf{A}$ we declare $U^{V;\mathsf{A}}_{{\rm eq};\bm{\epsilon}} = \tilde{U}^{V;\mathsf{A}}_{{\rm eq};\bm{\epsilon}} = +\infty$.

\begin{proposition}\label{restprop}  If $V$ is regular, confining, and uniformly to $V^{\{0\}}$ on $\mathsf{A}$. Then, for any closed set $\mathsf{F}$ and open set $\mathsf{O}$ of $\mathbb R$,
\begin{equation*}
\begin{split}
\limsup_{N\ra\infty}\frac{1}{N}\ln \mu^{V;\mathsf{A}}_{N,\beta;\bm{N}/N}\big(\big\{\exists i \in [N] \quad \lambda_i \in \mathsf{F}\big\}\big) & \leq -\frac{\beta}{2}\,\inf_{x \in \mathsf{F}} \tilde{U}^{V;\mathsf{A}}_{{\rm eq};\bm{\epsilon}}(x), \\
\liminf_{N\ra\infty}\frac{1}{N}\ln \mu^{V;\mathsf{A}}_{N,\beta;\bm{N}/N}\big(\big\{\exists i \in [N] \quad \lambda_i \in \mathsf{O}\big\}\big) & \geq -\frac{\beta}{2}\,\inf_{x \in \mathsf{O}} \tilde{U}^{V;\mathsf{A}}_{{\rm eq};\bm{\epsilon}}(x).
\end{split}
\end{equation*}
Moreover, Corollaries \ref{uuu2} and \ref{pro} also extend to  this setting.
\end{proposition}

We may omit the superscript $\mathsf{A}$ in the equilibrium measure, the effective potential, etc. when it is clear that we work with the compact set $\mathsf{A}$.

\subsection{Sketch of the proof of Lemma~\ref{uuu3}}
\label{Spr} We only sketch  the proof, since it is similar to \cite{BG11} as well as \cite[section 2.6.2]{AGZ}. The only technical difference is that the lower bound is achieved here by introducing the functions $H_{x,\varepsilon}$ and $\phi_{x,K}$ below, rather than localising $L_{N-1}$ to probability measures on some smaller sets than $\mathsf{B}$ in \cite{BG11}. We first give the proof for the initial model and at the end of the proof precise the necessary changes to deal with the model with fixed filling fractions. 

Recall that $L_N = N^{-1}\sum_{i = 1}^N\delta_{ \lambda_i}$ denotes the normalised empirical measure. We observe that:
\beq\label{proba}
 \frac{\Upsilon_{N,\beta}^{V;\mathsf{B}}(\mathsf{F})}{\Upsilon_{N,\beta}^{V;\mathsf{B}}(\mathsf{B})} \le \mu_{N,\beta}^{V;\mathsf{B}}\big(\big\{\exists i \in [N]\,\,|\,\,\lambda_i \in \mathsf{F}\big\}\big) \le  N\,\frac{\Upsilon_{N,\beta}^{V;\mathsf{B}}(\mathsf{F})}{\Upsilon_{N,\beta}^{V;\mathsf{B}}(\mathsf{B})},
\eeq
where, for any measurable set $\mathsf{X}$:
$$
\Upsilon_{N,\beta}^{V;\mathsf{B}}(\mathsf{X}) = \mu_{N - 1,\beta}^{\frac{NV}{N - 1};\mathsf{B}}\left[\int_{\mathsf{X}} \dd\xi\,\exp\Big\{-\frac{N\beta}{2}\,V(\xi) + 
(N - 1)\beta\int_{\mathsf{B}}\dd L_{N - 1}(\lambda)\ln|\xi - \lambda|
\Big\}\right] .
$$
We shall hereafter estimate $\frac{1}{N}\ln \Upsilon_{N,\beta}^{V;\mathsf{B}}(\mathsf{X})$. 

We first prove a lower bound for $\Upsilon_{N,\beta}^{V;\mathsf{B}}(\mathsf{X})$ with $\mathsf{X}$ open in $\mathsf{B}$. For any $x\in \mathsf{X}$ we can find $\varepsilon>0$ such that 
$(x-\varepsilon,x+\varepsilon) \cap \mathsf{B} \subset \mathsf{X}$.  Let
$$
\delta_{\varepsilon}^{V} =\max_{\substack{|x - y| \leq \varepsilon \\ x,y \in \mathsf{B}}} |V(x)-V(y)|
$$
Using twice Jensen inequality and the convention $V(\xi) = +\infty$ for $\xi \notin \mathsf{B}$, we get 
\begin{equation*}
\begin{split}
\Upsilon_{N,\beta}^{V;\mathsf{B}}(\mathsf{X}) & \geq \mu_{N - 1,\beta}^{\frac{NV}{N - 1};\mathsf{B}}\left[\int_{x - \varepsilon}^{x + \varepsilon}  \dd\xi
\exp\Big\{-\frac{N\beta}{2}\,V(\xi) + 
 (N - 1)\beta\int_{\mathsf{B}} \dd L_{N - 1}(\lambda)\ln|\xi - \lambda|\Big)\Big\}\right] \\
& \geq e^{-\frac{N\beta}{2}(V(x) + \delta_\varepsilon^V)}\,\mu_{N - 1,\beta}^{\frac{NV}{N - 1};\mathsf{B}}\left[\int_{x - \varepsilon}^{x + \varepsilon} \dd\xi \exp\Big\{(N - 1)\beta\,\int_{\mathsf{B}} \dd L_{N - 1}(\lambda)\ln|\xi - \lambda|\Big\}\right]  \\
& \geq {2\varepsilon}\ e^{-\frac{N\beta}{2}(V(x) + \delta_\varepsilon^V)}\,\exp\Big\{(N - 1)\beta\,\mu_{N - 1,\beta}^{\frac{NV}{N - 1};\mathsf{B}}\Big[\int_{\mathsf{B}} \dd L_{N - 1}(\lambda)\,H_{x,\varepsilon}(\lambda)\Big]\Big\} \\
& \geq 2\varepsilon\,e^{-\frac{N\beta}{2}(V(x) + \delta_{\varepsilon}^V)}\,\exp\Big\{(N - 1)\beta\,\mu_{N - 1,\beta}^{\frac{NV}{N - 1};\mathsf{B}}\Big[\int_{\mathsf{B}} \dd L_{N - 1}(\lambda)\,\phi_{x,K}(\lambda) H_{x,\varepsilon}(\lambda)\Big]\Big\},
\end{split}
\end{equation*}
where we have set:
$$
H_{x,\varepsilon}(\lambda) = \int_{x - \varepsilon}^{x + \varepsilon} \frac{\dd\xi}{2\varepsilon}\,\ln|\xi - \lambda|,
$$
and $\phi_{x,K}$ is a continuous function vanishing outside of a large compact $K$ that includes the support of $\mu_{\mathrm{eq}}^{V}$, is equal to $1$ on a ball around $x$ with radius $1+\varepsilon$ and on the support of $\mu_{\mathrm{eq}}^{V}$, and takes values in $[0,1]$.
For any fixed $\varepsilon > 0$, $\phi_{x,K} \cdot H_{x,\varepsilon}$ is bounded continuous, so we have by Theorem~\ref{th:1}:
$$
\Upsilon_{N,\beta}^{V;\mathsf{B}}(\mathsf{X}) \geq 2\varepsilon\, e^{-\frac{N\beta}{2}(V(x) + \delta_\varepsilon^V)}\exp\Big\{
 (N - 1)\beta \int_{\mathsf{B}} \dd\mu_{\mathrm{eq}}^{V}(\lambda)\,\phi_{x,K}(\lambda)\,H_{x,\varepsilon}(\lambda) + NR(\varepsilon,N)\Big\}
$$
with $\lim_{N \rightarrow \infty} R(\varepsilon,N) = 0$ for all $\varepsilon>0$. Letting $N \rightarrow \infty$, we deduce since
$$
\int_{\mathsf{B}} \dd\mu_{\mathrm{eq}}^{V}(\lambda)\,\phi_{x,K}(\lambda)\,H_{x,\varepsilon}(\lambda)=\int_{\mathsf{B}} \dd\mu_{\mathrm{eq}}^{V}(\lambda)\,H_{x,\varepsilon}(\lambda),
$$
and since $V$ converges uniformly towards $V^{\{0\}}$, that
$$
\liminf_{N \rightarrow \infty} \frac{1}{N}\,\ln \Upsilon_{N,\beta}^{V;\mathsf{B}}(\mathsf{X}) \geq -\frac{\beta}{2}\,\delta_\varepsilon^{V^{\{0\}}} - \frac{\beta}{2}\Big(V^{\{0\}}(x) - 2\int_{\mathsf{B}} \dd\mu_{\mathrm{eq}}^{V}(\lambda)\,H_{x,\varepsilon}(\lambda)\Big).
$$
Exchanging the integration over $\xi$ and $\lambda$, observing that $\xi\rightarrow\int_{\mathsf{B}} \dd\mu_{\mathrm{eq}}^{V}(\lambda)\,\ln|\xi-\lambda|$ is continuous and
then  letting $\varepsilon \rightarrow 0$,  we conclude that for all $x\in\mathsf{X}$,
\beq\label{qw}
\liminf_{N \rightarrow \infty} \frac{1}{N}\,\ln\Upsilon_{N,\beta}^{V;\mathsf{B}}(\mathsf{X}) \geq - \frac{\beta}{2}\, \tilde{U}_{{\rm eq}}^{V;\mathsf{B}}(x),
\eeq
where we have recognised the effective potential of Equation~\eqref{eq31}.  We finally optimise over $x\in\mathsf{X}$ to get the desired lower bound. To prove the upper bound, we note that for any $M>0$, 
$$
\Upsilon_{N,\beta}^{V;\mathsf{B}}(\mathsf{X}) \le  \mu_{N - 1,\beta}^{\frac{NV}{N - 1};\mathsf{B}}\left[\int_{\mathsf{X}} \dd\xi\,\exp\Big\{-\frac{N\beta}{2}\,V(\xi) + (N - 1)\beta\int_{\mathsf{B}} \dd L_{N - 1}(\lambda)\ln \mathrm{max}\big(|\xi - \lambda|,M^{-1}\big)\Big\}\right] \,.
$$

Observe that  there exists $C_0$ and $c>0$  and $d$ finite such that for $|\xi| \geq C_0$ and all probability measures $\mu$ on $\mathsf{B}$
$$
W_\mu(\xi)=V(\xi)-2 \int_{\mathsf{B}} \dd \mu(\lambda)\ln \mathrm{max}\big(|\xi - \lambda|,M^{-1}\big)\ge c\ln |\xi|+ d
$$
 by the confinement Hypothesis \ref{hmainc1}. As a consequence, if $\mathsf{X}\subset \mathsf{B} \setminus [-C,C]$ for some $C$ large enough, we deduce that:
\begin{equation}
\label{qw2}
\Upsilon_{N,\beta}^{V;\mathsf{B}}(\mathsf{X}) \leq  \int_{\mathsf{X}} \dd\xi\,e^{-\frac{\beta}{2}V(\xi)}\,e^{-(N-1)\frac{\beta}{2} (c\ln|\xi|+d)}\leq  e^{-N\frac{\beta}{4} c\ln C},
\end{equation}
where the last bound holds for $N$ large enough. 
Combining Equations~\eqref{qw}, \eqref{qw2} and \eqref{proba} shows that
$$\limsup_{C\ra\infty}\limsup_{N\ra\infty}\frac{1}{N}\ln \mu_{N,\beta}^{V;\mathsf{B}}\big(\big\{\exists i \in [N] \quad |\lambda_i|\ge C\big\}\big)=-\infty\,.$$
Hence, we may restrict ourselves to $\mathsf{X}$ bounded. Moreover,  the same bound extends to 
$ \mu_{N - 1,\beta}^{\frac{NV}{N - 1};\mathsf{B}}$ so that 
 we can restrict the expectation over $L_{N-1}$ to probability measures supported on $[-C,C]$ 
up to an arbitrary small error  $e^{-N e(C)}$, provided $C$ is large enough and where $\lim_{C \rightarrow +\infty} e(C) = +\infty$.
The confinement hypothesis also guarantees that $V(\xi)-2\int_{\mathsf{B}} \dd L_{N - 1}(\lambda)\ln \mathrm{max}\big(|\xi - \lambda|,M^{-1}\big)$ is uniformly bounded from below by a constant $D$. 
As $\lambda \ra \ln \mathrm{max}\big(|\xi - \lambda|,M^{-1}\big)$ is bounded continuous on compacts and  $M$-Lipschitz on $\mathbb{R}$, we can then use the large deviation principles of Theorem~\ref{th:1}   to deduce that for any $\varepsilon>0$, any $C\ge C_0$,
\begin{equation*}
\label{turlut}
\begin{split}
\Upsilon_{N,\beta}^{V;\mathsf{B}}(\mathsf{X}) &\leq e^{ N^2 \tilde{R}(\varepsilon,N,C)} +e^{-N(e(C)-\frac{\beta}{2}D)}
\\
& \quad +\int_{\mathsf{X}} \dd\xi
\,\exp\bigg(-\frac{N\beta}{2}\,V(\xi) +
 (N - 1)\beta\int_{\mathsf{B}} \dd\mu_{\mathrm{eq}}^{V}(\lambda) \ln \mathrm{max}\big(|\xi - \lambda|,M^{-1}\big) +N M \varepsilon \bigg).
 \end{split}
 \end{equation*}
with
$$
\limsup_{N\rightarrow\infty} \tilde{R}(\varepsilon,N,C)=
\limsup_{N \rightarrow \infty} \frac{1}{N^2}\ln\mu_{N - 1,\beta}^{\frac{NV}{N -1};\mathsf{B}}\big( \{L_{N-1}([-C,C])=1\}\cap \{\mathfrak{d}(L_{N-1},\mu_{\mathrm{eq}}^{V})>\varepsilon\}\big)<0.
$$
in terms of the Vaserstein distance between two probability measures:
$$
\mathfrak{d}(\mu,\nu) = \sup\bigg\{\Big|\int_{\mathbb{R}} f(\xi)\dd[\mu - \nu](\xi)\Big| \quad : \quad f: \mathbb{R} \rightarrow \mathbb{R} \,\,1\,\,\text{-Lipschitz}\bigg\}.
$$
Moreover, $\xi \rightarrow V(\xi) - \int_{\mathsf{B}} \dd\mu_{\mathrm{eq}}^{V}(\lambda) \ln \mathrm{max}\big(|\xi - \lambda|,M^{-1}\big)$ is bounded 
continuous so that a standard Laplace method yields, as $V$ goes to $V^{\{0\}}$, 
$$
\limsup_{N \rightarrow \infty} \frac{1}{N}\,\ln\Upsilon_{N,\beta}^{V;\mathsf{B}}(\mathsf{X}) \le \max\bigg\{ -\inf_{\xi\in \mathsf{X}}
\Big[\frac{\beta}{2}\Big(V^{\{0\}}(\xi) - \int_{\mathsf{B}} \dd\mu_{\mathrm{eq}}^{V}(\lambda) \ln \max\big(|\xi - \lambda|,M^{-1}\big)\Big) \Big], \frac{\beta D}{2} -e(C)\bigg\} \,.
$$
We finally choose  $C$ large enough so that the first term is larger than the second. Then, by the monotone convergence theorem we deduce that $\int_{\mathsf{B}} \dd\mu_{\mathrm{eq}}^{V}(\lambda) \ln \mathrm{max}\big(|\xi - \lambda|,M^{-1}\big)$  
increases as $M$ goes to infinity
towards $\int_{\mathsf{B}} \dd\mu_{\mathrm{eq}}^{V}(\lambda) \ln|\xi - \lambda|$. This completes the proof of the large deviation in the initial model.

For the fixed filling fractions model, we make the decomposition
$$
\mu_{N,\beta;\bm{\epsilon}}^{V;\mathsf{A}}\big[\big\{\exists i \in [N] \quad  \lambda_i\in \mathsf{X}\big\}\big]=\sum_{h=0}^g \mu_{N,\beta;\bm{\epsilon}}^{V;\mathsf{A}}\big[\big\{\exists i\in \ldbrack 1,N_h\rdbrack \quad \lambda_{h,i}\in \mathsf{X}\cap \mathsf{A}_h\big\}\big],
$$
with
$$
\frac{\Upsilon_{N,\beta,h}^{V;\mathsf{B}}(\mathsf{X}\cap\mathsf{A}_h)}
{\Upsilon_{N,\beta,h}^{V;\mathsf{B}}(\mathsf{A}_h)}\le  \mu_{N,\beta;\bm{\epsilon}}^{V;\mathsf{B}}\big[\big\{\exists i\in \ldbrack 1,N_h \rdbrack \quad \lambda_{h,i}\in \mathsf{X}\cap \mathsf{A}_h\big\}\big]\le N_h\frac{\Upsilon_{N,\beta,h}^{V;\mathsf{B}}(\mathsf{X}\cap\mathsf{A}_h)}
{\Upsilon_{N,\beta,h}^{V;\mathsf{B}}(\mathsf{A}_h)},
$$
and 
$$
\Upsilon_{N,\beta,h}^{V;\mathsf{B}}(\mathsf{X}\cap\mathsf{A}_h)=\mu_{N ,\beta;\bm{\epsilon}- 1_h/N}^{\frac{NV}{N - 1};\mathsf{A}}\left(\int_{\mathsf{X}\cap\mathsf{A}_h} \dd\xi\,\exp\Big\{-\frac{N\beta}{2}\,V(\xi) + 
(N - 1)\beta\int_{\mathsf{B}}\dd L_{N - 1}(\lambda)\ln|\xi - \lambda|\Big)\Big\}\right),
$$
where $\bm{\epsilon}-1_h/N$ corresponds to the filling fraction where one eigenvalue has been suppressed from $\mathsf{A}_h$. The estimates for $\Upsilon_{N,\beta,h}^{V;\mathsf{B}}(\mathsf{X}\cap\mathsf{A}_h)$ are done exactly as above and the result follows since the logarithm of a finite sum of exponentially small terms is asymptotically equivalent to the logarithm of the maximal term. \hfill $\Box$

\subsection{Concentration of measure and consequences}

\label{Scons}

We will need rough \textit{a priori} bounds on the correlators, which can be derived by purely probabilistic methods. This type of result first appeared in the work of \cite{Boutet,Johansson98} and more recently \cite{Shch2,MMEMS}. Given their importance, we find useful to prove independently the bound we need by elementary means.

Hereafter, we will say that a function $f\,:\,\mathbb R\rightarrow \mathbb{C}$ is $b$-H\"older if
$$
\kappa_b[f]= \sup_{x\neq y}\frac{|f(x)-f(y)|}{|x-y|^b}<\infty.
$$
Our final goal is to control 
 $\int_{\mathsf{A}} \varphi(x)\dd[L_N - \mu_{\mathrm{eq}}^{V}](x)$ for a class of functions $\varphi$ which is large enough, in particular contains analytic functions on a neighbourhood of the interval of integration $\mathsf{A}$. This problem can be settled by controlling the ``distance" between $L_N$ and $\mu_{\mathrm{eq}}^{V}$ for an appropriate notion of ``distance''. We introduce the pseudo-distance $\mathfrak{D}$  between probability measures $\mu,\nu$ given by: 
\beq
\label{dee}\mathfrak{D}[\mu,\nu] = \left(-\iint_{\mathbb{R}^2} \dd[\mu - \nu](x)\dd[\mu - \nu](y)\,\ln|x - y|\right)^{\frac{1}{2}}.
\eeq
It can be represented in terms of Fourier transform of the measures:
\beq
\label{pseu}\mathfrak{D}[\mu,\nu] = \left(\int_{0}^{\infty} \frac{\dd p}{|p|}\,\big|(\widehat{\mu} - \widehat{\nu})(p)\big|^2\right)^{\frac{1}{2}}.
\eeq
Since $L_N$ has atoms, its pseudo-distance to another measure is in general infinite. There are several methods to circumvent this issue, and one of them, that we borrow from \cite{MMEMS}, is to define a regularised measure $\widetilde{L}_N^{\mathrm{u}}$ (see the beginning of \S~\ref{compsa} below) from $L_N$. Then, the result of concentration  takes the form:

\begin{lemma}\label{theoconc} 
Let $V$ be regular, $\mathcal{C}^3$, confining, satisfying a control of large deviations on $\mathsf{A}$, and satisfying \eqref{expV} for $K=0$ (namely: $N(V-V^{\{0\}})$ is uniformly bounded by a constant $v^{\{1\}}$ on $\mathsf{A}$). There exists $C > 0$ so that, for $t$ small enough and $N$ large enough:
$$
\mu_{N,\beta}^{V;\mathsf{A}}\big(\mathfrak{D}[\widetilde{L}_N^{\mathrm{u}},\mu_{\mathrm{eq}}^{V}] \geq t\big) \leq e^{CN\ln N - N^2t^2}.
$$
Moreover, for any $\bm{N}=(N_1,\ldots,N_g)$ so that $\bm{\epsilon}=\bm{N}/N\in \mathcal{E}$,
\beq\label{concfix}
\mu_{N,\beta;\bm{\epsilon}}^{V; \mathsf{A}}\big(\mathfrak{D}[\widetilde{L}_N^{\mathrm{u}},\mu_{\mathrm{eq};\bm{\epsilon}}^{V}] \geq t\big) \leq e^{CN\ln N - N^2t^2}.
\eeq
\end{lemma}
We prove it in \S~\ref{compsa} below. The assumption $V$ of class $\mathcal{C}^3$ ensures that the effective potential \eqref{eq31} defined from the equilibrium measure is a $\frac{1}{2}$-H\"older function (and even
Lipschitz if all edges are soft) on the compact set $\mathsf{A}$, as one can observe on  Equation~\eqref{A17} given in Appendix~\ref{appA}. This lemma allows an \textit{a priori} control of expectation values of test functions:
\begin{corollary}
\label{Ls1} Let $V$ be regular, $\mathcal{C}^3$, confining, satisfying a control of large deviations on $\mathsf{A}$, and satisfying \eqref{expV} for $K=0$ (namely: $N(V-V^{\{0\}})$ is uniformly bounded by a constant $v^{\{1\}}$ on $\mathsf{A}$). Let $b > 0$, and assume $\varphi\,:\,\mathbb{R} \rightarrow \mathbb{C}$ is a $b$-H\"older function with constant $\kappa_{b}[\varphi]$, and such that:
$$
| \varphi |_{1/2} : = \Big(\int_{\mathbb{R}} \dd p\,|p|\,|\widehat{\varphi}(p)|^2\Big)^{\frac{1}{2}} < \infty.
$$
Then, there exists $C_3 > 0$ such that, for $t$ small enough and $N$ large enough:
$$
\mu_{N,\beta}^{V;\mathsf{A}}\Big[\Big|\int_{\mathsf{A}} \dd[L_N - \mu_{\mathrm{eq}}^{V}](x)\,\varphi(x)\Big| \geq \frac{2\kappa_{b}[\varphi]}{(b + 1)N^{2b}} + t\,|\varphi|_{1/2}\Big] \leq e^{C_3N\ln N - \frac{\beta}{2} N^2t^2}.
$$
and for any $\bm{N}=(N_1,\ldots,N_g)$ so that $\bm{\epsilon}=\bm{N}/N\in \mathcal{E}$,
$$ 
\mu_{N,\beta;\bm{\epsilon}}^{V;\mathsf{A}}\Big[\Big|\int_{\mathsf{A}} \dd[L_N - \mu_{\mathrm{eq};\bm{\epsilon}}^{V}](x)\,\varphi(x)\Big| \geq \frac{2\kappa_{b}[\varphi]}{(b + 1)N^{2b}} + t\,|\varphi|_{1/2}\Big] \leq e^{C_3N\ln N - \frac{\beta}{2}N^2t^2}.
$$
\end{corollary}
As a special case, we can obtain a rough \textit{a priori} control on the correlators. Recall the notation, for $\bm{\epsilon}\in\mathcal{E}$,
$$
W_{1;\bm{\epsilon}}^{\{-1\}}(x) =  \int_{\mathsf{A}} \frac{\dd\mu^{V;\mathsf{A}}_{{\rm eq};\bm{\epsilon}}(\xi)}{x - \xi}.
$$
\begin{corollary}
\label{apco} Let $V$ be regular, $\mathcal{C}^3$, confining and satisfying a control of large deviations on $\mathsf{A}$. Let $D' > 0$, and:
$$
w_N = \sqrt{N\ln N},\qquad f(\delta) = \frac{\sqrt{|\ln \delta|}}{\delta},
\qquad d(x,\mathsf{A}) = \inf_{\xi \in \mathsf{A}} |x - \xi| \geq \frac{D'}{\sqrt{N^2\ln N}}.
$$
There exists a constant $\gamma_1(\mathsf{A},D') > 0$ so that, for $N$ large enough, for any $\bm{N}=(N_1,\ldots,N_g)$ so that $\bm{\epsilon}=\bm{N}/N\in \mathcal{E}$:
\beq\label{apriori0}
\big|W_{1;\bm{\epsilon}}(x) - NW_{1;\bm{\epsilon}}^{\{-1\}}(x)\big| \leq \gamma_1(\mathsf{A},D')\,w_N\,f\big(d(x,\mathsf{A})\big).
\eeq
Similarly, for any $n \geq 2$, there exist constants $\gamma_n(\mathsf{A},D') > 0$ so that, for $N$ large enough:
\beq\label{apriori}
\big|W_{n;\bm{\epsilon}}(x_1,\ldots,x_n)\big| \leq \gamma_n(\mathsf{A},D')\,w_N^{n} \prod_{i = 1}^n f\big(d(x_i,\mathsf{A})\big).
\eeq
\end{corollary}
In the $(g + 1)$-cut regime with $g \geq 1$, we denote $(\mathsf{S}_h)_{0 \leq h \leq g}$ the connected components of the support of $\mu_{\mathrm{eq}}^{V}$, and we take $\mathsf{A} = \bigcup_{h = 0}^{g} \mathsf{A}_h$, where $\mathsf{A}_h = [a_h^-,a_h^+] \subseteq \mathsf{B}$ are pairwise disjoint bounded segments such that $\mathsf{S}_h \subset \mathring{\mathsf{A}}_h$. For any configuration $\lambda \in \mathsf{A}^N$, we denote $N_h$ the number of $\lambda_i$s in $\mathsf{A}_h$, and $\bm{N} = (N_h)_{1 \leq h \leq g}$. The following result gives an estimate for large deviations of $\bm{N}$ away from $N\bm{\epsilon}_{\star}$ in the large $N$ limit.
\begin{corollary}
\label{cosq}Let $\mathsf{A}$ be as above, and $V$ be $\mathcal{C}^3$, confining, satisfying a control of large deviations on $\mathsf{A}$, and leading to a $(g + 1)$-cut regime. There exists positive constants $C,C'$ such that, for $N$ large enough and uniformly in $t$:
\begin{equation}
\mu_{N,\beta}^{V;\mathsf{A}}\big(|\bm{N} - N\bm{\epsilon}_{\star}|_1 > t \sqrt{N\ln N}\big) \leq e^{N\ln N(C - C' t^2)}.
\end{equation}
\end{corollary}
As an outcome of this article, we will obtain in Section~\ref{Lare} a stronger large deviation statement for filling fractions when the potential satisfies the stronger Hypotheses~\ref{hmainc1}--\ref{hmainc4}.

\subsection{Concentration of \texorpdfstring{$L_N$}{LN}: Proof of Lemma~\ref{theoconc}}
\label{ProofC}
Throughout this section, proofs will be given for the initial model: they are exactly the same for the  fixed filling fractions model. 
\subsubsection{Regularisation of \texorpdfstring{$L_N$}{LN}}
\label{compsa}

We start by following an idea introduced by Ma\"{i}da and Maurel-Segala \cite[Proposition 3.2]{MMEMS}. Let $\sigma_N,\eta_N \rightarrow 0$ be two sequences of positive numbers. To any configuration of points $\lambda_1 \leq \ldots \leq \lambda_N$ in $\mathsf{A}$, we associate another configuration $\widetilde{\lambda}_1,\ldots,\widetilde{\lambda}_N$ by the formula:
\beq
\label{minia}\widetilde{\lambda}_1 = \lambda_1,\qquad \widetilde{\lambda}_{i + 1} = \widetilde{\lambda}_i + \mathrm{max}(\lambda_{i + 1} - \lambda_i,\sigma_N)\,.
\eeq
It has the properties:
\beq\label{gf}
\forall i \neq j,\qquad |\widetilde{\lambda}_i - \widetilde{\lambda}_j| \geq \sigma_N,\qquad |\lambda_i - \lambda_j| \leq |\widetilde{\lambda}_i - \widetilde{\lambda}_j|,\qquad |\widetilde{\lambda}_i - \lambda_i| \leq (i - 1)\sigma_N.
\eeq  
Let us denote  by $\widetilde{L}_N = N^{-1} \sum_{i = 1}^N \delta_{\widetilde{\lambda}_i}$ the new counting measure. Then, we define $\widetilde{L}_N^{\mathrm{u}}$ be the convolution of $\widetilde{L}_N$ with the uniform measure on $[0,\eta_N\sigma_N]$.

We are going to compare the (opposite of the) logarithmic energy of $L_N$ to that of $\widetilde{L}_N^{\mathrm{u}}$, which has the advantage of having no atom.
We first have:
\begin{equation}\label{lkjh}
\sum_{i \neq j} \ln|\lambda_i - \lambda_j| \leq \sum_{i \neq j} \ln\big|\widetilde{\lambda}_i - \widetilde{\lambda}_j\big|,
\end{equation}
because the logarithm is increasing and the spacings of $\tilde{\lambda}$s are larger than the spacings of $\lambda$s.
Let:
$$
\Sigma[\mu]=\iint_{\mathbb{R}^2} \ln|x-y|\dd \mu(x)\dd\mu(y)
$$
denote the (opposite of the) logarithmic energy of a probability measure $\mu$. Then
$$
N^2 \Sigma[\widetilde{L}_{N}^{{\rm u}}]
- \sum_{i \neq j} \ln|\widetilde{\lambda}_i - \widetilde{\lambda}_j| 
= \sum_{i\neq j} \iint_{[0,1]^2} \!\!\dd u \,\dd v \ln \Big| 1+\eta_N\sigma_N \frac{(u-v)}{\tilde \lambda_i-\tilde\lambda_j}\Big|
+ \sum_{i = 1}^{N} \iint_{[0,1]^2} \!\!\dd u \,\dd v \ln\big|\eta_N\sigma_N(u - v)\big|.
$$
Thanks to the minimal distance $\sigma_N$ enforced between the $\widetilde{\lambda}_i$s in Equation~\eqref{gf}, $\sigma_N\,\big|(u - v)/(\tilde \lambda_i-\tilde\lambda_j)\big|$ is bounded by $1$, so that for $\eta_N\le \frac{1}{2}$ (thus for $N$ large enough),  
$$
\bigg|  \sum_{i\neq j} \iint_{[0,1]^2} \dd u\, \dd v \ln \Big| 1+\eta_N\sigma_N \frac{(u-v)}{\widetilde \lambda_i-\widetilde\lambda_j}\Big|\bigg|\le 2 N(N-1)\eta_N\,.
$$
Since  $(u,v) \mapsto \ln|u - v|$ is integrable in $[0,1]^2$, we find for some constants $c_1,c_2 > 0$:
$$
\Big|\sum_{i \neq j} \ln|\widetilde{\lambda}_i - \widetilde{\lambda}_j| - N^2\Sigma[\widetilde{L}_{N}^{u}] \Big| \leq c_1N\big|\ln(\eta_N\sigma_N)\big| + c_2N^2\eta_N,
$$
so that finally, with Equation~\eqref{lkjh}, we have proved that for any $(\lambda_i)_{1\le i\le N}\in\mathbb R^N$:
\begin{equation}
\label{lkj}
\sum_{i \neq j} \ln|\lambda_i - \lambda_j| \leq  N^2\Sigma[\widetilde{L}_{N}^{u}]+ c_1N\big|\ln(\eta_N\sigma_N)\big| + c_2N^2\eta_N\,.
\end{equation}
Besides, if $b > 0$ and $\varphi\,:\,\mathsf{A} \rightarrow \mathbb{C}$ is a $b$-H\"older function with constant $\kappa_b[\varphi]$, we have by Equation\eqref{gf}:
\begin{equation}
\label{Ca3} 
\begin{split}
\Big|\int_{\mathsf{A}} \dd[L_N - \widetilde{L}_N^{\mathrm{u}}](x)\,\varphi(x)\Big| & \leq \frac{\kappa_{b}[\varphi]}{N} \sum_{i = 1}^N \big((i - 1)\sigma_N + \eta_N\sigma_N\big)^b \\
& \leq  \frac{\kappa_b[\varphi]}{N}\Big(\sigma_N^b\eta_N^b + \sum_{i = 2}^{N} (i - 1)^b \sigma_N^b(1 + \eta_N)^b\Big) \leq \frac{2\kappa_b[\varphi]}{(1 + b)}\,(N\sigma_N)^b\,.
\end{split}
\end{equation}

\subsubsection{Deviations of \texorpdfstring{$\widetilde{L}_N^{\mathrm{u}}$}{LNu}}
\label{Sec342}
We would like to estimate the probability of  deviations of $\widetilde{L}_N^{\mathrm{u}}$ from the equilibrium measure $\mu_{\mathrm{eq}}^V$. We need first a lower bound on $Z^{V;\mathsf{A}}_{N,\beta}$ similar to that of \cite{BAG} obtained by localising the ordered eigenvalues at a distance $N^{-3}$ of the quantiles $\lambda_i^{\mathrm{cl}}$ of the equilibrium measure $\mu_{\mathrm{eq}}^{V}$, which are defined as:
$$
\lambda_i^{\mathrm{cl}} = \inf\Big\{x \in \mathsf{A}\,\,:\,\, \mu_{\mathrm{eq}}^{V}\big([-\infty,x]\big)\geq \frac{i}{N}\Big\},\qquad i \in \ldbrack 1,N \rdbrack.
$$
Since $V$ is $\mathcal{C}^2$, $\dd\mu_{\mathrm{eq}}^{V}$ has a continuous density on the interior of its support, which diverges only at hard edges, where it blows at most like the inverse of a squareroot, and vanishes only at soft edges. Therefore, there exists a constant $C > 0$ such that, for $N$ large enough:
\beq
\label{LB40} \forall i \in \ldbrack 2,N \rdbrack,\qquad  \frac{c}{N^2} \leq |\lambda^{\mathrm{cl}}_i-\lambda^{\mathrm{cl}}_{i-1}|.
\eeq
Then, since $V$ is a fortiori $\mathcal{C}^1$ on $\mathsf{A}$ compact,
\begin{equation*}
\begin{split}
Z^{V;\mathsf{A}}_{N,\beta}&\geq  N!\int_{|\delta_i|\leq N^{-3}}\prod_{1 \leq i< j \leq N}|\lambda_i^{\mathrm{cl}}-\lambda_j^{\mathrm{cl}} + \delta_i - \delta_j|^\beta\,\prod_{i = 1}^N e^{-\frac{\beta N}{2} V(\lambda_i^{\mathrm{cl}}+\delta_i)} \dd\delta_i \\
&\geq  N!\,N^{-3N}e^{-C_1\,N} \prod_{1 \leq i< j \leq N}|\lambda_i^{\mathrm{cl}} -\lambda_j^{\mathrm{cl}}|^\beta\,\prod_{i = 1}^N e^{-\frac{N\beta}{2}\sum_{i = 1}^N V(\lambda_i^{\mathrm{cl}})},
\end{split}
\end{equation*}
for some constant $C_1 > 0$. Then:
\begin{equation}
\label{nunudnsaifzb}\begin{split}
\sum_{1 \leq i < j \leq N} \ln |\lambda_i^{{\rm cl}} - \lambda_j^{{\rm cl}}| & = \sum_{\substack{1 \leq i,j \leq N \\ i + 1 < j}} \ln|\lambda_{i}^{{\rm cl}} - \lambda_j^{{\rm cl}}| + \sum_{i = 1}^{N - 1} \ln|\lambda_i^{{\rm cl}} - \lambda_{i + 1}^{{\rm cl}}| \\
& \geq \sum_{1 \leq i < j \leq N - 1} \ln|\lambda_{i}^{{\rm cl}} - \lambda_{j + 1}^{{\rm cl}}| + (N - 1)\ln\big(\tfrac{C}{N^2}\big) \\
& \geq N^2 \iint_{\lambda_1^{{\rm cl}} \leq x < y \leq \lambda_N^{{\rm cl}}} \ln|x - y| \dd \mu_{{\rm eq}}^{V}(x)\dd \mu_{{\rm eq}}^{V}(y) + (N - 1)\ln\big(\tfrac{c}{N^2}\big) \\
& \geq \frac{N^2}{2} \iint_{[\lambda_1^{{\rm cl}},\lambda_N^{{\rm cl}}]^2} \ln|x - y| \dd \mu^{V}_{{\rm eq}}(x) \dd \mu_{{\rm eq}}^{V}(y) + (N - 1)\ln\big(\tfrac{c}{N^2}\big) \\
& \geq \frac{N^2}{2} \iint_{\mathsf{A}^2} \ln|x - y| \dd\mu_{{\rm eq}}^{V}(x)\dd\mu_{{\rm eq}}^{V}(y) - \frac{N^2}{2}  \iint_{\substack{x \in \mathsf{A} \\y < \lambda_1^{{\rm cl}}}} \ln|x - y|\dd \mu_{{\rm eq}}^{V}(x)\dd\mu_{{\rm eq}}^{V}(y) \\
& \quad - \frac{N^2}{2} \iint_{\substack{x > \lambda_1^{{\rm cl}} \\ y < \lambda_1^{{\rm cl}}}} \ln|x - y|\dd\mu_{{\rm eq}}^{V}(x)\dd\mu_{{\rm eq}}^V(y) + (N - 1)\ln\big(\tfrac{c}{N^2}\big) \\
& \geq  \frac{N^2}{2} \iint_{\mathsf{A}^2} \ln|x - y| \dd\mu_{{\rm eq}}^{V}(x)\dd\mu_{{\rm eq}}^{V}(y) - \frac{N^2}{4}  \int_{y < \lambda_1^{{\rm cl}}} \big(C + V^{\{0\}}(y)\big)\dd\mu_{{\rm eq}}^V(y) \\
& \quad - \frac{N^2}{2} \iint_{\substack{x > \lambda_1^{{\rm cl}} \\ y < \lambda_1^{{\rm cl}}}} \ln|x - y|\dd\mu_{{\rm eq}}^V(x)\dd\mu_{{\rm eq}}^V(y) + (N - 1)\ln\big(\tfrac{c}{N^2}\big).
\end{split} 
\end{equation}
Between the first two lines, we used Equation~\eqref{LB40} to get a lower bound for the second term. In the third line we have used the fact that the logarithm is an increasing function. In the fourth line, we have symmetrised the integral. In the fifth line, we observed that the definition of $\lambda_N^{{\rm cl}}$ implies that $\mu_{{\rm eq}}^V$ has support included in $\mathsf{A}_- := \mathsf{A} \cap (-\infty,\lambda_N^{{\rm cl}}]$ and completed the square domain $[\lambda_1^{{\rm cl}},\lambda_N^{{\rm cl}}]$ to $\mathsf{A}^2$ while subtracting the extra contributions coming from this procedure. In the last line, we used the equality case of the characterisation of the equilibrium measure. Since $\dd \mu_{{\rm eq}}^V$ has a continuous density possibly blowing up like $\alpha$ an inverse squareroot at the endpoints of its support, $y \mapsto \int_{x > \lambda_1^{{\rm cl}}} \ln|x - y| \dd \mu_{{\rm eq}}^V(x)$ is uniformly $O(\ln N)$ for $y \in \mathsf{A}_-$ (recall that $\mathsf{A}_-$ is compact since $\mathsf{A}$ is). Since $\mu_{{\rm eq}}^V\big((-\infty,\lambda_1^{{\rm cl}}]\big) \leq \frac{1}{N}$ by definition of $\lambda_1^{{\rm cl}}$, we deduce that the first term in the last line of Equation~\eqref{nunudnsaifzb} is $O(N\ln N)$. Besides, $V^{\{0\}} + C$ is continuous, hence bounded on $\mathsf{A}$ compact, showing for a similar reason that the last term in the penultimate line of  Equation~\eqref{nunudnsaifzb} is $O(N)$. All in all, this shows the existence of a constant $C_2$ such that for $N$ large enough:
$$
\sum_{1 \leq i < j \leq N} \ln |\lambda_i^{{\rm cl}} - \lambda_j^{{\rm cl}}| \geq \sum_{1 \leq i < j \leq N} \ln |\lambda_i^{{\rm cl}} - \lambda_j^{{\rm cl}}| - C_2 N \ln N.
$$

Next, we have
\begin{equation*} 
\begin{split}
& \quad \left|\frac{1}{N}\sum_{i=1}^N V(\lambda_i^{\mathrm{cl}})-\int_{\mathsf{A}} V(x)\dd\mu_{\mathrm{eq}}^{V}(x)\right|  \\
& = \frac{V(\lambda_N^{{\rm cl}})}{N} - \int_{x < \lambda_1^{{\rm cl}}} V(x) \dd\mu_{{\rm eq}}^V(x) + \sum_{i = 1}^{N - 1} \int_{\lambda_i^{{\rm cl}}}^{\lambda_{i + 1}^{{\rm cl}}} \big(V(\lambda_i^{{\rm cl}}) - V(x)\big)\dd\mu_{{\rm eq}}^V(x) \\
& \leq \frac{2\p V \p_{\infty}^{\mathsf{A}}}{N} + \p V' \p_{\infty}^{\mathsf{A}} \bigg(\sum_{i = 1}^{N - 1} \int_{\lambda_i^{{\rm cl}}}^{\lambda_{i +1}^{{\rm cl}}} |x - \lambda_i^{{\rm cl}}| \dd \mu_{{\rm eq}}^V(x)\bigg) \\ 
& \leq \frac{2\p V \p_{\infty}^{\mathsf{A}}}{N} + \frac{\p V' \p_{\infty}^{\mathsf{A}}}{N} \sum_{i = 1}^{N - 1} (\lambda_{i + 1}^{{\rm cl}} - \lambda_i^{{\rm cl}}) \leq \frac{2 \p V \p_{\infty}^{\mathsf{A}} + C_3 \p V' \p_{\infty}^{\mathsf{A}}}{N}
\end{split}
\end{equation*}
for some constant $C_3 > 0$. Then, as $N^{-1} \sum_{i = 1}^N \delta_{\lambda_i^{{\rm cl}}}$ is a sequence of measures converging to the minimiser $\mu_{{\rm eq}}^V$ of the energy functional $E$ introduced in Equation~\eqref{Enf}, we find:
\beq
\label{lowe}Z^{V;\mathsf{A}}_{N,\beta} \geq \exp\Big\{- \frac{\beta}{2}\,C_4\,N\ln N - N^2\,E[\mu_{\mathrm{eq}}^{V}]\Big\}
\eeq
for some positive constant $C_4$.

Now, consider the event $\mathcal{S}_{N}(t) = \big\{\mathfrak{D}[\widetilde{L}_N^{\mathrm{u}},\mu_{\mathrm{eq}}^{V}] \geq t\big\}$.  Observing that 
$$ 
\mu_{N,\beta}^{V;\mathsf{A}}\big(\mathcal{S}_N(t)\big) = \frac{1}{Z_{N,\beta}^{V;\mathsf{A}}} \int_{\mathcal{S}_N(t)} 
e^{\frac{\beta }{2}( \sum_{i\neq j} \ln |\lambda_i-\lambda_j| - N^2\int_{\mathsf{A}} \dd L_N(x)\,V(x))}\prod_{i = 1}^N \dd\lambda_i,
$$
and using the comparison \eqref{lkj} of \S~\ref{compsa}, we find, with the notations of Theorem \ref{th:10}:
$$
\mu_{N,\beta}^{V;\mathsf{A}}\big(\mathcal{S}_N(t)\big) \leq \frac{e^{\frac{\beta}{2}\,R_N}}{Z_{N,\beta}^{V;\mathsf{A}}} \int_{\mathcal{S}_N(t)} e^{\frac{\beta N^2}{2}(\Sigma[\widetilde{L}_N^{\mathrm{u}}] - \int_{\mathsf{A}} \dd\widetilde{L}_N^{\mathrm{u}}(x)\,V^{\{0\}}(x))} \prod_{i = 1}^N\dd\lambda_i,
$$
with:
$$
R_N = N^3\sigma_N\,\kappa_1[V] + c_2 N^2\eta_N +c_1 N|\ln(\sigma_N\eta_N)|+Nv^{\{1\}}\,.
$$
We then decompose:
\begin{equation*}
\begin{split}
E[\widetilde{L}_N^{{\rm u}}] & = \frac{\beta}{2}\Big(-\Sigma[\widetilde{L}_N^{{\rm u}}] + \int_{\mathsf{A}} \dd\widetilde{L}_N^{{\rm u}}(x)\,V^{\{0\}}(x)\Big) \\
& =  E[\mu_{{\rm eq}}^{V}] + \frac{\beta}{2}\Big(\int_{\mathsf{A}} U_{{\rm eq}}^{V}(x)\dd[\widetilde{L}_{N}^{{\rm u}} - \dd\mu_{{\rm eq}}^{V}](x) + \mathfrak{D}^2[\widetilde{L}_{N}^{{\rm u}},\mu_{{\rm eq}}^{V}]\Big).
\end{split}
\end{equation*}
 The effective potential $U_{{\rm eq}}^{V}$ is defined in Equation~\eqref{eq31}, and since it is integrated against a signed measure of zero mass, we can add to it a constant and thus replace it with $\tilde{U}_{{\rm eq}}^{V}$. According to the characterisation of the equilibrium measure, $\widetilde{U}_{{\rm eq}}^{V}$ vanishes $\mu_{{\rm eq}}^{V}$-everywhere, hence:
$$
E[\widetilde{L}_N^{{\rm u}}] = E[\mu_{{\rm eq}}^{V}] + \frac{\beta}{2}\Big(\mathfrak{D}^2[\widetilde{L}_N^{{\rm u}},\mu_{{\rm eq}}^{V}] + \int_{\mathsf{A}} \widetilde{U}^{V}_{{\rm eq}}(x)\,\dd\widetilde{L}_N^{{\rm u}}(x)\Big),
$$
and we obtain:
$$
\mu_{N,\beta}^{V;\mathsf{A}}\big(\mathcal{S}_N(t)\big)  \leq  \frac{e^{\frac{\beta}{2}R_N - N^2\,E[\mu_{\mathrm{eq}}^{V}]}}{Z_{N,\beta}^{V;\mathsf{A}}}\int_{\mathcal{S}_N(t)} e^{-\frac{\beta N^2}{2}\big(\mathfrak{D}^2[\widetilde{L}_N^{\mathrm{u}},\mu_{\mathrm{eq}}^{V}] + \int_{\mathsf{A}} \dd\widetilde{L}_N^{\mathrm{u}}(x)\,\widetilde{U}^{V}_{{\rm eq}}(x)\big)} \prod_{i = 1}^N \dd\lambda_i\,.
$$
Since $\widetilde{U}^{V;\mathsf{A}}$ is at least $\frac{1}{2}$-H\"older on $\mathsf{A}$ (and even Lipschitz if all edges are soft),we find by Equation~\eqref{Ca3}:
$$
\mu_{N,\beta}^{V;\mathsf{A}}\big(\mathcal{S}_N(t)\big) \leq \frac{e^{\frac{\beta}{2}(R_N + \kappa_{1/2}[\widetilde{U}_{{\rm eq}}^{V}]\,N^{\frac{5}{2}}\sigma_N^{\frac{1}{2}}) -  N^2\,E[\mu_{\mathrm{eq}}^{V}]}}{Z_{N,\beta}^{V;\mathsf{A}}}\int_{\mathcal{S}_N(t)} e^{-\frac{\beta N^2}{2}\,\mathfrak{D}^2[\widetilde{L}_N^{\mathrm{u}},\mu_{\mathrm{eq}}^V]}\,\prod_{i = 1}^N e^{-\frac{\beta N}{2}\,\widetilde{U}_{{\rm eq}}^{V}(\lambda_i)}\,\dd\lambda_i.
$$
We now use the lower bound \eqref{lowe} for the partition function, and the definition of the event $\mathcal{S}_N(t)$, in order to obtain:
\begin{equation*}
\begin{split}
\mu_{N,\beta}^{V;\mathsf{A}}\big(\mathcal{S}_N(t)\big) & \leq e^{\frac{\beta}{2}(R_N + \kappa_{1/2}[\widetilde{U}_{{\rm eq}}^{V}]\,N^{\frac{5}{2}}\sigma_N^{\frac{1}{2}}  + C_2\,N\ln N - N^2t^2)}\Big(\int_{\mathsf{A}} \dd\lambda\,e^{-\frac{\beta N}{2}\,\widetilde{U}_{{\rm eq}}^{V}(\lambda)}\Big)^N \\
& \leq e^{\frac{\beta}{2}(\tilde{R}_N + C_2\,N\ln N - N^2t^2)},
\end{split}
\end{equation*} 
with:
\beq
\label{LB346} \tilde{R}_N = R_N + \kappa_{1/2}[\widetilde{U}_{{\rm eq}}^{V}]\,N^{\frac{5}{2}}\sigma_N^{\frac{1}{2}} + \frac{2N}{\beta}\,\ln \ell(\mathsf{A}).
\eeq
Indeed, since $\widetilde{U}^{V;\mathsf{A}}$ is nonnegative on $\mathsf{A}$, we observed that the integral in bracket is bounded by the total length $\ell(\mathsf{A})$ of the range of integration, which is here finite. We now choose:
\beq
\label{chois1}\sigma_N = \frac{1}{N^3},\qquad \eta_N = \frac{1}{N},
\eeq
which guarantees that $\tilde{R}_N = O(N\ln N)$. Thus, there exists a positive constant $C_3$ such that, for $N$ large enough:
$$
\mu_{N,\beta}^{V;\mathsf{A}}\big(\mathcal{S}_N(t)\big) \leq e^{\frac{\beta}{2}(C_3\,N\ln N - N^2t^2)},
$$
which concludes the proof of Proposition~\ref{theoconc}. We may rephrase this result by saying that the probability of $\mathcal{S}_N(t)$ becomes small for $t$ larger than $\sqrt{\frac{2C_3\ln N}{N}}$.

The proof of \eqref{concfix} for fixed filling faction is similar since the same algebra holds, \textit{cf.}  Equation~\eqref{appveff} (with measures with same mass on the $\mathsf{A}_h$). 

\hfill $\Box$

\subsection{Large deviations for test functions}

\label{skp}

\subsubsection{Proof of Corollary~\ref{Ls1}}
\label{subsusb}
Since $\varphi$ is $b$-H\"older, we can use the comparison \eqref{Ca3} with $\sigma_N = N^{-3}$ chosen in Equation~\eqref{chois1}:
\beq\label{bnm}
\Big|\int_{\mathsf{A}} \dd[L_N - \widetilde{L}_N^{\mathrm{u}}](x)\,\varphi(x)\Big| \leq \frac{2\kappa_{b}[\varphi]}{(b + 1)N^{2b}}.
\eeq
Then, we compute in Fourier space and using Cauchy--Schwarz inequality:
$$
\Big|\int_{\mathsf{A}} \dd[\widetilde{L_N^{\mathrm{u}}} - \mu_{\mathrm{eq}}^V](x)\,\varphi(x)\Big| = \Big|\int_{\mathbb{R}} \dd p\big(\widehat{\widetilde{L}}_N^{\mathrm{u}} - \widehat{\mu_{\mathrm{eq}}^V}\big)(p)\,\overline{\widehat{\varphi}(p)} \Big| \leq |\varphi|_{1/2}\Big(\int_{\mathbb{R}} \frac{\dd p}{|p|}\,\big|(\widehat{\widetilde{L}}_N^{\mathrm{u}} - \widehat{\mu_{\mathrm{eq}}^V})(p)|^2\Big)^{\frac{1}{2}}, 
$$
and we recognise in the last factor the definition \eqref{pseu} of the pseudo-distance:
\beq
\label{346}\Big|\int_{\mathsf{A}} \dd[\widetilde{L}_N^{\mathrm{u}} - \mu_{\mathrm{eq}}^V](x)\,\varphi(x)\Big| \leq \sqrt{2}\,|\varphi|_{1/2}\,\mathfrak{D}[\widetilde{L}_N^{\mathrm{u}},\mu_{\mathrm{eq}}^V].
\eeq
Corollary~\ref{Ls1} then follows from this inequality combined with Lemma~\ref{theoconc}.

\subsubsection{Bounds on correlators and filling fractions (Proof of Corollary~\ref{apco} and \ref{cosq})}
\label{corbounds}

Let $\mathsf{A}_{\eta} = \{x \in \mathbb{R},\,\,\,:\,\,\, d(x,\mathsf{A}) \leq \eta\}$. As we chose $\sigma_N = N^{-3}$ and $\eta_N = N^{-1}$, the support of $\widetilde{L}_N^{\mathrm{u}}$ is included in $\mathsf{A}_{2/N^3}$. If $\mu$ is a probability measure, let $\mathcal{W}_{\mu}$ denote its Stieltjes transform. We have:
\beq
\label{firstine} \big(\mathcal{W}_{L_N} - \mathcal{W}_{\mu_{\mathrm{eq}}^V}\big)(x) = \int_{\mathsf{A}} \dd[L_N - \mu_{\mathrm{eq}}^V](\xi)\,\psi_{x}(\xi),\qquad \psi_{x}(\xi) = \psi^{R}_{x}(\xi) + {\rm i}\psi^{I}_{x}(\xi) = \frac{1}{x - \xi}\,.
\eeq
Since $\psi_{x}$ is Lipschitz on $\mathsf{A}_{2/N^3}$ with constant $\kappa_{1}[\psi_{x}] = d^{-2}(x,\mathsf{A}_{2/N^3})$, we have for  $d(x,\mathsf{A}) \geq \frac{3}{N^3}$:
\beq
\label{nji}\big|\mathcal{W}_{L_N}(x) - \mathcal{W}_{\widetilde{L}_N^{\mathrm{u}}}(x)\big| \leq \frac{3}{N^2 d^2(x,\mathsf{A})}\,.
\eeq
We focus on estimating $\mathcal{W}_{\widetilde{L}_N^{\mathrm{u}}} - \mathcal{W}_{\mu_{\mathrm{eq}}^V}$. We have the freedom to replace $\psi_{x}^{\bullet}$ by any function $\phi_{x}^{\bullet}$ which coincides with $\psi_{x}^{\bullet}$ on $\mathsf{A}_{2/N^3}$ since the support  of $\widetilde{L}_N^{\mathrm{u}}$ and $\mu_{\mathrm{eq}}^V$
are included in $\mathsf{A}_{2/N^3}$, and then :
\beq
\label{boundin}\big|\mathcal{W}_{\widetilde{L}_N^{{\rm u}}}(x) - \mathcal{W}_{\mu_{{\rm eq}}^V}(x)\big| \leq \sqrt{2}\big(|\phi_{x}^{R}|_{1/2} + |\phi_{x}^{I}|_{1/2}\big)\mathfrak{D}[\widetilde{L}_N^{{\rm u}},\mu_{{\rm eq}}^V]\,.
\eeq
We wish to choose $\phi_x^\bullet$ so that our estimates depend on the distance to $\mathsf{A}_{2/N^3}$ (whereas the choice of the function $\psi_x^\bullet$ would 
only gives bounds in terms of the distance to the real line, and therefore would not allow bounds for $x\in\mathbb R\backslash\mathsf{A}_{2/N^3}$). We now explain a suitable choice of $\phi_{x}^{\bullet}$. Let $a_{x,h,2/N^3} \in \mathsf{A}_{h,2/N^3}$ the point such that $d(x,\mathsf{A}_{h,2/N^3}) = |x - a_{x,h,2/N^3}|$. Then, for $\xi \in \mathsf{A}_{h,2/N^3}$, we have:
$$
\big|(\psi^{\bullet}_{x})'(\xi)\big| \leq \frac{1}{d(x,\mathsf{A}_{h,2/N^3})^2 + (\xi - a_{x,h,2/N^3})^2},
$$
and therefore:
\beq
\label{psirhs}\forall \xi \in \mathsf{A}_{2/N^3},\qquad \big|(\psi^{\bullet}_{x})'(\xi)\big| \leq \sum_{h = 0}^{g} \frac{1}{d(x,\mathsf{A}_{h,1/N^2})^2 + (\xi - a_{x,h,2/N^3})^2}\,.
\eeq
Then, we take a function $(\phi^{\bullet}_{x})'$ which coincides with $(\psi^{\bullet}_{x})'$ on $\mathsf{A}_{1/N^2}$, extends it continuously on $\mathbb{R}$, with compact support included in $\big[-\frac{M}{2},\frac{M}{2}\big]$ for some $M$ large enough, independent of $N$,  and such that:
\beq
\label{phirhs} \forall \xi \in \mathbb{R},\qquad \big|(\phi^{\bullet}_{x})'(\xi)\big| \leq \sum_{h = 0}^{g} \frac{1}{d(x,\mathsf{A}_{h,1/N^2})^2 + (\xi - a_{x,h,1/N^2})^2}\,.
\eeq
We denote $\phi^{\bullet}_{x}$ an antiderivative of this function, and use it in Equation~\eqref{boundin}. We compute:
\begin{equation}
\label{bineq}
\begin{split}
|\phi_{x}^{\bullet}|_{1/2}^2 & = \int_{\mathbb{R}} |p|\,|\widehat{\phi_{x}^{\bullet}}(p)|^2 \dd p = \int_{\mathbb{R}} \frac{1}{|p|}\,|\widehat{(\phi_{x}^{\bullet})'}(p)|^2 \dd p \\
& =  -2\int_{\mathbb{R}^2} \ln|\xi_1 - \xi_2|\,(\phi_x^{\bullet})'(\xi_1)(\phi_x^{\bullet})'(\xi_2) \dd\xi_1\dd\xi_2 \\
 & \leq 2 \int_{\mathbb{R}^2} \big|\ln|\xi_1 - \xi_2|\big|\,|(\phi^{\bullet}_x)'(\xi_1)|\,|(\phi_x^{\bullet})'(\xi_2)|\dd\xi_1\dd\xi_2.
\end{split}
\end{equation}
We note that, for any $a_1,a_2\in [-M,M]$, $b_1,b_2\in \mathbb R$, we can find a finite constant $C$ (depending only on $M$) such that 
$$\int_{\mathbb{R}^2} \big|\ln|\xi_1-\xi_2|\big|\frac{\dd\xi_1}{(\xi_1-a_1)^2+b_1^2}\frac{\dd\xi_2}{(\xi_2-a_2)^2+b_2^2}\le \frac{C}{b_1 b_2} (1+|\ln|b_1||+|\ln|b_2||)\,.$$
So, after we insert the bounds of \eqref{phirhs} in \eqref{bineq}, we obtain
$$
\big|\phi_{x}^{\bullet}\big|_{1/2}^{2} 
\leq \frac{D\,\ln d(x,\mathsf{A}_{2/N^3})}{d^2(x,\mathsf{A}_{2/N^3})}
$$
for some constant $D > 0$ depending only on $\mathsf{A}_{2/N^3}$. If $d(x,\mathsf{A}) \geq \frac{3}{N^3}$ and for $N$ large enough, we can also write with a larger constant $D$:
$$
\big|\phi_{x}^{\bullet}\big|_{1/2}^{2} \leq \frac{D\,\ln d(x,\mathsf{A})}{d^2(x,\mathsf{A})}.
$$

Then, with Equations~\eqref{bnm}, \eqref{firstine} and \eqref{346}:
\begin{equation*}
\begin{split} 
\Big|\frac{1}{N}W_1(x) - 
\mathcal{W}_{\mu_{\mathrm{eq}}^V}(x)\Big| & = \Big|\mu_{N,\beta}^{V;\mathsf{A}}\big(\mathcal{W}_{L_N}(x) - \mathcal{W}_{\mu_{\mathrm{eq}}^V}(x)\big)\Big|  \\
& \leq \frac{3}{N^2 d^2(x,\mathsf{A})} + 2D\,\sqrt{\frac{\ln N}{N}}\,\frac{\sqrt{|\ln d(x,\mathsf{A})}|}{d(x,\mathsf{A})}.
\end{split} 
\end{equation*} 
If we restrict ourselves to $x \in \mathbb{C}\setminus\mathsf{A}$ such that
$$
d(x,\mathsf{A}) \geq \frac{D'}{\sqrt{N^{2} \ln N}} 
$$
for some constant $D' > 0$, then:
$$
\Big|\frac{1}{N}\, W_1(x) - \mathcal{W}_{\mu_{\mathrm{eq}}^V}(x)\Big| \leq (2D + D'')\,\sqrt{\frac{\ln N}{N}}\,\frac{\sqrt{|\ln d(x,\mathsf{A})}|}{d(x,\mathsf{A})}
$$
for some constant $D'' > 0$.

Now let us consider  the higher correlators. For any $n \geq 2$,   the same arguments show that there exists a finite constant $c_n$ so that for
any $x_i$ such that $d(x_i,\mathsf{A}) \geq \frac{D'}{\sqrt{N^{2} \ln N}} $,
$$
m_n(x_1,\ldots,x_n)=\mu_{N,\beta}^{V;\mathsf{A}}\Big[\prod_{i=1}^n (\mathcal{W}_{L_N} - \mathcal{W}_{\mu_{\mathrm{eq}}^V})(x_i)\Big]
$$
satisfies 
$$
|m_n(x_1,\ldots,x_n)|\le c_n (N\ln N)^{\frac{n}{2}}\,\prod_{i = 1}^n \frac{\sqrt{|\ln d(x_i,\mathsf{A})|}}{d(x_i,\mathsf{A})}.
$$
As 
$W_n^{V;\mathsf{A}}$ is an homogeneous polynomial of degree $n$ in the moments $(m_k)_{1 \leq k \leq n}$, we conclude that:
$$
|W_n(x_1,\ldots,x_n)| \leq \gamma_n\,(N\ln N)^{\frac{n}{2}}\,\prod_{i = 1}^n \frac{\sqrt{|\ln d(x_i,\mathsf{A})|}}{d(x_i,\mathsf{A})}.
$$
for some constant $\gamma_n > 0$, which depends only on $\mathsf{A}$. This concludes the proof of Corollary~\ref{apco}.

Similarly, to have a control on filling fractions, we write:
$$
N_h - N\epsilon_{\star,h} = N \int_{\mathsf{A}} \dd[L_N - \mu_{\mathrm{eq}}^V](\xi)\,\mathbf{1}_{\mathsf{A}_h}(\xi).
$$
Following the same steps to extend the function $x \mapsto \mathbf{1}_{\mathsf{A}_h}(x)$ initially defined on $\mathsf{A}$ by a function defined on $\mathbb{R}$ and with finite $|\cdot|_{1/2}$ norm, we can apply Corollary~\ref{Ls1} to deduce Corollary~\ref{cosq}. \hfill $\Box$

\section{Dyson--Schwinger equations for \texorpdfstring{$\beta$}{beta} ensembles}
\label{S31}
Let $\mathsf{A} = \bigcup_{h = 0}^{g} \mathsf{A}_h$ be a finite union of pairwise disjoint bounded segments, and $V$ be a $\mathcal{C}^1$ function of $\mathsf{A}$. Dyson--Schwinger equations for the initial model $\mu_{N,\beta}^{V;\mathsf{A}}$ can be derived by integration by parts. Since the derivation does not use any information on the location of the $\lambda$s, it is equally valid for the model with fixed filling fractions $\mu_{N,\beta;\bm{\epsilon}}^{V;\mathsf{A}}$, in which $N\epsilon_h = N_h$ are integers.

Since these equations are well-known (and have been reproved in \cite{BG11}), we state them without proof. They can be written in several equivalent forms, and here we recast them in a way which is convenient for our analysis. We assume that $V$ extends to a holomorphic function in a neighbourhood of $\mathsf{A}$, so that they can be written in terms of contour integrals of correlators --- an extension to $V$ harmonic will be mentioned in \S~\ref{S62}. We introduce (arbitrarily for the moment) a partition $\partial \mathsf{A} = (\partial \mathsf{A})_+ \cup (\partial \mathsf{A})_-$ of the set of edges of the range of integration, and
let 
\beq
\label{L2def} L(x) = \prod_{a \in (\partial \mathsf{A})_-} (x - a),\qquad L_1(x,\xi) = \frac{L(x) - L(\xi)}{x - \xi},\qquad L_2(x;\xi_1,\xi_2) = \frac{L_1(x,\xi_1) - L_1(x,\xi_2)}{\xi_1 - \xi_2}.
\eeq

\begin{theorem}\label{dystheo}
Dyson--Schwinger equation in one variable. For any $x \in \mathbb{C}\setminus \mathsf{A}$, we have:
\begin{equation}
\label{234}
\begin{split}
0 & = W_2(x,x) + \big(W_1(x)\big)^2 + \Big(1 - \frac{2}{\beta}\Big)\partial_x W_1(x)  \\
& \quad - N\oint_{\mathsf{A}} \frac{\dd\xi}{2{\rm i}\pi}\,\frac{L(\xi)}{L(x)}\,\frac{V'(\xi)\,W_1(\xi)}{x - \xi} - \frac{2}{\beta} \sum_{a \in (\partial \mathsf{A})_+} \frac{L(a)}{x - a}\,\partial_{a}\ln Z_{N,\beta}^{V;\mathsf{A}} \\
& \quad + \Big(1 - \frac{2}{\beta}\Big) \oint_{\mathsf{A}} \frac{\dd\xi}{2{\rm i}\pi}\,\frac{L_2(x;\xi,\xi)}{L(x)}\,W_1(\xi) \\
& \quad - \oiint_{\mathsf{A}^2} \frac{\dd\xi_1\dd\xi_2}{(2{\rm i}\pi)^2}\,\frac{L_2(x;\xi_1,\xi_2)}{L(x)}\big(W_2(\xi_1,\xi_2) + W_1(\xi_1)W_1(\xi_2)\big). \end{split}
\end{equation}
\end{theorem}
And similarly, for higher correlators:
\begin{theorem} 
\label{SDN}Dyson--Schwinger equation in $n \geq 2$ variables. For any $x,x_2,\ldots,x_n \in \mathbb{C}\setminus \mathsf{A}$, if we denote $I = \ldbrack 2,n \rdbrack$, we have:
\begin{equation}
\label{235}
\begin{split}
 0 & = W_{n + 1}(x,x,x_I) + \sum_{J\subseteq I} W_{|J| + 1}(x,x_J)W_{n - |J|}(x,x_{I\setminus J}) + \Big(1 - \frac{2}{\beta}\Big)\partial_x W_n(x,x_I) \\
& \quad  - N\oint_{\mathsf{A}} \frac{\dd\xi}{2{\rm i}\pi}\,\frac{L(\xi)}{L(x)}\,\frac{V'(\xi)\,W_n(\xi,x_I)}{x - \xi} - \frac{2}{\beta}\sum_{a \in (\partial \mathsf{A})_+} \frac{L(a)}{x - a}\,\partial_{a} W_{n - 1}(x_I)  \\
& \quad + \frac{2}{\beta} \sum_{i \in I} \oint_{\mathsf{A}} \frac{\dd\xi}{2{\rm i}\pi}\,\frac{L(\xi)}{L(x)}\,\frac{W_{n - 1}(\xi,x_{I\setminus\{i\}})}{(x - \xi)(x_i - \xi)^2} + \Big(1 - \frac{2}{\beta}\Big)\oint_{\mathsf{A}} \frac{\dd\xi}{2{\rm i}\pi}\,\frac{L_2(x;\xi,\xi)}{L(x)}\,W_n(\xi,x_I)  \\
& \quad - \oiint_{\mathsf{A}^2} \frac{\dd\xi_1\dd\xi_2}{(2{\rm i}\pi)^2}\,\frac{L_2(x;\xi_1,\xi_2)}{L(x)}\Big(W_{n + 1}(\xi_1,\xi_2,x_I) + \sum_{J \subseteq I} W_{|J| + 1}(\xi_1,x_J)W_{n - |J|}(\xi_2,x_{I\setminus J})\Big).
\end{split}
\end{equation}
\end{theorem}
The last line in Equation~\eqref{234} or \eqref{235} is a rational fraction in $x$, with poles at $a \in \partial \mathsf{A}$, whose coefficients are linear combination of moments of $\lambda_i$.

We stress that the Dyson--Schwinger equations are exact for any finite $N$, and hold for any choice of splitting $\partial \mathsf{A} = (\partial \mathsf{A})_+ \cup (\partial†\mathsf{A})_-$. Note here that $L,L_{1},L_{2}$ depend on $A_{-}$ so that in fact all the terms  except those in the first line of Dyson--Schwinger equations  depend a priori on this splitting. Later, when we perform a large $N \rightarrow \infty$ asymptotic analysis, we are led to distinguish soft edges and hard edges (this is a property of the equilibrium measure). It will be then be convenient to declare $\partial\mathsf{A}_-$ to be the set of hard edges and $\partial\mathsf{A}_+$ the set of soft edges. This will have for consequence that the simple poles in \eqref{234}-\eqref{235} at $x = a \in \partial\mathsf{A}_+$ have exponentially small residues and therefore can be neglected to any order $O(N^{-K})$ in the asymptotic analysis.

\section{Fixed filling fractions: expansion of correlators}
\label{S3}
\label{secloop}

\subsection{Notations, assumptions and operator norms}

The model with fixed filling fractions corresponds to the case where we condition the number of eigenvalues in each segment $\mathsf{A}_h$ to be a given integer $N_{h}$. We set $\epsilon_h = N_h/N$ for $h \in \llbracket 0,g \rrbracket$ and $\bm{\epsilon} = (\epsilon_1,\ldots,\epsilon_{g})$. Throughout this section, the equilibrium measure, the correlators $W_{n} = W_{n;\bm{\epsilon}}$, etc. all depend on $\bm{\epsilon}$. The vector $\bm{\epsilon}$ itself could also depend on $N$, but this dependence will remain implicit. Accordingly, all coefficients we will find in the asymptotic expansion of the correlators will implicitly be functions of $\bm{\epsilon}$.

As explained in Section~\ref{S31}, the correlators in the fixed filling fractions model satisfy the same Dyson--Schwinger equation as in the initial model. We analyse them under the following assumptions:
\begin{hypothesis}
\label{hyu222}
\begin{itemize}
\item[] \phantom{sss}
\item[$\bullet$] $\mathsf{A}$ is a disjoint finite union of bounded segments $\mathsf{A}_h = [a_h^-,a_h^+]$.
\item[$\bullet$] (Real-analyticity) $V\,:\,\mathsf{A} \rightarrow \mathbb{R}$ extends to a holomorphic function in a neighbourhood $\mathsf{U} \subseteq \mathbb{C}$ of $\mathsf{A}$.
\item[$\bullet$] (Expansion for the potential) There exists a sequence $(V^{\{k\}})_{k \geq 0}$ of holomorphic functions in $\mathsf{U}$ and constants $(v^{\{k\}})_{k \geq 1}$, so that, for any $K \geq 0$:
$$
\sup_{\xi \in \mathsf{U}} \Big| V(\xi) - \sum_{k = 0}^{K} N^{-k}\,V^{\{k\}}(\xi)\Big| \leq v^{\{K + 1\}}\,N^{-(K + 1)}.
$$
\item[$\bullet$] ($(g + 1)$-cut regime) The probability measure $\mu_{\rm{eq};\bm{\epsilon}}^V$  is supported on $\mathsf{S}$ which is a disjoint union of $(g + 1)$ segments $\mathsf{S}_h = [\alpha_h^-,\alpha_h^+] \subseteq \mathsf{A}_h$. We set $W_{1}^{\{-1\}}$ to be its Stieltjes transform and recall 
 that 
  $$
  \lim_{N \rightarrow \infty} (N^{-1}\,W_{1}(x)-W_1^{\{-1\}}(x))=0,
  $$   uniformly  for $x$ in any compact of $\mathbb{C}\setminus \mathsf{A}$.
 \item[$\bullet$] (Off-criticality) $y(x) = \frac{(V^{\{0\}})'(x)}{2} - W_1^{\{-1\}}(x)$ takes the form:
\beq
\label{ydq}y(x) = S(x)\,\prod_{h = 0}^{g} \sqrt{(x - \alpha_{h}^+)^{\rho_{h}^+}(x - \alpha_{h}^-)^{\rho_{h}^-}},
\eeq
where $S$ does not vanish on $\mathsf{A}$, $\alpha_{h}^{\bullet}$ are all pairwise distinct, and $\rho_{h}^{\bullet} = -1$ if $\alpha_{h}^{\bullet} \in \partial \mathsf{A}$, and $\rho_{h}^{\bullet} = 1$ otherwise.
\end{itemize}
\end{hypothesis}
Later in Section~\ref{musqo}, we will come back to the analysis of the initial model, which has $\mu_{{\rm eq}}^V = \mu_{{\rm eq};\bm{\epsilon}_{\star}}^V$ as equilibrium measure. We will show in Lemma~\ref{Lhq2} that the initial Hypotheses~\ref{hmainc1}--\ref{hmainc4} imply the present Hypotheses~\ref{hyu222} for $\bm{\epsilon}$ in some neighbourhood of $\bm{\epsilon}_{\star}$, in particular the off-criticality assumption \eqref{ydq}  is verified, 
making the results of the present Section applicable.

\begin{definition} 
If $\delta > 0$, we introduce the norm $\p \cdot \p_{\delta}$ on the space $\mathcal{H}_{m_1,\ldots,m_n}^{(n)}(\mathsf{A})$ of holomorphic functions on $(\mathbb{C}\setminus \mathsf{A})^n$ which behave like $O(\frac{1}{x_i^{m_i}})$ when $x_i \rightarrow \infty$:
$$
\p f \p_{\delta} = \sup_{\min_i d(x_i,\mathsf{A}) \geq \delta} |f(x_1,\ldots,x_n)| = \max_{\min_i d(x_i,\mathsf{A}) = \delta} |f(x_1,\ldots,x_n)|,
$$
the last equality following from the maximum principle. If $n \geq 2$, we denote $\mathcal{H}_m^{(n)} = \mathcal{H}_{m,\ldots,m}^{(n)}$.
\end{definition}
From Cauchy residue formula, we have a naive bound on the derivatives of a function $f \in \mathcal{H}_{1}^{(1)}$ in terms of $f$ itself:
$$
\p \partial_x^m f(x) \p_{\delta} \leq \frac{2^{m + 1}C}{\delta^{m + 1}}\,\p f \p_{\delta/2}.
$$
In practice, we will take $\delta$ independent of $N$, and therefore the constants depending on $\delta$ will not matter. 

Our goal in the next section is to establish under Hypothesis~\ref{hyu222} below an asymptotic expansion for the correlators when $N \rightarrow \infty$, exploiting the Dyson--Schwinger equations. We already notice that it is convenient to choose
$$
(\partial \mathsf{A})_{\pm} = \{a_{h}^{\bullet} \in (\partial \mathsf{A})\quad | \quad \rho_{h}^{\bullet} = \pm 1\},
$$
as bipartition of $\partial \mathsf{A}$ to write down the Dyson--Schwinger equation, since the terms involving $\partial_a \ln Z$ and $\partial_a W_{n - 1}$ for $a \in (\partial \mathsf{A})_+$ will be exponentially small according to Corollary~\ref{pro}. If $a = a_{h}^{\bullet}$, we denote $\alpha(a) = \alpha_{h}^{\bullet}$.

To perform the asymptotic analysis to all order, we need a rough \textit{a priori} estimate on the correlators. We have established in \S~\ref{Scons} (actually under weaker assumptions than Hypothesis~\ref{hyu222})  that for any $\delta>0$:
\beq
\label{59}\|W_1 - N\,W_1^{\{-1\}}\|_\delta \le C_1(\delta)\,\sqrt{N \ln N},
\eeq
and for any $n \geq 2$:
\beq
\label{59b}\|{W}_{n} \|_\delta \leq C_n(\delta)\,(N\ln N)^{\frac{n}{2}}\,.
\eeq

\subsection{Some relevant linear operators}
\label{S33}

In this subsection, we give the list of linear operators that are used in \S~\ref{parso} to recast the Dyson--Schwinger equations in a form suitable for the asymptotic analysis. 
The precise expression of these operators is not essential, but we  establish  
 bounds on suitable operator norms that are needed later in the analysis.

\subsubsection{Periods}

We fix once for all a neighbourhood $\mathsf{U}$ of $\mathsf{A}$ so that $S$ has no zeroes in $\mathsf{U}$, and pairwise non-intersecting contours $\bm{\mathcal{A}} = (\mathcal{A}_h)_{1 \leq h \leq g}$ surrounding $\mathsf{A}_h$ in $\mathsf{U}$. It is not necessary to introduce a contour surrounding $\mathsf{A}_0$, since it is homologically equivalent to $-\sum_{h = 1}^g \mathcal{A}_{h}$ in $\widehat{\mathbb{C}}\setminus\mathsf{A}$. We define the period operator $\mathcal{L}_{\bm{\mathcal{A}}}\,:\,\mathcal{H}^{(1)}_{1} \rightarrow \mathbb{C}^{g}$ by the formula:
\begin{equation}\label{perioddef}
\mathcal{L}_{\bm{\mathcal{A}}}[f] = \Big(\oint_{\mathcal{A}_{1}} \frac{\dd\xi}{2{\rm i}\pi}\,f(\xi)\,,\,\ldots\,,\,\oint_{\mathcal{A}_{g}} \frac{\dd\xi}{2{\rm i}\pi}\,f(\xi)\Big).
\end{equation}
By Cauchy residue formula, the periods of the Stieltjes transform of the empirical measure are the filling fractions:
$$
\mathcal{L}_{\bm{\mathcal{A}}}\Big[x \mapsto \int_{\mathsf{A}} \frac{\dd L_N(\xi)}{x - \xi}\Big] = \bm{\epsilon}.
$$
Since the $(W_n)_{n \geq 1}$ are cumulants and the $\bm{\epsilon}$ are fixed (see the remark in Section~\ref{fixfil}), we have:
\begin{equation}\label{lkh}
\mathcal{L}_{\bm{\mathcal{A}}}[ W_n(\bullet,x_2,\ldots,x_n)] = \delta_{n,1}\,N\bm{\epsilon}\,.
\end{equation}
In other words, we know that in the model with fixed filling fractions, the correlators (as functions of one of their variables) have to satisfy the $g$ constraints \eqref{lkh}.

\begin{definition}
If $\bm{X}$ is an element of $(\mathbb{C}^{g})^{\otimes n}$, we define its $L^1$-norm:
$$
|\bm{X}|_{1} = \sum_{1 \leq h_1,\ldots,h_n \leq g} |X_{h_1,\ldots,h_n}|.
$$
\end{definition}

\subsubsection{The operator \texorpdfstring{$\mathcal{K}$}{K}}

\label{opK}

We introduce an operator $\mathcal{K}$ which is the linearisation around the equilibrium measure of the generator of Dyson--Schwinger equations. It is defined on functions $f \in \mathcal{H}^{(1)}_{2}(\mathsf{A})$ by the formula:
\begin{equation}
\label{Kdeff}
\mathcal{K}[f](x) = 2W_1^{\{-1\}}(x)f(x) - \frac{1}{L(x)}\oint_{\mathsf{A}}\frac{\mathrm{d}\xi}{2{\rm i}\pi}\Big[\frac{L(\xi)\,(V^{\{0\}})'(\xi)}{x - \xi} + P^{\{-1\}}(x;\xi)\Big]f(\xi),
\end{equation}
where $x$ is outside the contour of integration, and:
$$
P^{\{-1\}}(x;\xi) = \oint_{\mathsf{A}} \frac{\dd\eta}{2{\rm i}\pi}\,2L_2(x;\xi,\eta)\,W_1^{\{-1\}}(\eta).
$$
We remind that $L(x) = \prod_{a \in (\partial \mathsf{A})_-} (x - \alpha(a))$ and $L_2$ was defined in Equation~\eqref{L2def}.
Notice that $W_1^{\{-1\}}(x) \sim \frac{1}{x}$ when $x \rightarrow \infty$, and $P^{\{-1\}}(x,\xi)$ is a polynomial in two variables, of maximal total degree $|(\partial \mathsf{A})_{-}| - 2$ (and it is zero if $|(\partial \mathsf{A})_-| < 2$). Hence, we have at least $\mathcal{K}[f](x) = O(\frac{1}{x})$ when $x \rightarrow \infty$. This gives us a linear operator:
$$
\mathcal{K}\,:\,\mathcal{H}_{2}^{(1)}(\mathsf{A}) \rightarrow \mathcal{H}_{1}^{(1)}(\mathsf{A}).
$$
Notice also that:
\beq
\label{514} y(x) = \frac{\big(V^{\{0\}}\big)'(x)}{2} - W_1^{\{-1\}}(x) = S(x)\,\sqrt{\frac{\tilde{L}(x)}{L(x)}},
\eeq
where $\tilde{L}(x) = \prod_{a \in (\partial \mathsf{A})_+} (x - \alpha(a))$, and by the off-criticality assumption the zeroes of $S$ are away from $\mathsf{A}$. If we define
$$
\sigma(x) = \sqrt{\prod_{a \in (\partial \mathsf{A})} (x - \alpha(a))} = \sqrt{\tilde{L}(x)L(x)},
$$
we can rewrite
\begin{equation}
\label{themissin}
\frac{\sigma(x)}{y(x)} = \frac{L(x)}{S(x)}.
\end{equation}
Then:
\beq
\label{420}\mathcal{K}[f](x) = -2y(x)f(x) + \frac{\mathcal{Q}[f](x)}{L(x)},
\eeq
where:
$$
\mathcal{Q}[f](x) = - \oint_{\mathsf{A}} \frac{\dd\xi}{2{\rm i}\pi}\,\Big[\frac{L(\xi)\,(V^{\{0\}})'(\xi) - L(x)\,(V^{\{0\}})'(x)}{x - \xi} + P^{\{-1\}}(x;\xi)\Big]\,f(\xi).
$$
For any $f \in \mathcal{H}_{2}^{(1)}(\mathsf{A})$, $x \mapsto \mathcal{Q}[f](x)$ is holomorphic in a neighbourhood of $\mathsf{A}$. It is clear from  Equation~\eqref{Kdeff} that $\mathrm{Im}\,\mathcal{K} \subseteq \mathcal{H}_{1}^{(1)}(\mathsf{A})$. Let $\varphi \in \mathrm{Im}\,\mathcal{K}$, and $f \in \mathcal{H}_{2}^{(1)}(\mathsf{A})$ such that $\varphi = \mathcal{K}[f]$. We can write:
$$
\sigma(x)\,f(x) = \Res_{\xi = x} \frac{\dd\xi}{\xi - x}\,\sigma(\xi)\,f(\xi) = \psi(x) - \oint_{\mathsf{A}} \frac{\dd\xi}{2{\rm i}\pi} \frac{\sigma(\xi)\,f(\xi)}{\xi - x},
$$
where:
\beq\label{eqpsi}
\psi(x) = - \Res_{\xi = \infty} \frac{\dd\xi}{\xi - x}\,\sigma(\xi)\,f(\xi).
\eeq
Since $f(x) = O(\frac{1}{x^2})$, $\psi(x)$ is a polynomial in $x$ of degree at most $g  - 1$.  Recall that $\mathcal{K}[f]=\varphi$.
We then compute:
\begin{equation}
\label{519}
\begin{split} 
\sigma(x)f(x) & = \psi(x) - \oint_{\mathsf{A}} \frac{\dd\xi}{2{\rm i}\pi}\,\frac{1}{\xi - x}\,\frac{\sigma(\xi)}{2y(\xi)}\Big(- \varphi(\xi) + \frac{\mathcal{Q}[f](\xi)}{L(\xi)}\Big)  \\
& = \psi(x) + \oint_{\mathsf{A}} \frac{\dd\xi}{2{\rm i}\pi}\,\frac{1}{\xi - x}\,\frac{1}{2S(\xi)}\,\Big[L(\xi)\,\varphi(\xi) - \mathcal{Q}[f](\xi)\Big]\\
& = \psi(x) + \oint_{\mathsf{A}} \frac{\dd\xi}{2{\rm i}\pi}\,\frac{1}{\xi - x}\,\frac{L(\xi)}{2S(\xi)}\,\varphi(\xi),
\end{split}
\end{equation}
using the fact that $S$ has no zeroes on $\mathsf{A}$ and $\mathcal{Q}[f]$ is analytic in a neighbourhood of $\mathsf{A}$. Let us denote $\mathcal{G}\,:\,\mathrm{Im}\,\mathcal{K} \rightarrow \mathcal{H}_{2}^{(1)}(\mathsf{A})$ the linear operator defined by:
\beq
\label{Gdeq}\mathcal{G}[\varphi](x) = \frac{1}{\sigma(x)}\,\oint_{\mathsf{A}} \frac{\dd\xi}{2{\rm i}\pi}\,\frac{1}{\xi - x}\,\frac{L(\xi)}{2S(\xi)}\,\varphi(\xi).
\eeq
One deduces:  
\beq
\label{426f}f(x) = \frac{\psi(x)}{\sigma(x)} + (\mathcal{G}\circ\mathcal{K})[f](x).
\eeq

\subsubsection{The extended operator \texorpdfstring{$\widehat{\mathcal{K}}$}{Khat} and its inverse}
\label{opK222}
It was observed in \cite{Ake96} that $\psi(x)\dd x/\sigma(x)$ defines a holomorphic one-form on the compactification $\Sigma$ of the Riemann surface of equation $\sigma^2 = \prod_{a \in (\partial \mathsf{A})} (x - \alpha(a))$. The space $H^1(\Sigma)$ of holomorphic one-forms on $\Sigma$ has dimension $g$ if all $\alpha(a)$ are pairwise distinct (which is the case by off-criticality) and the number of cuts is $(g + 1)$. So, if $g \geq 1$, $\mathcal{K}$ is not invertible. But we can define an extended operator:
\bea
\widehat{\mathcal{K}}\,:\,\mathcal{H}_{2}^{(1)}(\mathsf{A}) & \longrightarrow & \mathrm{Im}\,\mathcal{K} \times \mathbb{C}^g \nonumber \\
\label{eq522} f & \longmapsto & \big(\mathcal{K}[f],\mathcal{L}_{\bm{\mathcal{A}}}[f]\big).
\eea
Since $\big(x^{j - 1}\dd x/\sigma(x)\big)_{0 \leq j \leq g - 1}$ are linearly independent over $\mathbb{C}$ and holomorphic one-forms on $\Sigma$, it forms a basis of $H^1(\Sigma)$ which can be thus identified with $\sigma(x)^{-1}\cdot\mathbb{C}_{g - 1}[x]$, where $\mathbb{C}_{g - 1}[x]$ is the set of polynomials in $x$ of degree $\leq (g-1)$. On the other hand, the family of linear forms $\mathcal{L}_{\bm{\mathcal{A}}}$ defined in Equation~\eqref{perioddef} are linearly independent (see \textit{e.g.} \cite{FarkasKra}), so they determine a unique basis
\beq
\label{varpiphi} \varpi_h (x)= \frac{\psi_h(x)}{\sigma(x)} \in \sigma(x)^{-1}\cdot\mathbb{C}_{g - 1}[x]
\eeq
such that:
\beq
\label{sqfd}\forall h,h' \in \ldbrack 1,g \rdbrack,\qquad \oint_{\mathcal{A}_h} \varpi_{h'}(x)\,\dd x = \delta_{h,h'}.
\eeq
Therefore, we can define an operator $\mathcal{L}_{\bm{\mathcal{A}}}^{-1}\,:\,\mathbb{C}^{g} \rightarrow \sigma(x)^{-1}\cdot \mathbb{C}_{g - 1}[x] \subseteq \mathcal{H}_2^{(1)}$ by the formula:
\beq
\label{jujus}\mathcal{L}_{\bm{\mathcal{A}}}^{-1}[\bm{w}] = \sum_{h = 1}^g w_h\,\varpi_{h}(x).
\eeq

We deduce that $\widehat{\mathcal{K}}$ is an isomorphism. Indeed,  $\widehat{\mathcal{K}}[f]=(\varphi, \bm{w})$ iff we have, according to Equation~\eqref{Gdeq}:
\begin{equation}\label{define}f(x)=\frac{\psi(x)}{\sigma(x)} +(\mathcal{G}\circ\mathcal{K})[f](x),\qquad {\rm and}\qquad  \mathcal{L}_{\bm{\mathcal{A}}} [f]=\bm{w}\,.\end{equation}
Plugging the first equality into the second, we deduce
$$
\mathcal{L}_{\bm{\mathcal{A}}}\Big[\frac{\psi}{\sigma} +(\mathcal{G}\circ\mathcal{K})[f]\Big]=\bm{w},
$$
which is equivalent to
$$
\frac{\psi(x)}{\sigma(x)}=   \mathcal{L}_{\bm{\mathcal{A}}}^{-1} \big[\bm{w}- \mathcal{L}_{\bm{\mathcal{A}}}\big[(\mathcal{G}\circ\mathcal{K})[f]\big]\big]=
 \mathcal{L}_{\bm{\mathcal{A}}}^{-1} \left[\bm{w}- \mathcal{L}_{\bm{\mathcal{A}}}\big[\mathcal{G}[\varphi]\big]\right]\,.
 $$
Plugging this back into Equation~\eqref{define},  we deduce that $\widehat{\mathcal{K}}$ is invertible, with   inverse  given by:
\beq
\label{inveJ}\widehat{\mathcal{K}}^{-1}[\varphi,\bm{w}](x) = \mathcal{L}_{\bm{\mathcal{A}}}^{-1}\left[\bm{w} - \mathcal{L}_{\bm{\mathcal{A}}}\big[\mathcal{G}[\varphi]\big]\right](x) + \mathcal{G}[\varphi](x),
\eeq
where $ \mathcal{G}$ is defined in Equation~\eqref{Gdeq}. We will use the notation $\widehat{\mathcal{K}}^{-1}_{\bm{w}}[\varphi] = \widehat{\mathcal{K}}^{-1}[\varphi,\bm{w}]$. In other words, $\widehat{\mathcal{K}}_{\bm{w}}^{-1}[\varphi] = f$ is the unique solution of $\mathcal{K}[f] = \varphi$  with $\bm{\mathcal{A}}$-periods equal to $\bm{w}$. It is equal to $\psi(x)\sigma(x)^{-1}+\mathcal{G}[\varphi](x)$ for some polynomial $\psi(x)$ of degree smaller than $g-1$ so that the $\bm{\mathcal{A}}$-periods equal to $\bm{w}$.
 The continuity of this inverse operator is the key ingredient of our method:
\begin{lemma}
\label{ImK} $\mathrm{Im}\,\mathcal{K}$ is closed in $\mathcal{H}_{2}^{(1)}(\mathsf{A})$, and for $\delta > 0$ small enough, there exist constants $C,C',C'' > 0$ such that:
\beq\label{binv}
\forall (\varphi,\bm{w}) \in \mathrm{Im}\,\mathcal{K}\times \mathbb{C}^g,\qquad \p \widehat{\mathcal{K}}^{-1}_{\bm{w}}[\varphi] \p_{\delta} \leq \delta^{-\kappa}\Big\{\big(CD_{c}(\delta) + C'\big)\p \varphi \p_{\delta} + C'' |\bm{w}|_1\Big\},
\eeq
with exponent $\kappa = \frac{1}{2}$ and $D_c(\delta)$ defined in Equation~\eqref{Dcdelta}. When the potential is off-critical, $D_c(\delta)$ remains bounded.
\end{lemma}
\begin{remark}
In the analysis of the model with fixed filling fractions, we will only make use of $\widehat{\mathcal{K}}_{\bm{0}}^{-1}$.
\end{remark}
\noindent\textbf{Proof.} If one is interested in controlling the large $N$ expansion of the correlators explicitly in terms of the distance of $x_i$s to $\mathsf{A}$, it is useful to give an explicit bound on the norm of $\widehat{\mathcal{K}}^{-1}_{\bm{w}}$. Let $\delta_0 > 0$ be small enough but fixed once for all, and let us  move the contour in Equation~\eqref{Gdeq} to a contour  staying at distance larger than $\delta_0$ from $\mathsf{A}$. If we choose now a point $x$ so that $d(x,\mathsf{A}) < \delta_0$, we can write:
$$
\mathcal{G}[\varphi](x) = \frac{\varphi(x) L(x)}{2S(x)\sigma(x)} - \frac{\varphi(x)}{\sigma(x)}\oint_{d(\xi,\mathsf{A}) = \delta_0} \frac{\dd\xi}{2{\rm i}\pi}\,\frac{L(\xi)}{2S(\xi)}\,\frac{1}{x - \xi} + \frac{1}{\sigma(x)}\oint_{d(\xi,\mathsf{A}) = \delta_0} \frac{\dd\xi}{2{\rm i}\pi}\,\frac{L(\xi)}{2S(\xi)}\,\frac{\varphi(\xi)}{x - \xi}.
$$
Hence, there exist constants $\tilde{C},\tilde{C}' > 0$ depending only on the position of the pairwise disjoint segments $\mathsf{A}_h$ such that, for any $\delta > 0$ smaller than $\frac{\delta_0}{2}$:
\beq
\label{5313G}\p\mathcal{G}[\varphi ]\p_{\delta} \leq (\tilde{C} D_c(\delta) + \tilde{C}')\,\delta^{-\frac{1}{2}}\,\p \varphi \p_{\delta}, \eeq
where
\begin{equation}
\label{Dcdelta}
D_c(\delta) = \sup_{d(\xi,\mathsf{A}) = \delta} \Big|\frac{L(\xi)}{S(\xi)}\Big|\,.
\end{equation}
For $\delta$ small enough but fixed, $D_c(\delta)$ blows up when the parameters of the model are tuned to achieve a critical point, \textit{i.e.} it measures a distance to criticality. Besides, we have for the operator $\mathcal{L}_{\bm{\mathcal{A}}}$:
\beq
\label{5313L}\big|\mathcal{L}_{\bm{\mathcal{A}}}[f]\big|_{1} \leq \tilde{C}\,\p f \p_{\delta}
\eeq
and for  $\mathcal{L}_{\bm{\mathcal{A}}}^{-1}$ written in Equation~\eqref{jujus} we find:
\beq 
\label{532D} \p \mathcal{L}_{\bm{\mathcal{A}}}^{-1}[\bm{w}] \p_{\delta} \leq \frac{\max_{1 \leq h \leq g} \p \psi_{h} \p_{\mathsf{U}}^{\infty}}{\mathrm{inf}_{d(\xi,\mathsf{A}) = \delta} |\sigma(x)|}\,|\bm{w}|_{1},
\eeq
and the denominator behaves like $\delta^{-\frac{1}{2}}$ when $\delta \rightarrow 0$. We then deduce from Equation~\eqref{inveJ} the existence of constants $C,C',C'' > 0$ so that:
\begin{equation}
\label{530}
\begin{split}
\p \widehat{\mathcal{K}}^{-1}_{\bm{w}}[\varphi] \p_{\delta} &\leq (\tilde{C}D_c(\delta) + \tilde{C}')\delta^{-\frac{1}{2}}\p \varphi \p_{\delta} + \delta^{-\frac{1}{2}}\,|\bm{w}- \mathcal{L}_{\bm{\mathcal{A}}}\big[\mathcal{G}[\varphi]\big]|_1 \\
&\le  (CD_c(\delta) + C')\delta^{-\frac{1}{2}}\p \varphi \p_{\delta} + C''\delta^{-\frac{1}{2}}\,|\bm{w}|_1
\end{split}
\end{equation}
\hfill $\Box$

\vspace{0.2cm}
\label{oda}
\noindent \textbf{Remark.} From the expression \eqref{inveJ} for the inverse, we observe that, if $\varphi$ is holomorphic in $\mathbb{C}\setminus\mathsf{S}$, so is $\widehat{\mathcal{K}}^{-1}_{\bm{w}}[\varphi]$ for any $\bm{w} \in \mathbb{C}^g$, in other words $\widehat{\mathcal{K}}^{-1}_{\bm{w}}(\mathrm{Im}\,\mathcal{K}\cap \mathcal{H}_{1}^{(1)}(\mathsf{S})) \subseteq \mathcal{H}_{2}^{(1)}(\mathsf{S})$.

\subsubsection{Other linear operators}
\label{opN}

Some other linear operators appear naturally in the Dyson--Schwinger equation. We collect them below. Let us first define,
with the notations of Equation~\eqref{L2def}:
\begin{equation}
\label{defd1} 
\begin{split}
\Delta_{-1} W_1(x) & =  N^{-1}\,W_1(x) - W_1^{\{-1\}}(x), \\
\Delta_{-1}P(x;\xi) & = \oint_{\mathsf{A}} \frac{\dd\eta}{2{\rm i}\pi}\,2L_2(x;\xi,\eta) \Delta_{-1} W_1 (\eta), \\
\Delta_0 V(x) & = V(x) - V^{\{0\}}(x).
\end{split}
\end{equation}
Let also $h_1,h_2$ be two holomorphic functions in $\mathsf{U}$. We define:
\bea
\label{alltheop}
\mathcal{L}_1\,:\,\mathcal{H}_1^{(1)}(\mathsf{A}) \rightarrow \mathcal{H}_2^{(1)}(\mathsf{A}) & \quad & \mathcal{L}_1[f](x) = \oint_{\mathsf{A}} \frac{\dd\xi}{2{\rm i}\pi}\,\frac{L_2(x;\xi,\xi)}{L(x)}\,f(\xi)\,, \nonumber \\
\mathcal{L}_{2}\,:\,\mathcal{H}_{1}^{(2)}(\mathsf{A}) \rightarrow \mathcal{H}_{1}^{(1)}(\mathsf{A}) & \quad & \mathcal{L}_2[f](x) = \oint_{\mathsf{A}} \frac{\dd\xi_1\dd\xi_2}{(2{\rm i}\pi)^2}\,\frac{L_2(x;\xi_1,\xi_2)}{L(x)}\,f(\xi_1,\xi_2)\,, \nonumber \\
\mathcal{M}_{x'}\,:\,\mathcal{H}_{1}^{(1)}(\mathsf{A}) \rightarrow \mathcal{H}^{(2)}_{1}(\mathsf{A}) & \quad & \mathcal{M}_{x'}[f](x) = \oint_{\mathsf{A}} \frac{\dd\xi}{2{\rm i}\pi}\,\frac{L(\xi)}{L(x)}\,\frac{f(\xi)}{(x - \xi)(x' - \xi)^2}\,, \nonumber \\
\mathcal{N}_{h_1,h_2}\,:\,\mathcal{H}_1^{(1)}(\mathsf{A}) \rightarrow \mathcal{H}_1^{(1)}(\mathsf{A}) & \quad & \mathcal{N}_{h_1,h_2}[f](x) = \frac{1}{L(x)} \oint_{\mathsf{A}} \frac{\dd\xi}{2{\rm i}\pi}\Big(\frac{L(\xi)h_1(\xi)}{x - \xi} + h_2(\xi)\Big)f(\xi)\,, \nonumber \\
\Delta\mathcal{K}\,:\,\mathcal{H}_{1}^{(1)}(\mathsf{A}) \rightarrow \mathcal{H}_1^{(1)}(\mathsf{A}) & \quad & \Delta\mathcal{K}[f](x) = -\mathcal{N}_{(\Delta_0 V)',\Delta_{-1}P(x;\bullet)}[f](x) + 2\Delta_{-1}W_1(x)\,f(x) \nonumber \\
& & \phantom{\Delta\mathcal{K}[f](x) = } + \frac{1}{N}\Big(1 - \frac{2}{\beta}\Big)(\partial_x + \mathcal{L}_1)[f](x). \label{defDeltaK} \nonumber \\
\Delta\mathcal{J}\,:\,\mathcal{H}_{1}^{(1)}(\mathsf{A}) \rightarrow \mathcal{H}_{1}^{(1)}(\mathsf{A}) & \quad & \Delta\mathcal{J}[f](x) = -\mathcal{N}_{(\Delta_0 V)',\Delta_{-1}P(x;\bullet)/2}[f](x) + \Delta_{-1}W_1(x)\,f(x) \nonumber \\
\label{3454}& & \phantom{(\Delta\mathcal{J})f(x) = } + \frac{1}{N}\Big(1 - \frac{2}{\beta}\Big)(\partial_{x} + \mathcal{L}_{1})[f](x).
\eea
We shall encounter $\Delta\mathcal{K}$ as a correction to the operator $\mathcal{K}$ of \S~\ref{opK}, which appears in the Dyson--Schwinger equations with $n \geq 2$ variables. For $n = 1$ variable equation, we shall need the modified version denoted $\Delta\mathcal{J}$, which only differs from $\Delta\mathcal{K}$ by some symmetry factors $\frac{1}{2}$.

All those operators are continuous for appropriate norms, since we have the bounds, for $\delta_0$ small enough but fixed, and $\delta<\delta_0$ small enough:
\begin{equation}
\label{525}
\begin{split}
\p \mathcal{L}_1[f] \p_{\delta} & \leq \frac{C\,\p L'' \p^{\mathsf{U}}_{\infty}}{D_{L}(\delta)}\,\p f \p_{\delta_0}\,, \\
\p \mathcal{L}_2[f] \p_{\delta} & \leq \frac{C^2\,\p L'' \p^{\mathsf{U}}_{\infty}}{D_L(\delta)}\,\p f \p_{\delta_0}\,, \\
\sup_{d(x',\mathsf{A})\ge\delta} \p\mathcal{M}_{x'} f \p_{\delta} & \leq \frac{C \p L \p^{\mathsf{U}}_{\infty}}{D_L(\delta)\,\delta^3}\,\p f \p_{\delta/2}\,,  \\
\p \mathcal{N}_{h_1,h_2}[f]\p_{\delta} & \leq \p h_1 \p^{\mathsf{U}}_{\infty}\,\p f \p_{\delta} + C\,\frac{\p Lh_1 \p^{\mathsf{U}}_{\infty} + \p h_2 \p^{\mathsf{U}}_{\infty}}{\delta_0\,D_L(\delta)}\,\p f \p_{\delta_0}\,, \\
 \max\big\{\p \Delta\mathcal{K}[f] \p_{\delta},\p \Delta\mathcal{J}[f] \p_{\delta}\big\} & \leq \big(\p (\Delta_0 V)' \p^{\mathsf{U}}_{\infty} + 2\,\p \Delta_{-1} W_1 \p_{\delta} \big) \p f \p_{\delta} + \Big|1 - \frac{2}{\beta}\Big|\,\frac{2C}{N\delta^2}\,\p f \p_{\delta/2} \\
 & \quad + C\,\frac{\p L\,(\Delta_0 V)' \p^{\mathsf{U}}_{\infty} + \p \Delta_{-1} P \p^{\mathsf{U}^2}_{\infty}}{D_L(\delta)\,\delta_0}\,\p f \p_{\delta_0} 
\end{split}
\end{equation} 
for any $f$ in the domain of definition of the corresponding operator, and:
\beq 
\label{539} C = \ell(\mathsf{A})/\pi + (g + 1),\qquad D_L(\delta) = \inf_{d(x,\mathsf{A}) \geq \delta} |L(x)|.
\eeq
If all edges are soft, $D_L(\delta) \equiv 1$, whereas if there exists at least one hard edge, $D_L(\delta)$ scales like $\delta$ as $\delta \rightarrow 0$.

\subsection{Recursive expansion of the correlators}
\label{S523}
\subsubsection{Rewriting Dyson--Schwinger equations}
\label{sec:subs}
\label{parso}

For $n \geq 2$ variables, we can organise the Dyson--Schwinger equation of Theorem~\ref{SDN} as follows:
\beq
\label{437f}(\mathcal{K} + \Delta\mathcal{K})[W_{n}(\bullet,x_I)](x) = A_{n + 1}(x;x_I) + B_{n}(x;x_I) + C_{n - 1}(x;x_I) + D_{n - 1}(x;x_I),
\eeq
where:
\begin{equation}
\label{eqBAN}
\begin{split}
A_{n + 1}(x;x_I) & = N^{-1}(\mathcal{L}_2 - \mathrm{id})[W_{n + 1}(\bullet_1,\bullet_2,x_I)](x),  \\
B_{n}(x;x_I) & =  N^{-1}(\mathcal{L}_2 - \mathrm{id})\Big[\sum_{\substack{J \subseteq I \\ J \neq \emptyset,I}} W_{|J| + 1}(\bullet_1,x_J)W_{n - |J|}(\bullet_2,x_{I\setminus J})\Big](x),  \\ 
C_{n - 1}(x;x_I) & =  - \frac{2}{\beta N} \sum_{i \in I} \mathcal{M}_{x_i}[W_{n - 1}(\bullet,x_{I\setminus\{i\}})](x), \\
D_{n - 1}(x;x_I) & = \frac{2}{\beta N}\,\sum_{a \in (\partial \mathsf{A})_+} \frac{L(a)}{x - a}\,\partial_{a} W_{n - 1}(x_I).
\end{split}
\end{equation}

For $n = 1$, the equation has the same structure but some terms come with an extra symmetry factor. With the notation of \eqref{defd1}, and in view of Equation~\eqref{234}, we can write:
\beq
\label{AAAB} (\mathcal{K} + \Delta\mathcal{J})[\Delta_{-1}W_1](x) = \frac{A_{2}(x) + D_0}{N} - \frac{1 - 2/\beta}{N}(\partial_x + \mathcal{L}_1)[W_1^{\{-1\}}](x) + \mathcal{N}_{(\Delta_0 V)',0}[W_1^{\{-1\}}](x),
\eeq
where the operator $\Delta\mathcal{J}$ was introduced in \S~\ref{opN} and $D_0$ is given by formula \eqref{eqBAN} with the convention 
 $W_0=\ln Z^{V;\mathsf{A}}_{N,\beta;\bm{\epsilon}}$.

Since we are in the model with fixed filling fractions, the $\bm{\mathcal{A}}$-periods of $W_n(\bullet,x_I)$ for $n \geq 2$, and of $\Delta_{-1}W_1$, vanish --- cf. \eqref{lkh}. So, we are left with equations of the form:
$$
(\mathcal{K} + \Delta\mathcal{X})[\varphi] = f,\qquad \mathcal{X} = \mathcal{K}\,\,\,{\rm or}\,\,\,\mathcal{J},
$$
and the function $\varphi$ to determine satisfies $\mathcal{L}_{\bm{\mathcal{A}}}[\varphi] = 0$ by \eqref{lkh}. We can then invert $\mathcal{K}$ on the subspace of functions with zero periods and write:
$$
\varphi = \widehat{\mathcal{K}}^{-1}_{\bm{0}}\big[f - \Delta\mathcal{X}[\varphi]\big]
$$

We will need to check under which conditions the contribution of $\Delta\mathcal{X}$ is negligible compared to the contribution of $\mathcal{K}$ in Equation~\eqref{437f}. This is achieved with the following lemma.
\begin{lemma}
\label{neglinu} There exists a finite constant $C_3$ such that 
for  any $\delta > 0$, for $N$ large enough,  if $\Delta\mathcal{X}$ is any of the operator $\Delta\mathcal{K}$ or $\Delta\mathcal{J}$, for any function
$\varphi\in \mathcal{H}_1^{(1)}(\mathsf{A})$,  we have:
\beq
\label{thehnn}
 \frac{\p \widehat{\mathcal{K}}_{\bm{0}}^{-1}\big[\Delta \mathcal{X}[\varphi]\big] \p_{2\delta}}{\p \varphi \p_{\delta}} \le C_3\Big(\sqrt{\frac{\ln N}{N}}\,\frac{\sqrt{\ln\delta}}{\delta^{\kappa + \theta}}\,\frac{D_{c}(2\delta)}{D_{L}(2\delta)} \Big)\,.
\eeq
with $\kappa = \frac{1}{2}$ coming from the inversion of $\widehat{\mathcal{K}}$, and $\theta = 1$ coming from the \textit{a priori} bound \eqref{apriori0}.
\end{lemma}
\textbf{Proof.} We have from Equation~\eqref{binv}:
\beq
\label{mumCD}\p \widehat{\mathcal{K}}_{\bm{0}}^{-1}\big[\Delta\mathcal{X}[f]\big] \p_{2\delta} \leq (2\delta)^{-\kappa}\big(CD_{c}(2\delta) +  C'\big) \p \Delta \mathcal{X}[f] \p_{2\delta},\qquad \kappa = \tfrac{1}{2}\,.
\eeq
Since we have the same bound \eqref{525} for the operator norm of $\Delta\mathcal{X} = \Delta\mathcal{K}$ or $\Delta\mathcal{J}$, we can keep the generic letter $\mathcal{X}$ in the proof.  We have the \textit{a priori} bound from Corollary~\ref{apco}:
$$
\p \Delta_{-1} W_1 \p_{\delta} \leq C_1\,\sqrt{\frac{\ln N}{N}}\,\frac{\sqrt{\ln \delta}}{\delta^{\theta}},\qquad \theta = 1\,,
$$
which also implies:
$$
\p \Delta_{-1} P \p^{\mathsf{U}^2}_{\infty} \leq C_1'\,\sqrt{\frac{\ln N}{N}}
$$
with the notations of \S~\ref{opN}. We also remind that by Hypothesis~\ref{hyu222}, $\p \Delta_0 V \p_{\mathsf{U}}^{\infty} = O(\frac{1}{N})$. We insert these bounds in Equation~\eqref{525} and use $\p \varphi \p_{2\delta} \leq \p \varphi \p_{\delta}$ to find:
$$
\p \Delta\mathcal{X}[\varphi] \p_{2\delta} \leq C_2\Big(\sqrt{\frac{\ln N}{N}}\,\frac{\sqrt{\ln \delta}}{\delta^{\theta}\,D_L(2\delta)} + \Big|1 - \frac{2}{\beta}\Big|\frac{1}{N\delta^{\theta + 1}}\Big) \p \varphi \p_{\delta}.
$$
Together with Equation~\eqref{mumCD} this yields:
\beq
\label{utgia}\frac{\p \widehat{\mathcal{K}}_{\bm{0}}^{-1}\big[\Delta\mathcal{X}[\varphi]\big] \p_{2\delta}}{\p \varphi \p_{\delta}} \leq C_2'\Big(\sqrt{\frac{\ln N}{N}}\,\frac{\sqrt{\ln\delta}}{\delta^{\kappa + \theta}}\,\frac{CD_{c}(2\delta)+C'}{D_{L}(2\delta)} + \Big|1 - \frac{2}{\beta}\Big|\,\frac{CD_c(2\delta)+C'}{N\delta^{\kappa + \theta + 1}}\Big).
\eeq 
As we pointed out at the end of \S~\ref{opN}, the fact that the potential is off-critical ensures that $D_{c}(\delta)$ remains bounded when $\delta \rightarrow 0$, while we have in the worst case $\frac{1}{D_L(\delta)} = o(\frac{1}{\delta})$ --- see Equation~\eqref{539}. In any case, the second term in the above right-hand side is  negligible  with respect to the first one and we can replace $CD_{c}(2\delta)+C'$ by $D_c(\delta)$ up to a change in the constant.
  \hfill $\Box$

Hereafter, we shall not use the precise dependency of the constants on $\delta$, simply use the fact that they are finite when $\delta$ is positive independent of $N$. We will denote $c(\delta)$ for a generic finite constant 
depending only on $\delta$, which may change from line to line. 

\subsubsection{Initialisation and order of magnitude of \texorpdfstring{$W_n$}{Wn}}
\label{mang}
The goal of this section is to prove the following bounds, for
$\delta$ independent of $N$, and $N$ large enough. We know from Corollary~\ref{pro} that the $D$-terms in Equations~\eqref{437f}--\eqref{AAAB} are exponentially small, and remain so after application of $\mathcal{K}_{\bm{0}}^{-1}$, so they will never contribute to the order we are looking at, and we will not bother mentioning them again.

\begin{proposition}
\label{coJ1}\label{iniL} There exists a function $W_1^{\{0\}} \in \mathcal{H}_{2}^{(1)}(\mathsf{S})$ depending only on $W_1^{\{-1\}}, V^{\{0\}},V^{\{1\}}$  so that:
\beq
\label{W1cor} W_1 = NW_1^{\{-1\}} + W_1^{\{0\}} + \Delta_0 W_1,
\eeq
so that for all $\delta>0$, there exists a finite constant $C(\delta)$ such that  for $N$ large enough
$$
\|\Delta_{0}W_1\|_\delta\le C(\delta) \frac{(\ln N)^{\frac{3}{2}}}{N^{\frac{1}{2}}}\,.
$$
It is given by:
\beq 
\label{W1c} W_1^{\{0\}}(x) = 
 \widehat{\mathcal{K}}^{-1}_{\bm{0}}\left[\Big(-\big(1 - \tfrac{2}{\beta}\big)(\partial_x + \mathcal{L}_1) + \mathcal{N}_{(V^{\{1\}})',0}\Big)[W_1^{\{-1\}}]\right](x)
\eeq
\end{proposition}
\begin{proposition}
\label{coJ2} For any $n \geq 1$, 
we have :
\beq
\label{weo}W_{n} = N^{2 - n}(W_{n}^{\{n - 2\}} + \Delta_{n - 2}W_{n}),
\eeq
where  for $n\ge 2$, we have defined
\begin{equation}
\label{foram}
\begin{split}
W_{n}^{\{n - 2\}}(x,x_I) & = \widehat{\mathcal{K}}^{-1}_{\bm{0}}\Big[- \frac{2}{\beta}\sum_{i \in I} \mathcal{M}_{x_i}[W_{n - 1}^{\{n - 3\}}(\bullet,x_I)](x) \\
& \quad + (\mathcal{L}_2 - \mathrm{id})\Big[\sum_{\substack{J \subseteq I \\ J\neq \emptyset,I}} W_{|J| + 1}^{\{|J| - 1\}}(\bullet_1,x_J)W_{n - |J|}^{\{n - |J| - 2\}}(\bullet_2,x_{I\setminus J})\Big](x),
\end{split}
\end{equation}
and  for any $\delta>0$, there exists a finite constant $C_n(\delta)$ such that for $N$ large enough
$$
\|\Delta_{n - 2}W_n \|_{\delta}\le C_n(\delta) \frac{(\ln N)^{2n -\frac{1}{2}}}{{N}^{\frac{1}{2}}}\,.
$$
\end{proposition}
In this result, the main information about the error is its order of magnitude. Prior to those results, we are going to prove:
\begin{lemma}
\label{lemi} Denote $r^*_n = 3n - 4$. For any integers $n \geq 2$ and  $\delta>0$,
there exists a finite constant $C_n(\delta)$ such that for $N$ large enough 
\beq
\label{recusa0}\|W_{n}\|_\delta \le C_n(\delta)
N^{\frac{n - r_n^*}{2}}(\ln N)^{\frac{n + r_n^*}{2}}\,.\eeq
\end{lemma}
\textbf{Proof.} We shall prove by induction that
for any integers $n \geq 2$ and $r \geq 0$ such that $r \leq r^*_n$, for any $\delta>0$,
there exists a finite constant $C_{n,r}(\delta)$ such that for $N$ large enough 
\beq
\label{recusa}\|W_{n}\|_\delta \le C_{n,r}(\delta)
N^{\frac{n - r}{2}}(\ln N)^{\frac{n + r}{2}}\,.
\eeq 
The \textit{a priori} control of correlators  \eqref{apriori} provides the result for $r = 0$. Let $s$ be an integer, and assume the result is true for any $r \in \ldbrack 0,s \rdbrack$. Let $n$ be such that $s + 1 \leq r^*_n = 3n - 4$. We consider Equation~\eqref{437f} which gives after application of $\widehat{\mathcal{K}}^{-1}_{\bm{0}}$ that if $x_I=(x_2,\ldots,x_n)$
\begin{equation}\label{qwe} 
W_n(x,x_I)= \widehat{\mathcal{K}}^{-1}_{\bm{0}}\big[A_{n+1}(\bullet,x_I)+B_n(\cdot,x_I)+C_{n-1}(\bullet,x_I)+D_{n-1}(\bullet,x_I)-\Delta\mathcal{K}[W_n(\bullet,x_I)]\big](x)\,.\end{equation}
It is understood that all linear operators appearing here (and defined in \S~\ref{S33}) act on the variables which at the end are assigned the value  $x$. 
This formula gives 
the correlator $W_{n}$ in terms of $W_{n + 1}$ and $W_{n'}$ for $n' < n$. We systematically use the control \eqref{530} on the operator norm of $\widehat{\mathcal{K}}^{-1}_{\bm{0}}$, and the fact that $\Delta\mathcal{K}$ only gives negligible contributions compared to the latter (Lemma~\ref{neglinu}). At each step of application of Lemma~\ref{neglinu}, we have to use the operator norm with smaller $\delta$, namely $\delta\rightarrow \frac{\delta}{2}$. This is fine since our induction hypothesis holds for all $\delta>0$ and we use these bounds only a finite number of times (in fact  at most $r$ times  to get the bound at  step $r$).  Note here that this reduction \textit{a priori} holds only on the variable $x$ as $x_I$ is kept fixed, but this is bounded above by the norm where all are greater are equal to $\frac{\delta}{2}$. 

We obtain the following bound on the $A$-term, by using the induction at $(n+1,s)$,  and Equation~\eqref{525},
\begin{eqnarray*}
\|\widehat{\mathcal{K}}_{\bm{0}}^{-1}[A_{n + 1}] \|_\delta&\le&c(\delta) \|A_{n+1}\|_{\delta/2}\\
&\le&\frac{c(\delta)}{N}\|W_{n+1}\|_{\delta/2}\le \frac{c(\delta)}{N} C_{n + 1,s}(\delta/2)
N^{\frac{n+1 - s}{2}}(\ln N)^{\frac{n+1 + s}{2}}\,
\end{eqnarray*}
so that rearranging terms yields a finite constant $c^A_{n,s + 1}(\delta)$ such that
\beq
\label{orde}\|\widehat{\mathcal{K}}_{\bm{0}}^{-1}[A_{n + 1}] \|_\delta\le  c^A_{n,s+1}(\delta)
N^{\frac{n - (s + 1)}{2}}(\ln N)^{\frac{n + s + 1}{2}}\,.
\eeq
Let us consider the $B$ term.  It  involves linear combinations of $W_{j + 1}W_{n - j}$. Notice that:
$$
s \leq r_{n}^* - 1 = r_{j + 1}^* + r_{n - j}^*
$$
Thus, it is always possible to decompose (arbitrarily) $s = s' + s''$ such that $s' \leq r_{j + 1}^*$ and $s'' \leq r_{n - j}^*$, and we can use the induction hypothesis with $r = s'$ for $W_{j + 1}$ and with $r = s''$ for $W_{n - j}$. Multiplying the bounds and using the control \eqref{530} on $\widehat{\mathcal{K}}^{-1}_{\bm{0}}$ and Equation~\eqref{525}, we obtain:\
$$
\|\widehat{\mathcal{K}}_{\bm{0}}^{-1}[B_{n}]\|_\delta  \le \frac{c(\delta)}{N} \sum_{J}\|W_{|J|+1}\|_{\delta/2} \|W_{n-|J|}\|_{\delta/2} \le c^B_{n,s+1}(\delta)
N^{\frac{n - (s + 1)}{2}}(\ln N)^{\frac{n + s + 1}{2}}\,.
$$
 The $C$-term involves $W_{n - 1}$. If $s \leq r^*_{n - 1}$, we can use the induction hypothesis with $r = s$ to find by Equation~\eqref{525} that
$$
\|\widehat{\mathcal{K}}_{\bm{0}}^{-1}[C_{n - 1}]\|_\delta\le \frac{c(\delta)}{N}\sup_{d(x,\mathsf{A})\geq \delta} \|\mathcal M_{x}[W_{n-1}]\|_{\delta/2}
\le \frac{c(\delta)}{N}\|W_{n-1}\|_{\delta/4} \le  \frac{c^C_{n,s+1}(\delta)}{N\ln N} N^{\frac{n - (s + 1)}{2}}(\ln N)^{\frac{n + s + 1}{2}}\,.
$$
If $s > r^*_{n - 1}$, we can only use the induction hypothesis for $r = r^*_{n - 1}$, and find the bound:
$$
\|\widehat{\mathcal{K}}^{-1}_{\bm{0}}[C_{n - 1}]\|_{\delta} \le c^{C}_{n,s+1}(\delta) N^{\frac{n-3-r_{n-1}^*}{2}}(\ln N)^{\frac{n - 1 + r_{n-1}^*}{2}}.
$$
Using that $r_n^*=r_{n-1}^*+3$ and $s\in \ldbrack r_{n-1}^*+1, r_n^* \rdbrack$, we see that the above right-hand side 
is of the same order than  the $A$-term. 
Finally, by Equation~\eqref{thehnn},  and the induction hypothesis at $s$, we find the bound:
$$
\| \widehat{\mathcal{K}}_{\bm{0}}^{-1}\big[\Delta\mathcal{K}[W_n]\big] \|_\delta\le 
 c(\delta)\sqrt{\frac{\ln N}{N}}\,\|W_n\|_{\delta/2}\le c^{\Delta\mathcal K}_{n,s+1}(\delta)\sqrt{\frac{\ln N}{N}}
N^{\frac{n-s}{2}}(\ln N)^{\frac{n+s}{2}},
$$
which is of the same order that the bound on the $A$-term. Equation~\eqref{qwe} and summing all our bounds on the error terms  proves the bound \eqref{recusa} for $r=s+1$ and  we can conclude by induction. \hfill $\Box$

\vspace{0.2cm}

\noindent \textbf{Proof of Proposition~\ref{coJ1}.} 
It appears in Equation~\eqref{AAAB} that $N\Delta_{-1}W_1=W_1-NW_1^{\{-1\}}$ is given by 
\begin{eqnarray}
N\Delta_{-1}W_1 &=& W_1^{\{0\}} + \Delta_0 W_1\label{es1}\end{eqnarray}
where 
\begin{equation}
\begin{split}
\label{5588} W_1^{\{0\}}(x) &= \widehat{\mathcal{K}}_{\bm{0}}^{-1}\Big[-\Big(1 - \frac{2}{\beta}\Big)(\partial_{x} + \mathcal{L}_{1})[W_1^{\{-1\}}] + \mathcal{N}_{(V^{\{1\}})',0}[W_1^{\{-1\}}]\Big](x), \\
\Delta_0 W_1(x) &= 
 \widehat{\mathcal{K}}^{-1}_{\bm{0}}\Big[\mathcal{N}_{(N(\Delta_0 V)'-(V^{\{1\}})'),0}[W_1^{\{-1\}}] +A_2+D_0- \Delta\mathcal{J}[N\Delta_{-1}W_{1}]\Big](x).\end{split}
\end{equation}
Recalling Remark page \pageref{oda}, $W_1^{\{0\}}$  belongs to $\mathcal{H}_{2}^{(1)}(\mathsf{S})$. 
To bound the norm of the first  term in $\Delta_0 W_1$ observe that by Hypothesis~\ref{hyu222}:
$$
N\Delta_{0}V = V^{\{1\}} + \Delta_{1}V,\qquad \p \Delta_{1} V \p_{\infty}^{\mathsf{U}} = O\bigg(\frac{1}{N}\bigg)\,,
$$ 
so that Equation~\eqref{525} yields
$$
\| \widehat{\mathcal{K}}^{-1}_{\bm{0}}\big[\mathcal{N}_{(\Delta_1 V)',0}[W_1^{\{-1\}}]\big]\|_\delta \leq \frac{c(\delta)}{N}\|W_1^{\{-1\}}\|_\delta\le \frac{c(\delta)}{N}\,.
$$
For the second term,  note that Lemma~\ref{lemi} for $n=2$ gives the bound:
$$
\|W_2\|_\delta \le C_2(\delta) (\ln N)^4\,.
$$
Equations~\eqref{binv} and \eqref{525} imply
 $$
 \|A_2 \|_\delta \le \frac{c(\delta)}{N} \|W_2\|_{\delta} \le  \frac{c(\delta)}{N} (\ln N)^4 \,.
 $$
Moreover, $D_0$ is exponentially small by Proposition \ref{restprop}.
By Lemma~\ref{neglinu} and the \textit{a priori} bound \eqref{apriori0} on $\Delta_{-1}W_1$:
$$
\|\widehat{\mathcal{K}}^{-1}_{\bm{0}}\big[\Delta\mathcal{J}[N\Delta_{-1}W_{1}]\big]\|_\delta \le c(\delta)
\ln N\,.
$$ 
This already shows that $\|\Delta_0 W_1\|_\delta$ is at most of order $\ln N$. To improve this bound observe that
$$
\widehat{\mathcal{K}}^{-1}_{\bm{0}}\big[\Delta\mathcal{J}[N\Delta_{-1}W_{1}]\big]=\widehat{\mathcal{K}}^{-1}_{\bm{0}}\big[\Delta\mathcal{J}[W_{1}^{\{0\}}]\big]+\widehat{\mathcal{K}}^{-1}_{\bm{0}}\big[\Delta\mathcal{J}[\Delta_{0}W_{1}]\big]\,.
$$
From  Lemma~\ref{neglinu} we deduce that:
\beq
\label{Ddeki}
\|\widehat{\mathcal{K}}_{\bm{0}}^{-1}\big[\Delta\mathcal{J}[W_1^{\{0\}}]\big] \|_\delta\le c(\delta) \sqrt{\frac{\ln N}{N}}\,,\qquad 
\|\widehat{\mathcal{K}}^{-1}_{\bm{0}}\big[\Delta\mathcal{J}[\Delta_{0}W_{1}]\big]\|_\delta \le  c(\delta)\frac{(\ln N)^{\frac{3}{2}}}{N^{\frac{1}{2}}}\,.
\eeq
We finally deduce from Equation~\eqref{Ddeki} and the fact that the other error terms are  smaller,  the error bound:
$$
\|
\Delta_{0}W_1\|_\delta \le  c(\delta)  \frac{(\ln N)^{\frac{3}{2}}}{N^{\frac{1}{2}}}\,.
$$

\noindent\textbf{Proof of Proposition~\ref{coJ2}.} We already know the result for $n = 1$ by Proposition \ref{iniL}. Let $n \geq 2$, and assume the result holds for all $n' \in \ldbrack 1, n - 1 \rdbrack$. We want to use Equation~\eqref{437f}   once more to compute $W_{n}$.
We have $W_{n} = N^{2 - n}(W_{n}^{\{n - 2\}} + \Delta_{n - 2}W_{n})$ with $W_n^{\{n-2\}}$ as in Equation~\eqref{foram}. 
The error term $\Delta_{n - 2}W_{n}$ receives contributions from
\begin{itemize}
\item The term in $\Delta\mathcal K$. It can be estimated by applying Lemma~\ref{lemi} which yields the bound $W_n = O\big(N^{2 - n}(\ln N)^{2n - 2}\big)$, and Lemma \ref{neglinu}  to show that 
$$\|\widehat{\mathcal{K}}^{-1}_{\bm{0}}\big[\Delta\mathcal K[W_n]\big]\|_\delta \le  c(\delta)\sqrt{\frac{\ln N}{N}} \|W_n\|_{\delta/2}\le 
 c(\delta) N^{\frac{3}{2} -n}\,(\ln N)^{2n - \frac{3}{2}}\,.$$

\item The $A$-term.
 Applying Lemma~\ref{lemi} for $W_{n+1}$ and Equation~\eqref{525}, we find: 
\beq
\label{Ajm}\| \widehat{\mathcal{K}}^{-1}_{\bm{0}}[A_{n + 1}]\|_\delta \le \frac{c(\delta)}{N}\|W_{n+1}\|_{\delta}\le 
N^{ - n}\,(\ln N)^{2n }.
\eeq
\item The $B$-term contributes to the second term in the definition of $W_n^{\{n-2\}}$, and also from
errors $\Delta_{n' - 2}W_{n'}$ with $n'\le n-1$ to this limiting term.
They are by the induction hypothesis of order 
$ N^{2-n - \frac{1}{2}} (\ln N)^{2n-\frac{1}{2}}$.
\item The $C$-term yields the first contribution in $W_n^{\{n-2\}}$ 
and  the remaining term from 
$C_{n - 1}$ is   of the same order than the error coming from the  $B$-term, divided by $(\ln N)^2$.

Hence, we deduce by subtracting $W_{n}^{\{n - 2\}}$ and applying $\widehat{\mathcal{K}}_{\bm{0}}^{-1}$ that:
$$
\|\Delta_{n - 2}W_{n} \|_\delta \le c(\delta) \frac{(\ln N)^{2n -\frac{1}{2}}}{{N^{\frac{1}{2}}}},
$$
which is the desired result for the $n$-point correlator. We conclude by induction.
\end{itemize}  \hfill $\Box$

\subsection{Recursive expansion of the correlators}
\label{secini}

\begin{proposition}
\label{expoqq}For any   $n \geq 1$ and $k_0 \geq n-2$, we have an expansion of the form:
$$
W_{n}(x_1,\ldots,x_n) = \sum_{k = n - 2}^{  k_0} N^{-k}\,W_{n}^{\{k\}}(x_1,\ldots,x_n) + N^{-k_0}(\Delta_{k_0} W_{n})(x_1,\ldots,x_n),
$$
where:
\begin{itemize}
\item[$(i)$] for any $n \geq 1$ and any $k \in \ldbrack n-2,k_0 \rdbrack$, $W_{n}^{\{k\}}$  in $\mathcal{H}_{2}^{(n)}(\mathsf{S})$ are specified by the data of $W_1^{\{-1\}}$ and $V^{\{j\}}$ for $0 \leq j \leq k + 3 - n$. More precisely, they are defined inductively  by Equation~\eqref{foram} and the equation:
\beq
\label{fdiex}W_{n}^{\{k + 1\}}(x,x_I) = \widehat{\mathcal{K}}^{-1}_{\bm{0}}\big[E_{n}^{\{k\}}(\bullet,x_I)\big](x),
\eeq
with  for $n=1$
\begin{equation}
\label{490b} 
\begin{split}
 E_{1}^{\{k\}}(x) & = (\mathcal{L}_2 - \mathrm{id})\big[W_{2}^{\{k\}}(\bullet_1,\bullet_2)\big](x) \\
& \quad +  (\mathcal{L}_{2} - \mathrm{id})\Big[\sum_{l=0}^{k} W_1^{\{k-l \}}(\bullet_1)  W_1^{\{l\}}(\bullet_2)\Big](x)  \\
& \quad  - \Big(1 - \frac{2}{\beta}\Big)(\partial_x + \mathcal{L}_1)[W_{1}^{\{k\}}](x) +\sum_{\ell=1}^{k+2}   \mathcal{N}_{(V^{\{\ell \}})',0}[W_{1}^{\{k+1-\ell\}}](x) \,,
\end{split}
\end{equation}
whereas for $n\ge 2$
\begin{equation}
\label{490}
\begin{split}
E_{n}^{\{k\}}(x;x_I) & = (\mathcal{L}_2 - \mathrm{id})\big[W_{n + 1}^{\{k\}}(\bullet_1,\bullet_2,x_I)\big](x) \\
& \quad + \sum_{\substack{0 \leq \ell \leq k \\ J \subseteq I}}  (\mathcal{L}_{2} - \mathrm{id})\big[W_{|J| + 1}^{\{\ell\}}(\bullet_1,x_J)W_{n - |J|}^{\{k - \ell\}}(\bullet_2,x_{I\setminus J})\big](x) \\
& \quad - \Big(1 - \frac{2}{\beta}\Big)(\partial_x + \mathcal{L}_1)\big[W_{n}^{\{k\}}(\bullet,x_I)\big](x)  + \sum_{\ell = n - 2}^{k} \mathcal{N}_{(V^{\{k + 1 - \ell\}})',0}\big[W_{n}^{\{\ell\}}(\bullet,x_I)\big](x) \\
& \quad - \frac{2}{\beta}\sum_{i \in I} \mathcal{M}_{x_i}\big[W_{n - 1}^{\{k\}}(\bullet,x_{I\setminus\{i\}})\big](x). 
\end{split}
\end{equation}
In the above formula $W_p^{\{\ell\}}$ vanishes if $\ell\le p-1$.

\item[$(ii)$] for any $n \geq 1$, $\Delta_{k_0}W_{n} \in \mathcal{H}_{2}^{(n)}(\mathsf{A})$ and there exists a finite constant $c_{n,k_0}(\delta)$ so that  for any $\delta>0$ for $N$ 
large enough:
\beq
\label{uhqa}\|\Delta_{k_0} W_{n}\|_\delta \le  c_{n, k_0}(\delta) \frac{(\ln N)^{2n-\frac{1}{2}+2(k_0-n+2)}}{{N^{\frac{1}{2}}}}\,.
\eeq
\end{itemize}
\end{proposition}

\noindent \textbf{Proof.} The case $k_0 = n-2$ follows from \S~\ref{mang}, and we prove the general case by induction on $k_0$, which can be seen as the continuation of the proof of Proposition~\ref{coJ2}. Assume the result holds for all $n\ge 1$ and all  $k \le  n-2+j=:k_n-1 $ for some $j\ge 0$. We prove it by induction  for all $n$ and $k_n$.  Let us decompose:
$$
V = \sum_{k = 0}^{j + 2} N^{-k}\,V^{\{k\}} + N^{-(j + 2)}\Delta_{j + 2}V.
$$
We already know that the Dyson--Schwinger equation for $W_n$ is satisfied up to order $N^{1 - k_n}$ for all $n$.  We first show that it holds at $k_1$ for $n=1$.
Returning to Equation~\eqref{234},  we see that
\begin{equation*}
\begin{split}
N\Delta_{k_1-1}W_1 (x) &=  W_{1}^{\{k_1\}}(x)+  \widehat{\mathcal{K}}^{-1}_{\bm{0}}[R_1^{\{k_1\}}](x) \\
 R_1^{\{k_1\}}(x)&= (\mathcal{L}_2 - \mathrm{id})\big[\Delta_{k_1-1}W_{2}(\bullet_1,\bullet_2)\big](x)  -\Big(1 - \frac{2}{\beta}\Big)(\partial_{x} + \mathcal{L}_{1})[\Delta_{k_1-1} W_1] \\
& \quad +2(\mathcal{L}_{2} - \mathrm{id})\big[\Delta_{k_1-1} W_1(\bullet_1)  (\Delta_{-1} W_1)(\bullet_2)](x) \\
& \quad - \Big(1 - \frac{2}{\beta}\Big)(\partial_x + \mathcal{L}_1)[\Delta_{k_1-1} W_{1}](x) +  \mathcal{N}_{(\Delta_{0} V)',0}[\Delta_{k_1-1} W_{1}](x).
\end{split}
\end{equation*}
Strictly speaking, we should also add the $D$-terms, but since they are always exponentially small, we will systematically omit them.
But we have bounded by induction the $\delta$-norms of 
\begin{itemize}
\item $\Delta_{k_1-1} W_1$  by $ c_{1,k_1-1}(\delta)(\ln N)^{2-\frac{1}{2}+2 k_1}N^{-\frac{1}{2}}$,
\item   $\Delta_{k_1-1} W_2$ (notice that $k_2\ge k_1$) by $c_{2, k_1-1}(\delta) (\ln N)^{4-\frac{1}{2}+2(k_1-1)} N^{-\frac{1}{2}}$,
\item $\Delta_{-1} W_1$ has norm of order $\frac{1}{N}$ by Proposition \ref{coJ1} and 
$(\Delta_0 V)'$ has also  norm of order $\frac{1}{N}$ by hypothesis,  
\end{itemize}
Hence, the continuity of $ \widehat{\mathcal{K}}^{-1}_{\bm{0}}$ implies that
$$
\| \widehat{\mathcal{K}}^{-1}_{\bm{0}}[R^{\{k_1\}}]\|_\delta\le c_{1,k_1}(\delta) \frac{(\ln N)^{2-\frac{1}{2}+2 k_1}}{N^{\frac{1}{2}}},
$$
which is our inductive bound.

This proves the induction hypothesis for $n=1$ and $k_1$. Let us assume that it was proved for all $n$ and $k_n-1$, and for $n\le n_0$ at $k_n$. Let us prove it at
$n=n_0+1$ and $k_0$ with $k_0=k_{n_0}$.
We can decompose the remainder for $n\ge 2$  as:
$$
N\Delta_{k_0-1}W_{n}(x,x_I) =\widehat{\mathcal{K}}^{-1}_{\bm{0}}[E_{n}^{\{k_0\}}(\bullet;x_I) + R_{n}^{\{k_0\}}(\bullet;x_I)](x)\,.
$$
Where $E_n^{\{k\}}$ was defined in Proposition \ref{expoqq}, we have set
\begin{equation*}
\begin{split}
R_{n}^{\{k_0\}}(x;x_I) & = (\mathcal{L}_2 - \mathrm{id})\big[\Delta_{k_0-1}W_{n + 1}(\bullet_1,\bullet_2,x_I)\big](x)  \\
& \quad + \sum_{J \subseteq I} (\mathcal{L}_2 - \mathrm{id})\big[\Delta_{k_{|J|+1}}W_{|J| + 1}(\bullet_1,x_J)W_{n  - |J|}^{\{k_0-k_{|J|+1}\}}(\bullet_2,x_{I\setminus J})\big](x) \\
& \quad +  \sum_{J \subseteq I} (\mathcal{L}_2 - \mathrm{id})\big[W_{|J| + 1}^{\{ k_0- k_{n-|J|}-1\} }(\bullet_1,x_J)\Delta_{ k_{n-|J|}} W_{n  - |J|}(\bullet_2,x_{I\setminus J})\big](x) \\
& \quad + \mathcal{N}_{N[( V'-V^{\{0\}})'],0}\big[\Delta_{k_0-1} W_{n}(\bullet,x_I)\big](x) - \frac{2}{\beta}\sum_{i \in I} \mathcal{M}_{x_i}\big[\Delta_{k_0-1}W_{n - 1}(\bullet,x_{I\setminus\{i\}})\big](x). 
\end{split}
\end{equation*}
Again, by the continuity of the involved operators,  and because $k_0-k_{|J|+1}-1\le k_{n-|J|}+1$ so that the induction hypothesis can be used, we get the announced bound. Again, the largest error comes from the first term and is by induction of order $(\ln N)^{2(n+1)-\frac{1}{2}+2(k_0-1-n+2)}N^{-\frac{1}{2}}$ which is of the announced order. \hfill $\Box$

This proves the first part of Theorem~\ref{th:3} for real-analytic potentials (\textit{i.e.} the stronger Hypothesis~\ref{hmainc3} instead of \ref{hmainc4}).
For given $n$ and $k$, the bound on the error $\Delta_{k}W_n$ depends only on a finite number of constants $v^{\{k'\}}$ appearing in Hypothesis~\ref{hyu222}.

\subsection{Central limit theorem}
\label{S5clt}
With Proposition~\ref{iniL} at our disposal, we can already establish a central limit theorem for linear statistics of analytic functions in the fixed filling fractions model.  \begin{proposition}
\label{CLTT}
Let $\varphi\,:\,\mathsf{A} \rightarrow \mathbb{R}$ extending to a holomorphic function in a neighbourhood of $\mathsf{S}$. Let $\bm{N}=(N_1,\ldots, N_g)$ be a sequence (indexed by $N$) of $g$-tuples of integers such that $\sum_{h = 1}^g N_h \leq N$, denote $\bm{\epsilon}=\bm{N}/N$, and assume all limit points of $\bm{\epsilon}$ are in $\mathcal{E}$. Assume Hypothesis~\ref{hyu222}. Then, when $N \rightarrow \infty$:
\beq
\label{ooe}\mu_{N,\beta;\bm{\epsilon}}^{V;\mathsf{A}}\Big[\exp\Big(\sum_{i = 1}^N \varphi(\lambda_i)\Big)\Big] = \exp\Big(N\int_{\mathsf{A}} \dd \mu_{{\rm eq};\bm{\epsilon}}^{V}(x) \varphi(x) + M_{\beta;\bm{\epsilon}}[\varphi] + \frac{1}{2}\,Q_{\beta;\bm{\epsilon}}[\varphi,\varphi]\Big) + o(1).
\eeq
where:
\begin{equation*}
\begin{split}
 M_{\beta;\bm{\epsilon}}[\varphi] & = \oint_{\mathsf{A}} \frac{\dd\xi}{2{\rm i}\pi}\,\varphi(\xi)\,W_{1;\bm{\epsilon}}^{\{0\}}(\xi), \\
Q_{\beta;\bm{\epsilon}}[\varphi,\varphi] & = \oiint_{\mathsf{A}} \frac{\dd\xi_1\,\dd\xi_2}{(2{\rm i}\pi)^2}\,\varphi(\xi_1)\varphi(\xi_2)\,W_{2;\bm{\epsilon}}^{V;\{0\}}(\xi_1,\xi_2).
\end{split}
\end{equation*}
Here $W_1^{\{0\}} = W_{1;\bm{\epsilon}}^{\{0\}}$ is the term of order $1$ (subleading correction) in $W_1$ --- \textit{cf.} Equation~\eqref{W1cor} --- and $W_2^{\{0\}} = W_{2;\bm{\epsilon}}^{\{0\}}$ is the leading order of $W_2$ --- \textit{cf.} Equation~\eqref{foram}. Observe above that $\bm{\epsilon}$ may depend on $N$, and therefore so does the right-hand side of Equation~\eqref{ooe}.
\end{proposition}

\noindent \textbf{Proof.} Let us define $V_{t} = V - \frac{2t}{\beta N}\,\varphi$. Since the equilibrium measure is the same for $V_{t}$ and $V$, we still have the result of Proposition~\ref{iniL} for the model with potential $V_{t}$ for any $t \in [0,1]$, with uniform errors. We can thus write:
\begin{equation}
\label{toto}
\begin{split}
\ln \mu_{N,\beta;\bm{\epsilon}}^{V;\mathsf{A}}\Big[\exp\Big(\sum_{i = 1}^N \,\varphi(\lambda_i)\Big)\Big] & = \int_{0}^{1}\dd t \oint_{\mathsf{A}} \frac{\dd\xi}{2{\rm i}\pi}\,W_{1;\bm{\epsilon}}^{V_{t}}(\xi)\,\varphi(\xi) \\
& =  \int_{0}^1 \dd t\oint_{\mathsf{A}} \frac{\dd\xi}{2{\rm i}\pi}\,\varphi(\xi)\big[N\,W_{1;\bm{\epsilon}}^{V_t;\{-1\}}(\xi) + W_{1;\bm{\epsilon}}^{V_t;\{0\}}(\xi)\big] + o(1).
\end{split}
\end{equation}
As already pointed out, $W_{1;\bm{\epsilon}}^{V_t;\{-1\}} = W_{1;\bm{\epsilon}}^{V;\{-1\}}$, and from Equation~\eqref{W1c}:
$$
W_{1;\bm{\epsilon}}^{V_t;\{0\}} = W_{1;\bm{\epsilon}}^{V;\{0\}} - \frac{2t}{\beta}\big(\widehat{\mathcal{K}}^{-1}_{\bm{0}}\circ\mathcal{N}_{\varphi',0}\big)[W_{1;\bm{\epsilon}}^{V;\{-1\}}].
$$
Hence, Equation~\eqref{toto} yields Equation~\eqref{ooe} with:
\beq
\label{qhh}Q_{\beta;\bm{\epsilon}}[\varphi,\varphi] = -\frac{1}{\beta}\oint_{\mathsf{A}} \frac{\dd\xi}{2{\rm i}\pi}\,\varphi(\xi)\big(\widehat{\mathcal{K}}^{-1}_{\bm{0}}\circ\mathcal{N}_{\varphi',0}\big)[W_{1;\bm{\epsilon}}^{V;\{-1\}}](\xi)\,.
\eeq
This expression can be transformed by comparing  with \eqref{foram} for $n = 2$, but we can cut this short by observing that $Q_{\beta;\bm{\epsilon}}[\varphi,\varphi]$ must also be the limiting covariance of $\sum_{i=1}^N \varphi(\lambda_i)$. Hence:
\beq
\label{qhh2} Q_{\beta;\bm{\epsilon}}[\varphi,\varphi] = \oiint_{\mathsf{A}} \frac{\dd\xi_1\,\dd\xi_2}{(2{\rm i}\pi)^2}\,\varphi(\xi_1)\varphi(\xi_2)\,W_{2;\bm{\epsilon}}^{V;\{0\}}(\xi_1,\xi_2),
\eeq
where $W_{2;\bm{\epsilon}}^{V;\{0\}}$ has been introduced in Equation~\eqref{weo}. From the proof of Proposition~\ref{iniL}, we observe that the $o(1)$ in \eqref{ooe} is uniform in $\varphi$ such that $\sup_{d(\xi,\mathsf{A}) \geq\delta} |\varphi(\xi)|$ is bounded by a fixed constant. \hfill $\Box$

\vspace{0.2cm}
 
In other words, if $\lim_{N \rightarrow \infty} \bm{\epsilon} = \bm{\epsilon}_{\infty}$, the random variable $\Phi_N = \sum_{i = 1}^N \varphi(\lambda_i) - N \int_{\mathsf{A}}\varphi(\xi) \dd\mu_{{\rm eq};\bm{\epsilon}}^V(\xi)$ converges in law to a Gaussian variable with mean $M_{\bm{\epsilon}_{\infty}}[\varphi]$ and variance $Q_{\bm{\epsilon}_{\infty}}[\varphi,\varphi]$ when $N \rightarrow \infty$. This is a generalisation of the central limit theorem already known in the one-cut regime \cite{Johansson98,BG11}. A similar result was recently obtained in \cite{Smulticut}.
 In the next Section, we are going to extend it to holomorphic $\varphi$ which could be complex-valued on $\mathsf{A}$ (Proposition~\ref{coms}).
\section{Fixed filling fractions: refined results}
\label{refine}
In this section, we show how to extend our results to the case of harmonic potentials, and potentials containing a complex-valued term of order $O(\frac{1}{N})$. The latter is performed by using fine properties of analytic functions (the two-constants theorem) as was recently proposed in \cite{Smulticut}.

\subsection{Extension to harmonic potentials}
\label{S62}
The main use of the assumption that $V$ is analytic came from the representation \eqref{repw} of $n$-linear statistics described by a holomorphic function, in terms of contour integrals of the $n$-point correlator. If $\varphi$ is holomorphic in a neighbourhood of $\mathsf{A}$, its complex conjugate $\overline{\varphi}$ is anti-holomorphic, and we can also represent:
\beq
\label{antioce}\mu_{N,\beta;\bm{\epsilon}}^{V;\mathsf{A}}\Big[\sum_{i = 1}^N \overline{\varphi(\lambda_i)}\Big] = \overline{\oint_{\mathsf{A}} \frac{\dd x}{2{\rm i}\pi}\,\varphi(x)\,W_{1;\bm{\epsilon}}(x)}.
\eeq
In this paragraph, we explain how to use a weaker set of assumptions than Hypothesis~\ref{hmainc3}, where ``analyticity" and ``$\frac{1}{N}$ expansion of the potential"  are weakened as follows.
\begin{hypothesis}
\label{hmain4} 
\begin{itemize}
\item[\phantom{$\bullet$}]
\item[$\bullet$] (Harmonicity) $V\,:\,\mathsf{A} \rightarrow \mathbb{R}$ can be decomposed $V = \mathcal{V}_1 + \overline{\mathcal{V}_2}$, where $\mathcal{V}_1,\mathcal{V}_2$ extends to holomorphic functions in a neighbourhood $\mathsf{U}$ of $\mathsf{A}$.
\item[$\bullet$] ($\frac{1}{N}$ expansion of the potential) For $j = 1,2$, there exists a sequence of holomorphic functions $(\mathcal{V}_j^{\{k\}})_{k \geq 0}$ and constants $(v_{j}^{\{k\}})_{k}$ so that, for any $K \geq 0$:
\beq
\label{ve20}\sup_{\xi \in \mathsf{U}} \Big|\mathcal{V}_j(\xi) - \sum_{k = 0}^{K} N^{-k}\,\mathcal{V}_j^{\{k\}}(\xi)\Big| \leq v_{j}^{\{K\}}\,N^{-(K + 1)}.
\eeq
\end{itemize}
\end{hypothesis}
In other words, we only assume $\mathcal{V}$ to be harmonic. ``Analyticity" corresponds to the special case $\mathcal{V}_2 \equiv 0$. The main difference lies in the representation \eqref{antioce} of expectation values of antiholomorphic statistics, which come into play at various stages, but does not affect the reasoning. Let us enumerate the small changes to take into account in the order they appear in Section~\ref{S3}.

In \S~\ref{S31}, in the Dyson--Schwinger equations (Theorem~\ref{234} and \ref{SDN}), we encounter a term:
\beq
\label{therem}\mu_{N,\beta;\bm{\epsilon}}^{V;\mathsf{A}}\Big[\sum_{i = 1}^N \frac{L(\lambda_i)}{L(x)}\,\frac{V'(\lambda_i)}{x - \lambda_i} \prod_{j = 2}^{n}\Big(\sum_{i_j = 1}^{N} \frac{1}{x_j - \lambda_{i_j}}\Big)\Big]_{c}.
\eeq
It is now equal to:
\beq
\label{therem2}\frac{1}{L(x)}\oint_{\mathsf{A}} \frac{\dd\xi}{2{\rm i}\pi}\,L(\xi)\,\frac{\mathcal{V}'_1(\xi)}{x - \xi}\,W_{n;\bm{\epsilon}}(\xi,x_I) - \frac{1}{L(x)}\overline{\oint_{\mathsf{A}}\frac{\dd\xi}{2{\rm i}\pi}\,L(\xi)\,\frac{\mathcal{V}'_2(\xi)}{\overline{x} - \xi}\,W_{n;\bm{\epsilon}}(\xi,x_I)}.
\eeq
Remark that Equation~\eqref{therem} or \eqref{therem2} still defines a holomorphic function of $x$ in $\mathbb{C}\setminus \mathsf{A}$. In \S~\ref{S33}, we can define the operator $\mathcal{K}$ by Equation~\eqref{420} with $\mathcal{Q}(x)$ now given by:
\begin{equation*}
\begin{split}
\mathcal{Q}[f](x) & = - \oint_{\mathsf{A}} \frac{\dd\xi}{2{\rm i}\pi} P^{\{-1\}}_{\bm{\epsilon}}(\xi)(x;\xi)\,f(\xi)  \\
& \quad + \oint_{\mathsf{A}} \frac{\dd\xi}{2{\rm i}\pi}\,\frac{L(\xi)(\mathcal{V}_1^{\{0\}})'(\xi) - L(x)(\mathcal{V}_1^{\{0\}})'(x)}{\xi - x}\,f(\xi)  \\
& \quad + \overline{\oint_{\mathsf{A}} \frac{\dd\xi}{2{\rm i}\pi}\,\frac{L(\xi)(\mathcal{V}_2^{\{0\}})'(\xi) - L(x)(\mathcal{V}_2^{\{0\}})'(x)}{\xi - \overline{x}}\,f(\xi)}.
\end{split}
\end{equation*}
It is still a holomorphic function of $x$ in a neighbourhood of $\mathsf{A}$, thus it disappears in the computation leading to Equation~\eqref{426f} for the inverse of $\mathcal{K}$, which still holds. In \S~\ref{opN}, the expression \eqref{defDeltaK}
 for the operator $\Delta\mathcal{K}$ used in Equation~\eqref{437f} should be replaced by:
\begin{equation*}
\begin{split}
\Delta\mathcal{K}[f](x) & = 2\Delta_{-1}W_{1;\bm{\epsilon}}(x)\,f(x) + \frac{1}{N}\Big(1 - \frac{2}{\beta}\Big)\mathcal{L}_1[f](x)  \\
& \quad - \mathcal{N}_{(\Delta_0 \mathcal{V}_1)',\Delta_{-1} P_{\bm{\epsilon}}(x;\bullet)}[f](x) - \overline{\mathcal{N}_{(\Delta_0 \mathcal{V}_2)',0}[f](\overline{x})},
\end{split}
\end{equation*}
and the bound of the form \eqref{525} still holds, and involves the constants $v_{1}^{{\{1}\}}$ and $v_{2}^{\{1\}}$ introduced in Equation~\eqref{ve20}. $\Delta\mathcal{J}$ is defined and bounded similarly. In \S~\ref{sec:subs}--\ref{secini}, all occurrences of $\mathcal{N}_{V',0}[f](x)$  should be replaced by $\mathcal{N}_{(\mathcal{V}_1)',0}[f](x) + \overline{\mathcal{N}_{(\mathcal{V}_2)',0}[f](\overline{x})}$ (and similarly for $\mathcal{N}_{(\Delta_{k} V)',0}$ or $\mathcal{N}_{(V^{\{k\}})',0}$). The key remark is that the terms where $\overline{\mathcal{V}_2}$ appear involve complex conjugates of contour integrals of the type $f(\xi)\,W_{n;\bm{\epsilon}}^{\{k\}}(\xi,x_I)$ or $f(\xi)\,\Delta_k W_{n;\bm{\epsilon}}(\xi,x_I)$ where $f$ is some holomorphic function in a neighbourhood of $\mathsf{A}$. Their norm can be controlled in terms of the norms of $W_{n;\bm{\epsilon}}^{\{k\}}$ or $\Delta_k W_{n;\bm{\epsilon}}$ on contours, as were the terms involving $\mathcal{V}_1$, so the inductive control of errors in the large $N$ expansion of correlators for the fixed filling fractions model is still valid, leading to the first part of Theorem~\ref{th:3}, and to the central limit theorem (Proposition~\ref{CLTT}) for harmonic potentials in a neighbourhood of $\mathsf{A}$, which are still real-valued on $\mathsf{A}$.

\subsection{Complex perturbations of the potential}
\label{cxplane} 

\begin{proposition}
\label{coms} The central limit theorem \eqref{ooe} holds for $\varphi\,:\,\mathsf{A} \rightarrow \mathbb{C}$, which can be decomposed as $\varphi = \varphi_1 + \overline{\varphi_2}$, where $\varphi_1,\varphi_2$ are holomorphic functions in a neighbourhood of $\mathsf{A}$.
\end{proposition}
\textbf{Proof.} We present the proof for $\varphi = t\,f$, where $t \in \mathbb{C}$ and $f\,:\,\mathsf{A} \rightarrow \mathbb{R}$ extends to a holomorphic function in a neighbourhood of $\mathsf{A}$. Indeed, the case of $f\,:\,\mathsf{A} \rightarrow \mathbb{R}$ which can be decomposed as $f = f_1 + \overline{f_2}$ with $f_1,f_2$ extending to holomorphic functions in a neighbourhood of $\mathsf{A}$, can be treated similarly with the modifications pointed out in \S~\ref{S62}. Then, if $\varphi\,:\,\mathsf{A} \rightarrow \mathbb{C}$ can be decomposed as $\varphi = \varphi_1 + \overline{\varphi_2}$ with $\varphi_1,\varphi_2$ holomorphic, we may decompose further $\varphi_j = \varphi_j^{R} + {\rm i}\varphi_j^{I}$, then write $\tilde{V} = V - \frac{2}{\beta N}(\varphi_1^{R} + \varphi_2^{R})$ and $f = (\varphi_1^{I} - \varphi_2^{I})$, and:
$$
\mu_{N,\beta;\bm{\epsilon}}^{V;\mathsf{A}}\Big[\exp\Big(\sum_{i = 1}^N \varphi(\lambda_i)\Big)\Big] = \mu_{N,\beta;\bm{\epsilon}}^{V;\mathsf{A}}\Big[\exp\Big(\sum_{i = 1}^N (\varphi_1^{R} + \varphi_2^{R})(\lambda_i)\Big)\Big]\,\mu_{N,\beta;\bm{\epsilon}}^{\tilde{V};\mathsf{A}}\Big[\exp\Big(\sum_{i = 1}^N {\rm i}f(\lambda_i)\Big)\Big].
$$
The first factor can be treated with the initial central limit theorem (Proposition~\ref{CLTT}), while an equivalent of the second factor for large $N$ will be deduced from the following proof applied to the potential $\tilde{V}$.

This proof is inspired from the one of \cite[Lemma 1]{Smulticut}. From Theorem~\ref{th:3} applied to $V$ up to $o(1)$, we introduce $W_{n;\bm{\epsilon}}^{\{k\}}$ for $(n,k) = (1,-1),(2,0),(1,0)$ (see~\eqref{weo}--\eqref{W1cor}). If $t \in \mathbb{R}$, the central limit theorem (Proposition~\ref{CLTT}) applied to $\varphi = t\,f$ implies:
\beq
\label{rns} \mu_{N,\beta;\bm{\epsilon}}^{V;\mathsf{A}}\Big[\Big(\sum_{i = 1}^N t\,f(\lambda_i)\Big)\Big] = G_N(t)(1 + R_N(t)),\qquad G_N(t) = \exp\Big(Nt\,\Lambda_{\beta;\bm{\epsilon}}[f] + t\,M_{\beta;\bm{\epsilon}}[f] + \frac{t^2}{2}\,Q_{\beta;\bm{\epsilon}}[f,f]\Big),
\eeq
where $\sup_{t \in [-T_0,T_0]} |R_N(t)| \leq C(T_0)\,\eta_N$ and $\lim_{N \rightarrow \infty} \eta_N = 0$. Let $T_0 > 0$, and introduce the function:
$$
\tilde{R}_N(t) = \frac{1}{C(T_0)\eta_N}\,R_N(t).
$$
For any fixed $N$, it is an entire function of $t$, and by construction
\beq
\label{born1}\sup_{t \in [-T_0,T_0]}\,|\tilde{R}_N(t)| \leq 1.
\eeq
Besides, for any $t \in \mathbb{C}$, we have 
$$
\Big|\mu_{N,\beta;\bm{\epsilon}}^{V;\mathsf{A}}\Big[\exp\Big(\sum_{i = 1}^N t\,f(\lambda_i)\Big)\Big]\Big| \leq \mu_{N,\beta;\bm{\epsilon}}^{V;\mathsf{A}}\Big[\exp\Big(\sum_{i = 1}^N (\mathrm{Re}\,t)\,f(\lambda_i)\Big)\Big].
$$
Using that $f$ is real-valued on $\mathsf{A}$, we deduce that
\begin{equation}
\label{born2}
\begin{split}
\sup_{|t| \leq T_0}\,|\tilde{R}_N(t)| & \leq \frac{1}{C(T_0)\eta_N}\Big(1 + \sup_{|t|\le T_0}\frac{G_N(\mathrm{Re}\,t)}{|G_N(t)|}\Big) \\
& \leq \frac{1}{C(T_0)\eta_N}\sup_{|t|\le T_0} \exp\Big(\frac{(\mathrm{Im}\,t)^2}{2}\,Q_{\beta;\bm{\epsilon}}[f,f]\Big)  \\ 
& \leq \frac{1}{C'(T_0)\eta_N}
\end{split}
\end{equation}
for some constant $C'(T_0)$. By the two-constants lemma \cite{Nevan} (see \cite[p41]{Nevanbook} for a more recent reference), Equations~\eqref{born1}--\eqref{born2} imply
$$
\forall T \in (0,T_0),\qquad \sup_{|t| \leq T} |\tilde{R}_N(t)| \leq (C'(T_0)\eta_N)^{-2\phi(T,T_0)/\pi},\qquad \phi(T,T_0) = \mathrm{arctan}\Big(\frac{2T/T_0}{1 - (T/T_0)^2}\Big).
$$
In particular, for any compact $\mathsf{K} \subset \mathbb{C}$, we can find an open disk of radius $T_0$ containing $\mathsf{K}$, and thus show \eqref{rns} with $R_N(t) = o(1)$ uniformly in $\mathsf{K}$. \hfill $\Box$

We observe from the proof that Proposition~\ref{coms} cannot be easily extended to $|t|$ going to $\infty$ with $N$. Indeed, the ratio $G_N(T_N(\mathrm{Re}\,t))/|G_N(T_N t)|$ in Equation~\eqref{born2} will not be bounded when $N \rightarrow \infty$, hence applying the two-constants lemma as above does not show $R_N(t) \rightarrow 0$.

\begin{corollary}
\label{coc}In the model with fixed filling fractions $\bm{\epsilon}$, assume the potential $V_0$ satisfies Hypotheses~\ref{hyu222}. If $\varphi\,:\,\mathsf{A} \rightarrow \mathbb{C}$ can be decomposed as $\varphi = \varphi_1 + \overline{\varphi_2}$ with $\varphi_1,\varphi_2$ extending to holomorphic functions in a neighbourhood of $\mathsf{A}$, then the model with fixed filling fractions $\bm{\epsilon}$ and potential $V = V_0 + \varphi/N$ satisfies Hypotheses~\ref{hyu222}. Therefore, the result of Proposition~\ref{expoqq} also holds.
More generally, if  there exists a sequence of holomorphic functions $\mathcal{V}_i^{\{k\}}, k\ge 0, i=1,2$ on a neighbourhood $\mathsf{U}$ of $\mathsf{A}$
so that 
$$
\limsup_{N\ge 1} N^{K+1} \sup_{\xi\in \mathsf{U}} \Big|
\varphi(\xi)-\sum_{k=0}^K N^{-k} [\mathcal{V}_1^{\{k\}}+\overline{\mathcal{V}_1^{\{k\}}}](\xi)\Big|<\infty,
$$
the result of Proposition~\ref{expoqq} also holds with $V=V_0+\varphi/N$.
\end{corollary}

\noindent\textbf{Proof.} Hypothesis~\ref{hyu222} constrains only the leading order of the potential, \textit{i.e.} it holds for $(V_0,\bm{\epsilon})$ iff it holds for $(V = V_0 + \varphi/N,\bm{\epsilon})$. Proposition~\ref{coms} implies a fortiori the existence of constants $C_+,C_- > 0$  and $C=\exp\big(-\Re( \int_{\mathsf{A}} \varphi (x) \dd\mu_{\mathrm{eq};\bm{\epsilon}}^{V}(x))\big)$,
such that:
$$
C_-\,C^{N} \leq \frac{ |Z_{N,\beta;\bm{\epsilon}}^{V;\mathsf{A}}|}{ |Z_{N,\beta;\bm{\epsilon}}^{V_0;\mathsf{A}}|} \leq C_+\,C^N.
$$
Using this inequality as an input, we can repeat the proof of the large deviation principles 
given in Section~\ref{concea} to check  Lemma \ref{uuu3}  (i.e the restriction to the vicinity of the support) and Corollary~\ref{apco} (\textit{i.e.} the \textit{a priori} control reminded in \eqref{59}--\eqref{59b}) for the potential $V$. Then, in the recursive analysis of the Dyson--Schwinger equation of Section~\ref{S3} for the model with fixed filling fractions, the fact that the potential is complex-valued does not matter: we have established the expansion of the correlators. \hfill $\Box$

\vspace{0.2cm}

This proves Theorem~\ref{th:3} in full generality.

\subsection{\texorpdfstring{$\frac{1}{N}$}{1/N} expansion of \texorpdfstring{$n$}{n}-point kernels}
\label{K63}
We can apply Corollary~\ref{coc} to study potentials of the form:
$$
V_{\bm{x},{\bf c}}(\xi) = V - \frac{2}{\beta N}\sum_{j = 1}^n c_j \ln(x_j - \xi),
$$
where $x_j \in \mathbb{C}\setminus \mathsf{A}$, and thus derive the asymptotic expansion of the kernels in the complex plane, \textit{i.e.} Corollaries~\ref{th:4} and \ref{th:33}.

First, let us choose a simply connected domain $\mathsf{D} \subset \mathbb{C}^*$ in which the complex logarithm is an analytic function. Choose $x_1,\ldots,x_n$ and an extra reference point $p$ such that all $x_j - \xi$ for $\xi$ in a complex neighborhood $\mathsf{A}_{\mathbb{C}}$ of $\mathsf{A}$ and $x_j - p$ belong to $\mathsf{D}$. Then, we can write for $\xi \in \mathsf{A}_{\mathbb{C}}$
$$
\ln(x_j - \xi) - \ln(p - \xi) = \int_{p}^{x_j} \frac{\dd z}{z - \xi}.
$$
Recalling the notation $\mathbb{L} = {\rm diag}(\lambda_1,\ldots,\lambda_N)$ for the random matrix, we have for $r \geq 1$
\begin{eqnarray}
\label{cumlog}
\partial_{c_{j_{1}}}\cdots\partial_{c_{j_{r}}}\ln Z^{V_{\bm{x},{\bf c}};\mathsf{A}}_{N,\beta}
&=&
\mu_{N,\beta}^{V_{\bm{x},{\bf c}};\mathsf{A}}\Big[{\rm Tr}\big(\ln(x_{j_1} - \mathbb{L}) - \ln(p - \mathbb{L})\big),\ldots,{\rm Tr}\big(\ln(x_{j_r} - \mathbb{L}) - \ln(p - \mathbb{L})\big)\Big]_c \nonumber\\
&=& \int_{p}^{x_{j_1}} \cdots \int_{p}^{x_{j_r}} W_{r;\bm{\epsilon}}(\xi_1,\ldots,\xi_r) \prod_{i = 1}^r \dd \xi_i,
\end{eqnarray}
where we pick the unique relative homology class of path between $p$ and $x_{j_r}$ in $\mathsf{D}' := \bigcap_{\xi \in \mathsf{A}_{\mathbb{C}}} (\xi + \mathsf{D})$ to perform the integration. Here the subscript $c$ refers to the  cumulant expectation value as in \eqref{defcor}. Since $\ln(p - \mathbb{L})$ is deterministic up a $o(1)$ when $p \rightarrow \infty$, we can take the limit $p \rightarrow \infty$ in \eqref{cumlog} for $r \geq 2$ and find
\begin{equation}
\label{EcTr}
\mathbb{E}_c\big[{\rm Tr}\,\ln(x_{j_1} - \mathbb{L}),\ldots,{\rm Tr}\,\ln(x_{j_r} - \mathbb{L})\big] = \int_{\infty}^{x_{j_1}} \cdots \int_{\infty}^{x_{j_r}} W_{r;\bm{\epsilon}}(\xi_1,\ldots,\xi_r) \prod_{i = 1}^r \dd \xi_i.
\end{equation}
Since $W_{r;\bm{\epsilon}}(\xi_1,\ldots,\xi_r)$ for $r \geq 2$ behaves as $O(1/\xi_i^2)$ when $\xi_i \rightarrow \infty$ and has vanishing periods around $\mathsf{A}_h$ for any $h$, the left-hand side of \eqref{EcTr} is a well-defined single-valued analytic function of $x_1,\ldots,x_n \in \mathbb{C} \setminus \mathsf{A}$ which does not depend on the choice of path from $\infty$ to $x_{j_i}$ in this domain. For $r = 1$ the situation is different. We have indeed
$$
\oint_{\mathsf{A}_h} W_{1;\bm{\epsilon}}(x) = N \epsilon_h,\qquad W_{1;\bm{\epsilon}}(x) \mathop{\sim}_{N \rightarrow \infty} \frac{N}{x}.
$$
By taking $|p|$ large enough, we may assume that $p \in \mathsf{D}$, and we can always choose $\mathsf{D}$ in conjunction with $x_1,\ldots,x_n$ such that $x_1,\ldots,x_n \in \mathsf{D}$. Then we can write
$$
\mathbb{E}\big[{\rm Tr}\,\ln(x_j - \mathbb{L})\big] = \mathbb{E}\big[{\rm Tr}\,\ln(p - \mathbb{L})\big] + N\big(\ln x_{j} - \ln p\big) + \int_{p}^{x_j} \Big(W_{1;\bm{\epsilon}}(\xi) - \frac{N}{\xi}\Big)\dd \xi,
$$
where the path from $p$ to $x_j$ remains in $\mathsf{D}'$. Choosing a continuous path $\tilde{\ell} \subseteq \mathsf{D}$ going to $\infty$, the limit
$$
\lim_{\substack{p \rightarrow \infty \\ p \in \tilde{\ell}}} \ln x_j - \ln p + \ln(p - \xi) = 2{\rm i}\pi \chi_j\qquad \chi_j \in \mathbb{Z}
$$
exists, is independent of $\xi \in \mathsf{A}_{\mathbb{C}}$ but depends on the choice of path $\tilde{\ell}$ and domain $\mathsf{D}$. Therefore
\begin{equation}
\label{thecorlon2}\mathbb{E}\big[{\rm Tr}\,\ln(x_j - \mathbb{L})\big]  = \int_{\infty}^{x_j} \Big(W_{1;\bm{\epsilon}}(\xi) - \frac{N}{\xi}\Big) \dd \xi + N \ln x_j + 2{\rm i}\pi N \chi_j.
\end{equation}
We stress that the ambiguity $\chi_j$ only appears for $r = 1$ and with a prefactor $N$. It depends on various choices pertaining to the determination of the logarithm, also restricting the allowed domain of $x_j$, but in such a way that $(\mathbb{C} \setminus \mathsf{A})^n$ can be covered by finitely many opens in which various domains $\mathsf{D}$ and determinations of the logarithm can be chosen to fulfil our needs.

Let us now introduce the random variable $H_{\bm{x},\mathbf{c}} =  \sum_{j = 1}^n c_j {\rm Tr}\,\ln(x_j - \mathbb{L})$. We know from Proposition~\ref{coms} that $\ln\mu_{N,\beta;\bm{\epsilon}}^{V;\mathsf{A}}\big(e^{tH_{\bm{x},\bf{c}}}\big)$ is an entire function. Therefore, its Taylor series is convergent for any $t \in \mathbb{C}$, and we have at $t = 1$:
$$
\mathsf{K}_{n,{\bf c};\bm{\epsilon}}(\boldsymbol{x}) = \bigg[\prod_{j = 1}^{n} (x_j - p)^{Nc_j}\bigg]\exp\Big(\ln \mu_{N,\beta;\bm{\epsilon}}^{V;\mathsf{A}}\big(e^{H_{\bm{x},\bf{c}}}\big)\Big).
$$
The right-hand side can be computed via the cumulants and thus \eqref{EcTr}-\eqref{thecorlon2}. We arrive to
\begin{equation}
\label{Lncunu}
\mathsf{K}_{n,{\bf c};\bm{\epsilon}}(\boldsymbol{x}) = \exp\bigg(\sum_{j = 1}^{n} Nc_j\big(\ln x_j + 2{\rm i}\pi \chi_j\big) + \sum_{r \geq 1} \frac{1}{r!} \mathcal{L}_{\bm{x},\mathbf{c}}^{\otimes r}[W_{r;\bm{\epsilon}}]\bigg),
\end{equation}
where
\begin{equation}
\label{Lxnc}\mathcal{L}_{\bm{x},\bf{c}}[f](x) = \sum_{j = 1}^n c_j \int_{\infty}^{x_j} \check{f}(\xi)\dd \xi,\qquad \check{f}(x) = f(x) + \frac{1}{x} \Res_{x = \infty} f(\xi)\dd \xi.
\end{equation}
 The difference between $\check{f}$ and $f$ in \eqref{Lxnc} is only relevant for the $r = 1$ term in \eqref{Lncunu}, and $\check{f}(x) = O(1/x^2)$ as $x \rightarrow \infty$, so is integrable near $\infty$. For this $r = 1$ term the integral does depend on choices of paths from $\infty$ to $x_j$ because the $W_{1;\bm{\epsilon}}$ has non-vanishing $\mathsf{A}_h$-periods, but these ambiguities are the same as the ambiguities in the definition of the kernels before taking any asymptotics (see \S~\ref{Kdefsec}), they only appear in the leading order term of the asymptotics. As we have explained above, they can be resolved if we work with a fixed $x_1,\ldots,x_n$ and fixed domain $\mathsf{D}$ of definition of the logarithm.

 As a consequence of Proposition~\ref{expoqq}, $W_{n;\bm{\epsilon}} = O(N^{2 - n})$ and has a $\frac{1}{N}$ expansion. Therefore, only a finite number of terms contribute to each order in the $n$-point kernels, and we find:
\begin{proposition}\label{propker}
Assume Hypothesis~\ref{hmainc3}. Then, for any given $K \geq -1$ and $\delta > 0$, we have a uniform asymptotic expansion for $\min_{1 \leq j \leq n} d(x_j,\mathsf{A}) \geq \delta$:
$$
\mathsf{K}_{n,{\bf c};\bm{\epsilon}}(\bm{x}) = \exp\left\{\sum_{j = 1}^{n} Nc_j(\ln x_j + 2{\rm i}\pi \chi_j\big) + \sum_{k = -1}^{K} N^{-k} \Big(\sum_{r = 1}^{k + 2} \frac{1}{r!} \mathcal{L}_{{\bm x},{\bm c}}^{\otimes r}[W_{r;\bm{\epsilon}}^{\{k\}}]\Big) + O(N^{-(K + 1)})\Big)\right\}\,.
$$
\hfill $\Box$
\end{proposition}

\section{Fixed filling fractions: \texorpdfstring{$\frac{1}{N}$}{1/N} expansion of the partition function}
\label{PartitionSec}
In this Section, we continue to work within the fixed filling fractions model: $\bm{N} = (N_1,\ldots,N_{g})$ is a sequence of integer vectors, we set $\bm{\epsilon} = \bm{N}/N$ (which may depend implicitly on $N$), and we assume Hypothesis~\ref{hyu222}.

\subsection{First step: one-cut interpolation}\label{firstsection}

\subsubsection{The result}

We remind that in the one-cut case $g = 1$, the main Theorem~\ref{th:22} was proved in \cite{BG11} and ensures that the partition function has an asymptotic expansion of the form, for any $K \geq 0$:
\beq
\label{exp1cut} Z_{N,\beta}^{V} = N^{\frac{\beta}{2} N + \varkappa} \exp\Big(\sum_{k = -2}^{K} N^{-k}\,F^{\{k\};V}_{\beta} + O(N^{-(K + 1)}) \Big).
\eeq
The leading term is of order $N^2$, and given by potential theory
\beq
\label{Fmoins2} F^{\{-2\};V}_{\beta} = \frac{\beta}{2}\bigg(\iint_{\mathsf{A}^2} \dd\mu_{{\rm eq}}^{V}(x)\dd\mu_{{\rm eq}}^{V}(y)\,\ln|x - y| - \int_{\mathsf{A}} V^{\{0\}}(x)\dd\mu_{{\rm eq}}^{V}(x)\bigg).
\eeq
It is well-known --- and we reprove below with Lemma~\ref{LemEnt} and  Equation~\eqref{Entref} --- that the terms of order $N$ is related to the entropy of the equilibrium measure:
\begin{proposition}
\label{LemEntV} We have
$$
F^{\{-1\};V}_{\beta} = - \frac{\beta}{2} \int_{\mathsf{A}} V^{\{1\}}(x) \dd\mu_{{\rm eq}}^V(x) +
 \Big(1 - \frac{\beta}{2}\Big)\Big({\rm Ent}(\mu_{{\rm eq}}^{V}) - \ln\big(\tfrac{\beta}{2}\big)\Big) + \frac{\beta}{2}\ln\big(\tfrac{2\pi}{e}\big) - \ln \Gamma\big(\tfrac{\beta}{2}\big),
 $$
where
$$
{\rm Ent}[\mu] = -\int_{\mathsf{S}} \ln\Big(\frac{\dd\mu}{\dd x}\Big)\dd\mu(x).
$$
\end{proposition}
In \cite{BG11} the potential was assumed independent of $N$, but it is straightforward to include a $V$ having a $\frac{1}{N}$ expansion and this results in $F^{\{-1\};V}_{\beta}$ in the extra term involving $V^{\{1\}}$. The exponent $\varkappa$ describing the $O(\ln N)$ correction is identified in \cite{BG11} as
\beq
\label{valuee} \varkappa = \left\{\begin{array}{lll} \frac{3 + \beta/2 + 2/\beta}{12} & & {\rm if}\,\,{\rm both}\,\,{\rm edges}\,\,{\rm are}\,\,{\rm soft}, \\ \frac{\beta/2 + 2/\beta}{6} & & {\rm if}\,\,{\rm one}\,\,{\rm edge}\,\,{\rm is}\,\,{\rm soft}\,\,{\rm and}\,\,{\rm the}\,\,{\rm other}\,\,{\rm is}\,\,{\rm hard}, \\
\frac{-1 + \beta/2 + 2/\beta}{4} && {\rm if}\,\,{\rm both}\,\,{\rm edges}\,\,{\rm are}\,\,{\rm hard}. \end{array}\right.
\eeq
This exponent can be compactly rewritten:
\begin{equation}
\label{varkappafull}
\varkappa = \frac{1}{2} + (\# {\rm soft} + 3 \# {\rm hard})\frac{-3 + \beta/2 + 2/\beta}{24}.
\end{equation}

\subsubsection{Strategy to prove  this result and computation of coefficients}

As we now review, this theorem was proved by interpolating, for fixed location of the cut $\gamma = [\gamma_-,\gamma_+]$ and nature of the edges, the partition function $Z^{V;\mathsf{A}}_{N,\beta}$ with a partition function $Z_{N,\beta}^{{\rm ref}}$ which is exactly computable by Selberg integrals. We denote $V_{{\rm ref}}$ the potential of these reference models. The choice of reference models will be made explicit in Section~\ref{refmod}, and depends only on the position of edges $\gamma_{\pm}$ and of their nature (soft or hard). For the moment, it is enough to mention that its associated equilibrium measure $\mu_{{\rm eq}}^{{\rm ref}}$ has same support $[\gamma_-,\gamma_+]$ as $\mu_{{\rm eq}}^{V}$, and $\gamma_{+}$ (resp. $\gamma_{-}$) have same nature --- hard or soft --- in $\mu_{{\rm eq}}^{{\rm ref}}$ and $\mu_{{\rm eq}}^{V}$. Moreover, $V_{{\rm ref}}$ will satisfy Hypothesis~\ref{hyu222}. Then, we observe that the measure:
$$
\mu^{t}_{{\rm eq}} = (1 - t)\mu_{{\rm eq}}^{V} + t\mu_{{\rm eq}}^{{\rm ref}}
$$
satisfies the characterisation of the equilibrium measure for the potential:
\begin{equation}
\label{Vtdepend}
V_{t} = (1 - t)V + t\Vref.
\end{equation}
Thus, by uniqueness, $\mu_{{\rm eq}}^{t}$ must be the equilibrium measure for $V_{t}$. It is then clear that, if $V$ satisfies Hypothesis~\ref{hyu222}, so does $V_{t}$ uniformly for $t \in [0,1]$. Proposition~\ref{expoqq} guarantees that the one-point correlator $W_{1}^{t}$ for the model with potential $V_{t}$ on $\mathsf{A}$ has an asymptotic expansion, for all $K \geq 0$:
\beq
\label{W1texp} W_{1}^{t} = \sum_{k = -1}^{K}N^{-k}\,W_{1}^{\{k\};t} + O(N^{-(K + 1)}),
\eeq
and the error is uniform for $t \in [0,1]$. Therefore, the exact formula:
$$
\ln\Big(\frac{Z_{N,\beta}^{V;\mathsf{A}}}{Z_{N,\beta}^{{\rm ref}}}\Big) = - \frac{N\beta}{2} \oint_{\mathsf{A}} \frac{\dd x}{2{\rm i}\pi} (V(x) - \Vref(x))\Big(\int_{0}^{1} W_{1}^{t}(x)\dd t\Big)
$$
turns into an asymptotic expansion:
\begin{lemma}
For any $K \geq -2$, we have:
\beq
\label{interpol1cut} \ln\Big(\frac{Z_{N,\beta}^{V;\mathsf{A}}}{Z_{N,\beta}^{{\rm ref}}}\Big) = \frac{\beta}{2} \sum_{k = - 2}^{K} N^{-k}\,F_{\beta}^{\{k\};V \rightarrow {\rm ref}} + O(N^{-(K + 1)}),
\eeq
where
\beq
\label{interpol1cuts} F_{\beta}^{\{k\};V \rightarrow {\rm ref}} = \frac{\beta}{2} \oint_{\mathsf{A}} \frac{\dd x}{2{\rm i}\pi}\,(\Vref(x) - V(x))\Big(\int_{0}^{1} W_{1}^{\{k + 1\};t}(x)\,\dd t\Big) .
\eeq
\end{lemma}
\hfill $\Box$

Let us explain the principles giving more explicit computations of $F_{\beta}^{\{k\};V \rightarrow {\rm ref}}$. As $W_{1}^{\{-1\};t}$ is the Stieltjes transform of $\mu_{{\rm eq}}^{t}$, we have:
$$
W_{1}^{\{-1\};t} = (1 - t)W_{1}^{\{-1\};V} +t W_{1}^{\{-1\};{\rm ref}} 
$$
with obvious notations. In this one-cut case, we remind the notations:
$$
\sigma(x) = \sqrt{(x - \gamma_+)(x - \gamma_-)},\qquad L(x) = \prod_{\gamma = {\rm hard}\,\,{\rm edge}} (x - \gamma),
$$
and the decomposition (see Equation~\eqref{514}):
\begin{equation}
\label{5142}W_{1}^{\{-1\};t}(x) = \frac{V'_t(x)}{2} - S_{t}(x)\,\frac{\sigma(x)}{L(x)}.
\end{equation}
By construction, we have:
\begin{equation}\label{Slinear}
S_{t}(x) = (1 - t)S^{V}(x) + tS^{{\rm ref}}(x),
\end{equation}
and it is a property of our choice of  reference models that $S^{{\rm ref}}(x) = S^{{\rm ref}}$ is a constant only depending on $\gamma_{\pm}$ and the nature of the edges. The proof of the expansion \eqref{W1texp} --- either in \cite{BG11} or here in Section~\ref{S3} specialised to the one-cut case --- also provides a recursive computation of the coefficients $W_{1}^{\{k\};t}$ for $k \geq 0$. The only place where $t$ is involved is via the initial data $W_{1}^{\{-1\};t}$, as well as the inverse operator $\mathcal{K}_{t}^{-1}$, which reads in the present one-cut case (see Equation~\eqref{519} with $g=0$):
\begin{equation}
\label{Ktinverse}
\mathcal{K}^{-1}_{t}[f](x) = \frac{1}{2\sigma(x)}\oint_{\gamma} \frac{\dd \xi}{2{\rm i}\pi}\,\frac{L(\xi)\,f(\xi)}{S_{t}(\xi)(\xi - x)}.
\end{equation}
Therefore, the integral over $t$ of the $k$-th term in Equation~\eqref{interpol1cut} is \textit{a priori} a rational function of $t$, and can be in principle explicitly performed.

In the present one-cut case, $L_2(x;\xi_1,\xi_2)$ defined in Equation~\eqref{L2def} is equal to $1$ if the two edges are hard, $0$ otherwise. One can then check using that $(W_{1} - NW_{1}^{\{0\}})(\xi) = O(\frac{1}{\xi^2})$ when $\xi \rightarrow \infty$ and for $n \geq 2$ using that $W_{n}(\xi_1,\ldots,\xi_n) = O(\frac{1}{\xi_i^2})$ uniformly for $(\xi_j)_{j \neq i}$ away from $\mathsf{A}$, that the terms involving the operators $\mathcal{L}_{1}$ and $\mathcal{L}_{2}$ in the Dyson--Schwinger equations vanish in the recursive computation of $W_{n}^{\{k\}}$s, independently of the nature of the edges.

We can easily check that $F_{\beta}^{\{-2\};V \rightarrow {\rm ref}}$ given by Equation~\eqref{interpol1cuts} is indeed the difference of \eqref{Fmoins2} for $V$ and for $\Vref$, since $W_{1}^{\{-1\};t}$ being a convex combination with respect to $t$ implies
$$
\int_{0}^1 W_{1}^{\{-1\};t}(x)\,\dd t = \frac{W_{1}^{\{-1\};V}(x)+W_{1}^{\{-1\};{\rm ref}}(x)}{2}.
$$

To obtain the order $N$, we need to compute $W_{1}^{\{0\};t}$ given by Equation~\eqref{W1c} taking into account the disappearance of $\mathcal{L}$s:
$$
W_{1}^{\{0\};t} = \mathcal{K}_{t}^{-1}\Big[-\Big(1 - \frac{2}{\beta}\Big)\partial_{x}W_{1}^{\{-1\};t}\Big]\,.
$$ 
Using Equation~\eqref{514} and the analyticity of $V$, we find
$$
W_1^{\{0\};t}(x)= \Big(\frac{2}{\beta} - 1\Big)\oint_{\gamma} \frac{\dd \xi}{2{\rm i}\pi}\,\frac{1}{\xi-x}\frac{\sigma(\xi)}{2\sigma(x)}\partial_\xi \ln \left(S_t(\xi)\frac{\sigma(\xi)}{L(\xi)}\right).
$$
Some algebra reveals
\begin{lemma}
\label{LemEnt}
$$
F_{\beta}^{\{-1\};V \rightarrow {\rm ref}} = \Big(1 - \frac{\beta}{2}\Big)\big({{\rm Ent}}[\mu_{{\rm eq}}^{V}] - {{\rm Ent}}[\mu_{{\rm eq}}^{{\rm ref}}]\big).
$$
\end{lemma}
\textbf{Proof.} We first make some preliminary remarks. If we denote $G_t(x) = S_t(x)\sigma(x)/L(x)$, the density of the equilibrium measure is given by
\beq
\label{densss} \rho_t(x) = - \frac{G_t(x - {\rm i}0) - G_t(x + {\rm i}0)}{2{\rm i}\pi}.
\eeq
in particular the total mass is
$$
1 = -\oint_{\gamma} G_t(x)\,\frac{\dd x}{2{\rm i}\pi}.
$$
Therefore, $x \mapsto \partial_{t} G(x)$ has zero period around $\gamma$. This implies that, for an arbitrary choice of $o \in \mathbb{C}\setminus \gamma$, the function
$$
H_t(x) = \int^{x}_{o} \partial_t G_t(y) \dd y
$$
is analytic for $x$ in a neighbourhood of $\gamma$ in $\mathbb{C}\setminus\gamma$. As $G_t(x)$ has at most inverse squareroot singularities, we conclude that $H_{t}(x)$ remains bounded when $x$ approaches $\gamma$. Besides, applying $\int^{x}\partial_{t}$ to $G_t(x + {\rm i}0) + G_t(x - {\rm i}0) = 0$ and taking into account that $\oint_{\gamma} \partial_t G_t(x)\,\dd x = 0$,  we deduce that $H_{t}(x + {\rm i}0) + H_{t}(x - {\rm i}0) = 0$ as well.

We can now start the computation of
$$
F_{\beta}^{\{-1\};V \rightarrow {\rm ref}} = - \frac{\beta}{2}\int_0^1 \dd t \oint_{\gamma} \frac{\dd x}{2{\rm i}\pi} 
\,\partial_t V_t(x)\,W_1^{\{0\};t}(x).
$$
We substitute, and find by Equation~\eqref{5142}
$$
\frac{\partial_t V_t(x)}{2} = C_{t} + \Big(\int^{x}_{o} \partial_{t} W_{1}^{\{-1\};t}(x')\dd x'\Big) + H_{t}(x),
$$
where $C_t$ is independent of $x$. Since $W_1^{t}(x) = \frac{N}{x} + O(\frac{1}{x^2})$ we have $W_1^{\{-1\};t}(x) = \frac{1}{x} + O(\frac{1}{x^2})$ and $W_1^{\{0\};t}(x) = O(\frac{1}{x^2})$ when $x \rightarrow \infty$. This implies as well $\partial_t W_1^{\{-1\}}(x) = O(\frac{1}{x^2})$ and $\int^{x}_o \partial_t W_{1}^{\{-1\};t}(\xi)\dd \xi$. Then, as we can transform the contour integral into a residue at infinity,  only $H_{t}(x)$ contributes to the contour integral. We then substitute $W_{1}^{\{0\};t}(x)$ for its expression to deduce
$$
F_{\beta}^{\{-1\};V \rightarrow {\rm ref}}  = \Big(1 - \frac{\beta}{2}\Big)\int_{0}^{1} \dd t\oint_{\gamma} \frac{\dd x}{2{\rm i}\pi}\,\frac{H_{t}(x)}{\sigma(x)}\,\oint_{\gamma} \frac{\dd\xi}{2{\rm i}\pi}\,\frac{\sigma(\xi)}{\xi - x}\,\partial_{\xi}\ln G_t(\xi),
$$ 
where the contour for $x$ surrounds the contour for $\xi$. If we exchange the two contours, we receive an extra term picking up the residue at $x = \xi$ and contour integrating $\xi$
\begin{equation}
\label{FbetaVmoins1}
\begin{split}
F_{\beta}^{\{-1\};V \rightarrow {\rm ref}} & = \Big(1 - \frac{\beta}{2}\Big) \int_{0}^{1}\,\dd t \bigg\{- \oint_{\gamma} \frac{\dd \xi}{2{\rm i}\pi}\,H_{t}(\xi)\,\partial_{\xi} \ln G_t(\xi)  \\
&\quad \qquad\qquad\qquad\qquad + \oint_{\gamma} \frac{\dd\xi}{2{\rm i}\pi}\,\sigma(\xi)\,\partial_{\xi}\ln G_t(\xi) \oint_{\gamma} \frac{\dd x}{2{\rm i}\pi}\,\frac{H_{t}(x)}{\sigma(x)(\xi - x)}\bigg\} ,
\end{split}
\end{equation}
where in the second term $\xi$ is now outside the contour of integration for $x$. The properties of $H_{t}$ imply that $\frac{H_{t}(x)}{\sigma(x)(\xi - x)}$ is integrable on $(\gamma \pm {\rm i}0)$. We can then squeeze the contour of integration to the union of $(\gamma - {\rm i}0)$ from left to right and $(\gamma + {\rm i}0)$ from right to left, and as $H_t(x)$ and $\sigma(x)$ both take a minus sign when $x$ crosses $\gamma$, the contribution of the upper and lower parts of the contour cancel each other. So, only remains the first term in Equation~\eqref{FbetaVmoins1}, which can be written after integration by parts
$$
F_{\beta}^{\{-1\};V \rightarrow {\rm ref}}  = \Big(1 - \frac{\beta}{2}\Big) \int_{0}^{1}\dd t \oint_{\gamma} \frac{\dd\xi}{2{\rm i}\pi} \partial_{t} G_t(\xi)\,\ln G_t(\xi) .
$$
Squeezing the contour to $\gamma = [\gamma_-,\gamma_+]$ and using Equation~\eqref{densss}, we find
$$
F_{\beta}^{\{-1\};V \rightarrow {\rm ref}}  = -\Big(1 - \frac{\beta}{2}\Big) \int_{0}^{1} \dd t \bigg\{\int_{\gamma_-}^{\gamma_+} \dd \xi\,\partial_{t} \rho_t(\xi)\,\ln(\rho_t(\xi))\bigg\}.
$$
Here we recognise
$$ 
\partial_{t} {\rm Ent}[\mu_{{\rm eq}}^{t}] = \partial_{t}\bigg(-\int_{\gamma_-}^{\gamma_+} \rho_t(x) \ln(\rho_t(x))\dd x\bigg) = -\int_{\gamma_-}^{\gamma_+} \partial_{t}\rho_t(x)\,\ln(\rho_t(x))\,\dd x,
$$
given that $\int_{\gamma_-}^{\gamma_+} \rho_t(x)\dd x = 1$ is independent of $t$. Performing the integration over $t \in [0,1]$ entails the claim.
\hfill $\Box$

To obtain the order $1$, we need to compute the leading covariance $W_{2}^{\{0\};t}$ and use the formula \eqref{fdiex} taking into account the disappearance of $\mathcal{L}$s:
$$
W_{1}^{\{1\};t}(x) = \mathcal{K}_{t}^{-1}\Big[-\iota[W_{2}^{\{0\};t}] - \big(W_{1}^{\{0\};t}\big)^2 - \Big(1 - \frac{2}{\beta}\Big)\partial_{x} W_{1}^{\{0\};t}\Big](x),
$$
where $\iota[f](x) = f(x,x)$. The leading covariance is itself obtained from the formula \eqref{foram} for $n = 2$:
$$ 
W_{2}^{\{0\};t}(x_1,x_2) = \mathcal{K}_{t}^{-1}\Big[- \frac{2}{\beta} \mathcal{M}_{x_2}[W_{1}^{\{-1\};t}]\Big](x_1).
$$
It can be computed explicitly and only depends on $\beta,\gamma_{\pm}$ and is independent of $t$ and of the nature of the edges:
\begin{lemma}
We have
\begin{equation}
\label{universal02}
W_{2}^{\{0\};t}(x_1,x_2) = \frac{2/\beta}{2(x_1 - x_2)^2}\Big(-1 + \frac{x_1x_2 - (x_1 + x_2)(\gamma_- + \gamma_+)/2 + \gamma_-\gamma_+}{\sigma(x_1)\sigma(x_2)}\Big).
\end{equation}
and 
$$
\iota[W_{2}^{\{0\};t}](x) = \frac{2}{\beta}\,\frac{(\gamma_+ - \gamma_-)^2}{16 \sigma^4(x)}.
$$
\end{lemma} 
\noindent \textbf{Proof.} This is the well-known universal expression for the leading covariance in the $1$-cut situation. The derivation of \eqref{universal02} from \eqref{foram} is classical but we include it  for the reader convenience (the formula for $\iota[W_2^{\{0\};t}]$ is then a direct consequence). We use the formula \eqref{Ktinverse} for $\mathcal{K}_t^{-1}$ and the definition \eqref{alltheop} of $\mathcal{M}_{x_2}$ to rewrite \eqref{foram} as
\begin{equation}
\label{W20integrale}
W_2^{\{0\};t}(x_1,x_2) = -\frac{2}{\beta}\,\frac{1}{2\sigma(x_1)} \oint_{\gamma} \frac{\dd \xi}{2{\rm i}\pi}\,\frac{1}{S_t(\xi_1)(\xi_1 - x_1)} \oint_{\gamma} \frac{\dd†\xi_2}{2{\rm i}\pi}\,\frac{L(\xi_2)W_1^{\{-1\};t}(\xi_2)}{(x_2 - \xi_2)^2(\xi_1 - \xi_2)},
\end{equation}
Here it is understood that the $\xi_2$-integration contour is closer to the cut $\gamma$ and the $\xi_1$-integration contour, and that both $x_1,x_2$ are kept outside those contours. We are going to prove the desired formula \eqref{universal02} for $x_2$ in the domain $\mathsf{U}$ of analyticity of $V$ (which is a complex neighborhood of the cut). By uniqueness of analytic continuation this implies the formula without this restriction  on $x_2$. We can always assume that the contours in \eqref{W20integrale} remain inside $\mathsf{U}$.

From the decomposition \eqref{514}-\eqref{themissin} with our specific potential \eqref{Vtdepend} we have
$$
W_1^{\{-1\};t}(\xi_2) = \frac{(V_t^{\{0\}}(\xi_2))'}{2} - \frac{S_t(\xi_2)}{L(\xi_2)}\sigma(\xi_2).
$$
Since $V_t^{\{0\}}$ is analytic in $\mathsf{U}$, its contribution to the $\xi_2$-integration in \eqref{W20integrale} vanishes. We can then rewrite
$$
W_2^{\{0\};t}(x_1,x_2)  = \frac{2}{\beta}\,\frac{1}{2\sigma(x_1)} \partial_{x_2}\bigg( \oint_{\gamma} \frac{\dd \xi_1}{2{\rm i}\pi}\,\frac{1}{S_t(\xi_1)(\xi_1 - x_1)} \oint_{\gamma} \frac{\dd \xi_2}{2{\rm i}\pi}\,\frac{S_t(\xi_2)\sigma(\xi_2)}{(\xi_2 - x_2)(\xi_1 - \xi_2)}\bigg).
$$
We push the $\xi_2$-integration contours towards the exterior while staying in the neighborhood $\mathsf{U}$. This picks up residues (with a minus sign) at $\xi_2 = x_2$ and $\xi_1$, while the new $\xi_2$-integration contour is now larger than the $\xi_1$-integration contour. The latter gives a vanishing contribution, as the integrand is an analytic function with respect to $\xi_1$ inside the $\xi_1$-integration contour. It remains only to evaluate the two residues, which gives
$$
W_2^{\{0\};t}(x_1,x_2) = \frac{2}{\beta}\,\frac{1}{2\sigma(x_1)} \partial_{x_2} \bigg(\oint_{\gamma} \frac{\dd \xi_1}{2{\rm i}\pi}\,\frac{1}{S_t(\xi_1)(\xi_1 - x_1)} \frac{S_t(\xi_1)\sigma(\xi_1) - S_t(x_2)\sigma(x_2)}{\xi_1 - x_2}\bigg).
$$
We split the numerator of the ratio in two. The second term is holomorphic inside the integration contour, thus gives zero, and remains the first term:
$$
W_2^{\{0\};t}(x_1,x_2) = \frac{2}{\beta}\,\frac{1}{2\sigma(x_1)}\partial_{x_2} \bigg(\oint_{\gamma} \frac{\dd \xi_1}{2{\rm i}\pi}\,\frac{\sigma(\xi_1)}{(\xi_1 - x_1)(\xi_1 - x_2)}\bigg).
$$
We now see that all the peculiarities of the model have disappeared and the answer only involves
$$
\sigma(\xi_1) = \sqrt{(\xi_1 - \gamma_-)(\xi_1 - \gamma_+)}.
$$
Moving the integration contour towards $\infty$ we pick (with a minus sign) the residues at $\xi_1 = x_1, x_2,\infty$, but the residue at $\infty$ gives a contribution independent of $x_2$ so disappears when we apply the derivative. This yields
$$
W_2^{\{0\};t}(x_1,x_2) = -\frac{2}{\beta}\,\frac{1}{2\sigma(x_1)} \partial_{x_2} \bigg(\frac{\sigma(x_1) - \sigma(x_2)}{x_1 - x_2}\bigg) .
$$
With $\sigma'(x) = (2x - \gamma_- - \gamma_+)/\sigma(x)$ we get
$$
W_2^{\{0\};t}(x_1,x_2) = \frac{2}{\beta}\,\frac{1}{2(x_1 - x_2)^2} \bigg(-1 + \frac{\sigma^2(x_2) + (x_1 - x_2)(x_2 - \frac{\gamma- + \gamma_+}{2})}{\sigma(x_1)\sigma(x_2)}\bigg)
$$
and this evaluates to \eqref{universal02}.
\hfill $\Box$ 

These are all ingredients necessary to compute $W_{1}^{\{1\};t}$ and thus the term of order $1$ in Equation~\eqref{interpol1cut}. We do not push the computation further.

\subsection{The reference partition functions}
\label{refmod}
To complete the description of the asymptotic expansion of $Z_{N,\beta}^{V;\mathsf{A}}$ in the one-cut regime, we describe as promised the reference potentials, and the asymptotic expansion of $Z_{N,\beta}^{{\rm ref}}$.

\subsubsection{Preliminaries}

The result builds on the properties of the double Gamma function $\Gamma_2$, that we now review following \cite{Sprea}. The Barnes double Zeta function is defined for $b_1,b_2 > 0$ by:
$$
\zeta_2(s;x;b_1,b_2) = \frac{1}{\Gamma(s)}\int_{0}^{\infty} \frac{e^{-tx}t^{s - 1}\dd t}{(1 - e^{-b_1t})(1 - e^{-b_2t})},
$$
for ${\rm Re}\,s > 2$, and admits a meromorphic analytic continuation to $s \in \mathbb{C}$. Barnes double Gamma function is then defined by:
$$
\Gamma_2(x;b_1,b_2) = \exp\Big(\frac{\dd}{\dd s}\Big|_{s = 0} \zeta_2(s;b_1,b_2,x)\Big),\qquad 
$$
In particular, it satisfies the functional equation:
\beq
\label{Gamma2id} \Gamma_2(x + b_2;b_1,b_2) = \frac{\Gamma_2(x;b_1,b_2)}{\Gamma\big(\frac{x}{b_1}\big)}\,\sqrt{2\pi}\,b_1^{\frac{1}{2} - \frac{x}{b_1}},\qquad \Gamma_2(1;b_1,b_2) = 1.
\eeq
We will only need the specialisation to $b_1 = \frac{2}{\beta}$ and $b_2 = 1$. It admits the asymptotic expansion, for any $K \geq 1$:
\begin{equation}
\label{Gamma2exp}
\begin{split}
\ln\Gamma_{2}\big(x;\tfrac{2}{\beta},1\big) & =  -\frac{\beta x^2 \ln x}{4} + \frac{3\beta x^2}{8} + \frac{1}{2}\Big(1 + \frac{\beta}{2}\Big)(x\ln x - x) - \frac{3 + \beta/2 + 2/\beta}{12}\ln x \\
 & \quad  - \chi'\big(0;\tfrac{2}{\beta},1\big) + \sum_{k = 1}^{K} (k - 1)!\,E_{k}\big(\tfrac{2}{\beta},1\big)\,x^{-k} + O(x^{-(K + 1)}),
\end{split}
\end{equation} 
where $E_{k}(b_1,b_2)$ are the polynomials in two variables defined by the series expansion, for any $K \geq 0$:
$$
\frac{1}{(1 - e^{-b_1t})(1 - e^{-b_2t})} \mathop{=}_{t \rightarrow 0} \sum_{k = -2}^K E_k(b_1,b_2)\,t^{k} + O(t^{K + 1}),
$$
and $\chi(s;b_1,b_2)$ is the analytic continuation to the complex plane of the series defined for ${\rm Re}\,s > 2$:
$$
\chi(s;b_1,b_2) = \sum_{\substack{m_1,m_2 \geq 0 \\ (m_1,m_2) \neq (0,0)}} \frac{1}{(m_1b_1 + m_2b_2)^{s}}.
$$
For instance:
$$
\chi'(0;1,1) = -\frac{\ln(2\pi)}{2} + \zeta'(-1)
$$
in terms of the Riemann zeta function. We also remind Stirling formula for the asymptotic expansion of the Gamma function, for any $K \geq 0$:
\beq
\label{Stirlingexp} \ln \Gamma(x) \mathop{=}_{x \rightarrow \infty} x\ln x - x - \frac{\ln x}{2} + \frac{\ln(2\pi)}{2} + \sum_{k = 1}^{K} \frac{B_{k + 1}}{k(k + 1)x^{k}} + O(x^{-(K + 1)}),
\eeq
where $B_{k}$ are the Bernoulli numbers: $B_{2} = \frac{1}{6}$, $B_{4} = -\frac{1}{30}$, $B_6 = \frac{1}{42}$, etc. and $B_{2j + 1} = 0$ for $j \geq 1$.

\subsubsection{Two soft edges} 

We have $L(x) = 1$. We take as reference the Gaussian potential:
$$
\Vref(x) = \frac{8}{(\gamma_+ - \gamma_-)^2}\Big(x - \frac{\gamma_- + \gamma_+}{2}\Big)^2.
$$
Its equilibrium measure is the semi-circle law, and its Stieltjes transform is:
$$
W_{1}^{\{-1\};{\rm ref}}(x) = \frac{(\Vref)'(x)}{2} - S^{{\rm ref}}\frac{\sigma(x)}{L(x)},\qquad S^{{\rm ref}} = \frac{8}{(\gamma_+ - \gamma_-)^2}.
$$
The partition function with potential $\Vref$ over $\mathbb{R}^N$ is equal to \cite{Mehtabook}:
\beq 
\label{SelbergGaussian} Z_{N,\beta}^{{\rm ref}} = \Big[\prod_{j = 1}^N \frac{\Gamma\big(1 + j\frac{\beta}{2}\big)}{\Gamma\big(1 + \frac{\beta}{2}\big)}\Big]\,(2\pi)^{\frac{N}{2}}\,\Big(\frac{(\gamma_+ - \gamma_-)^2}{16}\,\frac{2/\beta}{N}\Big)^{\frac{\beta}{4}N^2 + (1 - \frac{\beta}{2})\frac{N}{2}},
\eeq
and it differs from the partition function on $\mathsf{A}$ by exponentially small corrections (see Corollary~\ref{uuu2}). Equation~\eqref{SelbergGaussian} can be rewritten in terms of Barnes double Gamma function: if we express the Gamma function using Equation~\eqref{Gamma2id} with $b_1 = \frac{2}{\beta}$ and $b_2 = 1$, the product becomes telescopic. The result is:
\begin{equation*}
\begin{split}
Z_{N,\beta}^{{\rm ref}} & = \frac{N!\,(2\pi)^N\,\big(\frac{\beta}{2}\big)^{(\frac{\beta}{2} - 1)N}}{\Gamma^N\big(\frac{\beta}{2}\big)\,\Gamma_2\big(N + 1;\frac{2}{\beta},1\big)} \Big(\frac{(\gamma_+ - \gamma_-)^2}{16N}\Big)^{\frac{\beta}{4} N^2 + (1 - \frac{\beta}{2})\frac{N}{2}},
\end{split}
\end{equation*}
and its asymptotic expansion can be computed with help of Equations~\eqref{Gamma2exp}--\eqref{Stirlingexp}. It yields an expansion of the form \eqref{exp1cut} with
\begin{equation*} 
\begin{split}
F_{\beta}^{\{-2\};{\rm ref}} &  = \frac{\beta}{2}\Big[-\frac{3}{4} + \ln\Big(\frac{\gamma_+ - \gamma_-}{4}\Big)\Big], \\
F_{\beta}^{\{-1\};{\rm ref}}& = \Big(1 - \frac{\beta}{2}\Big)\ln\Big(\frac{\gamma_+ - \gamma_-}{4}\Big) -\frac{1}{2} - \frac{\beta}{4} + \ln(2\pi) - \ln\Gamma\big(\tfrac{\beta}{2}\big) + \big(\tfrac{\beta}{2} - 1\big)\ln\big(\tfrac{\beta}{2}\big), \\
F_{\beta}^{\{0\};{\rm ref}} & = \chi'\big(0;\tfrac{2}{\beta},1\big) + \frac{\ln(2\pi)}{2}, \\
\varkappa^{\rm ref} & = \frac{3 + \beta/2 + 2/\beta}{12},
\end{split}
\end{equation*}
and explicitly computable higher $F^{\{k\};{\rm ref}}_{\beta}$.

\subsubsection{One soft edge, one hard edge}

Up to exchanging the role of $\gamma_{\pm}$, we can assume that $\gamma_+$ is hard and $\gamma_-$ is soft. Then $L(x) = (x - \gamma_+)$. We take as reference the linear potential:
$$
\Vref(x) = \frac{4(\gamma_+ - x)}{\gamma_+ - \gamma_-}.
$$
Its equilibrium measure is the Mar\v{c}enko--Pastur law, whose Stieltjes transform is:
$$
W_{1}^{\{-1\};{\rm ref}}(x) = \frac{(\Vref)'(x)}{2} - S^{{\rm ref}}\frac{\sigma(x)}{L(x)},\qquad S^{{\rm ref}} = \frac{2}{\gamma_+ - \gamma_-}.
$$
The partition function for $\Vref$ over $(-\infty,\gamma_+]$ is the Laguerre Selberg integral:
$$
Z_{N,\beta}^{{\rm ref}} = \prod_{j = 1}^{N} \frac{\Gamma\big(1 + (j - 1)\frac{\beta}{2}\big)\Gamma\big(1 + j\frac{\beta}{2}\big)}{\Gamma\big(1 + \frac{\beta}{2}\big)}\,\Big(\frac{2/\beta}{N}\,\frac{\gamma_+ - \gamma_-}{4}\Big)^{\frac{\beta}{2}N^2 + (1 - \frac{\beta}{2})N},
$$
and it differs from the partition function over $\mathsf{A}$ by exponentially small corrections. We transform it using Barnes double Gamma function:
\begin{equation*}
\begin{split}  
Z_{N,\beta}^{{\rm ref}} & = \frac{N!^2\,(2\pi)^N\, \big(\frac{\beta}{2}\big)^{(\beta - 1)N}}{\Gamma\big(1 + N\frac{\beta}{2}\big)\,\Gamma^N\big(\frac{\beta}{2}\big)\, \Gamma_2^2\big(N + 1;\frac{2}{\beta},1\big)}\Big(\frac{\gamma_+ - \gamma_-}{4N}\Big)^{\frac{\beta}{2}N^2 + (1 - \frac{\beta}{2})N} 
\end{split}
\end{equation*}
We then deduce the asymptotic expansion with coefficients
\begin{equation*}
\begin{split}
F_{\beta}^{\{-2\};{\rm ref}} & = \frac{\beta}{2}\Big[-\frac{3}{2} + \ln\Big(\frac{\gamma_+ - \gamma_-}{4}\Big)\Big], \\
F_{\beta}^{\{-1\};{\rm ref}} & =  \Big(1 - \frac{\beta}{2}\Big)\ln\Big(\frac{\gamma_+ - \gamma_-}{4}\Big) -1 + \ln(2\pi) - \ln \Gamma\big(\tfrac{\beta}{2}\big) + (\tfrac{\beta}{2} - 1)\ln\big(\tfrac{\beta}{2}\big),  \\
F_{\beta}^{\{0\};{\rm ref}} &=  2\chi'\big(0;\tfrac{2}{\beta},1\big) - \frac{\ln(\beta/2)}{2} + \frac{\ln(2\pi)}{2}, \\
\varkappa^{\rm ref} & = \frac{\beta/2 + 2/\beta}{6},
\end{split}
\end{equation*}
and explicitly computable higher $F^{\{k\};{\rm ref}}_{\beta}$.

\subsubsection{Two hard edges}

We have $L(x) = (x - \gamma_+)(x - \gamma_-)$. We take as reference potential $\Vref = 0$ on $[\gamma_-,\gamma_+]$. The equilibrium measure is the arcsine law, and its Stieltjes transform is:
$$
W^{\{-1\};{\rm ref}}_{1} = \frac{1}{\sigma(x)} = \frac{(\Vref)'(x)}{2} - S^{{\rm ref}}\,\frac{\sigma(x)}{L(x)},\qquad S_{{\rm ref}} = -1.
$$
The partition function for the zero potential on $[\gamma_-,\gamma_+]$ is the Jacobi Selberg integral:
$$
Z_{N,\beta}^{{\rm ref}} = \frac{1}{\Gamma^2\big(1 + N\frac{\beta}{2}\big)} \prod_{j = 1}^N \frac{\Gamma^3\big(1 + j\frac{\beta}{2}\big)}{\Gamma\big(2 + (N + j - 2)\frac{\beta}{2}\big)\Gamma\big(1+ \frac{\beta}{2}\big)}\,(\gamma_+ - \gamma_-)^{\frac{\beta}{2}N^2 + (1 - \frac{\beta}{2})N}
$$
We rewrite it in terms of the Barnes double Gamma function:
\begin{equation*}
\begin{split}
Z_{N,\beta}^{{\rm ref}} & = \frac{(2\pi)^N\,N!^3\, (N - 2)!\,\Gamma\big(\frac{2}{\beta} + N - 1\big)\,\big(\frac{\beta}{2}\big)^{(\frac{3\beta}{2} - 1)N}}{(2N - 2)!\,\Gamma\big(\frac{2}{\beta} + 2N - 1)\,\Gamma^N\big(\frac{\beta}{2}\big)\, \Gamma^2\big(1 + N\frac{\beta}{2}\big)}\,\frac{\Gamma_2\big(2N - 1;\frac{2}{\beta},1\big)}{\Gamma_{2}\big(N - 1;\frac{2}{\beta},1\big)\Gamma_2^3\big(N + 1;\frac{2}{\beta},1\big)} \\
& \quad \times (\gamma_+ - \gamma_-)^{\frac{\beta}{2}N^2 + (1 - \frac{\beta}{2})N},
\end{split}
\end{equation*}
and we find the asymptotic expansion with coefficients:
\begin{equation*}
\begin{split}
F_{\beta}^{\{-2\};{\rm ref}} & = \frac{\beta}{2}\ln\Big(\frac{\gamma_+ - \gamma_-}{4}\Big), \\ 
F_{\beta}^{\{-1\};{\rm ref}}& = \Big(1 - \frac{\beta}{2}\Big)\ln\Big(\frac{\gamma_+ - \gamma_-}{8}\Big) - \tfrac{\beta}{2} + \ln(2\pi) - \ln \Gamma\big(\tfrac{\beta}{2}\big) + \big(\tfrac{\beta}{2} - 1\big)\ln\big(\tfrac{\beta}{2}\big), \\
F_{\beta}^{\{0\};{\rm ref}}& =   3\chi'\big(0;\tfrac{2}{\beta},1\big) + \frac{27 - 13(\beta/2 + 2/\beta)}{12}\ln(2) - \ln\big(\tfrac{\beta}{2}\big) + \frac{\ln(2\pi)}{2}, \\
\varkappa^{\rm ref} & = \frac{-1 + \beta/2 + 2/\beta}{4},
\end{split}
\end{equation*}
with explicitly computable higher $F^{\{k\};{\rm ref}}_{\beta}$.

\vspace{0.2cm}

\subsubsection{Non decaying terms}

The asymptotic expansion of the reference partition takes the form, for any $K \geq 0$:
$$
\ln Z_{N,\beta}^{{\rm ref}} = \sum_{k = - 2}^{K} N^{-k}\,F_{\beta}^{\{k\};{\rm ref}} + \frac{\beta}{2}\ln N + \varkappa^{{\rm ref}} \ln N + O(N^{-(K + 1)}).
$$
As the reference equilibrium measures are explicit, we can check by explicit computation that the potential-theoretic formula \eqref{Fmoins2} holds. Using the change of variables $x = \frac{\gamma_+ + \gamma_-}{2}$, we can also also compute the entropy of the reference equilibrium measures. The result is
\beq
{{\rm Ent}}[\mu_{{\rm eq}}^{{\rm ref}}] = \left\{\begin{array}{lll} \vspace{2pt} -\frac{1}{2} + \ln(2\pi) + \ln\Big(\frac{\gamma_+ - \gamma_-}{4}\Big) && {\rm if}\,\,\gamma_{+}\,\,{\rm and}\,\,\gamma_{-}\,\,{\rm are}\,\,{\rm soft}, \\ \vspace{2pt} -1 + \ln(2\pi) + \ln\Big(\frac{\gamma_+ - \gamma_-}{4}\Big) & & {\rm if}\,\,\gamma_{\pm}\,\,{\rm is}\,\,{\rm soft}\,\,{\rm and}\,\,\gamma_{\mp}\,\,{\rm is}\,\,{\rm hard}, \\ -\ln(2) + \ln(2\pi) + \ln\Big(\frac{\gamma_+ - \gamma_-}{4}\Big) & & {\rm if}\,\,\gamma_{+}\,\,{\rm and}\,\,\gamma_{-}\,\,{\rm are}\,\,{\rm hard}. \end{array}\right.
\eeq
Collecting the previous expressions, we find that independently of the nature of the edges:
\beq
\label{Entref} F_{\beta}^{\{-1\};{\rm ref}} = \Big(1 - \frac{\beta}{2}\Big)\Big({\rm Ent}[\mu_{{\rm eq}}^{{\rm ref}}] - \ln\big(\tfrac{\beta}{2}\big)\Big) +  \frac{\beta}{2}\ln\big(\tfrac{2\pi}{e}\big) - \ln\Gamma\big(\tfrac{\beta}{2}\big).
\eeq
 Adding this contribution to the formula of Lemma~\ref{LemEnt} gives a proof of Proposition~\ref{LemEntV} relating $F_{\beta}^{\{-1\};V}$ to the entropy of the equilibrium measure for general potential $V$. We also remark from the previous expressions that
\begin{equation*}
\begin{split} 
\varkappa^{{\rm ref}} & = \frac{1}{2} + (\# {\rm soft} + 3\#†{\rm hard}) \frac{-3 + \beta/2 + 2/\beta}{24}, \\
F_{\beta}^{\{0\};{\rm ref}} & = \frac{\# {\rm soft} + 3 \# {\rm hard}}{2}\,\chi'\big(0;\tfrac{2}{\beta},1\big) + \frac{\ln(2\pi)}{2} - \frac{\#†{\rm hard}}{2}\,\ln\big(\tfrac{\beta}{2}\big) \\
& \quad + \delta_{\# {\rm hard},2} \frac{27 - 13(\beta/2 + 2/\beta)}{12}\,\ln(2).
\end{split} 
\end{equation*}

\subsection{Second step: decoupling the cuts}\label{sectiondecouple}

\subsubsection{General strategy}

This step is new compared to the one-cut situation treated in \cite{BG11}. We are going to interpolate between the partition function of a $(g + 1)$-cut model with fixed filling fractions, to a product of $(g + 1)$ partition functions of one-cut models. For this purpose, we introduce a slightly more general model
\begin{equation*}
\begin{split}
& Z_{N,\beta;\bm{\epsilon}}^{V;\mathsf{A}}(s)  \quad \\
& = \! \int_{\mathsf{A}_h^{N_h}} \Big[\prod_{h = 0}^{g} \prod_{i = 1}^{N_h} \dd\lambda_{h,i}\,e^{-N\frac{\beta}{2}V(\lambda_{h,i})}\Big] \Big[\prod_{0 \leq h < h' \leq g} \prod_{\substack{1 \leq i \leq N_h \\ 1 \leq i' \leq N_{h'}}}\!\! |\lambda_{h,i} - \lambda_{h',i'}|^{s\beta}\Big] \Big[\prod_{h = 0}^{g} \prod_{1 \leq i < j \leq N_{h}} \!\! |\lambda_{h,i} - \lambda_{h,j}|^{\beta}\Big],
\end{split}
\end{equation*}
which realises our interpolation for $s \in [0,1]$. Although this $s$-dependent model is not of the form of the $\beta$-ensemble announced in introduction, we justify in \S~\ref{decoupleproof} below that:
\begin{lemma}
\label{htem}
Assume Hypothesis~\ref{hmainc1}--\ref{hmainc4} for $V$, and consider the $s$-dependent model with $s$-dependent potential: 
\beq
\label{Vsfamily} T_s(x) = V(x) - 2(1 - s)\sum_{h' \neq h} \int_{\mathsf{A}_{h'}} \dd\mu_{{\rm eq};\bm{\epsilon}}^{V}(\xi)\,\ln|x - \xi|,\qquad x \in \mathsf{A}_{h}.
\eeq 
The correlators $W_{n;\bm{\epsilon}}^{s}$ of the  model  $
Z_{N,\beta;\bm{\epsilon}}^{T_{s};\mathsf{A}}(s)$ have a $\frac{1}{N}$ asymptotic expansion of the form:
$$
W_{n;\bm{\epsilon}}^{s} = \sum_{k = n - 2}^{K} N^{-k}\,W_{n;\bm{\epsilon}}^{\{k\};s} + O(N^{-(K + 1)}).
$$
for any $K \geq -2$, for some $N$-independent functions $W_{n;\bm{\epsilon}}^{\{k\};s}$. This expansion is uniform for $s \in [0,1]$. Besides, $W_{1;\bm{\epsilon}}^{\{-1\};s}$ is  independent of $s$, and  therefore equals the Stieltjes transform of the equilibrium measure $\mu_{{\rm eq};\bm{\epsilon}}^{V}$. It is
simply denoted by  $W_{1;\bm{\epsilon}}^{\{-1\}}$. Moreover, for any $K\ge 0$, we have
\begin{equation}\label{devfree}
\ln\bigg(\frac{Z_{N,\beta;\bm{\epsilon}}^{V;\mathsf{A}}}{Z_{N,\beta;\bm{\epsilon}}^{T_0;\mathsf{A}}(s = 0)}\bigg)= N^{2}F^{\{-2\};T}_{\beta;\bm{\epsilon}}
+\sum_{k=0}^{K} N^{-k}F^{\{k\};T}_{\beta;\bm{\epsilon}}+O(N^{-K-1})
\end{equation}
with
$$F^{\{-2\};T}_{\beta;\bm{\epsilon}}=\frac{\beta}{2} \sum_{0 \leq h \neq h' \leq g} \int_{\mathsf{A}_{h}} \int_{\mathsf{A}_{h'}} \ln|x - y|\,\dd\mu_{{\rm eq};\bm{\epsilon}}^{V}(x)\dd\mu_{{\rm eq};\bm{\epsilon}}^{V}(y) $$
and 
 some constants $F^{\{k\};T}_{\beta;\bm{\epsilon}}$ depending on $W_{n;\bm{\epsilon}}^{s}$ with $n = 1,2$ and $s\in [0,1]$.
\end{lemma}
The choice of our interpolation has two advantages. First,  at $s=1$ we get our initial model whereas at $s=0$ we get a product of one cut models which have already been analyzed, see section \ref{secdec}. On the other hand, our choice is such that the equilibrium measure for the model $\mu_{N,\beta;\bm{\epsilon}}^{T_{s};\mathsf{A}}(s)$ is independent of $s$ and equals $\mu_{{\rm eq};\bm{\epsilon}}^{V}$, see Section  \ref{secDS}. This implies clearly that $W_{1;\bm{\epsilon}}^{\{-1\};s}$ is independent of $s$. Integrating the log-derivative of $Z_{N,\beta;\bm{\epsilon}}^{T_s;\mathsf{A}}(s)$ along the family of potentials $(T_s)_{s \in [0,1]}$ given in  Equation~\eqref{Vsfamily}, we have the exact formula:
\begin{equation*}
\begin{split}
& \quad \ln\bigg(\frac{Z_{N,\beta;\bm{\epsilon}}^{V;\mathsf{A}}}{Z_{N,\beta;\bm{\epsilon}}^{T_0;\mathsf{A}}(s = 0)}\bigg)  \\
&= \beta \int_{0}^{1} \dd s\, \mu_{N,\beta;\bm{\epsilon}}^{T_{s};\mathsf{A}} \Bigg[\sum_{0 \leq h< h' \leq g}\sum_{{\substack{1 \leq i \leq N_h \\ 1 \leq i' \leq N_{h'}}}\!\! }\ln|\lambda_{h,i} - \lambda_{h',i'}|
-N\sum_{0 \leq h' \neq h \leq g}\sum_{{\substack{1 \leq i \leq N_h }}} \int_{\mathsf S_{h'}} \ln |\lambda_{h,i}-x| \dd\mu_{{\rm eq};\bm{\epsilon}}^{V}(x)\Bigg] \\
& =   -N\beta \sum_{0 \leq h \neq h' \leq g} \oint_{\mathsf{A}_{h}}\oint_{\mathsf{A}_{h'}} \frac{\dd x\,\dd x'}{(2{\rm i}\pi)^2}\,\ln[(x - x'){\rm sgn}(h - h')]\,W_{1;\bm{\epsilon}}^{\{-1\}}(x)\,\bigg(\int_{0}^{1} \dd s\, W_{1;\bm{\epsilon}}^{s}(x')\bigg) \\
&\quad + \sum_{0 \leq h' \neq h \leq g} \frac{\beta}{2} \oint_{\mathsf{A}_{h}}\oint_{\mathsf{A}_{h'}} \frac{\dd x\,\dd x'}{(2{\rm i}\pi)^2} \ln[(x - x'){\rm sgn}(h - h')]\bigg(\int_{0}^{1} \dd s \big[W_{2;\bm{\epsilon}}^{s}(x,x') + W_{1;\bm{\epsilon}}^{s}(x)W_{1;\bm{\epsilon}}^{s}(x')\big]\bigg),
\end{split}
\end{equation*}
and in the right-hand side, the uniformity of the asymptotic expansion when $N \rightarrow \infty$ of $W_{1;\bm{\epsilon}}^{s}$ and $W_{2;\bm{\epsilon}}^{s}$ with respect to $s$ allows integrating over $s \in [0,1]$ term by term. We obtain, for any $K \geq 0$:
\begin{equation*}
\begin{split}
& \quad \ln\bigg(\frac{Z_{N,\beta;\bm{\epsilon}}^{V;\mathsf{A}}}{Z_{N,\beta;\bm{\epsilon}}^{T_0;\mathsf{A}}(s = 0)}\bigg)  \\
& = - \frac{\beta N^2}{2} \sum_{0 \leq h \neq h' \leq g} \int_{\mathsf{A}_{h}} \int_{\mathsf{A}_{h'}} \ln|x - y|\,\dd\mu_{{\rm eq};\bm{\epsilon}}^{V}(x)\dd\mu_{{\rm eq};\bm{\epsilon}}^{V}(y)  \\
& \quad + \sum_{k = 0}^{K} N^{-k} \sum_{0 \leq h' \neq h \leq g} \frac{\beta}{2}\oint_{\mathsf{A}_{h}}\oint_{\mathsf{A}_{h'}} \frac{\dd x\,\dd x'}{(2{\rm i}\pi)^2} \ln[(x - x'){\rm sgn}(h - h')] \\
& \quad \qquad\qquad\qquad \times\bigg\{\int_{0}^1\Big(W_{2;\bm{\epsilon}}^{\{k\};s}(x,x') + \sum_{\substack{k',k'' \geq 0 \\ k' + k'' = k}} W_{1;\bm{\epsilon}}^{\{k'\};s}(x)W_{1;\bm{\epsilon}}^{\{k''\};s}(x')\Big)\dd s\bigg\} + O(N^{-(K + 1)})
\end{split}
\end{equation*}
where we noticed that the term depending linearly on $N$  vanishes since the two first terms of the expansion  reads  $W_{1;\bm{\epsilon}}^{s}=NW_{1;\bm{\epsilon}}^{\{-1\}}(x)+W_{1,\bm{\epsilon}}^{\{0\};s}(x)+....$ This proves \eqref{devfree} if the first part of the Lemma is granted.

\subsubsection{The decoupled partition function}\label{secdec}

For $s = 0$, we have:
\beq
\label{ZNorderedgroup} Z_{N,\beta;\bm{\epsilon}}^{T_0;\mathsf{A}}(s = 0) = \prod_{h = 0}^{g} Z_{N\epsilon_h,\beta}^{T_0/\epsilon_h;\mathsf{A}_h}
\eeq
and its asymptotic expansion follows from \eqref{exp1cut}. We remind that, in the partition function of the usual model \eqref{eqmes} where filling fractions are not fixed, the eigenvalues are not ordered, while in \eqref{ZNorderedgroup} the groups of eigenvalues are ordered. We shall therefore study the asymptotic expansion of $\frac{N!}{\prod_{h = 0}^{g} N_h!}\,Z_{N,\beta;\bm{\epsilon}}^{T_0;\mathsf{A}}(s = 0)$. Taking into account $\sum_{h = 0}^g \epsilon_h = 1$, Stirling expansion \eqref{Stirlingexp} yields:
\begin{equation*}
\begin{split}
\frac{N!}{\prod_{h = 0}^{g} N_h!} & = \Big[\prod_{h = 0}^{g} \epsilon_h^{-\frac{1}{2}}\Big]\,\exp\Big\{-N\Big(\sum_{h = 0}^g \epsilon_h\ln \epsilon_h\Big) - \frac{g\ln N}{2} - \frac{g\ln(2\pi)}{2} \\
&\quad + \sum_{k = 1}^K  \frac{N^{-k}\,B_{k + 1}}{k(k + 1)}\Big(1 - \sum_{h = 0}^{g} \epsilon_h^{-k}\Big)\Big\} + O(N^{-(K + 1)}).
\end{split}
\end{equation*}
As the equilibrium measure of the $s$-dependent model with potential $T_{s}$ is independent of $s$, the equilibrium measure corresponding to the $h$-th model in \eqref{ZNorderedgroup} is the restriction to $\mathsf{A}_{h}$ of $\epsilon_h^{-1}\mu_{{\rm eq};\bm{\epsilon}}^{T_0/\epsilon_h}$, and it has only one cut $\mathsf{S}_h$. Noticing that the entropy is additive for measures with disjoint support, we find the an asymptotic expansion:
\begin{equation}
\label{theeeP}
\begin{split}
& \quad \ln\bigg(\frac{N!\,Z_{N,\beta;\bm{\epsilon}}^{T_0;\mathsf{A}}(s = 0)}{\prod_{h = 0}^{g} N_h!}\bigg) \\
& = N^2\bigg\{-E[\mu^{V}_{{\rm eq};\bm{\epsilon}}] + \frac{\beta}{2}\sum_{0 \leq h \neq h' \leq g} \iint_{\mathsf{A}_h \times \mathsf{A}_{h'}} \ln|x - y|\dd\mu_{{\rm eq};\bm{\epsilon},h}^{V}(x)\dd\mu_{{\rm eq};\bm{\epsilon},h}^{V}(y)\bigg\}  \\
&\quad + \frac{\beta}{2}N\ln N  + N \Bigg\{- \frac{\beta}{2} \int_{\mathsf{A}} V^{\{1\}}(x)\dd\mu_{{\rm eq};\bm{\epsilon}}^{V}(x) \\
& \quad + \Big(1 - \frac{\beta}{2}\Big)\Big({\rm Ent}[\mu_{{\rm eq};\bm{\epsilon}}^{V}] - \ln\big(\tfrac{\beta}{2}\big)\Big) + \frac{\beta}{2}\ln\big(\tfrac{2\pi}{e}\big) - \ln\Gamma\big(\tfrac{\beta}{2}\big)\bigg\} \\
& \quad + \Big(\frac{1}{2} + (\#{\rm soft} + 3\#{\rm hard})\frac{-3 + \beta/2 + 2/\beta}{24}\Big)\ln N  + \sum_{h = 0}^{g} \Big(F^{\{0\};T_0/\epsilon_h,\mathsf{A}_h}_{\beta} +\big(-\tfrac{\epsilon_h}{2} + \varkappa_{h}\big)\ln \epsilon_h\Big)  \\
& \quad  + \sum_{k = 1}^K N^{-k}\bigg\{\frac{B_{k + 1}}{k(k + 1)} + \sum_{h = 0}^{g} \frac{1}{\epsilon_h^{k}}\Big(F_{\beta}^{\{k\};T_0/\epsilon_h,\mathsf{A}_h} - \frac{B_{k + 1}}{k(k + 1)}\Big)\bigg\} + O(N^{-(K + 1)}). 
\end{split}
\end{equation}
where:
$$
\varkappa_h = \left\{\begin{array}{lll} \vspace{2pt} \frac{3 + \beta/2 + 2/\beta}{12} && {\rm if} \,\,{\rm two}\,\,{\rm soft}\,\,{\rm edges}\,\,{\rm in}\,\,\mathsf{S}_h \\ \vspace{2pt} \frac{\beta/2 + 2/\beta}{6} && {\rm if}\,\,{\rm one}\,\,{\rm soft}\,\,{\rm and}\,\,{\rm one}\,\,{\rm hard}\,\,{\rm edge}\,\,{\rm in}\,\,\mathsf{S}_h \\ \frac{-3 + \beta/2 + 2/\beta}{4} && {\rm if}\,\,{\rm two}\,\,{\rm hard}\,\,{\rm edges}\,\,{\rm in}\,\,\mathsf{S}_h†\end{array}\right.
$$
We are going to use the notation ${\rm ref}(h)$ for the reference model that we associate to the one-cut model $Z_{N\epsilon_h,\beta}^{T_0/\epsilon_h;\mathsf{A}_h}$.  When we write the coefficients of the large $N$ asymptotic expansion of $\ln\big(Z_{N\epsilon_h,\beta}^{T_0/\epsilon_h;\mathsf{A}_h}/Z_{N\epsilon_h,\beta}^{{\rm ref}(h)}\big)$ as in Equation~\eqref{interpol1cut}, we find that two possible sources\footnote{By explicit dependence in $\epsilon_h$, we mean dependence in the first variable for functionals of $(\epsilon_h,\mu^{V}_{{\rm eq};\bm{\epsilon}})$.} of explicit dependence in $\epsilon_h$: either from $(N\epsilon_h)^{-k}$ which is the natural variable of expansion for the $h$-th model, and a factor of $\frac{1}{\epsilon_h}$ from each occurrence of $S_{s}$ (\textit{i.e.} each application of $\mathcal{K}^{-1}_{s}$) due to the normalisation of the equilibrium measure of the $h$-th model. We then obtain:
$$
\ln Z_{N\epsilon_h,\beta}^{T_0/\epsilon_h;\mathsf{A}_h} = \sum_{k = -2}^{K} N^{-k}\,F_{\beta}^{\{k\};T_0/\epsilon_h,\mathsf{A}_h} + O(N^{-(K + 1)}),
$$
with
\beq
\label{coefhd} F_{\beta}^{\{k\};T_0/\epsilon_h,\mathsf{A}_{h}} = F_{\beta}^{\{k\};{\rm ref}(h)} + \frac{\beta}{2} \oint_{\mathsf{S}_{h}} \frac{\dd x}{2{\rm i}\pi}\,\big(V^{\mathrm{ref}(h)}(x)- T_0(x)/\epsilon_h\big)\Big(\int_{0}^{1} W_{1;(h)}^{\{k + 1\};s}(x)\,\dd s\Big),
\eeq
where by convention $V^{{\rm ref}(h)}$ denotes the reference potential associated with the equilibrium measure of the $h$-th model --- it only depends on the edges of the support $\mathsf{S}_{h}$ and their nature, and not on the filling fractions $\bm{\epsilon}$. Besides, $W_{1;(h)}^{\{k + 1\};s}$ (here denoting the $1$-point correlator of the $h$-th model) is obtained by $k + 2$ successive applications of $\mathcal{K}_{s}^{-1}$ to a quantity involving $W_{1;(h)}^{\{-1\};s}$, the latter being proportional to $\epsilon_h^{-1}$. Therefore, $W_{1;(h)}^{\{k + 1\};s}$ is proportional to $\epsilon_h^{-1 + (k + 2)}$. As a result, the contributions from \eqref{coefhd} result in Equation~\eqref{theeeP} in affine functions of $\epsilon_h$, and the terms of degree $1$ in $\epsilon_h$ are the ones involving $V^{\mathrm{ref}(h)}(x)$.

\subsubsection{Comparison with decoupled partition function}

Note that there is no contribution of order $N$ in the right-hand side, and that the contribution of order $N^2$ reconstructs with that in $\ln Z_{N,\beta;\bm{\epsilon}}^{T_0;\mathsf{A}}\big(s = 0)$ the energy functional for $\mu_{{\rm eq}}^{V}$. Putting all results together (mainly Lemma \ref{htem} and \eqref{theeeP}), we find:
\begin{proposition} Assume Hypothesis~\ref{hyu222}. The partition function with fixed filling fractions admits an asymptotic expansion of the form, for any $K \geq 0$:
\label{propodnun}\begin{equation*}
\begin{split} 
\ln\bigg(\frac{N!\,Z_{N,\beta;\bm{\epsilon}}^{V;\mathsf{A}}}{\prod_{h = 0}^{g} N_h!}\bigg) & =  -N^2 E[\mu_{{\rm eq};\bm{\epsilon}}^{V}] + \frac{\beta}{2}N\ln N \\
& \quad  + N\bigg\{-\frac{\beta}{2} \int_{\mathsf{A}} V^{\{1\}}(x)\dd\mu_{{\rm eq};\bm{\epsilon}}^{V}(x) \\
& \quad + \Big(1 - \frac{\beta}{2}\Big)\Big({\rm Ent}[\mu_{{\rm eq};\bm{\epsilon}}^{V}] - \ln\big(\tfrac{\beta}{2}\big)\Big) + \frac{\beta}{2}\ln\big(\tfrac{2\pi}{e}\big) - \ln\Gamma\big(\tfrac{\beta}{2}\big)\bigg\} \\
& \quad  + \varkappa \ln N + \sum_{k = 0}^{K} N^{-k}\,F_{\beta;\bm{\epsilon}}^{\{k\};V} + O(N^{-(K + 1)}).
\end{split}
\end{equation*}
The coefficient of $\ln N$ is: 
\begin{equation}
\label{varkappauni} \varkappa = \frac{1}{2} + (\# {\rm soft} + 3\# {\rm hard})\frac{-3 + \beta/2 + 2/\beta}{24}
\end{equation}
The constant term is:
\begin{equation}
\label{cteF}
\begin{split}
F_{\beta;\bm{\epsilon}}^{\{0\};V} & =  \frac{\#{\rm soft} + 3\#{\rm hard}}{2}\,\chi'\big(0;\tfrac{2}{\beta},1\big) + \frac{\ln(2\pi)}{2} + \frac{\#{\rm hard}}{2}\,\ln(\beta/2)  \\
 & \quad + \#({\rm hard}\,\,{\rm cut})\frac{27 - 13(\beta/2 + 2/\beta)}{12}\,\ln(2) + \sum_{h = 0}^{g} \Big(F_{\beta}^{\{0\};T_0/\epsilon_h,\mathsf{A}_{h}} - \frac{\epsilon_h\ln\epsilon_h}{2}\Big) \\
& \quad +\!\!\! \sum_{0 \leq h \neq h' \leq g} \frac{\beta}{2} \oiint_{\mathsf{A}_{h} \times \mathsf{A}_{h'}}\!\! \frac{\dd x\,\dd x'}{(2{\rm i}\pi)^2}\,\ln[(x - x'){\rm sgn}(h - h')]\big(W_{2;\bm{\epsilon}}^{\{0\};V}(x,x') + W_{1;\bm{\epsilon}}^{\{0\};V}(x)W_{1;\bm{\epsilon}}^{\{0\};V}(x')\big),
\end{split}
\end{equation} 
The corrections for $k \geq 1$ are:
\begin{equation}
\label{higherF} 
\begin{split}   
F_{\beta;\bm{\epsilon}}^{\{k\};V} & = \sum_{h = 0}^{g} \epsilon_h^{-k} F_{\beta}^{\{k\};T_0/\epsilon_h,\mathsf{A}_{h}}  + \frac{B_{k + 1}}{k(k + 1)}\Big(1 - \sum_{h = 0}^{g} \epsilon_h^{-k}\Big)\\
& \quad + \sum_{0 \leq h \neq h' \leq g} \frac{\beta}{2} \oiint_{\mathsf{A}_h \times \mathsf{A}_{h'}} \frac{\dd x\,\dd x'}{(2{\rm i}\pi)^2}\,\ln[(x - x'){\rm sgn}(h - h')] \\
 & \quad \qquad\qquad\qquad\bigg\{\int_{0}^{1} \Big(W_{2;\bm{\epsilon}}^{\{k\};s}(x,x') + \sum_{k' + k'' = k} W_{1;\bm{\epsilon}}^{\{k'\};s}(x)W_{1;\bm{\epsilon}}^{\{k''\};s}(x')\Big)\dd s\bigg\}. 
\end{split}
\end{equation} 
\end{proposition}
\hfill $\Box$

To compute the last term in Equation~\eqref{cteF} at least in principle, we need  formulas for $W_{1;\bm{\epsilon}}^{\{0\};V}$ and $W_{2;\bm{\epsilon}}^{\{0\};V}$ in the multi-cut fixed filling fractions case. $W_{1;\bm{\epsilon}}^{\{0\};V}$ is computed by Equation~\eqref{5588}. Although we can use Equation~\eqref{fdiex} to compute $W_{2}^{\{0\};V}$, it is better expressed via its relation to the fundamental bidifferential of the second kind, see Equations~\eqref{Bebl}-\eqref{W2Beb}.

\subsection{Proof of Lemma~\ref{htem}: expansion of correlators in the \texorpdfstring{$s$}{s}-dependent model}
\label{decoupleproof}

We indicate how the arguments used so far in the article can be carried to the $s$-dependent model with fixed filling fractions without any difficulty. The interested reader can find all the details --- in the greater generality of arbitrary pairwise interactions --- in \cite{BGK15}. Let us take Hypotheses \ref{hmainc1} and \ref{hmainc4}, as the weakening of the latter to Hypothesis~\ref{hmainc3} can be done as  in Section~\ref{refine}. 

\subsubsection{Preliminary: the \texorpdfstring{$s$}{s}-dependent energy functional and associated pseudo-distance}
Hereafter, we study the energy functional associated with  $Z_{N,\beta;\bm{\epsilon}}^{V;\mathsf{A}}(s)$.
We introduce the matrix:
$$
\bm{\varsigma}^s = \big(s + (1 - s)\delta_{h,h'}\big)_{0 \leq h,h' \leq g}.
$$
It is positive semi-definite, has an eigenvector $(1)_{h = 0}^{g}$ with positive eigenvalue $(g + 1)$ and a $g$-dimensional nullspace orthogonal to it. We define the $s$-dependent energy functional :
$$
E_s^V[\mu] = \frac{\beta}{2}\bigg(-\sum_{0 \leq h,h' \leq g} \iint_{\mathsf{A}_h \times \mathsf{A}_{h'}}  \varsigma^s_{h,h'} \ln|x - y| \dd\mu_{h}(x) \dd\mu_{h'}(y) + \sum_{h = 0}^{g} \int_{\mathsf{A}_h} V^{\{0\}}(x) \dd \mu_h(x)\bigg)
$$
depending on a probability measure $\mu$ supported $\mathsf{A}$, which we decompose as $\mu = \sum_{h = 0}^{g} \mu_h$ where $\mu_h$ is supported on $\mathsf{A}_h$. We see that, with $E=E^{V}$ as defined in \eqref{Enf}:
\begin{equation}
E_s^{V}[\mu] = \sum_{h = 0}^{g} E^{V}[\mu_h] + s \iint_{\mathsf{A}^2}\Lambda(x,y)\dd \mu(x)\dd \mu(y)\,,
\end{equation}
where:
\begin{equation} 
\label{Lfrak}
\Lambda(\xi,\xi') = \left\{\begin{array}{lll} \ln|\xi - \xi'| & & {\rm if}\,\,(\xi,\xi') \in \mathsf{A}_{h}\times \mathsf{A}_{h'}\,\,{\rm and}\,\,h \neq h' \\ 0 & & {\rm otherwise} \end{array}\right.
\end{equation}
is a smooth bounded function on $\mathsf{A}^2$. Since $E[\mu_h]$ is well-defined in $\mathbb{R} \cup \{+\infty\}$, this shows that $E_s^{V}[\mu]$ is also well-defined in $\mathbb{R} \cup \{+\infty\}$.

In intermediate steps we will need the $s$-dependent analog of the pseudo-distance $\mathfrak{D}$, namely:
\begin{equation}
\label{Dssqu}\begin{split}
\mathfrak{D}_{s}[\mu,\nu] & =  \bigg(- s\sum_{0 \leq h \neq h' \leq g} \iint_{\mathsf{A}_{h} \times \mathsf{A}_{h'}} \ln|x - y|\dd[\mu - \nu](x)\dd[\mu - \nu](y) \\
& \qquad - \sum_{h = 0}^{g} \iint_{\mathsf{A}_{h}^2} \ln|x - y| \dd[\mu - \nu](x)\dd[\mu - \nu](y)\bigg)^{\frac{1}{2}} \\
& =  \bigg(\int_{0}^{\infty} \frac{\dd p}{p}\Big\{ \sum_{0 \leq h,h' \leq g} \varsigma_{h,h'}^s (\widehat{\mu_{h}} - \widehat{\nu_{h}})(p)\overline{(\widehat{\mu_{h'}} - \widehat{\nu_{h'}})(p)}\Big\}\bigg)^{\frac{1}{2}},
\end{split}
\end{equation}
We claim that it a well-defined in $[0,+\infty]$ for any two positive measures $\mu,\nu$ of finite mass on $\mathsf{A}$ such that $\mu(\mathsf{A}_h) = \nu(\mathsf{A}_h)$ for any $h \in \ldbrack 0,g \rdbrack$. This is also the setting in which we need it since we work with the $s$-dependent model for fixed filling fractions. To see this, we first remark that
$$
-\sum_{h = 0}^{g} \iint_{\mathsf{A}_h^2} \ln|x - y|\dd[\mu - \nu](x)\dd[\mu - \nu](y) = \sum_{h = 0}^{g} \mathfrak{D}^2[\mu_h,\nu_h]
$$
is well-defined in $[0,+\infty]$. Again, as $(x,y) \mapsto \Lambda(x,y)$ is continuous bounded for $(x,y) \in \bigcup_{h \neq h'} \mathsf{A}_h \times \mathsf{A}_{h'}$, we see that
$$
- \sum_{0 \leq h \neq h' \leq g}\iint_{\mathsf{A}_h \times \mathsf{A}_{h'}} \ln|x - y|\dd[\mu - \nu](x)\dd[\mu - \nu](y)
$$
is well-defined in $\mathbb{R}$. So the quantity under the squareroot in Equation~\eqref{Dssqu} is well-defined \textit{a priori} in $\mathbb{R} \cup \{+\infty\}$. Since $\bm{\varsigma}^{s}$ is positive semi-definite, we deduce that $\mathfrak{D}_s[\mu,\nu] \in [0,+\infty]$ is well-defined. If $\mathfrak{D}_s[\mu,\nu] = 0$, we must have $\sum_{h = 0}^{g} \big(\widehat{\mu_h}(p) - \widehat{\nu_h}(p)\big) = 0$ for $p$ almost everywhere (corresponding to projection on the eigenvector with positive eigenvalue), hence $\sum_{h = 0}^{g} (\mu_h - \nu_h) = 0$. Since the summands have pairwise disjoint supports, this implies $\mu_h = \nu_h$ for all $h \in \ldbrack 0,g \rdbrack$, that is $\mu = \nu$. So, $\mathfrak{D}_s$ is a pseudo-distance.


We now explain how to control linear statistics in terms of $\mathfrak{D}_s$, uniformly in $s$. Let $\mu,\nu$ be two positive measures on $\mathsf{A}$ such that $\mu(\mathsf{A}_h) = \nu(\mathsf{A}_h)$ for any $h \in \ldbrack 0,g \rdbrack$. We decompose $\rho = \mu - \nu = \sum_{h = 0}^g \rho_{h}$, where $\rho_h$ is a signed measure of zero mass supported on $\mathsf{A}_h$. Let $f$ be a smooth test function on $\mathsf{A}$. Let $\chi_{h}[f]$ be a smooth function on $\mathbb{R}$ which is equal to $f$ in $\mathsf{A}_{h}$, $0$ outside a compact neighbourhood of $\mathsf{A}_{h}$ and in particular $0$ on $\bigcup_{h' \neq h} \mathsf{A}_{h'}$. One can choose the extension procedure so that
$$
|\chi_h[f]|_{1/2} \leq C |f|_{1/2}
$$
for a constant $C > 0$ independent of $f$ and $h \in \ldbrack 0,g \rdbrack$ (it is controlled by the minimum distance between the segments $(\mathsf{A}_h)_{h'}$). We observe that for $s \in [0,1]$, the matrix $\tilde{\bm{\varsigma}}^s = (u^s + v^s \delta_{h,h'})_{0 \leq h,h' \leq g}$ squares to $\bm{\varsigma}^s$ when we choose
$$
u^s = \frac{\sqrt{1 + gs} - \sqrt{1 - s}}{g + 1},\qquad v^s = \sqrt{1 - s}
$$
On the diagonal this matrix has diagonal entries:
\begin{equation}
\label{loweruv}
u^s + v^s = \frac{g\sqrt{1 - s} + \sqrt{1 + gs}}{g + 1} \geq \frac{1}{g + 1}
\end{equation}
Let us write:
\begin{equation*}
\begin{split}
\bigg|\int_{\mathsf{A}} f(x)\dd[\mu - \nu](x)\bigg| & = \bigg| \sum_{h = 0}^{g} \int_{\mathbb{R}} \chi_h[f](x) \dd\rho_{h}(x) \Big| = \frac{1}{u^s + v^s} \bigg|\int_{\mathbb{R}} \sum_{h = 0}^{g}  \chi_h[f](x) \Big(\sum_{h' = 0}^{g} \tilde{\varsigma}^s_{h,h'} \dd \rho_{h'}(x)\Big) \bigg| \\
& \leq (g + 1)\bigg| \int_{\mathbb{R}}  \sum_{h = 0}^{g} \overline{\widehat{\chi_h[f]}(p)}\Big(\sum_{h' = 0}^{g} \tilde{\varsigma}^s_{h,h'} \widehat{\nu_{h'}}(p)\Big) \dd p\bigg|
\end{split}
\end{equation*}
where we have used the bound \eqref{loweruv} in the last line. We then use the Cauchy--Schwarz inequality:
\begin{equation}
\label{3462} \begin{split}
& \quad \bigg|\int_{\mathsf{A}} f(x)\dd[\mu - \nu](x)\bigg| \\
 & \leq (g + 1) \bigg(\int_{\mathbb{R}} \sum_{h = 0}^{g} \big|\widehat{\chi_h[f]}(p) \big|^2 |p|\,\dd p \bigg)^{\frac{1}{2}} \bigg(\int_{\mathbb{R}} \sum_{0 \leq h,h',h''  \leq g} \tilde{\varsigma}^{s}_{h,h'} \tilde{\varsigma}^s_{h,h''} \widehat{\nu_{h'}}(p)\,\overline{\widehat{\nu_{h''}}(p)} \frac{\dd p}{|p|}\bigg)^{\frac{1}{2}}  \\
& \leq \sqrt{2}(g + 1)\Big(\sum_{h = 0}^{g} |\chi_h[f]|_{1/2}\Big)\Bigg(\int_{0}^{\infty} \sum_{0 \leq h',h'' \leq g} \varsigma^{s}_{h',h''} \widehat{\nu_{h'}}(p) \overline{\widehat{\nu_{h''}}(p)} \frac{\dd p}{|p|}\bigg)^{\frac{1}{2}} \\
& \leq \sqrt{2}C(g + 1)^{2}|f|_{1/2} \mathfrak{D}_s[\mu,\nu],
\end{split}
\end{equation}
where we have used $\sqrt{X_0 + \cdots + X_g} \leq \sqrt{X_0} + \cdots + \sqrt{X_g}$ for nonnegative $X_i$ in the first squareroot factor to get the second line.

\subsubsection{Equilibrium measure}

The properties of $E_s^{V}$ and its quadratic part established in the previous paragraph allows to apply the standard potential theoretic arguments. This leads to an analog of Theorem~\ref{th:10} for the $s$-dependent model with fixed filling fractions $\bm{\epsilon}$ and potential $V$. It states the existence and uniqueness of the minimiser $\mu_{{\rm eq};\bm{\epsilon}}^{V;s}$ of $E_s^V$ among probability measures supported in $\mathsf{A}$ and having fixed filling fractions $\bm{\epsilon}$. The analog of \eqref{ina0}, i.e. the characterisation of the $s$-dependent equilibrium measure is as follows: for each $h \in \ldbrack 0,g \rdbrack$ there exists a constant $C_{\bm{\epsilon},h}^{V;s}$ such that:
\beq
\label{caract} 2\int_{\mathsf{A}_{h}} \dd\mu_{{\rm eq};\bm{\epsilon}}^{V;s}(\xi)\ln|x - \xi| + \sum_{h' \neq h} 2s\int_{\mathsf{A}_{h'}} \dd\mu_{{\rm eq};\bm{\epsilon}}^{V;s}(\xi)\ln|x - \xi| -V(x) \leq C^{V;s}_{\bm{\epsilon},h},
\eeq
with equality $\mu_{{\rm eq};\bm{\epsilon}}^{V;s}$ almost surely.

\subsubsection{Concentration estimates}
\label{conces}
The $s$-dependent model differs from the $\beta$-ensemble --- \textit{i.e.} $s = 1$ --- by multiplication of the weight by:
$$
\exp\Big((1 - s)\beta \sum_{0 \leq h < h' \leq g} \sum_{\substack{1 \leq i \leq N_{h} \\ 1 \leq i' \leq N_{h'}}} \ln|\lambda_{h,i} - \lambda_{h',i'}|\Big) = \exp\Big( \frac{(1 - s)\beta}{2} \iint_{\mathbb{R}^2} \dd L_N(\xi_1)\dd L_N(\xi_2) \Lambda(\xi_1,\xi_2)\Big),
$$
where $\Lambda$ was introduced in Equation~\eqref{Lfrak} and is smooth bounded on $\mathsf{A}^2$. This is a perturbation of the $\beta$-ensemble by a smooth functional of the empirical measure $L_{N}$. Therefore, using $E_s$ and $\mathfrak{D}_s$ instead of $E$ and $\mathfrak{D}$, we can estimate the error made by replacing $L_N$ with the regularised empirical measure $\widetilde{L}_N^{{\rm u}}$ as done in Section~\ref{compsa}, and estimate the large deviations of $\mathfrak{D}_s[\widetilde{L}_N^{{\rm u}},\mu_{{\rm eq};\bm{\epsilon}}^{T;s}]$ as in Section~\ref{Sec342} leading to an analog of  Lemma~\ref{theoconc} with $s$-independent constants. We can then proceed to estimate the large deviations of fluctuations of linear statistics like in Section~\ref{subsusb} --- using  the new Equation~\eqref{3462} instead of \eqref{346} --- and obtain an analog of Corollary~\ref{Ls1}, where the constants are chosen independent of $s$ and the only difference is that $|\varphi|_{1/2}$ should be replaced by $C^{-1}|\varphi|_{1/2}$ for some $C > 0$ independent of $s$. We also get the a priori bound of the $n$-point correlators of the $s$-dependent model with filling fractions $\bm{\epsilon}$ (analog of Corollary~\ref{apco}) and an estimate of the large deviations of the filling fractions (analog of Corollary~\ref{cosq} with $t$ replaced by $C t$ in the right-hand side) by a similar adaptation of Section~\ref{corbounds}. We conclude that all results of Section~\ref{concea} extend to the $s$-dependent model with constants that can be chosen independent of $s \in [0,1]$.

\subsubsection{Dyson--Schwinger equations}

If $f$ is a holomorphic function in $\mathbb{C}\setminus \mathsf{A}$ and decaying like $O(\frac{1}{x})$ at infinity, we may write:
$$
f(x) = \sum_{h = 0}^{g} \mathcal{P}_{h}[f](x),\qquad \mathcal{P}_{h}[f](x) = \oint_{\mathsf{A}_{h}} \frac{\dd\xi}{2{\rm i}\pi}\,\frac{f(\xi)}{x - \xi}.
$$ 
The operator $\mathcal{P}_{h}$ is a projector, and by construction $\mathcal{P}_{h}[f]$ is holomorphic in $\mathbb{C}\setminus\mathsf{A}_{h}$, continuous across $\mathsf{A}_{h'}$ for $h' \neq h$, and behaves like $O(\frac{1}{x})$ at infinity.

As in Section~\ref{S31}, we can derive the one-variable Dyson--Schwinger equation for the $s$-dependent model with potential $V$ by integration by parts. The result is a small modification of Equation~\eqref{234}:
\begin{equation}
\label{SDmoda}
\begin{split}
0 & = \sum_{0 \leq h \neq h' \leq g} s\Big(\mathcal{P}_{h}\otimes \mathcal{P}_{h'}[W_{2}^{s}](x,x) + \mathcal{P}_{h}[W_1^{s}](x)\cdot\mathcal{P}_{h'}[W_1^{s}](x)\Big) \\
& \quad + \sum_{h = 0}^{g} \Big(\mathcal{P}_{h}\otimes\mathcal{P}_{h}[W_2^{s}](x,x) + \mathcal{P}_{h}[W_1^{s}](x)\cdot \mathcal{P}_{h}[W_1^{s}](x)\Big)  \\
& \quad  + \Big(1 - \frac{2}{\beta}\Big)\partial_{x}W^{s}_1(x)  + \Big(1 - \frac{2}{\beta}\Big)\oint_{\mathsf{A}} \frac{\dd\xi}{2{\rm i}\pi}\,\frac{L_2(x;\xi,\xi)}{L(x)}\,W_1^{s}(\xi)  \\
& \quad - N \sum_{h = 0}^{g} \oint_{\mathsf{A_{h}}} \frac{\dd\xi}{2{\rm i}\pi}\,\frac{L(\xi)}{L(x)}\,\frac{V_h'(\xi) \mathcal{P}_{h}[W_{1}^{s}](\xi)}{x - \xi} - \frac{2}{\beta} \sum_{a \in (\partial \mathsf{A})_+} \frac{L(a)}{x - a}\partial_{a}\ln Z_{N,\beta;\bm{\epsilon}}^{T;\mathsf{A}}(s)  \\
& \quad - \sum_{0 \leq h \neq h' \leq g}  s\bigg(\oiint_{\mathsf{A}_h \times \mathsf{A}_{h'}}  \frac{\dd\xi_1\dd\xi_2}{(2{\rm i}\pi)^2}\,\frac{L_2(x;\xi_1,\xi_2)}{L(x)}\big(\mathcal{P}_{h}\otimes\mathcal{P}_{h'}[W_2^{s}](\xi_1,\xi_2) + \mathcal{P}_{h}[W_1^{s}](\xi_1)\cdot \mathcal{P}_{h'}[W_1^{s}](\xi_2)\big)\bigg) \\
& \quad - \sum_{h = 0}^{g} \oiint_{\mathsf{A}_{h}^2} \frac{\dd\xi_1\dd\xi_2}{(2{\rm i}\pi)^2}\,\frac{L_2(x;\xi_1,\xi_2)}{L(x)}\,\big(\mathcal{P}_{h}\otimes\mathcal{P}_{h}[W_{2}^{s}](\xi_1,\xi_2) + \mathcal{P}_{h}[W_1^{s}](\xi_1)\cdot\mathcal{P}_{h}[W_1^{s}](\xi_2)\big),
\end{split}
\end{equation}
For $n \geq 2$ a similar modification of Equation~\eqref{235} for the $n$-variables Dyson--Schwinger equations can be written down. 

\subsubsection{Analysis of the Dyson--Schwinger equations}\label{secDS}

Let $\mu_{{\rm eq};\bm{\epsilon}}^{V}$ be the equilibrium measure of the $\beta$-ensemble --- \textit{i.e.} $s = 1$ --- and fix $\mathsf{U}_{h}$ pairwise disjoint neighbourhoods of $\mathsf{A}_{h}$. We remark that the equilibrium measure $\mu_{{\rm eq};\bm{\epsilon}}^{T_{s};s}$  in the $s$-dependent model  with the choice of a $s$-dependent potential on the $h$-th segment ($h$ fixed):
$$
T_s(x) := V(x) - 2(1 - s)\sum_{0 \leq h' \neq h \leq g} \int_{\mathsf{A}_{h'}} \dd\mu_{{\rm eq};\bm{\epsilon}}^{V}(\xi)\ln[(x - \xi){\rm sgn}(h - h')]
$$
satisfies from \eqref{caract} with $T_{s}$ in place of $V$ the same characterisation as $\mu_{{\rm eq};\bm{\epsilon}}^{V}$, hence by uniqueness is equal to $\mu_{{\rm eq};\bm{\epsilon}}^{V}$ for any $s \in [0,1]$. This justifies the choice of $T_{s}$ in Lemma~\ref{htem}.

Let us study the $s$-dependent model with this choice of $s$-dependent potential. The correlators are still denoted $W_{k}^{s}$. The previous remark means that:
$$
W_{1}^{s} = N(W_{1}^{\{-1\}} + \Delta_{-1}W_{1}^{s}),\qquad \Delta_{-1}W_1^{s} = o(1),
$$
where $W_{1}^{\{-1\}}$ is the ($s$-independent) Stieltjes transform of $\mu_{{\rm eq};\bm{\epsilon}}^{V}$, and the error is uniform in $s \in [0,1]$. We now decompose the modified Dyson--Schwinger equations \eqref{SDmoda}  with $V_h=T_{s}$ and the many variables analogue as in  Equation~\eqref{437f}, Section~\ref{sec:subs}. Note that for $x$ near $\mathsf{A}_h$ for a fixed $h$, we have
\begin{equation}
\label{Tsprime}
T_s'(x) = V'(x) - 2(1-s)\sum_{0 \leq h' \neq h \leq g} \mathcal{P}_{h'}[W_1^{\{-1\}}](x)\,.
\end{equation}
The relevant operators $\mathcal{K}^{s}$ and $\Delta\mathcal{K}^{s}$ are now
\begin{equation*}
\begin{split}
\mathcal{K}^{s} & =  \mathcal{K} + \mathcal{D}^s,  \\
\Delta\mathcal{K}^{s} & =  \Delta\mathcal{K} + \Delta\mathcal{D}^s, \\
\Delta\mathcal{J}^s & =  \Delta\mathcal{J} + \tfrac{1}{2}\Delta\mathcal{D}^s,
\end{split}
\end{equation*}
where
\begin{equation*}
\begin{split}
\mathcal{D}^s[f](x) & = 2(s - 1) \sum_{0 \leq h \neq h' \leq g} \Big(\mathcal{P}_{h}[W_{1}^{\{-1\}}](x)\cdot \mathcal{P}_{h'}[f](x) - \mathcal{P}_{h'}\big[\mathcal{P}_{h}[W_1^{\{-1\}}] \cdot f\big](x)\Big),  \\
\Delta\mathcal{D}^s[f](x) & =  2(s - 1) \sum_{0 \leq h \neq h' \leq g} \Big(\mathcal{P}_{h}[\Delta W_{1}^{\{-1\};s}](x)\cdot \mathcal{P}_{h'}[f](x) - \mathcal{P}_{h'}\big[\mathcal{P}_{h}[W_1^{\{-1\}}] \cdot f\big](x)\Big).
\end{split}
\end{equation*}
The second term in $\mathcal{D}^s$ is the contribution of the extra term in the $s$-dependent potential \eqref{Tsprime} to the linearisation of the fourth line of the $s$-dependent Dyson--Schwinger equation \eqref{SDmoda}, while the first term is what remains from the linearisation of the two first lines of \eqref{SDmoda} after we isolate the contribution of the usual $s = 1$ operator $\mathcal{K}$.

In general $\mathcal{D}^s$ and $\Delta \mathcal{D}^s$ are non-zero operators. Indeed, if $g_h \in \mathcal{H}_1^{(1)}(\mathsf{A}_h)$ and $f \in \mathcal{H}_2^{(1)}(\mathsf{A})$, we have for $h \neq h'$
\begin{equation*}
\label{theprejum}\begin{split}
g_h(x) \cdot \mathcal{P}_{h'}[f](x) - \mathcal{P}_{h'}[g_h \cdot f](x) & = g_h(x) \oint_{\mathsf{A}_{h'}} \frac{\dd \xi}{2{\rm i}\pi}\,\frac{f(\xi)}{x - \xi} - \oint_{\mathsf{A}_{h'}} \frac{\dd \xi}{2{\rm i}\pi}\,\frac{g_h(\xi) f(\xi)}{x - \xi} \\
& = \oint_{\mathsf{A}_{h'}} f(\xi)\,\frac{g_h(x) - g_h(\xi)}{x - \xi}  \frac{\dd \xi}{2{\rm i}\pi}= - \oint_{\mathsf{A}_h} f(\xi)\,\frac{g_h(x) - g_h(\xi)}{x - \xi}  \frac{\dd \xi}{2{\rm i}\pi},
\end{split}
\end{equation*}
where the last expression comes from moving the contour away from $\mathsf{A}_h$, noticing that $\xi \mapsto \frac{g_h(x) - g_h(\xi)}{x - \xi}$ is holomorphic in $\mathbb{C} \setminus \mathsf{A}_h$, and that there is no contribution from $\infty$ since the integrands are $O(\frac{1}{\xi^2})$ as $\xi \rightarrow \infty$ (by definition of the spaces $\mathcal{H}^{(1)}_m$).  The nature of \eqref{theprejum} is better seen if we further assume that  $f$ and $g_h$ have upper/lower boundaries values on $\mathsf{A}$ (resp. $\mathsf{A}_{h'}$). Indeed, by computing the difference of upper and lower boundary values of \eqref{theprejum} for $x \in \mathsf{A}_h$, we find
\begin{equation}
\label{thejump}\big(g_h(x + {\rm i}0) - g_h(x - {\rm i}0)\big) \mathcal{P}_{h'}[f](x),
\end{equation}
while for $x \in \mathsf{A}_{k}$ for $k \neq h$ (including $k = h'$) we find $0$. Therefore, \eqref{theprejum} reconstructs the unique function in $\mathcal{H}_2^{(1)}(\mathsf{A}_{h})$ whose jump (from upper to lower boundary value) is \eqref{thejump}. The $(h,h')$-term in $\mathcal{D}^s[f]$ is $2(s-1)$ times \eqref{theprejum} with $g_h = \mathcal{P}_{h}[W_1^{\{-1\}}](x)$.

Unlike $\mathcal{K}$, the operator $\mathcal{K}^s$ cannot be explicitly inverted, but we can nevertheless prove the analogue of Lemma~\ref{ImK} and \ref{neglinu} by functional analysis arguments
\begin{proposition}
\label{propo} Assume Hypothesis~\ref{hmainc1}. ${\rm Im}\,\mathcal{K}^s$ is closed in $\mathcal{H}_{2}^{(1)}(\mathsf{A})$, and there exists an operator $(\widehat{\mathcal{K}}^s_{\bm{0}})^{-1}$, with domain ${\rm Im}\,\mathcal{K}^s$ and target the subspace of functions $\mathcal{H}_{2}^{(1)}(\mathsf{A})$ with zero $\bm{\mathcal{A}}$-periods,  providing the unique such solution $f(x) = (\widehat{\mathcal{K}}^s_{\bm{0}})^{-1}[\varphi](x)$ to the equation $\mathcal{K}^s[f](x) = \varphi(x)$. For any $\delta > 0$ independent of $N$, there exists $s$-independent constant $C(\delta)  > 0$ such that 
$$
\forall \varphi \in {\rm Im}\,\mathcal{K}^{s} \times \mathbb{C}^g,\qquad \p (\widehat{\mathcal{K}}^s_{\bm{0}})^{-1}[\varphi] \p_{\delta} \leq C(\delta) \p \varphi \p_{\delta}.
$$
Besides, 
\beq
\label{Ks9} \p (\widehat{\mathcal{K}}^{s}_{\bm{0}})^{-1}[\Delta\mathcal{X}^{s}] \p_{2\delta} \leq C'(\delta)\,\sqrt{\frac{\ln N}{N}} \p \varphi \p_{\delta},\qquad \mathcal{X} = \mathcal{K}\,\,{\rm or}\,\,\mathcal{J}.
\eeq
\end{proposition}
\noindent \textbf{Proof.} Given $\varphi$, let us try to solve the equation $\mathcal{K}^s[f](x) = \varphi(x)$ for a function $f$ such that $\oint_{\mathsf{A}_{h}} \frac{f(x)\dd x}{2{\rm i}\pi} = 0$ for any $h \in \ldbrack 1,g \rdbrack$. Following the computations of Section~\ref{opK}, we have
\beq
\label{Pari}\big({\rm id} + \mathcal{G}\circ\mathcal{D}^s + \Pi \big)[f](x) = \mathcal{G}[\varphi](x),
\eeq
where
$$
\Pi[f](x) = \Res_{\xi = \infty} \frac{\sigma(\xi)}{\sigma(x)}\,\frac{f(\xi)\dd\xi}{\xi - x}.
$$

We now prove that the operator $({\rm id} + \mathcal{G}\circ \mathcal{D}^s + \Pi)$ with domain the subspace of functions in $\mathcal{H}^{(1)}_{2}$ with zero $\bm{\mathcal{A}}$-periods, is injective. Assume we have an element $q$ in the kernel of this operator. The expression
\begin{equation}
\label{compq} q(x) = -(\mathcal{G}\circ \mathcal{D}^s)[q](x) - \Pi[q](x)
\end{equation}
and the fact that $\mathcal{P}_{h'}[q](x)$ is holomorphic in a neighbourhood of $\mathsf{A}_{h}$ for $h \neq h'$, shows that $\sigma(x)q(x)$ admits continuous upper and lower boundary values on $\mathsf{S}_{h}$, and is continuous across $\mathsf{A}_{h}\setminus\mathsf{S}_{h}$. Hence there exists an integrable measure $\nu^{q}$, supported on $\bigcup_{h = 0}^g \mathsf{S}_{h}$ such that
$$
q(x) = \int_{\mathsf{A}} \frac{\dd\nu^{q}(\xi)}{x - \xi}.
$$
As $q(x)$ has zero $\bm{\mathcal{A}}$-periods, we have $\nu^{q}(\mathsf{A}_{h}) = 0$ for every $h$. Besides, computation with Equation~\eqref{compq} shows that
$$
\forall h \in \ldbrack 0,g \rdbrack \quad \forall x \in \mathsf{S}_{h},\qquad \mathcal{P}_{h}[q](x + {\rm i}0) + \mathcal{P}_{h}[q](x - {\rm i}0) + 2s \sum_{\substack{h' = 0 \\ h' \neq h}} ^g \mathcal{P}_{h'}[q](x) = 0, 
$$
which means in terms of the measure $\nu^q$:
$$
\forall h \in \ldbrack 0,g \rdbrack ,\quad \forall x \in \mathsf{S}_{h},\qquad 2 \fint_{\mathsf{S}_{h}}  \frac{\dd\nu^{q}(\xi)}{x - \xi} + 2s\sum_{\substack{h' = 0 \\ h' \neq h}}^{g} \int_{\mathsf{S}_{h'}} \frac{\dd\nu^{q}(\xi)}{x - \xi} = 0.
$$
Integrating this equation from the left edge of $\mathsf{S}_{h}$ to $x$ in the segment $\mathsf{S}_{h}$ yields
$$
\forall h \in \ldbrack 0,g \rdbrack,\quad \forall x \in \mathsf{S}_{h},\qquad \sum_{h' = 0}^{g} 2\varsigma_{h,h'}^{s} \int_{\mathsf{S}_{h'}} \ln|x - \xi|\dd\nu^{q}(\xi) = c_{h}  
$$
for some constant $c_h$, where we remind that $\varsigma_{h,h'}^{s} = 1$ if $h = h'$, and $s$ if $h \neq h'$. Integrating this equation against the measure $\dd\nu^q$ over $\mathsf{S}_{h}$, the constant in the right-hand side disappears as $\nu^{q}(\mathsf{A}_{h}) = 0$. Then summing over $h$, we find
$$
\sum_{0 \leq h,h' \leq g} \iint_{\mathsf{S}_h \times \mathsf{S}_{h'}} \varsigma_{h,h'}^{s} \ln|x - \xi|\dd\nu^q_{h}(x)\dd\nu^q_{h'}(\xi) = 0,
$$
but we have shown that in \S~\ref{conces} that this equality implies $\nu^{q} = 0$, hence $q = 0$. This concludes the proof of injectivity.

Therefore, $({\rm id} + \mathcal{G}\circ \mathcal{D}^s + \Pi)$ is invertible on its image. We proceed to show the continuity of this inverse. For this purpose, we fix once for all contours $\gamma_h$  surrounding $\mathsf{A}_{h}$ and not $(\mathsf{A}_{h'})_{h' \neq h}$, and set $\gamma = \bigcup_{h = 1}^{g} \gamma_{h}$ and $\bm{\gamma} = (\gamma_h)_{h = 1}^{g}$. We equip $\gamma$ with a curvilinear measure. From the expression of these operators --- by moving the contour of integration to $\gamma$ --- one readily sees that $(\mathcal{G}\circ \mathcal{D}^s + \Pi)$ can be considered as endomorphisms of $L^2(\gamma)$, denote $\mathfrak{N}^s$, which is compact. Let $\tilde{\gamma}$ be the disjoint union of  the set $\ldbrack 1,g \rdbrack$ (equipped with the uniform measure) and $\gamma$ (equipped with the curvilinear measure), so $L^2(\tilde{\gamma}) = \mathbb{C}^{g} \oplus L^2(\gamma)$. We consider further the operator
$$
\widehat{\mathfrak{N}}^s\,:\,\begin{array}{rcl} L^2(\tilde{\gamma})  & \longrightarrow & L^2(\tilde{\gamma}) \\ \big(\bm{w}, \phi\big) & \longmapsto & \big(-\bm{w} + \oint_{\bm{\gamma}} \frac{\phi(\xi)\,\dd \xi}{2{\rm i}\pi}\,,\, \mathfrak{N}^s[\phi]\big) \end{array},
$$
and one can check as before that ${\rm id} + \widehat{\mathfrak{N}}^s$ is injective. As $\widehat{\mathfrak{N}}^s$ is compact, Fredholm alternative ensures that ${\rm id} + \widehat{\mathfrak{N}}^s$ is continuously invertible. Its inverse is ${\rm id} - \mathfrak{R}^s$ where $\mathfrak{R}^s$ is the resolvent operator of $\widehat{\mathfrak{N}}^s$, and it has a smooth integral kernel. This is enough to prove continuous invertibility of $\widehat{\mathcal{K}}^s$ and a bound for the norm of its inverse. The sought for inverse for $\widehat{\mathcal{K}}^s$ is
$$
f(x) = {\rm pr}_{2} \circ (\widehat{\mathcal{K}}_{\bm{0}}^s)^{-1}\circ \mathcal{G}[\varphi](x) = ({\rm id}- \mathfrak{R}^s)(\bm{0},\mathcal{G}[\varphi]),
$$
where ${\rm pr}_{2}$ is the projection on the second factor $L^2(\tilde{\gamma})$. The fact that this solution is actually in $\mathcal{H}_{2}^{(1)}(\mathsf{A})$ can be read from the equivalent versions of Equation~\eqref{Pari} that we have encountered earlier, namely \eqref{426f} where one takes into account that ${\rm Im}\,\mathcal{G} \subseteq \mathcal{H}_2^{(1)}(\mathsf{A})$ (manifest on \eqref{Gdeq}) and the fact that $\psi(x)$ is a polynomial of degree $g - 1$ while $\sigma(x)$ is the squareroot of a polynomial of degree $2g + 2$.

The very construction of $\widehat{\mathfrak{N}}^s$ guarantees that $\oint_{\bm{\gamma}} \frac{f(x)\dd x}{2{\rm i}\pi} = 0$ as desired, and the estimate on the norm of $(\widehat{\mathcal{K}}^{s}_{\bm{0}})^{-1}$ comes from the properties of the resolvent kernel. The proof of the estimate \eqref{Ks9} follows the steps of Lemma~\ref{neglinu} and is omitted. \hfill $\Box$

\vspace{0.2cm}

For $n \geq 2$ variables, the Dyson--Schwinger equations of the $s$-deformed model can be recast as
$$
(\mathcal{K}^s + \Delta\mathcal{K}^s)[W_{n}^{s}(\bullet,x_I)](x) = A_{n + 1}^{s}(x;x_I) + B_{n}^{s}(x;x_I) + C_{n - 1}^{s}(x;x_I) + D_{n - 1}^{s}(x;x_I),
$$
with modified expression for $A$ and $B$. For $n \geq 2$, we have:
\begin{equation*}
\begin{split}
A_{n + 1}^{s}(x;x_I) & = N^{-1}(\mathcal{L}_{2} - {\rm id})\bigg\{ s\Big(\sum_{0 \leq h \neq h' \leq g} \mathcal{P}_{h}\otimes \mathcal{P}_{h'}[W_{n + 1}^s(\bullet_1,\bullet_2,x_I)](x,x)\Big) \\
& \quad \qquad \qquad \qquad + \sum_{h = 0}^g \mathcal{P}_{h}\otimes \mathcal{P}_{h}[W_{n + 1}^{s}(\bullet_1,\bullet_2,x_I)](x,x)\Big)\bigg\}, \\
B_{n + 1}^{t}(x;x_I) & =  N^{-1}(\mathcal{L}_2 - {\rm id})\bigg\{\sum_{\substack{J \subseteq I \\ J \neq (\emptyset,I)}} \sum_{0 \leq h \neq h' \leq g} s\mathcal{P}_{h}[W_{|J| + 1}^{s}(\bullet,x_J)](x)\cdot \mathcal{P}_{h'}[W_{n - |J|}^{s}(\bullet,x_{I\setminus J})](x) \\
& \quad \qquad\qquad\qquad + \sum_{h = 0}^g \mathcal{P}_{h}[W_{|J| + 1}^{s}(\bullet,x_J)](x)\cdot\mathcal{P}_{h}[W_{n - |J|}^{s}(\bullet,x_{I\setminus J})](x)\bigg\}, \\
C_{n - 1}^{s}(x;x_I) & =  -\frac{2}{\beta N} \sum_{i \in I} \mathcal{M}_{x_i}[W_{n - 1}^{s}(\bullet,x_{I\setminus\{i\}})](x), \\
D_{n - 1}^{s}(x;x_I) & =  \frac{2}{\beta N} \sum_{a \in (\partial\mathsf{A})_+} \frac{L(a)}{x - a}\,\partial_{a} W_{n - 1}^{s}(x_I).
\end{split}
\end{equation*} 
And for $n = 1$ variable, we find the analogue of Equation~\eqref{AAAB}
$$
\big[\mathcal{K}^s + \Delta\mathcal{J}^{s}][\Delta_{-1}W_1^{s}](x) = \frac{A_2^{s}(x) + D_0^{s}}{N} - \frac{1 - 2/\beta}{N}(\partial_{x} + \mathcal{L}_{1})[W_{1}^{\{-1\}}](x) + \mathcal{N}_{(\Delta_0 V)',0}[W_{1}^{\{-1\}}](x),
$$
with:
\begin{equation*}
\begin{split}
\Delta_{-1}P^{s}(x;\xi) & = \oint_{\mathsf{A}} \frac{\dd\eta}{2{\rm i}\pi}\,2L_2(x;\xi,\eta)\,\Delta_{-1}W_1^{s}(\eta), \\
\Delta \mathcal{J}^{s}[f](x) & = - \mathcal{N}_{(\Delta_0 V)',\Delta_{-1}P^{s}(x;\bullet)/2}[f](x) + \sum_{0 \leq h \neq h' \leq g} s\mathcal{P}_{h}[\Delta_{-1}W_1^{s}](x)\,\mathcal{P}_{h'}f(x) \\
& \quad + \sum_{h} \mathcal{P}_{h}[\Delta_{-1}W_1^{s}](x)\,\mathcal{P}_{h}[f](x) + \frac{1}{N}\Big(1 - \frac{2}{\beta}\Big)(\partial_{x} + \mathcal{L}_{1})[f](x).
\end{split}
\end{equation*}

One can then repeat all the steps of Section~\ref{S523}, the key point being that we use the inverse $(\widehat{\mathcal{K}}^{s}_{\bm{0}})^{-1}$ of $\mathcal{K}^s$ and its norm estimate constructed in Proposition~\ref{propo}. This results in the proof of an asymptotic expansion, for any $K \geq 0$:
$$
W_{n}^{s}(x_1,\ldots,x_n) = \sum_{k = n - 2}^{K} N^{-k}\,W_n^{\{k\};s}(x_1,\ldots,x_n) + O(N^{-(K + 1)}),
$$
where the coefficients $W_n^{\{k\};s}$ are $N$-independent, are given by a $s$-dependent recursion which is a $s$-dependent modification of the recursions provided in Section~\ref{secini}.

\subsection{Regularity with respect to the filling fractions}

Let $\bm{\epsilon}_{\star}$ be the equilibrium filling fraction in the initial model $\mu_{N,\beta}^{V;\mathsf{A}}$. In order to finish the proof of Theorem~\ref{th:3}, it remains to show that the Hypotheses~\ref{hmainc1}--\ref{hmainc3} for $\mu_{N,\beta}^{V;\mathsf{A}}$ imply Hypothesis~\ref{hyu222} for the model $\mu_{N,\beta;\bm{\epsilon}}^{V;\mathsf{A}}$ with fixed filling fractions $\bm{\epsilon} \in \mathcal{E}$ close enough to $\bm{\epsilon}_{\star}$, that all coefficients of the expansion extend as smooth functions of $\bm{\epsilon}$, and that the Hessian of $F^{\{-2\}}_{\bm{\epsilon}}$ with respect to filling fractions is negative definite. These properties are proved in  the Appendix, see Propositions~\ref{Lhq2}--\ref{hess}. 
\begin{lemma}
\label{sqoi} If $V$ satisfies Hypotheses~\ref{hmainc1}--\ref{hmainc4}, then $(V,\bm{\epsilon})$ satisfies Hypotheses~\ref{hyu222} for $\bm{\epsilon} \in \mathcal{E}$ close enough to $\bm{\epsilon}_{\star}$. Besides, the soft edges $\alpha_h^{\bullet}$ and $W_{1;\bm{\epsilon}}^{\{-1\}}(x)$ (for $x$ away from the edges) extend as  $\mathcal{C}^{\infty}$ functions of $\bm{\epsilon}$, while the hard edges remain unchanged, at least for $\bm{\epsilon}$ close enough to $\bm{\epsilon}_{\star}$.
\end{lemma}
We observe that, once $W^{\{-1\}}_{1;\bm{\epsilon}}$ and the edges of the support $\alpha_{\bm{\epsilon},h}^{\bullet}$ are known, the $W_{n;\bm{\epsilon}}^{\{k\}}$ for any $n \geq 1$ and $k \geq 0$ are determined recursively by Equations~\eqref{weo}--\eqref{W1cor} and \eqref{490}--\eqref{fdiex}, where the linear operator $\widehat{\mathcal{K}}^{-1}$ is given explicitly in Equations~\eqref{Gdeq}--\eqref{inveJ}, and thus depends smoothly on $\bm{\epsilon}$ close enough to $\bm{\epsilon}_{\star}$. Similarly, $F^{\{k\}}_{\beta;\bm{\epsilon}}$ for $k \geq 0$ are obtained from Equation~\eqref{exp1cut} leading to Equations~\eqref{cteF}--\eqref{higherF}, which shows their smooth dependence for $\bm{\epsilon}$ close enough to $\bm{\epsilon}_{\star}$.

\begin{corollary}
\label{co73} If $V$ satisfies Hypotheses~\ref{hmainc1}--\ref{hmainc4}, then $W_{n;\bm{\epsilon}}^{\{k\}}(x_1,\ldots,x_k)$ (for $x_1,\ldots,x_k$ away from the support) and $F^{\{k\}}_{\beta;\bm{\epsilon}}$ extend as  $\mathcal{C}^{\infty}$ functions of $\bm{\epsilon} \in \mathcal{E}_{g}$ close enough to $\bm{\epsilon}_{\star}$. \hfill $\Box$
\end{corollary}

This concludes the proof of Theorem~\ref{th:2} announced in Section~\ref{fixfil}.

\section{Asymptotic expansion in the initial model in the multi-cut regime}
\label{musqo}

\subsection{The partition function (Proof of Theorem~\ref{th:22})}
\label{S8111}
We come back to the initial model $\mu_{N,\beta}^{V;\mathsf{A}}$, and we assume Hypotheses~\ref{hmainc1}--\ref{hmainc4} with number of cuts $(g + 1) \geq 2$. We remind the notation $\bm{N} = (N_h)_{1 \leq h \leq g}$ for the number of eigenvalues in $\mathsf{A}_h$, and the number of eigenvalues in $\mathsf{A}_0$ is $N_0 = N - \sum_{h = 1}^{g} N_h$. The $N_h$ are here random variables, which take the value $N\bm{\epsilon}$ with probability $Z_{N,\beta;\bm{\epsilon}}^{V;\mathsf{A}}/Z_{N,\beta}^{V;\mathsf{A}}$. We denote $\bm{\epsilon}_{\star}$ the vector of equilibrium filling fractions, and $\bm{N}_\star = N\bm{\epsilon}_{\star}$. Let us summarise five essential points:
\begin{itemize}
\item[$\bullet$] By concentration of measures, Corollary~\ref{cosq} yields the existence of a constant $c,c' > 0$ such that, for $N$ large enough,
\beq
\label{81}\mu_{N,\beta}^{V;\mathsf{A}}\Big(|\bm{N} - \bm{N}_{\star}|_1 > c\sqrt{N \ln N}\Big) \leq e^{-c'N\ln N}.
\eeq
\item[$\bullet$] We have established in Theorem~\ref{th:2} an expansion for the partition function with fixed filling fractions.
\item[$\bullet$] Thanks to the strong off-criticality assumption and Lemma~\ref{sqoi}, we can apply Proposition~\ref{propodnun}: there exists $c''>0$ small enough such that for $|\bm{\epsilon} - \bm{\epsilon}*|_1 \leq c''$, the model with fixed filling fractions $\bm{\epsilon}$ admits an asymptotic expansion of the form, for any $K \geq 0$:
\beq
\label{81eq}\frac{N!\,Z_{N,\beta;\bm{\epsilon}}^{V;\mathsf{A}}}{\prod_{h = 0}^g (N\epsilon_h)!} = N^{\frac{\beta}{2}N + \varkappa}\exp\Big(\sum_{k = -2}^{K} N^{-k}\,F_{\beta;\bm{\epsilon}}^{\{k\};V} + O(N^{-(K + 1)})\Big),
\eeq
with $\varkappa$ independent of $\bm{\epsilon}$ and given by Equation~\eqref{varkappauni} and an error depending only on $c''$.
\item[$\bullet$] As established later in Proposition~\ref{hess},  the Hessian $(F^{\{-2\}}_{\beta;\star})''$ is negative definite.
\item[$\bullet$] According to Lemma~\ref{sqoi}, $\bm{\epsilon} \mapsto F^{\{k\};V}_{\beta;\bm{\epsilon}}$ is smooth in the domain $|\bm{\epsilon} - \bm{\epsilon}_\star| < c''$. From there, we deduce that, for any $K,k \geq -2$, there exist a constant $C_{k,K} > 0$ and tensors $(F_{\beta;\star}^{\{k\}})^{(j)} = \partial_{\bm{\epsilon}}^{\otimes j} F_{\beta;\bm{\epsilon}}^{\{k\};V}|_{\bm{\epsilon} = \bm{\epsilon}_{\star}}$, such that:
\beq
\label{r8}\left| N^{-k}\,F^{\{k\};V}_{\beta;\bm{N}/N} - \sum_{j = 0}^{K - k} N^{-(k + j)}\frac{(F^{\{k\}}_{\beta;\star})^{(j)}}{j!}\cdot(\bm{N} - \bm{N}_{\star})^{\otimes j}\right| \leq C_{k,K}\,N^{-(K + 1)}|\bm{N}- \bm{N}_{\star}|^{K - k + 1}_1.
\eeq

\end{itemize} 
We now proceed with the proof of Theorem~\ref{th:22}.

\subsubsection{Taylor expansion around the equilibrium filling fraction}

Due to the large deviation estimates for filling fractions  \eqref{81}, we can write for $N$ large enough
$$
(Z_{N,\beta}^{V;\mathsf{A}})^{-1} 
\bigg(\sum_{\substack{0 \leq N_1,\cdots,N_g \leq N \\ |\bm{N} - \bm{N}_{\star}|_1 \leq c\sqrt{N \ln N}}} \frac{N!\,Z_{N,\beta;\bm{N}/N}^{V;\mathsf{A}}}{\prod_{h = 0}^{g} N_h!} \bigg)=
\mu_{N,\beta}^{V;\mathsf{A}}\Big(|\bm{N} - \bm{N}_{\star}|_1 \le  c\sqrt{N \ln N}\Big)= 1
 + O(e^{-c'N \ln N}).
$$
In other words
$$Z_{N,\beta}^{V;\mathsf{A}}=
\bigg(\sum_{\substack{0 \leq N_1,\cdots,N_g \leq N \\ |\bm{N} - \bm{N}_{\star}|_1 \leq c\sqrt{N \ln N}}} \frac{N!\,Z_{N,\beta;\bm{N}/N}^{V;\mathsf{A}}}{\prod_{h = 0}^{g} N_h!} \bigg)\big(1
 + O(e^{-c'N \ln N})\big)\,.$$
For the range of filling fractions appearing in the sum in the right hand side, we dispose of an  asymptotic expansion  of each term which are 
the partition functions of the model with fixed filling fractions by \eqref{81eq}.  Moreover, we can do the 
Taylor expansion of its coefficients with respect to $\bm{N}/N$ around $\bm{\epsilon}_{\star}$ by \eqref{r8}
 up to order $O(N^{-(2K + 1)})$ (and these errors are uniform for the range of filling fractions considered). This gives:
 
\begin{equation}
\label{peirugb}
\begin{split}
& \quad \sum_{\substack{0 \leq N_1,\ldots,N_g \leq  N \\ |\bm{N} - \bm{N}_{\star}|_1 \leq c\sqrt{N \ln N}}} \frac{N!\,Z_{N,\beta;\bm{N}/N}^{V;\mathsf{A}}}{\prod_{h = 0}^g N_h!}\, \\
& =  \sum_{\substack{0 \leq N_1,\ldots,N_g \leq N \\ |\bm{N} - \bm{N}_{\star}|_1 \leq c\sqrt{N} \ln N}} \exp\bigg(\sum_{k = -2}^{2K} \sum_{j = 0}^{2K - k} N^{-(k + j)}\,\frac{(F^{\{k\}}_{\beta;\star})^{(j)}}{j!}\cdot(\bm{N} - \bm{N}_{\star})^{\otimes j} + N^{-(2K + 1)}R_{2K}(\bm{N})\bigg),
\end{split}
\end{equation}
The error $N^{-(2K + 1)}R_{2K}(\bm{N})$ can be controlled according to Equation~\eqref{r8} using the constraint $|\bm{N} - \bm{N}_{\star}|_1 \leq c \sqrt{N \ln N}$, as follows: 
\begin{equation}
\begin{split}
|N^{-(2K + 1)}R_{2K}(\bm{N})| & \leq N^{-(2K+1)}\sum_{k=-2}^{2K} C_{k,2K}c^{2K - k}|\bm{N}-\bm{N}_{\star}|_1^{2K - k} \\
& \leq C_Kc^{2K} N^{-(2K+1)} N^{K}(\ln N)^{K}=C'_K N^{-K-1} (\ln N)^K\,.
\end{split}
\end{equation}
Note here that all sums are finite since we stop them up to an error term which is uniformly bounded.
Observing that $\exp(N^{-(K + 1)}R_K(\bm{N})) - 1 = O\big(N^{-K - 1}(\ln N)^{K}\big)$ when $N \rightarrow \infty$ uniformly over the range of filling fractions on which we sum in Equation~\eqref{peirugb}, we get:
\begin{equation}
\label{expnund} \frac{Z_{N,\beta}^{V;\mathsf{A}}}{1 + O(N^{-(K + 1)}(\ln N)^{K})} = \sum_{\substack{0 \leq N_1,\ldots,N_g \leq N \\ |\bm{N} - \bm{N}_{\star}|_1 \leq c \sqrt{N\ln N}}}  \exp\bigg(\sum_{k = -2}^{2K} \sum_{j = 0}^{2K - k} N^{-(k + j)}\,\frac{(F^{\{k\}}_{\beta;\star})^{(j)}}{j!}\cdot(\bm{N} - \bm{N}_{\star})^{\otimes j}.\bigg).\end{equation} 
Here the previous error $O(e^{-c'N \ln N})$ has been absorbed in the larger error $O(N^{-(K + 1)}(\ln N)^K)$.
 
Since $\bm{\epsilon}_{\star}$ is the equilibrium filling fraction, which is characterised as the filling fraction maximising $F^{\{-2\}}_{\beta;\bm{\epsilon}}$, we have $(F^{\{-2\}}_{\beta;\star})' = 0$. We can factor out the exponential containing the $F_{\beta;\star}^{\{-k\}}$ without derivative. We then expand the exponential of terms containing $(F_{\beta;\star}^{\{k\}})^{(j)}$ with $k + j > 0$, doing so up to an error of magnitude $O(N^{-(K + 1)}(\ln N)^{K})$. In this way only, remain in the exponential $(F_{\beta;\star}^{\{-2\}})' = 0$, $(F_{\beta;\star}^{\{-2\}})''$ and  $(F_{\beta;\star}^{\{-1\}})'$. The result is the following expansion (note here that all sums are finite, including the one on $r$)
 \begin{equation}
\label{splint} \begin{split}
& \quad \frac{Z_{N,\beta}^{V;\mathsf{A}}}{1 + O(N^{-(K + 1)}(\ln N)^{K})} \\
& = \exp\bigg(\sum_{k = -2}^{K} N^{-k} F_{\beta;\star}^{\{k\}}\bigg) \times \Bigg\{ \sum_{r \geq 0} \frac{1}{r!} \sum_{\substack{k_1,\ldots,k_r \geq -2 \\ j_1,\ldots,j_r \geq 1 \\ k_i + j_i > 0 \\ \sum_{i = 1}^{r} k_i + j_i \leq 2K}}  \bigotimes_{i = 1}^{r} \frac{(F_{\beta;\star}^{\{k_i\}})^{(j_i)}}{j_i!} \\
& \quad \cdot \bigg(\sum_{\substack{0 \leq N_1,\ldots,N_g \leq N \\ |\bm{N} - \bm{N}_{\star} |_1 \leq c \sqrt{N \ln N}}} N^{-\sum_{i = 1}^r (k_i + j_i)} \,(\bm{N} - \bm{N}_{\star})^{\otimes (\sum_{i = 1}^r j_i)} e^{\frac{1}{2}(F_{\beta;\star}^{\{-2\}})'' \cdot (\bm{N} - \bm{N}_{\star})^{\otimes 2} + (F_{\beta;\star}^{\{-1\}})' \cdot (\bm{N} - \bm{N}_{\star})}\bigg)\Bigg\}.
\end{split}
\end{equation}

\subsubsection{Waiving the constraint on the sum}

Our next task will be, for each of the finitely many tuples $(j_1,\ldots,j_r)$ involved in the sum in the last line, to replace the constrained sum over $\bm{N}$ such that $|\bm{N} - \bm{N}_{\star}|  \leq c \sqrt{N \ln N}$, with an unconstrained sum over $\bm{N} \in \mathbb{Z}^g$. This will be possible because $(F^{\{-2\}}_{\beta;\star})''$ is negative definite (Proposition~\ref{hess}), in other words because the minimum eigenvalue $q$ of the symmetric matrix $-(F^{\{-2\}}_{\beta;\star})''$ is positive. More precisely, set $J = \sum_{i = 1}^{r} j_i$ and notice Equation~\eqref{splint} only involves $J \leq 2K$. Let us equip $\mathbb{R}^{g}$ with the euclidean norm:
$$
\forall \bm{w} \in \mathbb{R}^g,\qquad |\bm{w}|_2 = \sqrt{\sum_{h = 1}^{g} w_h^2}.
$$
In particular we denote $r = \big|(F_{\beta;\star}^{\{-1\}})'|_2$. The tensor product $(\mathbb{R}^{g})^{\otimes J}$ is naturally equipped with a euclidean norm also denoted $|\cdot|_2$, such that:
$$
\forall \bm{w}_1,\ldots,\bm{w}_{J} \in \mathbb{R}^g \qquad |\bm{w}_1 \otimes \cdots \otimes \bm{w}_J|_2 = |\bm{w}_1|_2 \cdots |\bm{w}_J|_2 .
$$
Let $m$ be a positive integer. We shall estimate the contribution --- with respect of the aforementioned euclidean norm in $(\mathbb{R}^{g})^{\otimes J}$ --- that the $\bm{N} \in \mathbb{Z}^g$ in the shell between the euclidean balls of radius $m - 1$ and $m$ would give to the sum:
\begin{equation*}
\begin{split}
& \sum_{\substack{\bm{N} \in \mathbb{Z}^g \\   |\bm{N} - \bm{N}_{\star}|_{2} \geq m}} \big|\bm{N} - \bm{N}_{\star}\big|_{2}^J e^{\frac{1}{2}(F_{\beta;\star}^{\{-2\}})'' \cdot (\bm{N} - \bm{N}_{\star})^{\otimes 2} + (F_{\beta;\star}^{\{-1\}})' \cdot (\bm{N} - \bm{N}_{\star})} \\
& \leq \sum_{\substack{\bm{N} \in \mathbb{Z}^g \\ m - 1 \leq |\bm{N} - \bm{N}_{\star}|_{2} < m}} m^J e^{-\frac{q}{2}(m - 1)^2 + mr}, \\
& \leq C\,m^{J + g - 1} e^{-\frac{q}{2}m^2 + mr'}
\end{split}
\end{equation*}
for some constants $C > 0$ coming from the number of integer points in the spherical shell in $g$-dimensional space, and $r' = r + q$. Then, there exists $M_K > 0$ such that for $m \geq M_K$ we have $C\,m^{J + g - 1} e^{-qm^2 + mr'} \leq e^{-\frac{q}{4}m^2}$. Up to choosing a larger $M_K$ we can assume as well for $M > M_K$:
$$
e^{-\frac{q}{2}M} < 1,\qquad \frac{e^{-\frac{q}{4}M^2}}{1 - e^{-\frac{q}{2}M}} \leq e^{-\frac{q}{8}M^2}.
$$
Then:
\begin{equation*}
\label{iubg}\begin{split}
& \quad \sum_{\substack{\bm{N} \in \mathbb{Z}^g \\  |\bm{N} - \bm{N}_{\star}|_{2} \geq M}} \big|\bm{N} - \bm{N}_{\star}\big|_{2}^J \exp\Big(\frac{1}{2} (F_{\beta;\star}^{\{-2\}})'' \cdot (\bm{N} - \bm{N}_{\star})^{\otimes 2} + (F_{\beta;\star}^{\{-1\}})' \cdot (\bm{N} - \bm{N}_{\star})\Big)  \\
& \leq \sum_{m \geq M} e^{-\frac{q}{4}m^2} \leq \sum_{m \geq 0} e^{-\frac{q}{4}M^2 - \frac{q}{2}mM} = \frac{e^{-\frac{q}{4}M^2}}{1 - e^{-\frac{q}{2}M}} \leq e^{-\frac{q}{8}M^2}.
\end{split}
\end{equation*} 
By Cauchy--Schwarz inequality, we have $|\bm{N} - \bm{N}_{\star}|_1 \leq \sqrt{g}|\bm{N} - \bm{N}_{\star}|_2$. Therefore, the terms $\bm{N} \in \mathbb{Z}^{g}$ not included in Equation~\eqref{splint} can be bounded by Equation~\eqref{iubg} with the choice $M = \big\lceil c\sqrt{\frac{N}{g} \ln N} \big\rceil$:
\begin{equation*}
\begin{split}
& \quad \bigg| \sum_{\substack{\bm{N} \in \mathbb{Z}^g \\  |\bm{N} - \bm{N}_{\star}|_{1} \geq c \sqrt{N \ln N}}} (\bm{N} - \bm{N}_{\star})_{2}^{\otimes J} \exp\Big((F_{\beta;\star}^{\{-2\}})'' \cdot (\bm{N} - \bm{N}_{\star})^{\otimes 2} + (F_{\beta;\star}^{\{-1\}})' \cdot (\bm{N} - \bm{N}_{\star})\Big)\bigg|_{2}  \\
& = O\big(e^{-\frac{q}{8}(\lceil c\sqrt{\frac{N}{g}\ln N} \rceil)^2}\big) = O(e^{-q'N \ln N})
\end{split}
\end{equation*}
for some constant $q' > 0$ when $N \rightarrow \infty$. As a result:
\begin{equation}
\label{vubfivs}\begin{split}
& \quad \frac{Z_{N,\beta}^{V;\mathsf{A}}}{1 + O(N^{-(K + 1)}(\ln N)^{K})} \\
& = \exp\bigg(\sum_{k = -2}^{K} N^{-k} F_{\beta;\star}^{V}\bigg) \times \Bigg\{ \sum_{r \geq 0} \frac{1}{r!} \sum_{\substack{k_1,\ldots,k_r \geq -2 \\ j_1,\ldots,j_r \geq 0 \\ k_i + j_i > 0 \\ \sum_{i = 1}^{r} k_i + j_i \leq 2K}}  N^{-\sum_{i = 1}^r (k_i + j_i)}  \bigotimes_{i = 1}^{r} \frac{(F_{\beta;\star}^{\{k_i\}})^{(j_i)}}{j_i!} \\
& \quad \cdot \bigg(\sum_{\bm{N} \in \mathbb{Z}^g} (\bm{N} - \bm{N}_{\star})^{\otimes (\sum_{i = 1}^r j_i)} \exp\Big((F_{\beta;\star}^{\{-2\}})'' \cdot (\bm{N} - \bm{N}_{\star})^{\otimes 2} + (F_{\beta;\star}^{\{-1\}})' \cdot (\bm{N} - \bm{N}_{\star})\Big)\bigg)\Bigg\}.
\end{split}
\end{equation}
Note that the error may not be uniform when $K$ increases, due to the choice of $M_K$ in the intermediate steps.

Eventually, we recognize in the sum of the last line the $J$-th tensor of derivatives of the Theta function defined in Equation~\eqref{thetadeff}, with arguments:
$$
\tau_{\beta;\star} = \frac{(F^{\{-2\}}_{\beta;\star})''}{2{\rm i}\pi},\qquad \bm{v}_{\beta;\star} = \frac{(F_{\beta;\star}^{\{-1\}})'}{2{\rm i}\pi},
$$
More precisely:
$$
\sum_{\bm{N} \in \mathbb{Z}^g} (\bm{N} - \bm{N}_{\star})^{\otimes J} e^{\frac{1}{2}(F_{\beta;\star}^{\{-2\}})'' \cdot (\bm{N} - \bm{N}_{\star})^{\otimes 2} + (F_{\beta;\star}^{\{-1\}})' \cdot (\bm{N} - \bm{N}_{\star})} = \Big(\frac{\nabla_{\bm{v}}}{2{\rm i}\pi}\Big)^{\otimes J} \vartheta\!\left[\begin{array}{@{\hspace{-0.02cm}}l@{\hspace{-0.02cm}}} -\bm{N}_{\star} \\\,\,\,\, \bm{0} \end{array}\right]\!\!(\bm{v}|\bm{\tau}_{\beta;\star})\Big|_{\bm{v} = \bm{v}_{\beta;\star}},
$$
and this contribution is of order $1$, so that we only need to sum up to $\sum_{i = 1}^{r} k_i + j_i \leq K$ in  Equation~\eqref{vubfivs} to get the expansion up to $O(N^{-(K + 1)}(\ln N)^{K})$.  By looking at the expansion for $K \mapsto K + 1$, we know that this error done for the expansion with $K$ is in fact $O(N^{-(K + 1)})$. This concludes the proof of Theorem~\ref{th:22}.

\subsection{Deviations of filling fractions from their mean value (Proof of Theorem~\ref{th:33t})}
\label{Lare}
We now describe the fluctuations of the number of eigenvalues in each segment. Let $\bm{P} = (P_1,\ldots,P_g)$ be a vector of integers, depending on $N$ in such a way that $\bm{P} - N\epsilon_{\star,h} = o(N^{\frac{1}{3}})$ when $N \rightarrow \infty$. We set $P_0 = N - \sum_{h = 1}^{g} P_h$. The joint probability for $h \in \ldbrack 1,g \rdbrack$ to find $P_h$ eigenvalues in the segment $\mathsf{A}_h$ is:
$$
\mu_{N,\beta}^{V;\mathsf{A}}[\bm{N} = \bm{P}] = \frac{N!}{\prod_{h = 0}^g P_h!}\,\frac{Z_{N,\beta;\bm{P}/N}^{V;\mathsf{A}}}{Z_{N,\beta}^{V;\mathsf{A}}}.
$$
We recall that the coefficients of the large $N$ expansion of the numerator are smooth functions of $\bm{P}/N$. Therefore, we can perform a Taylor expansion in $\bm{P}/N$ close to $\bm{\epsilon}_{\star}$ with the method used in Section~\ref{musqo}. We leave out the details, and only state the result: provided $\bm{P} - N\bm{\epsilon}_{\star} = o(N^{\frac{1}{3}})$, only the quadratic term of the Taylor expansion remains when $N \rightarrow \infty$:
\begin{equation*}
\begin{split}
\mu_{N,\beta}^{V;\mathsf{A}}[\bm{N} = \bm{P}] & \sim \frac{e^{\frac{1}{2}\,(F^{\{-2\}}_{\beta;\star})^{''}\cdot(\bm{P} - N\bm{\epsilon}_{\star})^{\otimes 2} + (F^{\{-1\}}_{\beta;\star})'\cdot(\bm{P} - N\bm{\epsilon}_{\star})}}{\vartheta\big[\begin{smallmatrix} -\bm{N}_{\star}\\ \bm{0} \end{smallmatrix}\big](\bm{v}_{\beta;\star}|\bm{\tau}_{\beta;\star})}.
\end{split}
\end{equation*}
In other words, the random vector $\Delta\bm{N} = (\Delta N_1,\ldots,\Delta N_g)$ defined by:
$$
\Delta N_{h} = N_h - N\epsilon_{\star,h} + \sum_{h' = 1}^g [(F^{\{-2\}}_{\beta;\star})'']^{-1}_{h,h'}\,(F^{\{-1\}}_{\beta;\star})'_{h'}
$$
is approximated in law by a random Gaussian vector, with covariance $[(F^{\{-2\}}_{\beta;\star})'']^{-1}$ and conditioned to live in the shifted lattice
$$
\Delta\bm{N} \in \Big(\mathbb{Z}^{g} - \lfloor N\bm{\epsilon}_{\star} \rfloor +\sum_{h' = 1}^g [(F^{\{-2\}}_{\beta;\star})'']^{-1}_{h,h'}\,(F^{\{-1\}}_{\beta;\star})'_{h'}\Big)
$$
where for $\bm{w} \in \mathbb{R}^g$ we denote $\lfloor \bm{w} \rfloor = \big(\lfloor w_1 \rfloor,\cdots,\lfloor w_g \rfloor\big)$. Strictly speaking, we cannot say that we have a convergence in law to a discrete Gaussian because the shift of the lattice oscillates with $N$. We observe that, when $\beta = 2$ and the potential $V$ is independent of $N$, the vector $F^{\{-1\}}_{\beta;\star}$ vanishes so that $\bm{N} - N\bm{\epsilon}_{\star}$  is approximated in law by a centered Gaussian vector conditioned to live in the shifted lattice $\big(\mathbb{Z}^g - \lfloor N \bm{\epsilon}_{\star}†\rfloor\big)$.

\subsection{Fluctuations of linear statistics}
\label{S833}

With a strategy similar to \S~\ref{S5clt}, the result of Section~\ref{S8111} implies, for $\varphi$ a test function which is analytic in a neighbourhood of $\mathsf{A}$:
\begin{equation}
\label{clclc} 
\mu_{N,\beta}^{V;\mathsf{A}}\big(e^{{\rm i}s\big(\sum_{i = 1}^N \varphi(\lambda_i) - N\int_{\mathsf{S}} \varphi(\xi)\dd\mu_{{\rm eq}}^V(\xi)\big)}\big)  \sim e^{{\rm i}s\,M_{\beta;\star}[\varphi] - \frac{s^2}{2}\,Q_{\beta;\star}[\varphi,\varphi]}\,\frac{\vartheta\big[\begin{smallmatrix} -N\bm{\epsilon}_{\star} \\ \bm{0} \end{smallmatrix}\big]\big(\bm{v}_{\beta;\star} + {\rm i}s\,\bm{u}_{\beta;\star}[\varphi]|\bm{\tau}_{\beta;\star}\big)}{\vartheta\big[\begin{smallmatrix} -N\bm{\epsilon}_{\star} \\ \bm{0} \end{smallmatrix}\big]\big(\bm{v}_{\beta;\star}|\bm{\tau}_{\beta;\star}\big)}.
\end{equation}
This formula gives an equivalent when $N \rightarrow \infty$, which features an oscillatory behaviour. We have set:
\begin{equation}\label{linearterm}
\bm{u}_{\beta;\star}[\varphi] = \Big(\frac{1}{2{\rm i}\pi} \partial_{{\epsilon}_h}\int_{\mathsf{S}} \varphi(\xi)\,\dd\mu_{{\rm eq};\bm{\epsilon}}^V(\xi)\Big)_{1 \leq h \leq g}\Big|_{\bm{\epsilon} = \bm{\epsilon}_{\star}} = \Big(\frac{1}{2{\rm i}\pi} \oint_{\mathsf{S}_h} \dd \xi\,\varphi(\xi)\,\varpi_{h}(\xi)\dd \xi\Big)_{1 \leq h \leq g},
\end{equation}
where $\varpi_{h}(\xi)\dd\xi$ are the holomorphic one-forms introduced in Equation~\eqref{sqfd}. The linear (resp. bilinear) form $M_{\beta;\bm{\epsilon}}[\varphi]$ (resp. $Q_{\beta;\bm{\epsilon}}[\varphi,\varphi]$) are defined in \S~\ref{S5clt}, and in Equation~\eqref{clclc} it is evaluated at $\bm{\epsilon} = \bm{\epsilon}_{\star}$. We recognise that the right-hand side of Equation~\eqref{clclc} is the Fourier transform of the sum of two independent random variables: one of them being Gaussian, and the other being the scalar product with $2{\rm i}\pi\bm{u}_{\beta;\star}[\varphi]$ of the sampling of a $g$-dimensional Gaussian vector at points belonging to $-\bm{N}_{\star} + \mathbb{Z}^{g}$. Therefore, among a codimension $g$ subspace of test functions determined by the equation $\bm{u}_{\beta;\star}[\varphi] = \bm{0}$, the ratio of Theta functions is $1$ and we do find a central limit theorem for fluctuations of linear statistics, as in the one-cut regime. But, when $\bm{u}_{\beta;\star}[\varphi] \neq \bm{0}$, we only find subsequential convergence in law --- along subsequences so that $(-N\bm{\epsilon}_{\star}\,\mathrm{mod}\,\mathbb{Z}^{g})$ converges ---  to the sum of a random Gaussian vector and an independent random Gaussian vector conditioned to belong to a lattice with oscillating center. Accordingly, the probability distribution of those fluctuations display interference patterns varying with $N$.

\appendix{}

\section{Elementary properties of the equilibrium measure with fixed filling fractions}
\label{appA}

We now prove Theorem~\ref{sqoi} stating that, if $V$ is analytic in a neighbourhood of $\mathsf{A}$, if we denote $(g + 1)$ the number of cuts of the equilibrium measure $\mu_{\mathrm{eq}}^V$ in the initial model, and assume it is off-critical, then $\mu_{\mathrm{eq};\bm{\epsilon}}^V$ still has $(g + 1)$ cuts and remains off-critical for $\bm{\epsilon}$ close enough to $\bm{\epsilon}_{\star}$, and depends smoothly on such $\bm{\epsilon}$.

\subsection{Lipschitz property}

We may decompose:
\beq
\label{ijjij}\mu_{\mathrm{eq};\bm{\epsilon}}^{V} = \sum_{h = 0}^{g} \epsilon_h\,\mu_{\mathrm{eq};\bm{\epsilon},h}^{V}.
\eeq
where $\mu_{{\rm eq};\bm{\epsilon},h}^{V}$ are probability measures in $\mathsf{A}_h$, and we know that $\mu_{\mathrm{eq};\bm{\epsilon}}^{V}$ minimises the energy functional $E[\mu]$ --- see Equation~\eqref{Enf} --- among such choices of probability measures.
We first establish that linear statistics of the equilibrium measure in the fixed filling fraction model are Lipschitz in $\bm{\epsilon}$.
Let $\delta \in (0,1]$ and set:
$$
\mathcal{E}_{\delta} = \Big\{\bm{\epsilon} \in (\delta,1-\delta)^{g}\quad \Big|\quad  \delta < 1 - \sum_{h = 1}^{g} \epsilon_h < 1 - \delta\Big\}.
$$
If $\bm{\epsilon} \in \mathcal{E}_{\delta}$ we denote $\epsilon_0 = 1 - \sum_{h = 1}^{g} \epsilon_h$. If $(\kappa_0,\ldots,\kappa_g)$ is such that $\sum_{h = 0}^{g} \kappa_h = 1$, we denote $\bm{\kappa} = (\kappa_1,\ldots,\kappa_g)$. 

\begin{lemma}
\label{Lhq} For $\delta > 0$ small enough, there exists a finite constant $c(\delta)$ such that, for any $\bm{\epsilon} \in \mathcal{E}_{\delta}$, for any $\kappa_h \in (0,2\epsilon_h]$ such that $\sum_{h = 0}^g \kappa_{h} = 1$, we have for any test function $\varphi$:
$$ 
\Big|\int_{\mathsf{A}} \varphi(x)\,(\dd\mu_{\mathrm{eq};\bm{\kappa}}^{V} - \dd\mu_{\mathrm{eq};\bm{\epsilon}}^{V})(x)\Big| \leq c(\delta) |\varphi|_{1/2}\,\max_{0 \leq h \leq g} |\kappa_h - \epsilon_h|
$$
\end{lemma}
\textbf{Proof.} As we have seen in Theorem~\ref{th:10}, $\mu_{\mathrm{eq};\bm{\epsilon}}^{V}$ is also characterised by saying that for \eqref{ijjij}, there exists constants $(C_{\bm{\epsilon},h}^{V})_{0 \leq h \leq g}$ so that for any $h \in \ldbrack 0,g \rdbrack$ and $x \in \mathsf{A}_h$:
$$
2  \int_{\mathsf{A}} \ln |x-\xi|\dd\mu_{{\rm eq};\bm{\epsilon}}^{V}(\xi) - V(x) \leq C_{\bm{\epsilon},h}^{V},
$$
with equality $\mu_{\mathrm{eq};\bm{\epsilon},h}^{V}$ almost everywhere. Recall the definition of the effective potential (here including the constants for convenience):
$$
\tilde{U}_{{\rm eq};\bm{\epsilon}}^{V}(x) = V(x) - 2 \int_{\mathsf{A}} \ln|x - \xi|\dd\mu_{\mathrm{eq};\bm{\epsilon}}^{V}(\xi) - \sum_{h = 0}^g C_{\bm{\epsilon},h}^{V}\mathbf{1}_{\mathsf{A}_h}(x),
$$
and of the pseudo-distance between two probability measures $\mu$ and $\nu$:
\begin{equation}
\label{Dpes}
\mathfrak{D}^2[\mu,\nu] = -\iint_{\mathbb{R}^2} \ln|x - y|\dd[\mu - \nu](x)\dd[\mu - \nu](y) \in [0,+\infty].
\end{equation}
We have for all probability measures on $\mathsf{A} = \bigcup_{h = 0}^{g} \mathsf{A}_h$:
\beq
\label{poi}E[\mu] = \frac{\beta}{2}\Big(\mathfrak{D}^2[\mu,\mu_{\mathrm{eq};\bm{\epsilon}}^{V}] + \int_{\mathsf{A}} \tilde{U}^{V}_{{\rm eq};\bm{\epsilon}}(x)\dd\mu(x) + \sum_{h = 0}^{g} C_{\bm{\epsilon},h}^{V}\,\mu(\mathsf{A}_h) + I_{{\rm eq};\bm{\epsilon}}^{V}\Big),
\eeq
with
$$
I_{{\rm eq};\bm{\epsilon}}^{V} = \iint_{\mathsf{A}^2} \ln|x - y|\dd\mu_{\mathrm{eq};\bm{\epsilon}}^{V}(x)\dd\mu_{\mathrm{eq};\bm{\epsilon}}^{V}(y).
$$
Indeed, a simple algebra shows that
\begin{equation}
\label{appveff}  
\begin{split}
E[\mu]&= E[\mu_{\mathrm{eq};\bm{\epsilon}}^{V}]+ \frac{\beta}{2}\Big(\mathfrak{D}^2[\mu,\mu_{\mathrm{eq};\bm{\epsilon}}^{V}] +\int_{\mathsf{A}} \Bigg(V(x)-2\int_{\mathsf{A}}\ln|x-y|\dd\mu_{\mathrm{eq};\bm{\epsilon}}^{V}(y)\Big)\dd[\mu-\mu_{\mathrm{eq};\bm{\epsilon}}^{V}](x)\Big) \\
&= E[\mu_{\mathrm{eq};\bm{\epsilon}}^{V}]+\frac{\beta}{2}\Big(\mathfrak{D}^2[\mu,\mu_{\mathrm{eq};\bm{\epsilon}}^{V}] +\int_{\mathsf{A}} 
\tilde U_{{\rm eq};\bm{\epsilon}}^{V}(x)
\dd[\mu-\mu_{\mathrm{eq};\bm{\epsilon}}^{V}](x) + \sum_{h=0}^g C_{\bm{\epsilon},h}^{V}(\mu-\mu_{\mathrm{eq};\bm{\epsilon}}^{V})(\mathsf{A}_h)\Big).
\end{split}
\end{equation}
Using  the characterisation of $\mu_{\mathrm{eq};\bm{\epsilon}}^{V}$, one finds that
$$
E[\mu_{\mathrm{eq};\bm{\epsilon}}^{V}]- \frac{\beta}{2} \sum_{h=0}^g C_{\bm{\epsilon},h}^{V}\epsilon_h= \frac{\beta}{2}\,I_{{\rm eq};\bm{\epsilon}}^{V},
$$
which completes the proof of Equation~\eqref{poi}.
We next choose $\bm{\kappa} \neq \bm{\epsilon}$ and write that if $\mu_{\bm{\kappa}}$ is \emph{any} probability measure such that $\mu_{\bm{\kappa}}(\mathsf{A}_h)=\kappa_h$, we must have
$$
E[\mu_{\mathrm{eq};\bm{\kappa}}^{V}] \leq  E[\mu_{\bm{\kappa}}].
$$
Since $\mu_{\mathrm{eq};\bm{\kappa}}^{V}$ and $\mu_{\bm{\kappa}}$ put the same masses on the $\mathsf{A}_h$ we deduce  from  Equation~\eqref{poi} that 
$$
\mathfrak{D}^2[\mu_{\mathrm{eq};\bm{\kappa}}^{V},\mu_{{\rm eq};\bm{\epsilon}}^{V}] +\int_{\mathsf{A}} \tilde{U}_{{\rm eq};\bm{\epsilon}}^{V}(x)\dd\mu_{\mathrm{eq}; \bm{\kappa}}^{V}(x) \leq \mathfrak{D}^2[\mu_{\bm{\kappa}},\mu_{\mathrm{eq};\bm{\epsilon}}^{V}]+\int_{\mathsf{A}} \tilde{U}_{{\rm eq};\bm{\epsilon}}^{V}(x)\dd\mu_{\bm{\kappa}}(x).
$$
We next choose $\mu_{\bm{\kappa}}$ whose support is included in the support of $\mu_{{\rm eq};\bm{\epsilon}}^{V}$ so that since $\tilde{U}_{{\rm eq};\bm{\epsilon}}^{V}$ vanishes there and is nonnegative everywhere, we deduce
\begin{equation}\label{poiu}
\mathfrak{D}^2[\mu_{\mathrm{eq};\bm{\kappa}}^{V},\mu_{\mathrm{eq};\bm{\epsilon}}^{V}] \leq \mathfrak{D}^2[\mu_{\bm{\kappa}},\mu_{\mathrm{eq};\bm{\epsilon}}^{V}].
\end{equation}
We put $\mu_{\bm{\kappa}} = t\,\mu_{{\rm eq};\bm{\epsilon}}^{V} +(1-t)\nu$ for $t \in [0,1]$ and a probability 
measure $\nu$ on $\mathsf{A}$  whose support is included in the support $\mu_{{\rm eq};\bm{\epsilon}}^{V}$ and such that for all $h$
\begin{equation}\label{fil}t \epsilon_h +(1-t)\nu(\mathsf{A}_h)=\kappa_h\,.\end{equation}
We have from Equation~\eqref{poiu}:
$$
\mathfrak{D}^2[\mu_{{\rm eq};\bm{\kappa}}^{V},\mu_{{\rm eq};\bm{\epsilon}}^{V}] \leq (1-t)^2\mathfrak{D}^2[\nu,\mu_{\mathrm{eq};\bm{\epsilon}}^{V}].
$$
We take
$$
1-t= \big(\max_{0 \leq h \leq g} \epsilon_h^{-1}|\kappa_h-\epsilon_h|\big) \in [0,1).
$$ 
If $\kappa_h\in (0,2\epsilon_h]$ and is such that $\nu(\mathsf{A}_h)\ge 0$ for any $h$, as it should for $\nu$ to be a probability measure. We finally choose $\nu$ such that $\mathfrak{D}^2[\nu,\mu_{{\rm eq};\bm{\epsilon}}^{V}]$ is finite (for instance the renormalised Lebesgue measure on the support of $\mu_{\mathrm{eq};\bm{\epsilon}}^{V}$) to conclude that there exists a constant $\tilde{c}(\delta)$ valid for all $\bm{\epsilon} \in \mathcal{E}_{\delta}$ such that:
$$
\mathfrak{D}^2[\mu_{{\rm eq};\bm{\kappa}}^{V} ,\mu_{{\rm eq};\bm{\epsilon}}^{V}] \leq \tilde{c}(\delta) \max_{0 \leq h \leq g} |\epsilon_h - \kappa_h|^2.
$$
Recalling that:
$$
\mathfrak{D}^2[\mu_{{\rm eq};\bm{\kappa}}^{V},\mu_{{\rm eq};\bm{\epsilon}}^{V}]=\int_0^\infty \frac{\dd p}{p} |\widehat{\mu_{{\rm eq};\bm{\kappa}}^{V}}(p) -\widehat{\mu_{{\rm eq};\bm{\epsilon}}^{V}}(p)|^2,
$$
we deduce that for all $\varphi \in L^1(\mathsf{A})$,
$$
\int_{\mathsf{A}} \varphi(x) \dd[\mu_{{\rm eq};\bm{\kappa}}^{V}- \mu_{{\rm eq};\bm{\epsilon}}^{V}](x) = \int_{\mathsf{A}} \dd p\,\widehat{f}(p)(\widehat{\mu_{{\rm eq};\bm{\kappa}}^{V}} - \widehat{\mu_{{\rm eq};\bm{\epsilon}}^{V}})(p).
$$
This implies that for all $\varphi$ with  $|\varphi|_{1/2} < \infty$ we have:
$$
\Big|\int_{\mathsf{A}} \varphi(x)\dd[\mu_{{\rm eq};\bm{\kappa}}^{V}-\mu_{{\rm eq};\bm{\epsilon}}^{V}](x)\Big|\leq c(\delta)\,|\varphi|_{1/2}\, \max_{0 \leq h \leq g} |\kappa_h - \epsilon_h|.
$$
\hfill $\Box$

\vspace{0.2cm}

\begin{lemma}
\label{Lhq2} If $\mu_{{\rm eq};\bm{\epsilon}}^{V}$ is off-critical and its support has $g_{\bm{\epsilon}} + 1$ cuts denoted $[\alpha_{\bm{\epsilon},h}^-,\alpha_{\bm{\epsilon},h}^+]$, then for $\bm{\epsilon}'$ in a small enough neighborhood of $\bm{\epsilon}$, $\mu_{{\rm eq};\bm{\epsilon}'}^V$ is off-critical and has the same number of cuts, of the form $[\alpha_{\bm{\epsilon}',h}^{-},\alpha_{\bm{\epsilon}',h}^{+}]$, and $\alpha_{\bm{\epsilon}',h}^{\bullet}$ are Lipschitz functions of $\bm{\epsilon}'$. Moreover, for $\delta>0$ small enough, assume that $\mathsf{A}$ contains
$$
\bigcup_{\bm{\epsilon}'} \bigcup_{\alpha_{\bm{\epsilon}'}\,\,{\rm soft}\,\,{\rm edge}}\big\{x \quad | \quad d(x,\alpha_{\bm{\epsilon}'})\le \delta\big\}
$$
when the union ranges over a small enough neighbourhood of $\bm{\epsilon}$. Then in the same neighbourhood of $\bm{\epsilon}$, the function $\bm{\epsilon}' \mapsto W^{\{-1\}}_{1;\bm{\epsilon}'}(x)$ is Lipschitz uniformly for $x$ in any compact of $\mathbb{C} \setminus \mathsf{A}$.
\end{lemma}
\textbf{Proof.} Restricting to $x$ in the domain $\mathsf{U}$ where $V$ is analytic, let us rewrite the leading order of the one-variable Dyson--Schwinger equation 
\beq
\label{kiqp} \big(W_{1;\bm{\epsilon}'}^{\{-1\}}(x)\big)^2 - V'(x)\,W_{1;\bm{\epsilon}'}(x) + \frac{Q_{\bm{\epsilon}'}(x)}{L_0(x)}= 0,
\eeq
where:
\beq
\label{A16} Q_{\bm{\epsilon}'}(x) = \int_{\mathsf{A}} L_0(\xi)\,\frac{V'(x) - V'(\xi)}{x - \xi}\,\dd\mu_{{\rm eq};\bm{\epsilon}'}^{V}(\xi),
\eeq
and we have chosen $L_0(x) = \prod_{a \in \partial\mathsf{A}} (x - a)$. Solving the quadratic equation \eqref{kiqp} we find:
\beq
\label{A17} W_{1;\bm{\epsilon}'}^{\{-1\}}(x) = \frac{V'(x)}{2} - \sqrt{\frac{L_0(x)\,V'(x)^2 - 4Q_{\bm{\epsilon}'}(x)}{4L_0(x)}},
\eeq
where the dependence in $\bm{\epsilon}'$ only appears through $Q_{\bm{\epsilon}'}(x)$. Owing to Lemma~\ref{Lhq}, since $V'$ is analytic in a neighbourhood of $\mathsf{A}$, $Q_{\bm{\epsilon}'}(x)$ is analytic for $x$ in this neighbourhood, and is Lipschitz in $\bm{\epsilon}'$, uniformly for $x$ in any compact of this neighbourhood. The edges of the support of $\mu_{\mathrm{eq};\bm{\epsilon}'}^{V}$ are precisely the zeroes or poles of $R_{\bm{\epsilon}'}(x) = (L_0(x)V'(x)^2 - 4Q_{\bm{\epsilon}'}(x))/L_0(x)$ on $\mathsf{A}$. Since $\mu_{{\rm eq};\bm{\epsilon}}^{V}$ is off-critical, for $\bm{\epsilon}' = \bm{\epsilon}$ these zeroes and poles are all simple. By a classical theorem of complex analysis, it implies that the zeroes of $R_{\bm{\epsilon}'}$ in $\mathsf{A}$ occur as Lipschitz functions $\bm{\epsilon}' \mapsto a_{\bm{\epsilon}',h}^{\bullet}$, in particular $\mu_{{\rm eq};\bm{\epsilon}}^{V}$ keeps the same number of cuts. Lemma~\ref{Lhq} also implies that $W_{1;\bm{\epsilon}'}^{\{-1\}}(x)$ is a Lipschitz function of $\bm{\epsilon}'$ for any fixed $x \notin \mathsf{A}$, and this is in fact uniform away from $\mathsf{A}$. \hfill $\Box$

\subsection{Smooth dependence in the filling fractions}
\label{Sumo}
The following result allows the conclusion that $\dd\mu_{{\rm eq};\bm{\epsilon}}^{V}/\dd x$ (or $W_{1;\bm{\epsilon}}^{\{-1\}}$) is smooth with respect to $\bm{\epsilon}$ for $x$ away from the edges.
\begin{proposition}
Lemma~\ref{Lhq2} holds with $C^{\infty}$ regularity instead of Lipschitz.
\end{proposition}
\textbf{Proof.}  We first prove that the Stieltjes transform $W_{1;\bm{\epsilon}}^{\{-1\}}(z)$ is a differentiable function of the filling fractions, for any $z \in \mathbb{C}\setminus\mathsf{S}_{\bm{\epsilon}}$. We take $\bm{\epsilon},\bm{\kappa},\bm{\kappa'} \in \mathcal{E}_{\delta}$. We choose $z,z' \in \mathbb{C}$ at distance at least $\delta'$ of $\mathsf{A}$ for $\delta' > 0$ fixed but small enough. Let $\psi_{z}(x) = \frac{1}{z - x}$ and $\psi_{z,z'}(x) = \psi_{z}(x) - \psi_{z'}(x)$. As in \S~\ref{corbounds}, we can build functions $\varphi_{z}(x)$ and $\varphi_{z,z'}(x)$ defined for $x \in \mathbb{R}$, which coincide with $\psi_{z}$ and $\psi_{z,z'}$ for $x \in \mathsf{A}$, and for which:
$$
|\varphi_{z}|_{1/2} \leq C(\delta')\qquad |\varphi_{z,z'}|_{1/2} \leq C(\delta')|z - z'|
$$
After Lemma~\ref{Lhq}, we have:
\begin{equation}
\label{numero1}
\begin{split}
|W_{1;\bm{\kappa}}^{\{-1\}}(z) - W_{1;\bm{\kappa}'}^{\{-1\}}(z)\big| & \leq C\,|\bm{\kappa} - \bm{\kappa}'|_{1}, \\
\big|\big(W_{1;\bm{\kappa}}^{\{-1\}}(z) - W_{1;\bm{\kappa}'}^{\{-1\}}(z)\big) - \big(W_{1;\bm{\kappa}}^{\{-1\}}(z') - W_{1;\bm{\kappa}'}^{\{-1\}}(z')\big)\big| & \leq C'\,|z - z'|\,|\bm{\kappa} - \bm{\kappa}'|_{1}.
\end{split}
\end{equation}
We fix $\bm{\eta} \in \mathbb{R}^{g + 1}$ such that $\sum_{h = 0}^{g} \eta_h = 0$, and for a given $z$ and $\bm{\kappa}$, we consider the function 
$t \mapsto W_{1;\bm{\kappa}+t\bm{\eta}}^{\{-1\}}(z)$ defined over
$$
\mathcal{V}_{\bm{\kappa},\bm{\eta}} = \big\{t \in \mathbb{R}\quad |\quad  \bm{\kappa} + t\bm{\eta} \in \mathcal{E}_{\delta}\big\}.
$$
We deduce from Equation~\eqref{numero1} and Rademacher theorem (stating that Lipschitz functions are almost surely differentiable), that:
$$
\partial_{s}W_{1;\bm{\kappa}+s\bm{\eta}}^{\{-1\}}(z) = \lim_{t \rightarrow 0} \frac{W_{1;\bm{\kappa} + (s+t)\bm{\eta}}^{\{-1\}}(z) -  W_{1;\bm{\kappa}+s\bm{\eta}}^{\{-1\}}(z)}{t}
$$
exists for $s$ in a subset $\mathcal{U}^{z}_{\bm{\kappa},\bm{\eta}}$ with probability $1$ in $\mathcal{V}_{\bm{\kappa},\bm{\eta}}$. Let $\mathfrak{N}_{\delta'}^{[\zeta]}$ be a countable $\zeta$-net of
$$
\tilde{\mathsf{A}}_{\delta'} = \big\{z \in \mathbb{C}\,\,:\,\,d(z,\mathsf{A}) \geq \delta'\big\}\,.
$$
By the previous point, we find a subset $\mathcal{U}_{\bm{\kappa},\bm{\eta}}^{\delta',[\zeta]}$ with probability $1$ in $\mathcal{V}_{\bm{\kappa},\bm{\eta}}$, such that for any $s\in \mathcal{U}_{\bm{\kappa},\bm{\eta}}^{\delta',[\zeta]}$ and $z \in \mathfrak{N}_{\delta'}^{[\zeta]}$, $\partial_{s}W_{1;\bm{\kappa}+s\bm{\eta}}^{\{-1\}}$ exists. We then choose the $\zeta$-nets to be increasing when $\zeta$ decreases, and denote:
$$
\mathcal{U}_{\bm{\kappa},\bm{\eta}}^{\delta'} =  \bigcap_{n \geq 1}  \mathcal{U}_{\bm{\kappa},\bm{\eta}}^{\delta',[1/n]}.
$$
$\mathcal{U}_{\bm{\kappa},\bm{\eta}}^{\delta'}$ has still probability $1$ in $\mathcal{V}_{\bm{\kappa},\bm{\eta}}$, and for any $s \in \mathcal{U}_{\bm{\kappa},\bm{\eta}}^{\delta'}$ in this set, $\partial_{s}W_{1;\bm{\kappa}+s\bm{\eta}}^{\{-1\}}(z)$ exists for all $z \in \bigcup_{n \geq 1} \mathfrak{N}_{\delta'}^{[1/n]}$. By Equation~\eqref{numero1}, this implies the existence of a Lipschitz (with respect to $z$) differential (with respect to $s$) for all $z \in \tilde{\mathsf{A}}_{\delta'}$ and any $s \in \mathcal{U}_{\bm{\kappa},\bm{\eta}}^{\delta'}$. By Montel theorem and Equation~\eqref{numero1}, $z \mapsto \partial_{s}W_{1;\bm{\kappa} + s\bm{\eta}}^{\{-1\}}(z)$ is a holomorphic function in $z$ for any $s$ such that  it exists.

By Equation~\eqref{A16}, $Q_{\bm{\kappa} + s\bm{\eta}}$ is the expectation value of an analytic function under $\mu_{{\rm eq};\bm{\kappa} + s\bm{\eta}}^V$, therefore:
$$
Q_{\bm{\kappa} + s\bm{\eta}}(x) = \oint_{\mathcal{C}} \frac{\dd \xi\,L_0(\xi)}{2{\rm i}\pi}\,\frac{V'(x) - V'(\xi)}{x - \xi}\,W_{1;\bm{\kappa} + s\bm{\eta}}^{\{-1\}}(\xi)
$$
with a contour $\mathcal{C}$ included in $\tilde{\mathsf{A}}_{\delta'}$. Besides, $Q_{\bm{\kappa} + s\bm{\eta}}(x)$ is a holomorphic function of $x$ in a neighbourhood $\mathsf{U}$ of $\mathsf{A}$ in $\mathbb{C}$ as $V$ is. Hence, $s \mapsto Q_{\bm{\kappa} + s\bm{\eta}}(x)$ is differentiable for $s \in \mathcal{U}_{\bm{\kappa},\bm{\eta}}^{\delta'}$ for each $x \in \mathsf{U}$, and Lipschitz in $z$. By Montel theorem, its derivative -- where it exists -- is holomorphic in $z \in \mathsf{U}$. Then, Equation~\eqref{A17} implies that $s \mapsto W_{1;\bm{\kappa} + s\bm{\eta}}^{\{-1\}}(x)$ is differentiable for $s \in \mathcal{U}_{\bm{\kappa},\bm{\eta}}^{\delta'}$ and any $x \in \mathbb{C}\setminus \partial \mathsf{S}_{\bm{\kappa} + s\bm{\eta}}$.

Now, let us fix a compact neighbourhood of $\bm{\epsilon} \in \mathcal{E}_{\delta}$ such that the regularity result of Lemma~\ref{Lhq2} applies. When we intersect $\mathcal{V}_{\bm{\kappa},\bm{\eta}}$ with a small enough neighbourhood of an off-critical $\bm{\epsilon} \in \mathcal{E}_{\delta}$, Lemma~\ref{Lhq2} guarantees that $\mu_{{\rm eq};\bm{\kappa}}^V$ remains uniformly off-critical. Arguments already used in Lemma~\ref{Lhq2} for Lipschitz regularity implies that edges at which $W_{1;\bm{\kappa}+s\bm{\eta}}^{\{-1\}}$ has a squareroot behaviour are functions $s \mapsto \alpha_{\bm{\kappa} + s \bm{\eta},h}^{\bullet}$ which are differentiable for $s \in \mathcal{U}_{\bm{\kappa},\bm{\eta}}^{\delta'}$. And, by Equation~\eqref{A17}, we can write at a hard edge $\alpha$ --- necessarily independent of $s$:
$$
W_{1;\bm{\kappa}+ s\bm{\eta}}^{\{-1\}}(x) = \frac{M_{\bm{\kappa} + s\bm{\eta}}^{[\alpha]}(x)}{(x - \alpha)^{\frac{1}{2}}},
$$
and at a soft edge $\alpha_{\bm{\kappa} + s\bm{\eta}}$:
$$
W_{1;\bm{\kappa} + s\bm{\eta}}^{\{-1\}}(x) = M_{\bm{\kappa} + s\bm{\eta}}^{[\alpha]}(x)\cdot \big(x- \alpha_{\bm{\kappa} + s\bm{\eta}}\big)^{\frac{1}{2}}
$$
with functions $M_{\bm{\kappa} + s\bm{\eta}}^{[\alpha]}(x)$ differentiable in $s\in \mathcal{U}_{\bm{\kappa},\bm{\eta}}^{\delta'}$ and holomorphic in $x$ a neighbourhood of  the edge $\alpha$.
 Therefore, for $s$ in this set, we have the behaviours:
$$
\partial_{s}W_{1;\bm{\kappa} + s\bm{\eta}}^{\{-1\}}(x) = O\big((x - \alpha_{\bm{\kappa} + s\bm{\eta}})^{-\frac{1}{2}}\big)
$$
at any edge. Given  the properties of the Stieltjes transform, we also  know that:
\begin{itemize}
\item[$\bullet$] $\partial_s W_{1;\bm{\kappa}+s\bm{\eta}}^{\{-1\}}(x)$ 
 behaves  like $O(\frac{1}{x^2})$ when $x \rightarrow \infty$ --- recall that the term in $\frac{1}{x}$ in $W_{1;\bm{\kappa}+s\bm{\eta}}^{\{-1\}}$ has constant coefficient.
\item[$\bullet$] for any $x \in \mathring{\mathsf{S}}_{\bm{\kappa}+s\bm{\eta}}$, we have $\partial_s W_{1;\bm{\kappa}+s\bm{\eta}}^{\{-1\}} (x + {\rm i}0) + \partial_s W_{1;\bm{\kappa}+s\bm{\eta}}^{\{-1\}}(x - {\rm i}0) = 0$.
\item[$\bullet$] for any $h \in \ldbrack 0,g \rdbrack$, $\oint_{\mathsf{A}_h} \partial_s W_{1;\bm{\kappa}+s\bm{\eta}}^{\{-1\}}(x)\,\frac{\dd x}{2{\rm i}\pi} = \eta_h$.
\end{itemize}

These properties imply that  $\partial_s W_{1;\bm{\kappa} + s\bm{\eta}}^{\{-1\}}(x)\dd x$ can be analytically continued to a holomorphic one-form\footnote{Note that on this Riemann surface, a local holomorphic coordinate near the points $x = \alpha$ is given by $\sqrt{x - \alpha}$.} on the Riemann surface of genus $g$ specified by the equation $\sigma^2 = \prod_{\alpha \in \partial\mathsf{S}_{\bm{\kappa}+s\bm{\eta}}}(x - \alpha)$ with periods $\eta_h$ around the $h$-th cut. As holomorphic one-forms are characterized by their periods, we deduce that
\beq
\label{limie}\partial_s W_{1;\bm{\kappa}+s\bm{\eta}}^{\{-1\}}(x) = 2{\rm i}\pi\,\sum_{h = 1}^{g} \eta_h\,{\varpi_{h}(x)},
\eeq
where $(\varpi_h(x)\dd x)_{h = 1}^{g}$ is the basis of holomorphic one-forms on the Riemann surface introduced in Equation~\eqref{sqfd}. These are completely determined by the endpoints and depend smoothly on them. Since the right-hand side of Equation~\eqref{limie} is a continuous function of $s$, we deduce that $s \mapsto W_{1;\bm{\kappa}+s{\bm\eta}}^{\{-1\}}(x)$ is actually  $\mathcal{C}^1$ for $s$ such that $\bm{\kappa}+s\bm{\eta}$ is in a vicinity of $\bm{\epsilon}$.  These arguments holding for any $\bm{\eta}, \bm{\kappa}$, we deduce that $\bm{\kappa} \ra W_{1;{\bm\kappa}}^{\{-1\}}$ is G\^{a}teaux differentiable, and hence Fr\'echet differentiable,   in a neighbourhood of $\bm{\epsilon}$.
Therefore, all the reasoning of the proof of Lemma~\ref{Lhq2} can be extended to show that the edges are $\mathcal{C}^1$. The differential equation \eqref{limie} (for any fixed $x$ away from the edges) then implies $\mathcal{C}^2$, and inductively, $\mathcal{C}^{\infty}$. \hfill $\Box$

\subsection{Hessian of the energy with respect to filling fractions}

We are now in position to prove:
\begin{proposition}
\label{hess} If $\mu_{{\rm eq};\bm{\epsilon}}^{V}$ is off-critical, then $F^{\{-2\};V}_{\beta;\bm{\epsilon}'}$ is $\mathcal{C}^{2}$ with negative definite Hessian at least for $\bm{\epsilon}'$ in a vicinity of $\bm{\epsilon}$.
\end{proposition}
In other words, the $g \times g$ matrix $\bm{\tau}_{\beta;\bm{\epsilon}}$ with purely imaginary entries:
\beq
\forall h,h' \in \ldbrack 1,g \rdbrack,\qquad (\bm{\tau}_{\beta;\bm{\epsilon}})_{h,h'} = \frac{1}{2{\rm i}\pi}\,\frac{\partial^2 F_{\bm{\epsilon}}^{\{-2\};V}}{\partial \epsilon_h\partial\epsilon_{h'}}
\eeq
is such that $\mathrm{Im}\,\bm{\tau}_{\beta;\bm{\epsilon}} > 0$.

\noindent \textbf{Proof.} Let $\bm{\eta},\bm{\eta}' \in \mathbb{R}^{g + 1}$ so that $\sum_{h = 0}^g \eta_h = \sum_{h = 0}^{g} \eta'_h = 0$ and $\bm{\epsilon}'$ be 
in a vicinity of $\bm{\epsilon}$.
The last paragraph has shown the existence of a integrable, signed measure with $0$ total mass:
$$
\nu_{\bm{\epsilon}';\bm{\eta}}^{V} = \lim_{t \rightarrow 0} \frac{\mu_{{\rm eq};\bm{\epsilon}' + t\bm{\eta}}^{V} - \mu_{{\rm eq};\bm{\epsilon}'}^{V}}{t}.
$$
By Equation~ \eqref{appveff}, we have:
\begin{equation*}
\begin{split}
F^{\{-2\};V}_{\beta;\bm{\kappa}}-F^{\{-2\};V}_{\beta;\bm{\epsilon}'}&= -\big(E[\mu_{\mathrm{eq};\bm{\kappa}}^{V}]-E[\mu_{\mathrm{eq};\bm{\epsilon}'}^{V}]\big)\\
&= \frac{\beta}{2}\Big(-\mathfrak{D}^2[\mu_{\mathrm{eq};\bm{\kappa}}^{V},\mu_{\mathrm{eq};\bm{\epsilon}'}^{V}] +\int_{\mathsf{A}}
\tilde U_{{\rm eq};\bm{\epsilon}'}^{V}(x)
\dd[\mu_{\mathrm{eq};\bm{\kappa}}^{V}-\mu_{\mathrm{eq};\bm{\epsilon}'}^{V}](x) + \sum_{h=0}^g C_{h;\bm{\epsilon}'}^{V}(\kappa_h-\epsilon_h')\Big).
\end{split}
\end{equation*}
Since $\tilde U_{{\rm eq};\bm{\epsilon}'}^{V;\mathsf{A}}$ vanishes on $\mathsf{S}_{\bm{\epsilon}'}$ and the derivatives of
 $\bm{\epsilon}' \mapsto \mu_{\mathrm{eq};\bm{\epsilon}'}^{V}$ are smooth and supported in $\mathsf{S}_{\bm{\epsilon}'}$, we deduce that
 $F^{\{-2\},V}_{\bm{\epsilon}'}$ is a $\mathcal{C}^2$ function of $\bm{\epsilon}'$ and its Hessian is:
\beq
{\rm Hessian}(F^{\{-2\};V}_{\beta;\bm{\epsilon}'})[\bm{\eta},\bm{\eta}'] = -\frac{\beta}{2} \sum_{h = 0}^{g} \mathfrak{D}^2[\nu_{\bm{\epsilon}';\bm{\eta}}^{V}\,\mathbf{1}_{\mathsf{A}_h},\nu_{\bm{\epsilon}';\bm{\eta}'}^{V}\,\mathbf{1}_{\mathsf{A}_h}],
\eeq
where we recall that $\mathfrak{D}$ is the pseudo-distance from Equation~\eqref{Dpes}. Therefore, the Hessian is definite negative. \hfill $\Box$

\subsection*{Acknowledgments} 

We thank V.~Gorin and the anonymous referees for useful comments. The work of G.B. has been supported by Fonds Europ\'een S16905 (UE7 - CONFRA) and the Fonds National Suisse (200021-143434), and he would like to thank the ENS Lyon where part of this work was conducted. This research was supported by ANR GranMa ANR-08-BLAN-0311-01, Simons foundation and ERC Project LDRAM:  ERC-2019-ADG	Project 884584.

\newcommand{\etalchar}[1]{$^{#1}$}
\providecommand{\bysame}{\leavevmode\hbox to3em{\hrulefill}\thinspace}
\providecommand{\MR}{\relax\ifhmode\unskip\space\fi MR }

\providecommand{\MRhref}[2]{
  \href{http://www.ams.org/mathscinet-getitem?mr=#1}{#2}
}
\providecommand{\href}[2]{#2}

\end{document}